\documentclass[12pt,a4paper,openany]{book}
\usepackage{pdfpages}
\usepackage{graphicx}
\usepackage{geometry}
\usepackage{incgraph}
\usepackage[utf8]{inputenc}
\usepackage[english]{babel}
\usepackage{amsmath,amsfonts,bbm,mathrsfs,amssymb}
\usepackage{bigstrut}
\usepackage[bottom]{footmisc}
\usepackage{multirow}

\usepackage{cite}
\usepackage[DIV=13]{typearea}
\usepackage{ragged2e}
\usepackage{catchfile}
\usepackage{enumerate}

\usepackage[labelfont=bf,font=footnotesize]{caption}
\usepackage[normalem]{ulem}

\usepackage{afterpage}
\usepackage{upgreek}

\usepackage{romannum}
\usepackage{scalerel}
\usepackage[inline]{enumitem}
\usepackage{dcolumn}
\usepackage{placeins}
\usepackage{bm}
\usepackage{booktabs}

\renewcommand{\footnoterule}{%
  \kern -3pt
  \hrule width \textwidth 
  \kern 2pt
}

\usepackage{fancyhdr}
\usepackage{yfonts}
\usepackage{lastpage}

\usepackage{physics,slashed}

\usepackage{mathtools}

\usepackage[export]{adjustbox}

\usepackage{fontawesome5} 

\usepackage{nicefrac}
\usepackage{xcolor} 
\definecolor{myred}{RGB}{128,0,0}
\usepackage{hyperref}
\usepackage{bbold}
\usepackage{comment}
\usepackage[capitalise]{cleveref}
\hypersetup{
    colorlinks = true,
    citecolor = myred,
    urlcolor = myred,
    linkcolor = myred,
    linktocpage=true
}
\usepackage{subfigure}
\usepackage{float}
\usepackage{slashed,cancel}
\usepackage{dirtytalk}
\usepackage{vmargin}
\usepackage{wallpaper}

\allowdisplaybreaks
\numberwithin{table}{chapter}
\numberwithin{figure}{chapter}
\setpapersize{A4}
\setmarginsrb
{3cm}
{1.5cm}
{3cm}
{1.5cm}
{0pt}
{1cm}
{0pt}
{2cm}
\usepackage{afterpage}

\newcommand\PB{\newpage\null\thispagestyle{empty}\newpage}

\usepackage[explicit]{titlesec}
\usepackage[default]{cinzel}
\usepackage[T1]{fontenc}
\usepackage{ebgaramond}
\usepackage{utopia}
\usepackage{tgadventor}
\usepackage{tgpagella}
\usepackage{etoolbox}

\usepackage[toc,page]{appendix}
\usepackage{rotating}

\usepackage{etoolbox}
\appto\appendix{\addtocontents{toc}{\protect\setcounter{tocdepth}{1}}}

\appto\listoffigures{\addtocontents{lof}{\protect\setcounter{tocdepth}{1}}}
\appto\listoftables{\addtocontents{lot}{\protect\setcounter{tocdepth}{1}}}

\definecolor{OliveGreen}{rgb}{0,0.6,0}
\definecolor{morado}{rgb}{0.4706,0.2510,0.5882}
\definecolor{blue-violet}{rgb}{0.33, 0.17, 0.89}

\newcommand{\sign}[1]{\text{sgn}(#1)}

\newcommand{\gitlink}{\href{https://github.com/mhostert/Heavy-Neutrino-Limits}{\textsc{g}it\textsc{h}ub {\large\color{blue-violet}\faGithub}}}

\definecolor{vierde}{rgb}{0.0, 0.5, 0.0}
\definecolor{OliveGreen}{rgb}{0,0.6,0}

\newcommand{\dolor}[2]{\Huge#1\LARGE #2}

\newcommand{\be}{\begin{equation}}
\newcommand{\ee}{\end{equation}}
\newcommand{\ba} {\begin{equation}\begin{aligned}}
\newcommand{\ea} {\end{aligned}\end{equation}}
\newcommand{\bea}{\begin{eqnarray}}
\newcommand{\eea}{\end{eqnarray}}

\newcommand{\LL}{\mathcal{L}}

\newcommand{\OO}{\mathcal{O}}
\newcommand{\cG}{\mathcal{G}}
\newcommand{\cM}{\mathcal{M}}

\newcommand{\unity}{\mathbbm{1}}

\newcommand{\hc}{\text{h.c.}}
\newcommand{\ov}[1]{\overline{#1}}
\newcommand{\nn}{\nonumber}

\newcommand{\eV}{\ \text{eV}}

\newcommand{\TeV}{\ \text{TeV}}
\newcommand{\GeV}{\ \text{GeV}}
\newcommand{\MeV}{\ \text{MeV}}

\def\Tr{{\rm Tr}}

\newcommand{\sL}{\mathcal{L}}

\newcommand{\cH}{\mathcal{H}}

\newcommand{\cN}{\mathcal{N}}
\newcommand{\cR}{\mathcal{R}}
\usepackage{chngcntr}
\counterwithin*{equation}{chapter}

\usepackage{fix-cm}

\usepackage[all]{nowidow}
\usepackage{titletoc}

    \titlecontents{part}
    [0em] %
    {\medskip\bfseries}
    {\thecontentslabel\quad}
    {}
    {\hfill\contentspage}

\usepackage{lettrine}
\usepackage{Acorn}
\usepackage{Eileen}
\usepackage{EileenBl}
\usepackage{Elzevier}
\usepackage{Romantik}
\usepackage{Zallman}

\counterwithout{footnote}{chapter}

\def\Tr{{\rm Tr}}

\def\BR{{\rm BR}}

\def\vep{\varepsilon}
\def\tH{\widetilde{H}}
\def\UPQ{U(1)_\text{PQ}}

\usepackage{titletoc}

    \titlecontents{chapter}
    [0em] %
    {\medskip\bfseries}
    {\thecontentslabel\quad}
    {}
    {\hfill\contentspage}

    \titlecontents{section}[1.72em]{\smallskip}%
    {\thecontentslabel.\enspace}
    {}
    {\titlerule*[1.2pc]{.}\contentspage}%

\titleformat{\section}[block]
    {\bfseries\Large
    \fontfamily{put}\selectfont}
    {\thesection}
    {10pt}    {#1}%
\titleformat{\subsection}[block]
    {\bfseries\large
    \fontfamily{put}\selectfont
    }
    {\thesubsection}
    {10pt}    {#1}%

\makeatletter

\newcommand*{\setupnormalparts}{%

  \titleformat{\part}[block]
    {\centering\Huge\thispagestyle{empty}}
    {\textsc{Episode \thepart}\\{\color{myred} \rule{\textwidth}{1mm}}}
    {10pt}{\\##1}
    {}%
}

\newcommand*{\setupfakechapters}{%

\titleformat{\chapter}[block]
    {\normalfont\vspace{-50mm}\Huge\scshape\bfseries}
    {}
      {0pt}{\flushleft ##1\\\vspace{-5mm}{\color{myred} \rule{\textwidth}{0.5mm}}\vspace{-15mm}}

        }

\newcommand*{\setupappendixparts}{%

\titleformat{\part}[block]
    {\centering\normalfont\vspace{-10mm}\Huge\scshape\bfseries\thispagestyle{empty}}
    {}
      {0pt}{##1\\\vspace{-5mm}{\color{myred} \rule{\textwidth}{0.5mm}}}

        }

\newcommand*{\setupnormalchapters}{%
\titleformat{\chapter}[block]
    {\normalfont\vspace{-30mm}}
    {\parbox[b]{2cm}{\fontsize{90}{10}\selectfont\color{myred}\thechapter}}
      {0pt}{\parbox[b]{\dimexpr\textwidth-2cm\relax}{
    \raggedleft
    \hfill{\LARGE##1}\\
        \rule{\dimexpr\textwidth-2cm\relax}{0.4pt}}\vspace{-5mm}}
}

\newcommand*{\setuplargerchapters}{%
\titleformat{\chapter}[block]
    {\normalfont\vspace{-30mm}}
    {\parbox[b]{3.5cm}{\fontsize{90}{10}\selectfont\color{myred}\thechapter}}
      {0pt}{\parbox[b]{\dimexpr\textwidth-3.5cm\relax}{
    \raggedleft
    \hfill{\LARGE##1}\\
        \rule{\dimexpr\textwidth-3.5cm\relax}{0.4pt}}\vspace{-5mm}}
}

\newcommand*{\setupappendixchapters}{%
\titleformat{\chapter}[block]
    {\normalfont\vspace{-30mm}}
    {\parbox[b]{2.8cm}{\fontsize{90}{10}\selectfont\color{myred}\thechapter}}
      {0pt}{\parbox[b]{\dimexpr\textwidth-2.8cm\relax}{
    \raggedleft
    \hfill{\LARGE##1}\\
        \rule{\dimexpr\textwidth-2.8cm\relax}{0.4pt}}}
}

\newcommand*{\setupspecialchapters}{%
\titleformat{\chapter}[block]
    {\normalfont\vspace{-50mm}\Huge\scshape\bfseries}
    {}
      {0pt}{\flushleft ##1\\\vspace{-5mm}{\color{myred} \rule{\textwidth}{0.5mm}}\vspace{-15mm}}

        }

\newcommand*{\setupbibliochapters}{%
\titleformat{\chapter}[block]
    {\normalfont\vspace{-50mm}\Huge\scshape\bfseries}
    {}
      {0pt}{\flushleft ##1\\\vspace{-5mm}{\color{myred} \rule{\textwidth}{0.5mm}}\vspace{-15mm}}

        }
        
\makeatother
\setupspecialchapters

\begin{document}
\emergencystretch 3em
\pagenumbering{gobble}


\incmultigraph{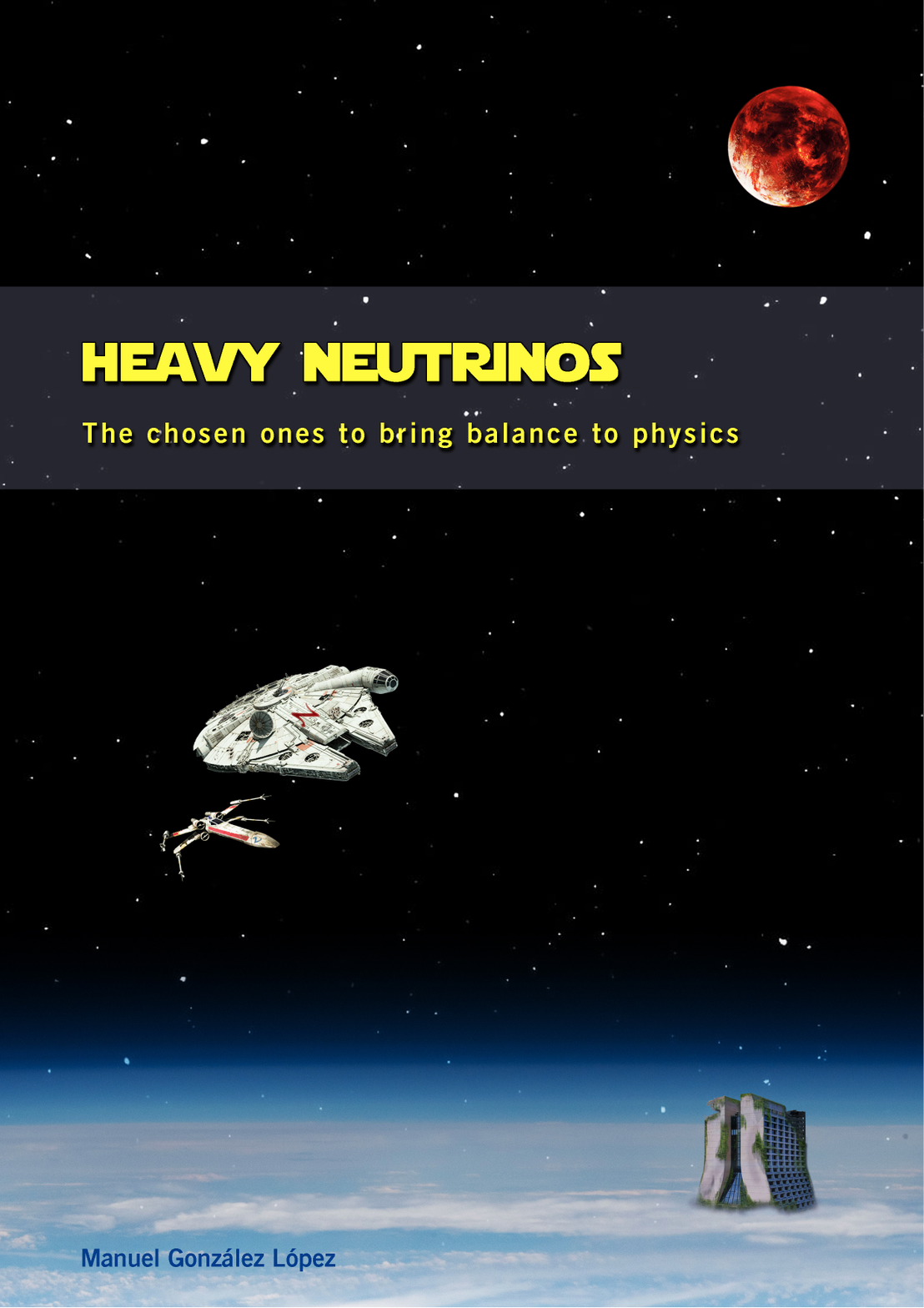}{1}
\begin{titlepage}
\begin{center}

{\scshape\LARGE \textcolor{myred}{Universidad Aut\'onoma de Madrid}\par}\vspace{1.5cm} 
\textsc{\Large Memoria de Tesis Doctoral}\\[0.5cm] 
\vspace{1cm}
\rule{\textwidth}{0.5mm} \\[0.4cm] 
{\huge  \cinzel\dolor{H}{eavy} \dolor{N}{eutrinos}: the \dolor{C}{hosen} \dolor{O}{nes} to bring \dolor{B}{alance} to \dolor{P}{hysics}

\par}\vspace{0.3cm} 
\rule{\textwidth}{0.5mm} \\[1.5cm] 
 
\vspace{1cm}
\begin{centering} \large
\textcolor{myred}{Manuel González López} 
\end{centering}

\vspace{15mm}

\large Dirigida por el Dr. Enrique Fern\'andez Mart\'inez
 
\vspace{15mm}

{\large Departamento de F\'isica Te\'orica\\
Instituto de F\'isica Te\'orica UAM-CSIC\\Madrid, 3 de noviembre de 2023}\\[4cm] 
\vfill
\vspace{-20mm}

\begin{figure}[htb!]
\centering
\includegraphics[width=0.35\textwidth]{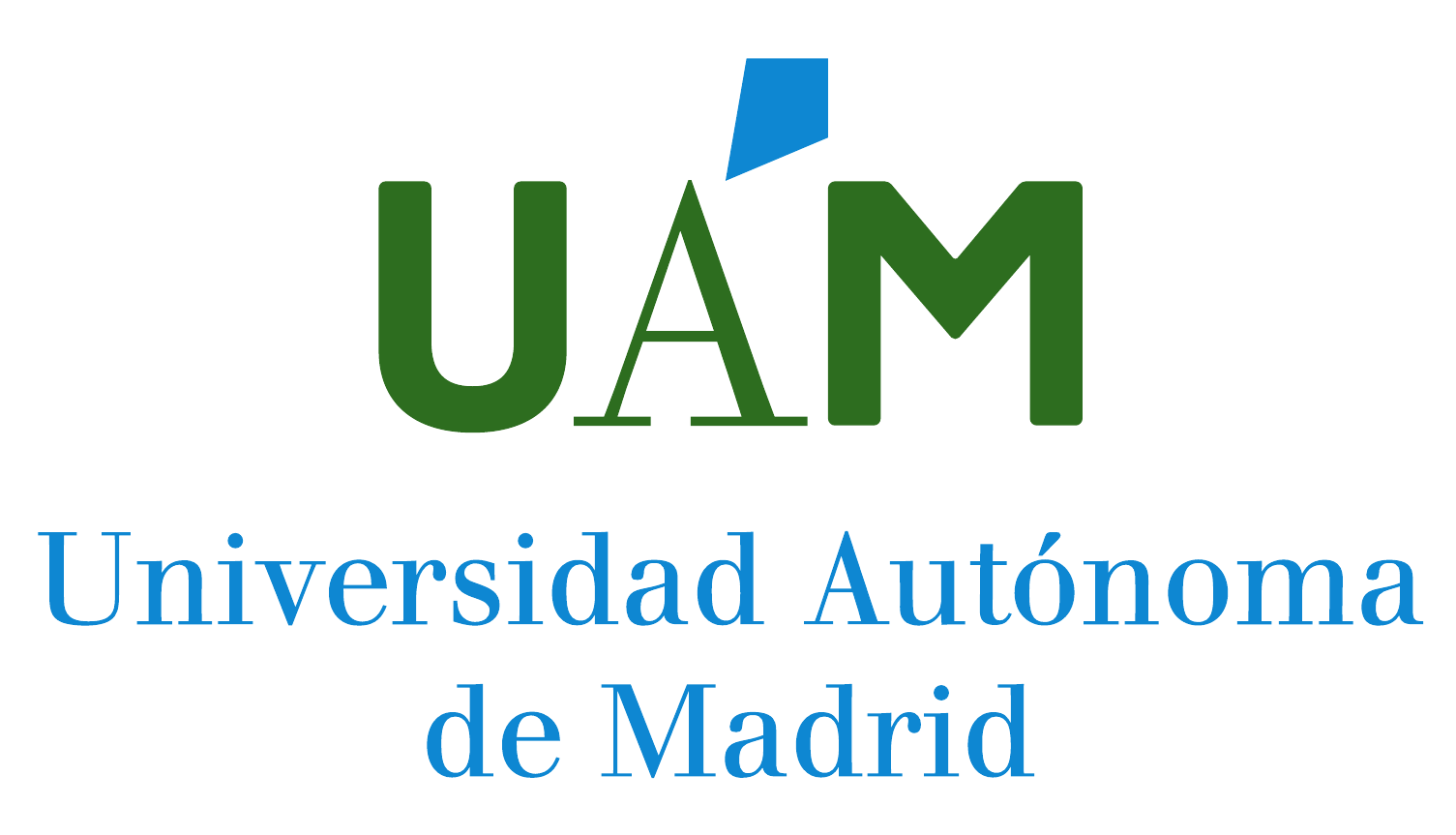}
\hfill
\includegraphics[width=0.35\textwidth]{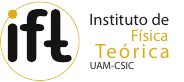}
\end{figure}

\end{center}
\end{titlepage}

\PB


\vspace*{\fill}
\begin{flushright}
\textit{Para mis padres.}
\vspace*{\fill}
\thispagestyle{empty}
\end{flushright}


\PB


\vspace*{\fill}
\begin{flushright}
\textit{Sé que no me queda\\
mucho más tiempo\\
de loop salvaje,\\
pero déjame,\\
mientras resista,\\
que me desangre.}
\end{flushright}

\vspace*{\fill}
\thispagestyle{empty}

\PB
\chapter*{Agradecimientos}
\addcontentsline{toc}{part}{Agradecimientos}
\fancyhead[RO]{}
\fancyhead[LE]{}
Sinceramente, llevo más de cuatro años pensando en estos agradecimientos, y era la parte de la tesis que más ganas tenía de escribir. Parece mentira que haya llegado el momento.

Mi madre siempre dice que soy una persona con suerte, y una prueba de que, como siempre, tiene razón, es toda la gente buena que me ha acompañado durante estos años. Es lo mejor que me llevo del doctorado, y espero no dejarme a nadie (pido perdón de antemano si es así!).

No podría empezar por nadie que no fuese Enrique. La frase típica de que sin él nada habría sido posible es completamente cierta en mi caso. Hace casi cinco años, parecía que eso de encontrar un director de tesis no iba a ser tan fácil como parecía. Después de visitar infructuosamente unos cuantos despachos, tuve la suerte de que Enrique tenía hueco para un nuevo estudiante, y casi sin conocerme aceptó. Desde el principio me ayudó un montón para pedir becas, hasta que tuve (otra vez) la suerte de ganar la ayudantía. Gracias Enrique por tratarme como uno más desde el principio, por tu paciencia con mi TOC, por saber tantísimo (y por saber enseñarlo), por tener paciencia cuando no me acordaba de cosas que nos dijiste en el máster, por tu sentido del humor y por escuchar mis dudas con respecto al futuro. Bajar a tu despacho a hablar de física, o simplemente a cotillear, siempre me hacía salir más contento de lo que entraba. Sé que entre pandemias, secretarías de departamento, filomenas y demás catástrofes, no ha sido el doctorado más canónico, pero aún así creo que no podría haber tenido un jefe mejor. Siempre estaré enormemente orgulloso de ser estudiante tuyo. 

También quiero acordarme de los otros dos tutores que tuve antes, Enrique Álvarez y María José. Con Enrique descubrí que quería ser físico teórico, y recuerdo la absoluta fascinación que sentía cuando hablaba con él en su despacho, y también el trato cercano y paciente con un chaval que no tenía ni idea. Con María José empecé en esto de la investigación, y en la física de partículas. En el TFM me empecé a sentir físico de verdad, y creo que si he llegado aquí es en parte gracias a esa etapa. También gracias a María José tuve la oportunidad de entrar en el IFT durante el máster, y estos años en el despacho 413 también se deben a ello. No puedo acabar estas palabras sin acordarme de lo que disfruté las clases de ambos en la carrera, y especialmente de su apoyo a la hora de sacar la ayudantía. Muchas gracias a los dos.

Yendo más atrás en el tiempo, he podido aprender física de mucha gente en estos años. Gracias a todos los profesores de la carrera, del máster, y a otras personas del IFT que me han hecho aprender. Me quiero acordar especialmente de dos personas con las que he trabajado. Gracias a Luca por hacerme parte de sus proyectos (pese a los files gordos), por hacer bromas, por las ayudas con la burocracia y por hacer piña con las cenas de sushi. Y gracias a Pilar, por ser tan detallista, por apoyarme con el TOC, y por tener siempre palabras de ánimo.

Por acabar con mis profesores, y aunque probablemente nunca lea esto, quiero darle las gracias a José Manuel. Sin sus clases de física en el cole es probable que no estuviera hoy escribiendo estas palabras.

No me quiero olvidar tampoco de otras personas que hacen del IFT un sitio estupendo para trabajar (limpieza, recepción, mantenimiento, etc). Gracias especialmente a Rebeca e Isabel por ser unas máquinas, y sobre todo por serlo siempre con una sonrisa.

Gracias también a la gente que me acogió en Boston, especialmente a Carlos, pero también a Iván, Alfonso, Will, Jeff, Elliot, Pavel y demás gente del grupo de los pingüinos. 

Me gustaría dedicar unas palabras a mi tribunal, por vuestra amabilidad y cercanía. Gracias a Nuria por ponerlo todo tan fácil desde el principio, y (enormes!) gracias a Ana por tener la paciencia de revisar mi tesis. A Xabi no sólo le tengo que agradecer aceptar ser mi secretario, sino mucho más. Siempre has estado ahí cuando iba a molestarte para hablar de física, del futuro, de fútbol o de cualquier otra cosa, o incluso para llevarme a urgencias.

También quiero aprovechar para acordarme de Ángela. Aunque me sigo sintiendo más como alumno que como profesor, poder enseñarte algo de este mundillo, aunque fuera un poquito, ha sido de las partes más agradecidas de esta etapa. Tienes un entusiasmo, unas ganas de aprender y un optimismo que cualquiera querríamos. Espero que te vaya genial en este camino.

Pasando a la parte personal, no sé muy bien que decir de la gente que siempre he tenido a mi lado: Alfon, Iñaki, Elena, Moflis, Dani, Dide y Moni. Probablemente no os he hablado mucho de la tesis (seguramente debería haberlo hecho más), pero sois lo que me mantiene en pie, lo que me hace feliz, y lo que me da ganas de seguir. Podría deciros muchísimas cosas a cada uno, recordar mil cosas divertidas y absurdas que nos han pasado, y también algunas no tan bonitas, en las que siempre habéis estado conmigo. Pero sólo diré que sois los mejores, y os quiero mucho. 

Y por supuesto tenía que estar presente Cuartel General. Si entré en esta cosa del doctorado fue probablemente por lo feliz que fui en la carrera, y en eso tenéis mucha culpa. Sois una fuente infinita de risas, buenos ratos, humor absolutamente absurdo y también de apoyo. Larga vida a Semanqui.

Aunque seguramente nunca leas esto, estuviste a mi lado durante esta tesis, así que gracias también a Jorge.

Y por fin llego a un grupo muy amplio, pero que ha sido fundamental para mí, el de mis compañeros del IFT. Siempre llego con miedo a los sitios nuevos, pero desde el principio me he sentido querido y arropado. Tengo que empezar por mis compañeros del 413 original, Claudia, Javi y Gallego. Me hicisteis sentir como en casa y, aunque no coincidimos durante tanto tiempo como me habría gustado, recuerdo esos días con mucho cariño. Gracias a Gallego por hacerme reír tanto en el despacho, por las remontadas del Madrid y por ser la bondad personificada. Gracias a Claudia, Javi y Roberto por descubrirme la pocha, aunque crearais un monstruo con ello. Gracias también a Fer, a Uga, a Salva, a Judit. Os echo mucho de menos en el IFT. Y, aunque no coincidimos, gracias especialmente al mayor de la familia, Josu, por rivalizar con mi TOC y por ayudarme siempre. 

Javi, aunque ya has salido antes, te mereces un párrafo aparte. Sé que no empezamos de la mejor forma, pero te has convertido en alguien importante para mí, aunque no te lo diga nunca. Gracias por tu paciencia \textit{infinita} en el pádel, por las referencias de Disney, por estar tan sordo, por aguantar el desesperanto, y por ser como tú eres. T\textquotesingle estimo moltissim.

Y llegando a la gente nueva del IFT, creo que dejamos el edificio en buenas manos, aunque el listón estaba alto. Gracias a Fer (siento que el mus no sea tan popular como la pocha), a los actuales integrantes del 413, y al resto del pacto. Mención especial merece mi querido hermano pequeño, Daniel Naredo-Tuero. Espero que nos quede mucho payasismo juntos.

Y siguiendo con la temática asturiana, gracias a David. He pensado mucho qué decirte para que la gente no te mire, y la verdad es que aún no sé que decir. Has sido lo mejor de este año, y no se me ocurre mejor manera de acabar la tesis que contigo. Por las recomendaciones musicales, por las red flags, por sacarme de fiesta, por no tener persianas funcionales, por la tarde en que me enseñaste el paraíso, y por ayudarme a ser mejor. Gracies:)

Como decía, siempre que llego a un sitio tengo un cierto miedo a estar solo, cosa que fue aún peor cuando, allá por 2019, parecía que iba a estar una semana aislado en una escuela en los Pirineos. Resulta que no iba solo, y que las dos personas que me acompañaban acabarían haciéndolo durante 4 años. Aunque ya les conocía de antes, Jesús y Jose han pasado de compañeros de clase a amigos de verdad, con los que he compartido muchísimas risas, anécdotas completamente absurdas (como olvidar Bachimala, París, el concesionario de Fuenlabrada o las cenas de pareja a las que yo iba solo), alguna que otra discusión de física, y también momentos no tan agradables, cuando vivíamos la peor parte de hacer una tesis. Habéis sido lo mejor de esta etapa, y solamente saber que estabais para comer y echar una pocha me hacía ver los días de otra manera. Me da mucha pena que esto se acabe, pero sé que nos queda mucho juntos. Os quiero mucho chicos. 

Y ya llegando al final de esta turra mítica, le toca a Claudia. Buena parte de todo esto es gracias a ti, y sinceramente nunca sé muy bien cómo agradecerlo. Desde el ordenador que me pusiste en tu despacho, hasta las revisiones de la tesis, pasando por el TFM, siempre he sentido que todos los problemas tendrían una solución si te los preguntaba a ti. Pero si te has convertido en una de las personas más importantes de mi vida no es por la física, sino por todo lo demás. No acabaría nunca si me pusiese a enumerar todos los recuerdos que tengo contigo. Los pub quiz (con sus respectivas infamias), el coche mal aparcado y demás risas absurdas, tu paciencia organizando viajes, y tu paciencia y comprensión cuando te he mandado audios de 8 minutos. Por desgracia, no todos los recuerdos son buenos, y sé que estos últimos años no han sido los mejores para ti. Pero como te he dicho alguna vez, tengo una fe absoluta en que puedes con una tesis, con una rodilla y con cualquier cosa que se te ponga por delante. Sé que lo mejor está por llegar, y puedo decir orgulloso que estaré ahí para vivirlo contigo. Te quiero mucho :)

Y por último llega la familia. Me gustaría dedicar también esta tesis a mi abuelo Jose. Te echo de menos, y ojalá pudieras estar aquí para ver el final de esta etapa.

Las últimas palabras de esta tesis tienen que ir para mis padres, porque todo, absolutamente todo, es gracias a ellos. Son las personas que más me quieren, y, aún así, las que ven lo peor de mí. Sé que soy injusto con vosotros, y que a veces no os trato como os merecéis. Pero sois lo más importante que tengo, y todo lo bueno que soy o que he conseguido es por cómo sois vosotros y porque me habéis querido siempre, incluso más que a vosotros mismos. Creo que no puede existir un amor desinteresado como el vuestro, y me siento la persona más afortunada del mundo por teneros. Os quiero mucho, muchísimo, y esta tesis es para vosotros.

\PB
\PB


\begin{center}
    \textbf{\Large\scshape Resumen}\\
{\color{myred} \rule{0.3\textwidth}{0.5mm}}
\end{center}
El Modelo Estándar de la Física de Partículas ha demostrado ser tremendamente exitoso como la teoría fundamental que describe las partículas elementales que componen nuestro universo, así como las interacciones entre ellas. Pese a los innumerables resultados experimentales que han ratificado las predicciones teóricas de este modelo, existen motivos de peso para pensar que ha de ser ampliado. Algunas de estas razones son de carácter teórico, mientras que otras aluden a la observación de ciertos fenómenos que el Modelo Estándar es incapaz de explicar. Entre estos últimos destaca el carácter masivo de los neutrinos, las partículas más elusivas del universo, a las que el Modelo Estándar adjudica una masa nula. Esta contradicción constituye el punto de partida de esta tesis doctoral. En ella se discute el papel de los neutrinos pesados (un hipotético nuevo tipo de neutrino) como responsables de la generación de la masa de los neutrinos ya conocidos. Se explorarán extensiones del Modelo Estándar que incluyen estas nuevas partículas, y ofrecen soluciones comunes a la masa de los neutrinos y a otros problemas abiertos. Además, se estudiarán las consecuencias que tendrían las interacciones de estos nuevos neutrinos con las partículas del Modelo Estándar, y cómo pueden ser caracterizados mediante observaciones experimentales presentes y futuras.
\thispagestyle{empty}


\PB


\begin{center}
    \textbf{\Large\scshape Abstract}\\
{\color{myred} \rule{0.3\textwidth}{0.5mm}}
\end{center}
    The Standard Model of Particle Physics has proven to be tremendously successful as the fundamental theory that describes the elementary particles that compose our Universe, as well as the interactions among them. Despite the countless experimental results that have ratified the theoretical predictions of this model, there are sound reasons to think it must be extended. Some are rather theoretical, while others are due to the observation of certain phenomena that the the Standard Model cannot explain. Chief among the latter is the massive nature of neutrinos, the most elusive particles in the Universe, which are rendered massless by the Standard Model. This contradiction establishes the starting point of this PhD thesis, which discusses the role of heavy neutrinos (a hypothetical new type of neutrino) as responsible of generating the mass of the observed neutrinos. Certain extensions of the Standard Model will be explored, which include these new particles and offer common solutions to neutrino masses and to other open problems. The consequences of the interactions of these new neutrinos with Standard Model particles will also be studied, as well as how they can be characterized by the means of present and future experimental observations. 

\thispagestyle{empty}


\PB

{
  \hypersetup{linkcolor=myred}
  \setcounter{tocdepth}{1}
  \tableofcontents
  \thispagestyle{empty}
}



\setupfakechapters
\PB
\chapter*{Introducción}
\fancyhead[RO]{\scshape \color{lightgray}Introducción}
\fancyhead[LE]{\scshape \color{lightgray}Introducción}
\addcontentsline{toc}{part}{Introducción}
\pagenumbering{gobble}
\pagenumbering{arabic}
\lettrine[depth=1, loversize=0.55,lraise=-0.4]
    {N}{}uestro conocimiento de las partículas fundamentales que componen el universo, sus características e interacciones, está contenido en el Modelo Estándar (SM, por sus siglas en inglés) de la física de partículas~\cite{Weinberg:1967tq,Glashow:1961tr,Salam:1968rm,Gell-Mann:1964ewy,Gross:1973id,Politzer:1973fx,Higgs:1964ia,Higgs:1964pj,Higgs:1966ev,Guralnik:1964eu,Englert:1964et}. Esta teoría es el resultado de numerosas aportaciones teóricas y experimentales desarrolladas a lo largo del siglo XX. Su última pieza, el bosón de Higgs, fue finalmente descubierta en 2012 en el Gran Colisionador de Hadrones del CERN~\cite{ATLAS:2012yve,CMS:2012qbp}.

El SM ha demostrado ser una de las teorías más exitosas en la historia de la ciencia. Contiene todas las partículas elementales que han sido observadas hasta la fecha, y proporciona una descripción cuántica de tres de las cuatro fuerzas fundamentales de la naturaleza. Esta teoría contiene algunos parámetros libres, que no pueden ser determinados por primeros principios. Sin embargo, una vez sus valores quedan fijados mediante información experimental, el SM es capaz de ofrecer predicciones para virtualmente cualquier fenómeno en física de partículas. Tales resultados se han visto ampliamente confirmados por un gran número de experimentos independientes, en algunos casos con una precisión sin precedentes.

Sin embargo, hay razones sólidas para pensar que el SM no puede ser la teoría definitiva de las partículas y las interacciones fundamentales. La primera de ellas es la ausencia de interacciones gravitatorias. Hasta el momento, no se conoce una descripción de la gravedad a nivel cuántico, al contrario que para las fuerzas electromagnética, nuclear débil y nuclear fuerte. Por tanto, el SM es incapaz de estudiar procesos en los que se espera que la gravedad cuántica juegue un papel relevante, es decir, a energías iguales o mayores que la masa de Planck.

Incluso a energías mucho menores que la masa de Planck, existen varios fenómenos, no relacionados con gravedad cuántica, que no pueden ser explicados satisfactoriamente por el SM. En algunos casos, esto se debe a motivos más bien teóricos, sugiriendo que el SM no es la teoría más elegante o sencilla posible. La mayoría de estos misterios se denominan problemas de \say{ajuste fino}. Se trata de situaciones en las que varios parámetros, que en principio describen una física similar, toman valores radicalmente diferentes, o en las cuales parámetros adimensionales han de ser extremadamente pequeños, incluso nulos. Esto contradice el principio de naturalidad de \textquotesingle t Hooft: \say{a cualquier escala de energía, un parámetro físico o conjunto de parámetros físicos sólo puede ser muy pequeño si así aumenta la simetría del sistema}.~\cite{tHooft:1979rat}. El ajuste fino también aparece cuando varios parámetros, a priori no relacionados, se cancelan a un nivel muy preciso.

Un ejemplo es el llamado problema CP fuerte. Las interacciones fuertes del SM podrían, en principio, romper la invariancia carga-paridad (CP); sin embargo, todos los resultados experimentales sugieren que la fuerza fuerte respeta tal simetría. En realidad, el parámetro adimensional que controla este efecto presenta dos contribuciones, que aparecen tanto del sector fuerte como del débil. Esto constituye un ejemplo excelente de problema de ajuste fino: para satisfacer los resultados experimentales, dos parámetros adimensionales deben ser extremadamente pequeños o cancelarse casi exactamente entre ambos, sin una razón teórica para ello.

El problema de la jerarquía pertenece a esta misma categoría. El descubrimiento del bosón de Higgs mostró que la masa de esta partícula está cercana a la escala electrodébil. Al contrario que para el resto de las partículas elementales, ninguna simetría protege la masa del Higgs frente a correcciones radiativas. En otras palabras, procesos cuánticos podrían contribuir a la masa del Higgs con términos proporcionales a la masa de Planck. Mantener esta partícula a la escala electrodébil requeriría una cancelación poco natural y extremadamente precisa de varias contribuciones. 

La estructura del contenido de partículas del SM también es un misterio. No hay una razón fundamental que explique por qué existen tres familias de fermiones, o por qué estas son progresivamente más pesadas. Además, es desconcertante que las masas de los fermiones sean tan diferentes, ya que la masa del más pesado, el quark top, es aproximadamente un millón de veces mayor que la del electrón. Los parámetros que controlan la masa de los fermiones deberían ser de la misma naturaleza física, y por tanto resulta sorprendente que presenten una jerarquía tan marcada. Se suele aludir a estas incógnitas como el puzle del sabor.

Uno de los problemas más graves de ajuste fino del SM es la energía oscura. Esta componente del universo, que supone un 70\% de su contenido energético~\cite{Planck:2018vyg}, es la responsable de su expansión acelerada~\cite{SupernovaCosmologyProject:1998vns,SupernovaSearchTeam:1998fmf}. En relatividad general, esta magnitud puede interpretarse como una constante cosmológica en las ecuaciones de Einstein; sin embargo, también puede calcularse como la energía del vacío desde el punto de vista de teoría cuántica de campos. Las observaciones cosmológicas han mostrado que la aceleración del universo requiere una constante cosmológica unos 53 órdenes de magnitud mayor que la contribución estimada del SM~\cite{Weinberg:1988cp}. Esta discrepancia puede resolverse ajustando el cero del potencial del SM, que no es físico en sí mismo (sólo las diferencias de energía lo son). Por tanto, reproducir los resultados cosmológicos requeriría un ajuste fino en el potencial de más de 50 órdenes de magnitud, sugiriendo un grave problema de naturalidad. 

Aunque se mantienen abiertos y constituyen una fuente de insatisfacción, estos enigmas teóricos no apuntan directamente a un fallo o ruptura del SM. Simplemente demuestran que, para que la teoría describa correctamente la naturaleza, algunas de sus características deben ser muy específicas, o estar misteriosamente fijadas. En ese sentido, no pueden compararse con otros problemas del SM, planteados por resultados experimentales que la teoría no puede explicar, o que están en contradicción directa con sus predicciones. Estas cuestiones claramente requieren una extensión del SM, y son las mejores motivaciones para la búsqueda de nueva física.

Una de ellas es la existencia de materia oscura (DM, por sus siglas en inglés). Si la teoría general de la relatividad es la descripción adecuada de la gravedad (y hay buenas razones para creer que así es), muchas observaciones astrofísicas y cosmológicas demuestran que el universo no puede estar principalmente compuesto de materia bariónica~\cite{Zwicky:1933gu,Zwicky:1937zza,Rubin:1970zza,Freeman:1970mx,Begeman:1991iy,Markevitch:2003at,Clowe:2006eq,WMAP:2012nax,Planck:2018vyg}. Las curvas de rotación de galaxias espirales, las lentes gravitacionales, la formación de estructuras a gran escala o las anisotropías del fondo cósmico de microondas son algunas de las evidencias que apuntan a la existencia de otro tipo de materia en el universo. Aunque la DM experimenta la gravedad del mismo modo que la materia bariónica, no interacciona con la luz ni con ninguna otra partícula conocida, al menos a un nivel detectable. Por tanto, el SM no contiene ningún candidato viable para DM. Asumiendo que este componente del universo esté formado por partículas, el SM ha de ser extendido para incluir la DM, que conforma el 25\% de la energía del cosmos~\cite{Planck:2018vyg}.

Continuando con la composición del universo, la asimetría materia-antimateria \cite{Planck:2018vyg} sigue siendo una pregunta sin respuesta. El hecho de que la materia bariónica esté formada por partículas (y no por antipartículas) requiere interacciones que violen conjugación de carga y carga-paridad, de condiciones fuera del equilibrio en el universo primitivo y de violación de número bariónico. El SM presenta todos estos requisitos, conocidos como condiciones de Sakharov~\cite{Sakharov:1967dj}. Sin embargo, la asimetría barión-antibarión que el SM puede generar no es lo suficientemente grande como para ser responsable del desequilibrio presente en el universo~\cite{Gavela:1993ts}. Se necesita, por tanto, un nuevo mecanismo que produzca la asimetría observada. 

Otra evidencia robusta que reclama una extensión del SM es la naturaleza masiva de los neutrinos. Estas partículas, las más esquivas del universo (visible), y sólo superadas en abundancia por los fotones, forman parte del SM, que predice una masa nula para ellas. Sin embargo, a finales del siglo XX, se confirmó la observación de oscilaciones de neutrinos~\cite{Super-Kamiokande:1998kpq,SNO:2002tuh}. Este fenómeno, que básicamente consiste en el cambio de sabor de los neutrinos según se propagan, sólo puede ocurrir si estas partículas tienen una masa no nula. Esta contradicción entre las predicciones de la teoría y los resultados empíricos es una clara prueba de que el SM está incompleto. Además, la presencia de masas de neutrinos deja sin respuesta varias preguntas que eran triviales en el SM, tales como la cantidad de violación de CP en el sector leptónico, la posible naturaleza de Majorana de los neutrinos, qué autoestados de masa son más pesados, o la escala absoluta de masa a la que realmente se encuentran.

Esta tesis está motivada por la evidencia de neutrinos masivos. Dentro de las partículas que ya conocemos, e independientemente de cualquier observación o modelo cosmológico, esta es la principal imperfección del SM, claramente requiriendo una explicación en términos de nueva física.

La predicción de neutrinos sin masa se debe a la ausencia de neutrinos dextrógiros (RH, por sus siglas en inglés) en el contenido de partículas del SM. Al contrario que el resto de los fermiones, los neutrinos levógiros (LH) carecen de su equivalente dextrógiro, impidiéndoles adquirir masa a través del mecanismo de Higgs. Parece natural extender el contenido del SM, introduciendo neutrinos dextrógiros y generando así masas de neutrinos. Resulta interesante que estas hipotéticas nuevas partículas no estarían afectadas por las interacciones \textit{gauge} del SM; por consiguiente, podrían presentar un término de masa de Majorana, al contrario que cualquier otro fermión. Esta nueva escala no estaría en principio relacionada con la electrodébil, posiblemente haciendo muy pesados estos neutrinos dextrógiros.

La forma más simple de explicar la masa de los neutrinos es el célebre mecanismo de \textit{seesaw} (balancín en castellano) tipo I~\cite{Yanagida:1979as,Mohapatra:1979ia,Gell-Mann:1979vob,Minkowski:1977sc}, que aumenta el contenido de partículas del SM, introduciendo neutrinos dextrógiros, y respeta sus simetrías originales. En este modelo, las masas de los neutrinos ligeros son proporcionales al acoplo de los neutrinos dextrógiros y levógiros con el campo de Higgs, e inversamente proporcionales a las masas de los nuevos neutrinos. Por tanto, proporciona una explicación natural a las medidas de las masas de los neutrinos, que han demostrado que estas partículas son mucho más ligeras que el resto de los fermiones del SM~\cite{Planck:2018vyg,KATRIN:2021uub}. Sin embargo, esto implica que sus posibles señales experimentales son muy difíciles de testar, debido bien a acoplos muy débiles o a masas extremadamente altas de los neutrinos dextrógiros. 

Este dilema ha llevado a la aparición de muchas otras extensiones del SM, que ofrecen mejores posibilidades a la hora de ser comprobadas (véase la Ref.~\cite{deGouvea:2016qpx} para un compendio de modelos de masas de neutrinos). Cada vez más barrocas, suelen plantear la existencia de más partículas nuevas (habitualmente diferentes tipos de nuevos neutrinos) o de otras interacciones fundamentales. A cambio, estas teorías más complejas suelen tener la ventaja de poder resolver otros problemas abiertos del SM, no directamente relacionados con el sector de los neutrinos. Por ejemplo, pueden proporcionar candidatos a DM, mecanismos que generan una asimetría bariónica o explicaciones al puzle del sabor.

Estos nuevos enfoques también son interesantes desde un punto de vista fenomenológico, ya que podrían ser potencialmente testados experimentalmente, bien detectando nuevas partículas, bien induciendo comportamientos nuevos y exóticos en las ya conocidas. Hasta ahora no se han encontrado evidencias concluyentes; sin embargo, la gran cantidad de datos experimentales disponible es muy valiosa para descartar algunas de las posibles nuevas teorías, o bien para señalar regiones favorables en el espacio de parámetros de estos modelos.

Los neutrinos pesados son el centro de esta tesis, que está organizada en tres episodios. En el primero se introducirá el SM, mostrando su incapacidad para explicar la naturaleza masiva de los neutrinos, y se resumirán las soluciones más simples y populares para este problema, que incluyen neutrinos pesados. En el segundo episodio se centrará la atención en modelos más complejos, que también ofrecen soluciones a otros problemas del SM, resaltando cómo los neutrinos pesados pueden ser la clave de varias preguntas aún por responder. El tercer episodio consistirá en un estudio de las interacciones de estas nuevas partículas desde un punto de vista independiente de modelos, tanto en un escenario mínimo como en una teoría efectiva más general, prestando atención a la información proporcionada por experimentos presentes y futuros. Por último, se finalizará con las conclusiones de esta tesis.

\chapter*{Introduction}
\fancyhead[RO]{\scshape \color{lightgray}Introduction}
\fancyhead[LE]{\scshape \color{lightgray}Introduction}
\addcontentsline{toc}{part}{Introduction}
\lettrine[depth=1, loversize=0.55,lraise=-0.4]
    {O}{}ur knowledge of the fundamental particles that compose the Universe, their characterstics and interactions, is comprised in the celebrated Standard Model (SM) of particle physics~\cite{Weinberg:1967tq,Glashow:1961tr,Salam:1968rm,Gell-Mann:1964ewy,Gross:1973id,Politzer:1973fx, Higgs:1964ia,Higgs:1964pj,Higgs:1966ev,Guralnik:1964eu,Englert:1964et}. This theory is the result of many theoretical and experimental efforts that took place during the 20th century. Its last piece, the Higgs boson, was finally discovered in 2012 at the Large Hadron Collider (LHC) at CERN~\cite{ATLAS:2012yve,CMS:2012qbp}. 

The SM has proven to be one of the most successful theories in the history of science. It contains all the elementary particles that have been observed up to date, and provides a quantum description of three of the four fundamental forces of Nature. Some of the parameters of the theory are a priori free, requiring experimental inputs to determine them; however, once those are fixed, the SM is able to offer predictions for virtually any phenomenon in particle physics. Such results have been widely confirmed by a great number of independent experiments, in some cases with an unprecedented precision.

Nevertheless, there are solid reasons to think that the SM cannot be the ultimate theory of fundamental particles and interactions. The first of them is the absence of gravitational interactions. So far, gravity cannot be described at the quantum level, in contrast to the electromagnetic, weak and strong forces. Thus, the SM cannot account for processes in which quantum gravity is expected to play a relevant role, that is, at energies at or above the Planck mass. 

Even at energies much below the Planck mass, there are several phenomena, unrelated to quantum gravity, that cannot be successfully explained by the SM. In some cases, this is due to rather theoretical arguments, suggesting that the SM is maybe not the most elegant, minimal or simple theory one could possibly build. Most of these puzzles are dubbed as \say{fine-tuning} problems. These are situations in which several parameters, that in principle describe similar physics, take wildly different values, or in which dimensionless parameters need to be extremely small, or even vanish. This is in contradiction to \textquotesingle t Hooft\textquotesingle s naturalness principle: \say{at any energy scale, a physical parameter or set of physical parameters is allowed to be very small only if it would increase the symmetry of the system}.~\cite{tHooft:1979rat}. Fine-tuning is also present when different, in principle uncorrelated parameters, cancel up to a very precise level.

Among this kind of puzzles is the so-called strong CP problem. The strong interactions of the SM are in principle allowed to break charge-parity (CP) invariance; however, all experimental results suggest that the strong force respects such a symmetry. Actually, the dimensionless parameter that controls this effect has two contributions, arising from both the strong and weak sectors. This is an excellent example of a fine-tuning problem: in order to satisfy the experimental results, two dimensionless parameters must be extremely small or cancel each other almost exactly, without a theoretical reason for it.

The hierarchy problem lies in the same category. The discovery of the Higgs boson showed that the mass of this particle is close to the electroweak scale. In contrast to the rest of elementary particles, no symmetry protects the Higgs mass against radiative corrections. In other words, quantum processes would potentially contribute to the Higgs mass with terms proportional to the highest physics scale, for instance the Planck mass. Keeping this particle at the electroweak scale would require an unnatural, extremely precise cancellation of several contributions.

The structure of the particle content of the SM is also a mystery. There is no fundamental reason that explains why the fermions are organized in three families, or why the latter are increasingly heavier. Furthermore, the very different masses of the fermions are quite puzzling, as the mass of the heaviest one, the top quark, is roughly a million times larger than that of the electron. The parameters that control the masses of the fermions are expected to be of a similar physical nature, and thus it is uncertain why they exhibit such a strong hierarchy. These issues are referred to as the flavour puzzle.

One of the worst fine-tuning problems in the SM is posed by dark energy. This component of the Universe, which accounts for almost the 70\% of its energy budget~\cite{Planck:2018vyg}, causes it to expand at an accelerated rate~\cite{SupernovaCosmologyProject:1998vns,SupernovaSearchTeam:1998fmf}. In general relativity, this quantity can be interpreted as a cosmological constant in Einstein\textquotesingle s equations; however, it also corresponds to the energy of the vacuum from the point of view of quantum field theory. Cosmological observations have shown that the acceleration of the Universe requires a cosmological constant around 53 orders of magnitude larger than the estimated contribution of the SM~\cite{Weinberg:1988cp}. This discrepancy can be addressed by fixing the zero of the potential of the SM, which is not physical by itself (only energy differences are). Thus, reproducing the cosmological results would require a fine-tuning in the potential of more than 50 orders of magnitude, suggesting a serious naturalness issue.

Although still open and a source of unease, these theoretical enigmas do not directly point to a failure or breaking of the SM. They just show that, in order for the theory to correctly describe Nature, some of its features need to be quite specific and mysteriously fixed. In that sense, they cannot be compared to other problems of the SM, which are posed by experimental results which cannot be explained by the theory, or which are directly in contradiction with its predictions. These issues clearly demand an extension of the SM, and are the best motivations to search for new physics.

One of them is the existence of dark matter (DM). If the General Theory of Relativity is the adequate description of gravity (and there are good reasons to believe so), many astrophysical and cosmological observations prove that the Universe cannot be mainly composed of baryonic matter~\cite{Zwicky:1933gu,Zwicky:1937zza,Rubin:1970zza,Freeman:1970mx,Begeman:1991iy,Markevitch:2003at,Clowe:2006eq,WMAP:2012nax,Planck:2018vyg}. Rotation curves of spiral galaxies, strong gravitational lensing, large scale structure formation and Cosmic Microwave Background (CMB) anisotropies are some of the evidences that point to the existence of another kind of matter in the Universe. Although the latter is sensitive to gravity in the same manner as baryonic matter, it does not interact with light, or with any other known particle, at least up to a testable level. Thus, none of the components of the SM is a suitable candidate for DM. Provided this component of the Universe is made of particles, the SM needs to be extended in order to accomodate DM, which comprises roughly the 25\% of the energy budget of the cosmos~\cite{Planck:2018vyg}.

Following up on the composition of the Universe, the matter-antimatter asymmetry of the Universe~\cite{Planck:2018vyg} is still an unanswered question. The fact that the baryonic matter is made of particles, and not antiparticles, requires charge conjugation and charge-parity violating interactions, out-of-equilibrium conditions in the early Universe and baryon number violation. The SM exhibits all these requisites, known as Sakharov conditions~\cite{Sakharov:1967dj}. However, the baryon-antibaryon asymmetry that the SM can generate is not large enough to account for the imbalance present in the Universe~\cite{Gavela:1993ts}. A new mechanism is thus needed to produce the observed asymmetry. 

Another sound evidence which calls for an extension of the SM is the massive nature of neutrinos. These particles, the most  elusive in the (visible) Universe, and only second in abundance to photons, are contained in the SM, which predicts a vanishing mass for them. However, in the late 20th century, the observation of neutrino oscillations was finally confirmed~\cite{Super-Kamiokande:1998kpq,SNO:2002tuh}. This phenomenon, which basically consists in neutrinos changing flavour as they propagate, can only take place if these particles have a non-zero mass. This contradiction between the predictions of the theory and the empirical results is a clear proof that the SM needs to be completed. Furthermore, the presence of non-vanishing neutrino masses leaves unanswered several questions that were trivial within the SM, such as the amount of CP violation in the lepton sector, the possible Majorana nature of neutrinos, which mass eigenstates are heavier, or the actual absolute scale at which they lie. 

This thesis is motivated by the evidence for non-zero neutrino masses. Within the particles we already know, and regardless of any cosmological model or observations, this constitutes the main flaw of the SM, blatantly calling for an explanation in terms of new physics.  

The prediction of massless neutrinos is due to the absence of right-handed (RH) neutrinos in the particle content of the SM. In contrast to the rest of the fermions, left-handed (LH) neutrinos lack their RH counterpart, preventing their acquisition of a mass via the Higgs mechanism. It seems natural to augment the particle content of the SM, introducing RH neutrinos, and thus generating neutrino masses. Interestingly, these hypothetical new particles would be unaffected by the gauge interactions of the SM; hence, a Majorana mass term for them would be allowed, unlike for any other known fermion. This new scale is not a priori related to the electroweak one, rendering the RH neutrinos possibly very heavy. 

The simplest way to account for massive neutrinos is the celebrated type I seesaw mechanism~\cite{Yanagida:1979as,Mohapatra:1979ia,Gell-Mann:1979vob,Minkowski:1977sc}, which enlarges the SM particle content, introducing RH neutrinos, and respects its original symmetries. In this setup, light neutrino masses are proportional to the coupling of LH and RH neutrinos to the Higgs field, and inversely proportional to the masses of the new neutrinos. Thus, it provides a natural explanation to the smallness of neutrino masses, which have been measured to be much lighter than the rest of the SM fermions~\cite{Planck:2018vyg,KATRIN:2021uub}. However, this comes at the cost of very unlikely experimental signatures, due to either very feeble couplings or extremely large masses of the RH neutrinos. 

This has led to the appearance of many other extensions of the SM, that offer better prospects to be tested (see Ref.~\cite{deGouvea:2016qpx} for a review on neutrino mass models). Increasingly baroque, they often hypothesize the existence of more new particles (usually different kinds of new neutrinos) or of new fundamental interactions. In turn, these more contrived theories usually come with the boon of being able to solve some other open problems of the SM, not directly related to the neutrino sector. For instance, they may provide DM candidates, mechanisms that generate a baryon asymmetry, or explanations to the flavour puzzle. 

These new frameworks are also interesting from a phenomenological point of view, as they could potentially be experimentally probed, either by detecting new particles or by inducing new, exotic behaviors in those that are already known. So far, no conclusive evidence has been found. However, the vast amount of available experimental data is very valuable in order to rule out some of the possible new theories, or to point to preferred regions in parameter space.

Heavy neutrinos are the center of this thesis, which is organized in three episodes. In the first one, we will introduce the SM, show how it cannot explain the massive nature of neutrinos, and summarize the simplest and most popular solutions to this problem, which include heavy neutrinos. In the second episode, we will focus on more complex models, which also offer solutions to other problems of the SM, highlighting heavy neutrinos as the key to several unanswered questions. The third episode will consist in a rather model-independent study of the interactions of these new particles, both in a minimal scenario and in a more general effective theory, paying attention to the information provided by current and future experiments. We will close the thesis summarizing its main conclusions.

\setupnormalparts
\setupnormalchapters

\part[Neutrinos in and beyond the Standard Model]
     {\cinzel The prophecy\\[\bigskipamount] 
      \normalfont\Large Neutrinos in and beyond the Standard Model}

\fancyhead[LE]{\scshape \color{lightgray}I. Neutrinos in and beyond the Standard Model}\PB

\chapter{The Standard Model}
\fancyhead[RO]{\scshape \color{lightgray}1. The Standard Model}
The SM is a quantum field theory based on the gauge symmetry group $SU(3)_C\times SU(2)_L\times U(1)_Y$, as well as on Lorentz symmetry. This direct product of three independent groups represents the three fundamental interactions: $SU(3)_C$ describes the strong interactions, or Quantum Chromodynamics (QCD), while $SU(2)_L$, or weak isospin, and $U(1)_Y$, or hypercharge, contain the unified electroweak interactions. The product of these two groups will be reduced to the Abelian group $U(1)_{\rm em}$, responsible for the electromagnetic force, once the electroweak symmetry is spontaneously broken (in a manner that will be described below).

The particle content of the SM includes the gauge bosons $-$spin 1 fields that mediate the fundamental interactions$-$, the fermions or matter fields $-$spin 1/2 particles$-$ and the Higgs doublet $-$a spinless field that is responsible for electroweak symmetry breaking (EWSB). All these particles are charged under the SM symmetry group, with quantum numbers summarized in Tab.~\ref{tab:SM_charges}.

\begin{table}[b!]
\centering

\begin{tabular}{cccc} \toprule\vspace{1mm}
    &$SU(3)_C$&$SU(2)_L$&$U(1)_Y$ \\ \toprule
    $Q_L$&\textbf{3}&\textbf{2}&1/3 \\
    $u_R$&\textbf{1}&\textbf{1}&4/3  \\
    $d_R$&\textbf{1}&\textbf{1}&-2/3 \\\midrule
    $L_L$&\textbf{1}&\textbf{2}&-1\\ 
    $e_R$&\textbf{1}&\textbf{1}&-2 \\\midrule
    $G_\mu$&\textbf{8}&\textbf{1}&0\\
    $W_\mu$&\textbf{1}&\textbf{3}&0  \\
    $B_\mu$&\textbf{1}&\textbf{1}&0\\ \midrule
    $H$&\textbf{1}&\textbf{2}&1 \\ \bottomrule
\end{tabular}
\caption{Transformation properties of the SM fields under the three gauge symmetry groups. Only the first generation fermions are shown here; those contained in the second and third families exhibit identical quantum numbers.}
\label{tab:SM_charges}
\end{table}

The Lagrangian of the SM can be decomposed into four main parts:
\begin{equation}
    \mathcal{L}_{\rm SM} = \mathcal{L}_{\psi} + \mathcal{L}_{\rm YM} + \mathcal{L}_{\rm Yuk} + \mathcal{L}_{\rm Higgs} \,.
\end{equation}
The fermionic part, $\mathcal{L}_{\psi}$, describes the matter fields, as well as their interactions with the gauge bosons, whose kinetic terms are contained in the Yang-Mills sector, $\mathcal{L}_{\rm YM}$. The Yukawa Lagrangian, $\mathcal{L}_{\rm Yuk}$, controls the interactions between the Higgs field and the fermions, being ultimately responsible for their masses. Finally, the Higgs sector, $\mathcal{L}_{\rm Higgs}$, describes the couplings of this field with the corresponding gauge bosons (eventually generating a mass for them) and is directly responsible for the breaking of electroweak symmetry. In the following we will describe each of these sectors separately.

\section{The fermion sector}
The SM contains three generations (also dubbed families or flavours) of fermions, which are almost identical copies. The first and lightest family contains the up and down quarks, the electron and the electron neutrino; the second one is composed of the charm and strange quarks, the muon and the muon neutrino; the third and heaviest generation includes the top and bottom quarks, the tau and the tau neutrino. The three families only differ from one another in the masses of the particles they contain, being identical from the point of view of the gauge interactions. In fact, all the \say{flavour copies} of a given particle have exactly the same quantum numbers. 

The SM fermions are chiral fields, exhibiting a left-handed and a right-handed part (save for neutrinos):
\begin{equation}
    \psi=\psi_L+\psi_R\,,
\end{equation}
where $\psi_{L(R)}=P_{L(R)}\psi$. $P_L$ and $P_R$ are the left and right projectors, which are defined as $P_{L(R)}=\left(\mathbb{1}\mp \gamma_5\right)/2$ and satisfy $P_{L(R)}^2=P_{L(R)}$ and $P_LP_R=0$. The LH and RH parts of each fermion are actually different fields, as they have different quantum numbers, and can be thought of as different particles. In fact, weak isospin only affects LH fermions, which transform as doublets under $SU(2)_L$, while leaving RH ones untouched (they are $SU(2)_L$ singlets). For instance, the first generation is composed of two weak doublets, $Q_L = \left( u_L, d_L\right)^T$ and $L_L = \left(\nu_{eL}, e_L \right)^T$, and three weak singlets, $u_R$, $d_R$ and $e_R$. The second and third families are structured in the same manner.

All these matter fields can be separated into two categories, quarks and leptons.
\begin{itemize}[topsep=-1pt,itemsep=0pt]
    \item[$\bullet$] Quarks are the only fermions that partake in strong interactions, transforming as triplets under $SU(3)_C$. In fact, each quark field actually has three copies, one for each QCD charge or \say{color}. There are two sets of quarks: the \say{up-type} (up, charm and top) have an electric charge of +2/3, while that of the \say{down-type} (down, strange and bottom) is -1/3.
    \item[$\bullet$] Leptons are not charged under QCD. They can also be subdivided in two categories, charged leptons (electron, muon and tau), with an electric charge of -1, and neutrinos, which are electrically neutral. 
\end{itemize}
The fermionic part of the SM Lagrangian contains the so-called covariant derivatives of the matter fields. These include the fermion kinetic terms, as well as their interactions with the corresponding gauge bosons, according to the quantum numbers of each fermion. We will only write the part concerning the first family, with identical terms for the other two.
\begin{equation}
    \mathcal{L}_\psi \supset \overline{Q_L}i\slashed{D}Q_L + \overline{L_L}i\slashed{D}L_L + \overline{u_R}i\slashed{D}u_R+ \overline{d_R}i\slashed{D}d_R + \overline{e_R}i\slashed{D}e_R\,,
\label{eq:ferm_lag}
\end{equation}
where
\begin{align}
D_\mu Q_L &= \left(\partial_\mu -i \frac{g_s}{2}\lambda_aG_\mu^a-i\frac{g}{2}\tau_aW_\mu^a-i\frac{g^\prime}{6}B_\mu\right) Q_L\,,
\\
D_\mu L_L &= \left(\partial_\mu -i\frac{g}{2}\tau_aW_\mu^a+i\frac{g^\prime}{2}B_\mu \right)L_L\,,
\\
D_\mu u_R &= \left(\partial_\mu -i \frac{g_s}{2}\lambda_aG_\mu^a-i\frac{2g^\prime}{3}B_\mu \right)u_R\,,
\\
D_\mu d_R &= \left(\partial_\mu-i \frac{g_s}{2}\lambda_aG_\mu^a+i\frac{g^\prime}{3}B_\mu\right) d_R\,,
\\
D_\mu e_R &= \left(\partial_\mu +ig^\prime B_\mu \right)e_R\,.
\end{align}
Here, $g_s$, $g$ and $g^\prime$ are the gauge couplings associated to the $SU(3)_C$, $SU(2)_L$ and $U(1)_Y$ groups, respectively. The eight $G_\mu$ fields represent the gluons, mediators of QCD, whereas the three $W_\mu$ are the gauge bosons associated to the weak isospin, and $B_\mu$ is the carrier of hypercharge. The Pauli and Gell-Mann matrices are denoted by $\tau$ and $\lambda$, respectively.

Although the distribution of the quantum numbers for the fermions may seem somewhat arbitrary, it is perfectly engineered to cancel the anomalies of the three gauge interactions. A symmetry is said to be anomalous if it is respected at the classical level, but is not conserved at the quantum level. This depends on the fermion fields that are charged under the symmetry, and on their corresponding quantum numbers; if a symmetry is anomalous, it cannot consistently describe a fundamental interaction. The transformation properties of the SM fermions ensure that the anomalies of QCD, weak isospin and hypercharge vanish exactly, rendering $SU(3)_C\times SU(2)_L\times U(1)_Y$ a good symmetry, also at the quantum level. Importantly, anomaly cancellations also require the SM to be composed of complete families (although they do not provide a rationale for the existence of exactly three generations).

Note that no explicit mass term appears for any SM fermion. This is a consequence of their chiral nature. A term proportional to $\overline{\psi}_{L(R)}\psi_{L(R)}$ would vanish, due to the ortogonality of the left and right projectors. A mass term proportional to $\overline{\psi}_{R(L)}\psi_{L(R)}$, dubbed Dirac mass, would in turn not cancel; however, such a bilinear would not be invariant under $SU(2)_L$ or $U(1)_Y$. Thus, such a term is forbidden in the SM Lagrangian. All in all, the matter fields of the SM remain massless at the Lagrangian level, acquiring a mass only via EWSB. 

\section{The Yang-Mills sector}
This part of the SM Lagrangian contains the kinetic terms for the gauge bosons associated to $SU(3)_C$, $SU(2)_L$ and $U(1)_Y$:
\begin{equation}
    \mathcal{L}_{\rm YM}=- \frac{1}{4}G_{\mu\nu}^aG^{\mu\nu a}- \frac{1}{4}W_{\mu\nu}^aW^{\mu\nu a}- \frac{1}{4}B_{\mu\nu}B^{\mu\nu}\,,
\end{equation}
where $G_{\mu\nu}$, $W_{\mu\nu}$ and $B_{\mu\nu}$ are the field strenght tensors associated to the corresponding vector fields:
\begin{align}
    G_{\mu\nu}^a&=\partial_\mu G_\nu^a-\partial_\nu G_\mu^a+g_sf^{abc}G_\mu^bG_\nu^c\,,
    \\
    W_{\mu\nu}^a&=\partial_\mu W_\nu^a-\partial_\nu W_\mu^a+g\epsilon^{abc}W_\mu^bW_\nu^c\,,
    \\
    B_{\mu\nu}&=\partial_\mu B_\nu-\partial_\nu B_\mu\,.
\end{align}
Here, $f^{abc}$ are the structure constants of $SU(3)$, whereas $\epsilon^{abc}$ is the fully antisymmetric Levi-Civita tensor. The appearance of quadratic terms in the field strengths of the gauge bosons of $SU(3)_C$ and $SU(2)_L$ is due to the non-Abelian nature of these groups, and has the physical consequence of the self-couplings of the corresponding bosons. As $U(1)_Y$ is Abelian, no such term appears in the field strength of the $B_\mu$ field.

Note that, at the Lagrangian level, the whole  $SU(3)_C\times  SU(2)_L\times U(1)_Y$ symmetry is exact. Thus, all the gauge bosons are exactly massless. The massive $W^\pm$ and $Z$ bosons, that are eigenstates of the Hamiltonian and thus dubbed \say{physical}, will acquire their masses after EWSB, while the photon remains massless. The gluons will still be physical and massless, as $SU(3)_C$ is unaffected by any symmetry breaking and stays exact.

Let us mention that the symmetries of the SM allow for an extra term concerning gluons,
\begin{equation}
    \mathcal{L}_{\cancel{\rm CP}}=\frac{g_s^2}{16\pi}\theta G_{\mu\nu}^a\widetilde{G}^{\mu\nu a}\,,
\label{eq:theta_term}
\end{equation}
where $\widetilde{G}^{\mu\nu a}=G_{\alpha\beta}^a\epsilon^{\alpha\beta\mu\nu}$ and $\theta$ is a dimensionless parameter. This so called $\theta$-term explicitly violates CP symmetry, and is the source of the strong CP problem, mentioned in the introduction. This is a serious fine-tuning issue, as $\theta$ is constrained to be extremely small, in contrast with what could be expected from naturalness arguments. 

\section{The Yukawa Lagrangian}
Yukawa interactions couple fermion fields to the Higgs doublet. These trilinear terms are controlled by dimensionless couplings, often dubbed Yukawas for short. Interestingly, this is the only part of the SM Lagrangian that treats the three generations of fermions in a distinct way. In order to reproduce their measured masses, the Yukawa couplings of the three generations must be very different. Let us write the terms corresponding to the first family:
\begin{equation}
    \mathcal{L}_{\rm Yuk}\supset -y_u\overline{Q_L}\widetilde{H}u_R-y_d\overline{Q_L}H d_R-y_e \overline{L_L}H e_R+\rm h.c.\,,
\label{eq:yuk_lag}
\end{equation}
where $H$ denotes the Higgs doublet field and the different $y_i$ are the Yukawa couplings of each fermion. $\widetilde{H}\equiv i\tau_2 H^*$ is the conjugate of the Higgs field. 

Once electroweak symmetry is broken, the Yukawa interactions will yield mass terms for the SM fermions, as well as couplings between them and the physical Higgs boson. Note that, once the three generations are considered, the Yukawa couplings are promoted to matrices, which are, in general, non-diagonal. The physical masses will correspond to the eigenvalues of such matrices, as will be described later. 

\section{The Higgs sector}
This last part of the SM Lagrangian describes the Higgs doublet, containing its kinetic term, its associated potential, and, as it is charged under the SM symmetry group, its interactions with the corresponding gauge bosons.
\begin{equation}
    \mathcal{L}_{\rm Higgs}=\left(D_\mu H\right)^\dagger\left(D^\mu H\right)-V\left(H\right)\,.
\end{equation}
The covariant derivative of the doublet is given by
\begin{equation}
    D_\mu H=\left(\partial_\mu -i\frac{g}{2}\tau_aW_\mu^a-i\frac{g^\prime}{2}B_\mu\right)H\,,
\label{eq:higgs_cov_der}
\end{equation}
while its potential takes the form
\begin{equation}
    V\left(H\right)=-\mu^2H^\dagger H+\lambda\left(H^\dagger H\right)^2\,,
\label{eq:higgs_pot}
\end{equation}
with $\mu$ and $\lambda$ being a priori free parameters. In order for the potential to be bounded from below, $\lambda$ must be positive. If $\mu^2$ is negative, the potential would be always positive, with a trivial, unique minimum at $\vert H\vert=0$. However, if $\mu^2>0$, the potential is minimized at a non-trivial configuration, $\vert H\vert=\sqrt{\mu^2/2\lambda}$. If this were the case, the electroweak vacuum would be non-trivial, and not invariant under $SU(2)_L\times U(1)_Y$. This situation, known as spontaneous symmetry breaking, is responsible for many features of the SM, such as the massive nature of the fermions and of the $W^\pm$ and $Z$ bosons, as well as for the appearance of a physical Higgs boson. Such a mechanism will be further detailed in the following. 

\section{The Higgs mechanism, EWSB and mass generation}
The Higgs field is a complex scalar doublet of $SU(2)_L$, also charged under $U(1)_Y$. Its components can be written as
  \begin{align}
    H &= \begin{pmatrix}
           \phi^+ \\
           \phi^0
         \end{pmatrix}\,,
  \end{align}
adding up to 4 degrees of freedom.

The parameters of the Higgs potential, $\mu$ and $\lambda$, are a priori free, so the theory alone is unable to distinguish if the electroweak vacuum is trivial, and the whole $SU(2)_L\times U(1)_Y$ symmetry is preserved, or whether it is spontaneously broken by a non-trivial vacuum. However, the SM fermions and the weak bosons behave as massive particles, a situation that could not be explained in the former scenario. Thus, EWSB is necessary, and the Higgs field must take a non-zero vacuum expectation value (vev). 

In principle, both the charged and the neutral components of the doublets could take vevs. However, if $\phi^+$ did, the vacuum would be electrically charged, implying that the electric charge would not be conserved. As that is not the case, only the neutral, lower component of the doublet takes a non-zero vev:
\begin{align}
    \left<H\right>=\begin{pmatrix}
           0 \\
           \frac{v}{\sqrt{2}}
         \end{pmatrix}\,,
         \label{eq:higgs_vev}
\end{align}
where $v=\sqrt{\mu^2/\lambda}$.

The minimization of the Higgs potential only depends on the modulus of the doublet, and not on the value of its angular component, as can be explictly seen in Eq.~\ref{eq:higgs_pot}, which is invariant under a rephasing of the $H$ field. This means that, even after $H$ takes a vev, there is a remaining, \say{rotational} symmetry.  This can be understood as the $U(1)$ that survives EWSB, and it is identified with the electromagnetic symmetry.

The way EWSB affects the gauge sector of the SM can be understood thanks to Goldstone\textquotesingle s theorem. This theorem states that, when a symmetry is spontaneously broken, a massless, spinless boson appears for each generator of the broken symmetry. In the breaking pattern $SU(2)_L\times U(1)_Y\to U(1)_{\rm em}$, 3 Goldstone bosons appear, associated to the 3 generators of $SU(2)$. These degrees of freedom are absorbed (\say{eaten}) by the corresponding gauge bosons, which acquire a longitudinal degree of freedom and become massive. In fact, if one introduces the vev of the Higgs field (Eq.~\ref{eq:higgs_vev}) into its covariant derivative (Eq.~\ref{eq:higgs_cov_der}), mass terms for the gauge bosons will appear. However, these will not be directly the physical masses, as mixing terms are also generated. The $W^1$ and $W^2$ can be easily combined into the physical $W_\mu^\pm$ bosons:
\begin{equation}
   W_\mu^\pm=\frac{W_\mu^1\mp iW_\mu^2}{\sqrt{2}}\,.
\end{equation}
The electrically neutral $W^3$ and $B$ need to be rotated into the physical basis. Upon diagonalization of the arising mass matrix, one massive state appears, the $Z$ boson, as well as a massless one, the photon, $A$. It is expected that the mass of the latter vanishes, as it is the gauge boson of the surviving 
$U(1)_{\rm em}$ symmetry, which remains preserved even by the non-trivial electroweak vacuum. The physical states are given in terms of the interaction basis by
\begin{align}
    Z_\mu =\cos{\theta_w}W_\mu^3-\sin{\theta_w}B_\mu\,,
    \\
    A_\mu =\sin{\theta_w}W_\mu^3+\cos{\theta_w}B_\mu\,.
\end{align}
Here, $\theta_w$ is the so-called weak mixing angle, that can be expressed in terms of the gauge couplings as
\begin{equation}
    \cos{\theta_w}=\frac{g}{\sqrt{g^2+g^{\prime 2}}}\,.
\end{equation}
The resulting mass terms for the weak bosons can be written as 
\begin{equation}
    \mathcal{L}_{\rm mass}^{W,Z}=-M_W^2W_\mu^-W^{\mu +}-\frac{1}{2}M_Z^2Z_\mu Z^\mu\,.
\end{equation}
It can easily be found that $M_W=\frac{gv}{2}$ and that $M_Z=\frac{gv}{2\cos{\theta_w}}$.

Note that EWSB cannot change the number of dynamical degrees of freedom. The Higgs doublet contains four of them, three of which are the Goldstone bosons, that become the longitudinal components of the $W$ and $Z$. The remaining one appears in the spectrum as a physical particle, the celebrated Higgs boson. It corresponds to the perturbations around the minimum of the potential, denoted by $h$ in the following. It acquires a mass, given by $M_h=\sqrt{2\lambda v^2}$, when introducing the perturbations into the Higgs potential (Eq.~\ref{eq:higgs_pot}). 

The masses of the SM fermions are also generated upon EWSB. If the Higgs doublet in the Yukawa Lagrangian (Eq.~\ref{eq:yuk_lag}) is replaced by its vev, Dirac mass terms are generated:
\begin{equation}
    \mathcal{L}_{\rm mass}\supset -m_u\overline{u_L}u_R-m_d\overline{d_L}d_R-m_e\overline{e_L}e_R\,.
\end{equation}
The mass of each fermion is controlled by its Yukawa coupling to the Higgs doublet, with the vev setting an overall scale:
\begin{equation}
    m_i=\frac{y_iv}{\sqrt{2}}\,.
\end{equation}
Experiments have shown no clear pattern for the masses of the SM fermions, which in fact range across six orders of magnitude. This large hierarchy is inherited by the Yukawa couplings. The SM does not provide an explanation for such a structure, which in fact involves serious fine-tuning issues in some of the couplings. Our lack of understanding of the origin of the very different fermion masses is the source of the so-called flavour puzzle. 

Aside from generating the masses of the fermions and the weak bosons, the couplings of the Higgs field are also responsible for the interactions of the physical Higgs boson (which we will just refer to as \say{Higgs} in the following). In the gauge sector, the covariant derivative of the doublet induces couplings between the Higgs and the $W_\mu^a$ fields. Once in the mass basis, this translates into vertices containing Higgses, $W^\prime$s and $Z^\prime$s. Importantly, no coupling between the scalar and the photon appears, as the Higgs is not electrically charged (a consequence of electromagnetism surviving EWSB). Similarly, the Yukawa Lagrangian will not only generate the masses of the fermions, but also their couplings to the Higgs, with an interaction strength also controlled by the corresponding Yukawa couplings.

\section{Other symmetries of the Standard Model}
Apart from Lorentz symmetry, the SM is built, as explained above, based on the gauge, local symmetry $SU(3)_C\times SU(2)_L\times U(1)_Y$, which is directly imposed and determines which terms are allowed in the Lagrangian and which are not, ultimately shaping the way in which elementary particles interact. Nevertheless, the SM exhibits other symmetries, either exact or only approximate.

This is the case of baryon ($B$) and  lepton number ($L$) symmetries. These are global, Abelian $U(1)$ symmetries, which are exact at the Lagrangian level, and have been widely confirmed by experiments. Baryon number only affects quarks, all of which are assigned a baryon number of 1/3, whereas all charged leptons and neutrinos have a lepton number of 1. Actually, there are three lepton number symmetries, $L_e$, $L_\mu$ and $L_\tau$, as the number of leptons of each family is conserved separately in the SM.

Baryon and lepton number are referred to as \say{accidental} symmetries of the SM, as they are not imposed when constructing the theory. In fact, there are good reasons to believe these are not good symmetries of Nature, as they are anomalous. Thus, they cannot be promoted to gauge symmetries, as they do not behave properly at the quantum level. 

Interestingly, the anomalies of baryon and lepton number cancel each other exactly. Thus, the difference of baryon and lepton number, $B-L$, is respected both at the classical and quantum levels. Although still accidental and global, $B-L$ is an exact symmetry of Nature, just as electric or color charge. This feature has led to many proposals in which $B-L$ is promoted to a new gauge symmetry (although that would require the addition of three RH neutrinos).

If one neglects the Yukawa interactions, the SM would also exhibit a $U(3)^5$ flavour symmetry, which is a consequence of the fact that the gauge interactions treat all fermion flavours equally. Each of the five $U(3)$\textquotesingle s corresponds to LH quarks, LH leptons, RH up-type quarks, RH down-type quarks and RH charged leptons, and connects particles of different generations; for instance, the electron, muon and tau transform under a triplet of their associated flavour symmetry. 

Yukawa interactions explicitly break flavour symmetry. The whole Yukawa Lagrangian, taking into account the three families, reads
\begin{equation}
    \mathcal{L}_{\rm Yuk} = -\overline{Q_L}^\prime\ Y_{u}\widetilde{H}u^\prime_{R}-\overline{Q_L}^\prime Y_{d}H d^\prime_{R}-\overline{L_L}^\prime
    Y_{e}H e^\prime_{R}+\rm h.c.\,
\end{equation}
where the fermion fields are now vectors that comprise all three flavour copies. Flavour indices are implicit and the Yukawa couplings are promoted to matrices in flavour space\footnote{The upper case Yukawas denote matrices, while the lower case ones shown before denote just couplings.}. The primes denote that these fields are not in the mass basis, as they do not describe physical particles.

The symmetries of the SM allow for non-diagonal Yukawa terms, with no reason for the Yukawa matrices to be diagonal. The physical masses of the fermions are obtained once these matrices are diagonalized. This change of basis requires different unitary rotations for up-type quarks, down-type quarks and charged leptons, and also for LH and RH fields:
\begin{equation}
    \psi_{L(R)}=V_{L(R)}^\psi\psi^\prime_{L(R)}\,,
\end{equation}
where these new, un-primed fields now correspond to the mass basis. The different rotation matrices are defined by
\begin{align}
    \left(V_L^d\right)^\dagger Y_dV_R^d=\frac{\sqrt{2}}{v}\mathrm{diag}\left(m_d\,,m_s\,,m_b\right)\,,
    \\
    \left(V_L^u\right)^\dagger Y_uV_R^u=\frac{\sqrt{2}}{v}\mathrm{diag}\left(m_u\,,m_c\,,m_t\right)\,,
    \\
    \left(V_L^e\right)^\dagger Y_eV_R^e=\frac{\sqrt{2}}{v}\mathrm{diag}\left(m_e\,,m_\mu\,,m_\tau\right)\,.
\end{align}
The interactions mediated by gluons, photons or $Z$ bosons, as well as those involving the Higgs boson, do not mix upper and lower components of weak doublets; thus, these matrices will always appear on the form $\left(V_{L(R)}^\psi\right)^\dagger V_{L(R)}^\psi$. As these are rotation matrices, they are by definition unitary, so this product always equals the identity matrix. This has the important physical consequence that strong, electromagnetic, neutral weak and Yukawa interactions are flavour diagonal and universal: no flavour-changing neutral currents (FCNCs) are allowed in the SM at tree level. 

Charged interactions, mediated by $W^\pm$ bosons, behave differently, as they couple up-type and down-type quarks, as well as charged leptons and neutrinos:
\begin{equation}
    \mathcal{L}_{W}=-\frac{g}{\sqrt{2}}W_\mu^+\left(\overline{u_L}\gamma^\mu\left(V_L^u\right)^\dagger V_L^d d_L + \overline{\nu_L}\gamma^\mu\left(V_L^\nu\right)^\dagger V_L^e e_L\right)+\rm h.c.\,
\end{equation}
In the lepton sector, the absence of RH neutrinos (and thus of the $V_R^\nu$ matrix) allows to perform a non-physical rotation on the neutrino fields, setting $V_L^\nu=V_L^e$ with no loss of generality. This is not possible in the quark sector, as it is impossible to simultaneously align the rotations of the LH and RH fields. Thus, the rotation matrices in the quark sector have a physical effect, inducing mixings in the quark sector (in other words, interactions that couple particles of different families) mediated by the $W$ bosons. These mixings are controlled by the Cabibbo-Kobayashi-Maskawa (CKM) matrix~\cite{Cabibbo:1963yz,Kobayashi:1973fv}, defined as
\begin{equation}
    V_{\rm CKM}\equiv \left(V_L^u\right)^\dagger V_L^d\,,
    \label{eq:CKM}
\end{equation}
which is unitary by construction. This matrix contains four free parameters, which can be parametrized as three mixing angles and a CP-violating phase, although other different parametrizations are available. Interestingly, the CP phase of the CKM matrix constitutes the only known source of CP violation in the SM. The only other possible source is the mysterious $\theta$-term of QCD, which has not been observed, and for which very stringent constraints apply. The CKM is also characterized by a strong hierarchy; its diagonal entries are much larger than the off-diagonal ones, so it is close to the identity matrix. Besides, the mixing between the first and third generations is much smaller than that between the first and second or second and third families. The lack of an understanding for such a structure, which seems to favour flavour conservation instead of flavour violation, is also a part of the flavour puzzle.

\chapter{Neutrinos in the SM}
\fancyhead[RO]{\scshape \color{lightgray}2. Neutrinos in the SM}
Neutrinos are fairly unique among the SM fermions. First of all, they are the only ones that do not partake in electromagnetic interactions, as they have no electric charge. Obviously, they are not charged under $SU(3)_C$ either, so they are only affected by weak interactions. This makes neutrinos very feebly interacting particles, and quite difficult to detect. 

Neutrinos experience both charged-current (CC) interactions, which couple them to $W$ bosons and charged leptons, and neutral-current (NC) interactions, in which a $Z$ boson couples to a pair of neutrinos. These couplings are given by the following Lagrangian:
\begin{equation}
    \mathcal{L}_\nu=-\frac{g}{\sqrt{2}}\sum_\alpha \overline{\ell_\alpha}\gamma^\mu P_L\nu_\alpha W_\mu^- -\frac{g}{4c_w}\sum_\alpha\overline{\nu_\alpha}\gamma^\mu P_L\nu_\alpha Z_\mu +\rm h.c.\,,
\end{equation}
where $\ell$ generically denotes a charged lepton, $c_w\equiv\cos{\theta_w}$ and the index $\alpha$ runs through the three lepton flavours, $\alpha =e,\mu,\tau$. As mentioned above, the absence of RH neutrinos makes it possible to align the rotations in the charged lepton and in the neutrino sectors. Thus, the electron, muon and tau neutrinos are both physical particles and interaction eigenstates, and they are defined by the flavour of the charged lepton to which they couple via the $W$ bosons. 

The main difference between neutrinos and the rest of the SM fermions is precisely the fact that the former always appear as LH fields, lacking their RH counterpart. Apart from preventing flavour violation in the lepton sector, this has the very relevant physical consequence of rendering the neutrinos massless: as there are no $\nu_R$ fields, there is no way to couple the lepton doublet to the Higgs field in order to generate a mass term for neutrinos upon EWSB. Note that this is only true at the renormalizable level. Such a coupling can be written as an effective, dimension-5 term, known as Weinberg operator~\cite{Weinberg:1979sa}. As it will be discussed later, such an operator yields a Majorana mass for neutrinos after electroweak symmetry is broken. Apart from the non-renormalizability of such a term, this Majorana mass would explicitly violate lepton number, so it is forbidden by $B-L$, which, although accidental, is an exactly preserved symmetry in the SM. All in all, the SM predicts neutrinos to be massless at all orders, due to its exact chiral and $B-L$ symmetries.

\chapter{Neutrino oscillations: evidence for neutrino masses}
\fancyhead[RO]{\scshape \color{lightgray}3. Neutrino oscillations: evidence for neutrino masses}
Physical particles are those whose quantum states only change up to a phase when propagating freely. That phase is given by the energy of the particle, $E$, and the time they have been propagating, $t$:
\begin{equation}
    \vert \psi (t)\rangle = \vert \psi (0)\rangle e^{-iEt}\,.
\end{equation}
In order for this to happen, a particle needs to have a well-defined energy, or, in other words, to be an eigenstate of the Hamiltonian operator. If this is the case, the modulus of its wavefunction will be unaltered upon propagation, and the probability of the particle remaining in the same state will be equal to unity:
\begin{equation}
    P(t)=\left| \langle\psi(0)\vert\psi(t)\rangle\right|^2=1\,.
\end{equation}
Obviously, in order for a particle to exhibit a well-defined energy, it also needs to have a well-defined mass (that is why the physical and mass bases are equivalent).

This is the case for the SM neutrinos. As they have an exactly vanishing mass, they have a well-defined energy, so they are eigenstates of the Hamiltonian and propagate as physical particles. Thus, if an electron neutrino is produced, it will remain in that state, and be detected as an electron neutrino with a 100\% probability (and equally for muon and tau neutrinos).

This prediction of the SM entered in conflict with experimental data when neutrino oscillations were discovered. This phenomenon consists in the change of flavour of neutrinos during their propagation. 

The first observations of neutrino oscillations, that took place at the Homestake experiment~\cite{Davis:1968cp}, led to the so-called solar neutrino problem, that consisted in the observation of a deficit of events in the detection of electron neutrinos coming from the Sun.  These findings, compatible with oscillations of electron neutrinos into other flavours, were confirmed by gallium experiments~\cite{Cleveland:1998nv,Lande:1999cv,Abazov:1991rx,GALLEX:1998kcz,Kirsten:2003ev}, as well as water Cherenkov detectors, such as Kamiokande~\cite{Kamiokande:1994sgx} and Super-Kamiokande~\cite{Super-Kamiokande:2002ujc}. However, the expected electron neutrino fluxes were based on the Standard Solar Model; a flaw in this framework could have been behind the observed deficit of events, even in the absence of neutrino oscillations. This explanation was disfavoured by both the SNO and Super Kamiokande detectors. The former was able to measure neutrino NC interactions, which were compatible with the standard solar model, and, as neutral currents are flavour blind, ruled out the possibility of a misunderstanding of the solar fluxes~\cite{SNO:2001kpb,SNO:2002hgz,SNO:2002tuh,SNO:2003bmh}. 

On the other hand, Super Kamiokande detected a deficit of muon neutrinos from atmospheric sources, consistent with their oscillation into tau neutrinos~\cite{Super-Kamiokande:1998kpq}. Muon neutrino disappearance was also observed in accelerator experiments~\cite{MINOS:2006foh,K2K:2002icj}. The confirmation of the solar model, together with the observation of oscillations in two different channels, clearly established that neutrinos do not behave as the SM predicted, and that the electron, muon and tau neutrinos are not mass eigenstates. 

The picture was completed in the following years, by the observations of oscillations of antineutrinos produced in reactors~\cite{KamLAND:2004mhv}. In fact, this kind of experiment was the first to provide enough information to confirm that the three neutrino flavours mix with each other~\cite{RENO:2012mkc,DoubleChooz:2011ymz,DayaBay:2012fng}. This result was also supported by accelerator experiments~\cite{NOvA:2016kwd,T2K:2015sqm}. 

In this three-flavour oscillation paradigm, the flavour neutrino eigenstates, $\nu_e$, $\nu_\mu$ and $\nu_\tau$, are rotated into the physical fields, $\nu_1$, $\nu_2$ and $\nu_3$, by the Pontecorvo-Maki-Nakagawa-Sakata (PMNS) matrix~\cite{Pontecorvo:1957cp,Pontecorvo:1957qd,Pontecorvo:1967fh,Gribov:1968kq,Maki:1962mu}:
\begin{equation}
    \nu_i = \sum_\alpha U_{\alpha i}^*\nu_\alpha\,,
    \label{eq:PMNS}
\end{equation}
or, conversely,
\begin{equation}
    \nu_\alpha = \sum_i U_{\alpha i}\nu_i\,,
\end{equation}
where the Latin (Greek) indices refer to mass (interaction) eigenstates. A flavour state after a certain time of propagation is given by
\begin{equation}
    \vert \nu_\alpha (t)\rangle =\sum_{i=1}^3U^*_{\alpha i}e^{iE_it}\vert\nu_i\rangle\,.
\end{equation}
Although a proper computation would require the treatment of neutrinos as wave packets, with finite size and momentum widths~\cite{Rich:1993wu,Beuthe:2001rc,Giunti:2002xg,Akhmedov:2009rb, Akhmedov:2010ms}, the same results can be obtained by means of a simpler approach, in which neutrinos are dealt with as ultrarelativistic plane waves. Thus, $t\simeq L$, $E_i\simeq E+\frac{m_i^2}{2E}$ and
\begin{equation}
        \vert \nu_\alpha (t)\rangle =\sum_{i=1}^3U^*_{\alpha i}e^{iEt}e^{\frac{im_i^2L}{2E}}\vert\nu_i\rangle\,,
\end{equation}
where $m_i$ is the mass of the eigenstate $\nu_i$, $E$ is the common momentum of all the mass states and $L$ is the distance travelled by the neutrino in its propagation.

If a neutrino is produced as a flavour eigenstate $\nu_\alpha$, the probability of detecting it as a $\nu_\beta$ after it has travelled a distance $L$ is given by $P_{\alpha\beta}=\left|\langle\nu_\beta\vert\nu_\alpha(t)\rangle\right|^2$, which can be expressed as
\begin{align}
    \overset{(-)}{P}_{\alpha\beta}&=\delta_{\alpha\beta}-4\sum_{i>j}\mathrm{Re}\left[U_{\alpha i}^*U_{\beta i}U_{\alpha j}U_{\beta j}^*\right]\sin^2\left(\frac{\Delta m_{ij}^2L}{4E}\right)
    \nonumber
    \\
    &\pm 2\sum_{i>j}\mathrm{Im}\left[U_{\alpha i}^*U_{\beta i}U_{\alpha j}U_{\beta j}^*\right]\sin\left(\frac{\Delta m_{ij}^2L}{2E}\right)\,,
\label{eq:osc_prob}
\end{align}
where $E$ is the energy of the neutrino and $\Delta m_{ij}^2\equiv m_i^2-m_j^2$ is the (squared) mass splitting between the states $\nu_i$ and $\nu_j$. $P_{\alpha\beta}$ denotes the probability for neutrino oscillations, and $\overline{P}_{\alpha\beta}$ the antineutrino oscillation probability (note that they differ in the sign of the last term). The entries of the PMNS matrix determine the amplitude of the oscillations, while the two mass splittings, together with the $L/E$ ratio, control their frequency. 

The PMNS plays the same role in the lepton sector as the one played by the CKM in the quark sector. In both cases, the individual rotations performed on the upper and lower components of the doublets are not physical by themselves; only the product of the matrices is. Thus, the effects of the transformation can be assigned arbitrarily between up-type and down-type quarks, in the case of the CKM, or between neutrinos and charged leptons, for the PMNS. The usual conventions are that the up-type quarks and charged leptons are both in the interaction and mass bases, assigning the whole rotations of the CKM and PMNS matrices to the down-type quarks and neutrinos, respectively.

The PMNS matrix is usually parametrized in terms of three mixing angles, $\theta_{12}$, $\theta_{13}$ and $\theta_{23}$, and a CP-violating phase, $\delta_{\rm CP}$. The whole PMNS is expressed as the product of three individual rotation matrices, $ U_{\rm PMNS}= \mathcal{R}(\theta_{23})\mathcal{R}(\theta_{13},\delta)\mathcal{R}(\theta_{12})$, which results in:
\begin{align}
    U_{\rm PMNS}
         =\begin{pmatrix}
           c_{12}c_{13}&s_{12}c_{13}&s_{13}e^{-i\delta_{\rm CP}} \\
           -s_{12}c_{23}-c_{12}s_{23}s_{13}e^{i\delta_{\rm CP}}&c_{12}c_{23}-s_{12}s_{23}s_{13}e^{i\delta_{\rm CP}}&s_{23}c_{13}\\
           s_{12}s_{23}-c_{12}c_{23}s_{13}e^{i\delta_{\rm CP}}&-c_{12}s_{23}-s_{12}c_{23}s_{13}e^{i\delta_{\rm CP}}&c_{23}c_{13}
         \end{pmatrix}\,.
         \label{eq:pmns}
\end{align}
The neutrinos that can be detected on Earth are produced in different sources, which determine the energy of these particles and the distance they travel before being detected. These characteristics establish several regimes, making the different types of neutrino experiments best suited to probe particular oscillation parameters. 

The Sun is one of the furthest sources of neutrinos we can observe, so solar neutrinos exhibit the largest $L/E$ ratio. According to the expression of the oscillation probability (Eq.~\ref{eq:osc_prob}), this would mean that the oscillations controlled by large mass splittings are too fast to be resolved. Hence, the small mass splitting, $\Delta m^2_{21}$, is dubbed \say{solar}. However, the experiments that detect these neutrinos are not too sensitive to small mass splittings, due to the very intense matter effects inside the Sun. In that environment, where only electron neutrinos are produced, densities are extremely high; thus, the matter potential dominates over the vacuum contribution, so $\nu_e$\textquotesingle s behave as eigenstates of the Hamiltonian. As they are the eigenstates with larger energy, they evolve adiabatically, due to the smoothness of the potential in the stellar medium, and exit the Sun as $\nu_2$\textquotesingle s, and not $\nu_1$\textquotesingle s (the production of $\nu_3$\textquotesingle s is suppressed due to the smallness of the angle $\theta_{13}$). These neutrinos exhibit energies at the MeV range, so they can only be detected as $\nu_e$\textquotesingle s. Thus, solar neutrino experiments are sensitive to $\vert U_{e2}\vert^2$, hence being ideal to test $\theta_{12}$.

Another very relevant source of neutrinos is the atmosphere. Cosmic rays impinging on the nuclei of the atmosphere produce showers of particles, mainly pions and kaons, whose decay chains generate neutrinos abundantly. In contrast to solars, atmospheric neutrinos are much more energetic, and they are also produced much closer to Earth. Thus, the ratio $L/E$ is much smaller in this regime, so the oscillations controlled by $\Delta m^2_{21}$ are suppressed with respect to those driven by $\Delta m^2_{31}$ and $\Delta m^2_{32}$. This means that the detection of electron neutrinos is hindered by the smallness of $\theta_{13}$. The $\nu_\mu$ disappearance channel is however not suppressed, making atmospheric neutrinos ideal to probe $\theta_{23}$.

Reactor neutrinos exhibit different $L/E$ ratios, as the distance to the source can be adjusted and tuned to enhance the sensitivity to different mass splittings. Long baseline experiments, such as KamLAND, are sensitive to the small mass difference, as the oscillations controlled by $\Delta m^2_{31}$ and $\Delta m^2_{32}$ cannot be resolved due to their large frequency (as well as suppressed by the small $\theta_{13}$). As these neutrinos have low energies, these experiments can only detect the electron flavour. In fact, their measurement of $\theta_{12}$ in the $\overline{\nu_e}$ disappearance channel provided a complementary test of solar neutrino oscillations, with no need to rely on solar models. 

In contrast, short-baseline reactor experiments, such as Double Chooz and Daya Bay, tuned their $L/E$ so that oscillations controlled by the large mass splitting dominate. As the proximity to the source allows for great luminosities, these experiments were the first to measure the small $\theta_{13}$ angle in $\overline{\nu_e}$ disappearance, completing the three-neutrino oscillation paradigm.

Nowadays, this framework is very robustly established. The different sources of neutrinos have been probed by many independent experiments, providing compatible determinations of the mixing angles and mass splittings. The reference values are summarized in Tab.~\ref{tab:osc_params}, as obtained by the NuFiT collaboration by performing a global fit to oscillation data~\cite{Esteban:2020cvm}.
\begin{table}[b!]
\centering

\begin{tabular}{ccc} \toprule\vspace{1mm}
     &Normal ordering &Inverted ordering \\ \toprule\vspace{1.5mm}
    $\theta_{12}(^\circ)$&$33.41_{-0.72}^{+0.75}$&$33.41_{-0.72}^{+0.75}$ \\\vspace{1.5mm}
    $\theta_{23}(^\circ)$&$42.2_{-0.9}^{+1.1}$&$49.0_{-1.2}^{+1.0}$ \\\vspace{1.5mm}
   $\theta_{13}(^\circ)$&$8.58_{-0.11}^{+0.11}$&$8.57_{-0.11}^{+0.11}$\\\vspace{1.5mm}
    $\delta_{\rm CP}(^\circ)$&$232_{-26}^{+36}$&$276_{-29}^{+22}$\\ \vspace{1.5mm}
    $\frac{\Delta m_{21}^2}{10^{-5}\rm eV^2}$&$7.41_{-0.20}^{+0.21}$&$7.41_{-0.20}^{+0.21}$\\
$\frac{\Delta m_{3\ell}^2}{10^{-3}\rm eV^2}$&$2.507_{-0.027}^{+0.026}$&$-2.486_{-0.028}^{+0.025}$\\ \bottomrule
\end{tabular}
\caption{Three-flavour oscillation parameters, determined by the NuFiT collaboration~\cite{Esteban:2020cvm}. The best fit points are displayed, with their corresponding 1$\sigma$ intervals.}
\label{tab:osc_params}
\end{table}

\begin{figure}[t!]
\centering
\includegraphics[width=0.9\textwidth]{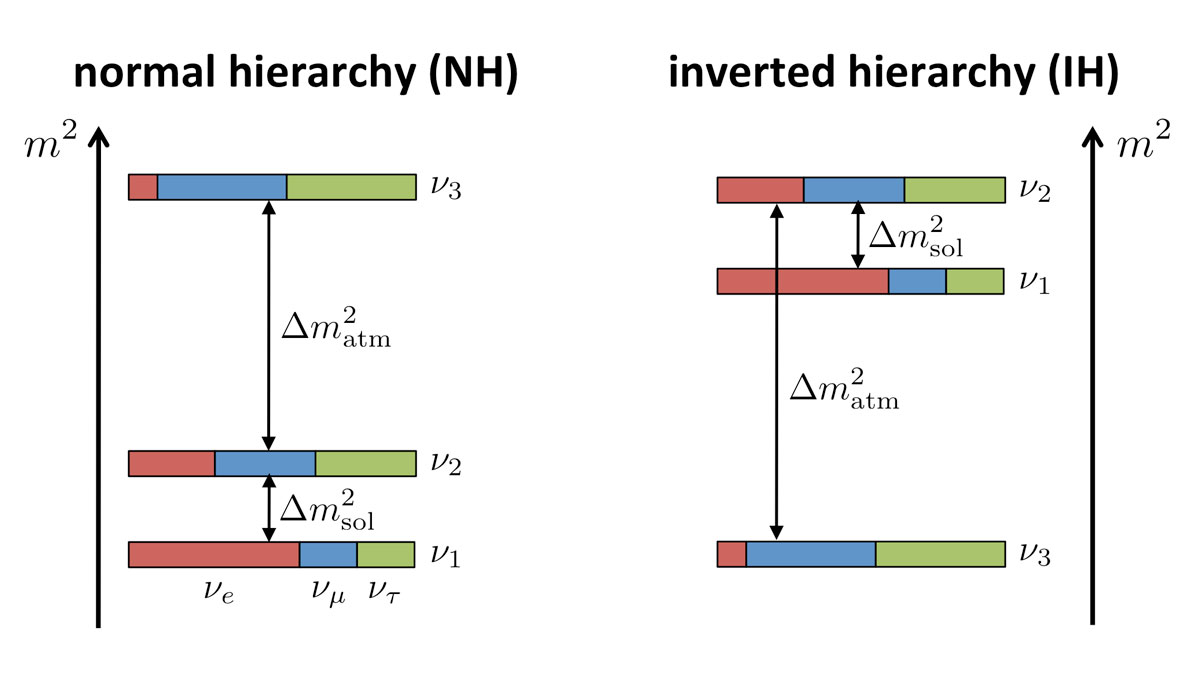}
\caption{Schematic representation of the ordering of the neutrino mass eigenstates, both for normal and inverted hierarchies~\cite{webjuno}.}
\label{fig:mass_orderings}
\end{figure}

Some questions remain unanswered, though. One of them is the mass hierarchy of neutrinos. The physical eigenstates $\nu_1$, $\nu_2$ and $\nu_3$ are defined according to their component of $\nu_e$, and not according to their masses. Thus, from a theoretical point of view, there is no rationale to determine which of these states are the heaviest. Oscillation data have shown that $\Delta m_{21}^2$ is positive and, thus, that $\nu_2$ is heavier than $\nu_1$. However, the sign of the atmospheric mass splitting is still unknown, so it is unclear whether $\nu_3$ is the heaviest or the lightest mass eigenstate. These two scenarios are dubbed normal (NO) and inverted ordering (IO), respectively. In the former, the largest mass splitting would be $\Delta m_{31}^2$ and would take positive values, whereas for inverted ordering, it would correspond to $\Delta m_{32}^2$, being negative. Both hierarchies are schematically depicted in Fig.~\ref{fig:mass_orderings}.

The other main puzzle in the oscillation paradigm concerns the CP-violating phase. Current experiments have not been precise enough to determine conclusively the value of this parameter, being unable to ascertain whether there is CP violation in the lepton sector or not. In fact, there is some tension between the NO$\nu$A~\cite{NOvA:2019cyt} and T2K~\cite{T2K:2019bcf} experiments, whose best fit points are opposite. However, the uncertainties in their measurements are still too large for this discrepancy to be too significant. 

The next generation of oscillation experiments, led by DUNE~\cite{DUNE:2015lol} and T2HK~\cite{Hyper-KamiokandeProto-:2015xww}, is being designed to determine how neutrinos are ordered in terms of masses, and whether CP is a good symmetry in the lepton sector.


\chapter{The absolute scale of neutrino masses}
\fancyhead[RO]{\scshape \color{lightgray}4. The absolute scale of neutrino masses}
As shown in the previous chapter, neutrino oscillations are determined by the (squared) mass differences between the neutrino mass eigenstates. However, they are not sensitive to the actual value of the masses of the neutrinos. Thus, some other input is needed in order to obtain information about the absolute mass scale of neutrinos.

A first insight could be gained from the theoretical perspective. The measured mass splittings are quite small (at the level of tens of meV), in fact extremely small when compared to the electroweak scale. Thus, naturalness arguments would point to neutrinos being very light particles. If that was not the case, and their absolute masses were much larger than the mass differences, neutrinos would be quasi-degenerate, pointing to some fine-tuning in the parameters controlling their masses. This reasoning is purely theoretical, and somehow biased by the expectation of naturalness, so it does not provide any definite proof of neutrinos being extremely light.

So far, experimental observations have been unable to measure the absolute masses of neutrinos. Only upper bounds are available, lying below the eV. Thus, these limits constrain neutrinos to be the lightest fermions known up to date, confirming what could be expected from a naturalness point of view. 

The strongest limits on neutrino masses arise from cosmology. Neutrinos are very abundant in the early Universe, playing a relevant role in the formation of large structures. The exact way in which this occurs depends on their massive or massless nature~\cite{Lesgourgues:2006nd}. So far, the combined astrophysical and cosmological observations are compatible with massless neutrinos, and only an upper bound on their mass is available. Note that all mass eigenstates play an equivalent role in structure formation, so the limits apply on the sum of the masses of the three neutrinos. The current constraint given by Planck data is
\begin{equation}
    \sum_im_{\nu_i}<0.12\rm\, eV
    \label{eq:cosmo_bound}
\end{equation}
at a 95\% confidence level (CL)~\cite{Planck:2018vyg}. 

This way of constraining neutrino masses strongly depends on the astrophysical and cosmological observables taken into account and, more importantly, on the assumed cosmological model. The bound shown above is computed assuming a standard $\Lambda$CDM cosmology, a framework that has proven very successful and is widely accepted. However, any deviations from the assumptions of $\Lambda$CDM, which are extrapolated through many orders of magnitude, could affect its predictions and, thus, its constraint on neutrino masses.

A more direct way to probe them is based on the kinematics of $\beta$-decay. In this process, in which a neutron decays into a proton, an electron and an electron antineutrino, the distribution of events with respect to the electron energy is affected by the value of the neutrino mass. The endpoint of this spectrum (the maximum energy the electron can carry) is given by the mass difference between the initial and final states; if neutrinos are massless, it is determined by the masses of the electron and of the isotopes involved in the $\beta$-decay. A modification in this endpoint would constitute an evidence of the massive nature of neutrinos, as the maximum energy the electron could carry would be smaller. 

No positive signal has been found so far. The strongest limit have been obtained by the KATRIN experiment, which analyzes the $\beta$-decay of tritium, and reads
\begin{equation}
    m_{\nu_e}=\sqrt{\sum_{i=1}^3\vert U_{ei}\vert^2m_{\nu_i}^2}<0.8\,\rm eV
\end{equation}
at a 90\% CL~\cite{KATRIN:2021uub}. As $\beta$-decays only produce electron neutrinos, KATRIN is only sensitive to the quantity $m_{\nu_e}$. This is the effective mass of the electron neutrino, and does not correspond to the physical mass of any of the neutrino mass eigenstates. However, it provides an insight on the absolute scale of neutrinos,  confirming that they lie below the eV scale.


Note that all experimental data is still compatible with a scenario in which the lightest neutrino ($\nu_1$ in NO and $\nu_3$ in IO) is exactly massless. Oscillations only provide the values of the two mass splittings, which could directly be the masses of the heavier neutrinos if the lightest one is massless. On the other hand, the upper limits from cosmology and KATRIN could be perfectly satisfied with only two massive states.

\chapter{Heavy neutrinos and generation of neutrino masses}
\fancyhead[RO]{\scshape \color{lightgray}5. Heavy neutrinos and generation of neutrino masses}
Neutrino oscillations clearly make manifest that the electron, muon and tau neutrino are not mass eigenstates. This is a blatant contradiction with the predictions of the SM, in which these particles have a well-defined, exactly vanishing mass. Thus, neutrino masses demand for an extension of the SM.

The simplest completion of the SM consists in the addition of RH neutrinos, which will be dubbed $N_R$ in the following. The rest of the fermions have both a LH and a RH counterpart, which couple to the Higgs doublet via Yukawa interactions, giving rise to a mass after EWSB. As the massless nature of neutrinos in the SM is due to the absence of their RH fields, it seems natural to add these states to the SM particle content, making neutrinos no different from the rest of fermions. This way, the Yukawa Lagrangian of the SM (Eq.~\ref{eq:yuk_lag}) would be extended with a Yukawa coupling for neutrinos,
\begin{equation}
    \mathcal{L}_{\rm Yuk}^\nu = -y_\nu\overline{L_L}\widetilde{H}N_R+\rm h.c.\,
\end{equation}
Once the Higgs field takes a vev, this term would give rise to a neutrino mass in the same fashion as for the other fermions, $m_\nu=y_\nu v/\sqrt{2}$. Note that this simplified Lagrangian only accounts for one generation of leptons.

This minimal setup has two important drawbacks. The first one is due to the extremely small mass of neutrinos. In order to generate a neutrino mass of 1 eV (which is even a bit too large), their Yukawa coupling should be of order $10^{-13}$, which is quite unnatural. Although the Yukawas of the SM already lie across several orders of magnitude, the lightness of neutrinos would worsen this problem considerably. However, once again, there is some theoretical bias in this rationale, as such a small Yukawa would not directly point to any failure of the theory.

The second issue is due to the unique properties of RH neutrinos, and their possible Majorana nature. Unless explicitly forbidden, these hypothetical particles could exhibit a Majorana mass term of the form~\cite{Majorana:1937vz}
\begin{equation}
    \mathcal{L}^\nu\supset -\frac{1}{2}\overline{N_R^c}M_MN_R+\rm h.c.\,,
\end{equation}
where $M_M$ is the Majorana mass and $N^c_R$ is the charge conjugate of the RH neutrino field.

Majorana mass terms are quite different to the usual, Dirac ones. Due to the transformation properties of the conjugated fermion fields, these terms explicitly break any symmetry that affects the corresponding particle. Thus, Majorana mass terms are forbidden for any SM fermion; as they are all charged under at least one gauge group, any such term would violate gauge invariance. In contrast, RH neutrinos are singlets of the SM gauge group. Being leptons, they have no color, and they have no weak isospin either, as they are right-handed. Taking into account that they are electrically neutral, their hypercharge also vanishes. This means that a Majorana mass for RH neutrinos is perfectly allowed, as it would not break any fundamental symmetry of the SM. 

The minimal extension of the SM described above, in which only the Yukawa term for neutrinos is added, would require to set the Majorana mass to zero by hand. Doing that without incurring in a fine-tuning problem would require a symmetry argument. The simplest option would be $B-L$, which, despite being accidental, is an exact symmetry of the SM. As a Majorana mass explicitly breaks $B-L$, promoting this symmetry to a fundamental one would be an argument to forbid such a term. Thus, if no extra symmetries are invoked, the simplest extension of the SM that accounts for neutrino masses includes both Dirac and Majorana terms: we give you the illustrious seesaw mechanism.

\section{The type I seesaw mechanism}
\label{sec:seesaw}
Considering a single leptonic family, the most general Lagrangian that describes neutrinos reads
\begin{equation}
    \mathcal{L}_{\rm seesaw}=-y_\nu\overline{L_L}\widetilde{H}N_R-\frac{1}{2}\overline{N_R^c}M_MN_R+\rm h.c.
    \label{eq:seesaw_lag}
\end{equation}
Once electroweak symmetry is broken, this mass Lagrangian can be compacted as
\begin{equation}
    \mathcal{L}_{\rm mass}=-\frac{1}{2}\overline{N_L^c}\mathcal{M}N_L+\rm h.c.\,,
\end{equation}
where $N_L= \left(\nu_L\,,N_R^c\right)^T$ is a vector containing all the neutrino fields, and the Majorana mass matrix reads
\begin{align}
    \mathcal{M}=\begin{pmatrix}
        0&m_D\\
        m_D&M_M
    \end{pmatrix}\,,
\end{align}
where $m_D=y_\nu v/\sqrt{2}$ is the Dirac mass generated via the Higgs mechanism. 

The fact that this mass matrix is not diagonal shows that neither the LH nor the RH neutrino fields are mass eigenstates. It is necessary to rotate to the mass basis, diagonalizing the mass matrix, to find the physical states:
\begin{equation}
    \mathcal{M}_{\rm diag}=\mathcal{R}^{T}\mathcal{M}R\,,
\end{equation}
where 
\begin{equation}
    \mathcal{M}_{\rm diag}=\begin{pmatrix}
        m_\nu&0\\0&M_N
    \end{pmatrix}\,,
\end{equation}
and 
\begin{equation}
    \mathcal{R}=\begin{pmatrix}
        \cos{\theta}&-\sin{\theta}\\\sin{\theta}&\cos{\theta}
\end{pmatrix}\,,
\label{eq:rot_matrix}
\end{equation}
with $\theta$ being the angle that controls the rotation.
 
Under the so-called seesaw approximation, that assumes that the Majorana mass is much larger than the Dirac one, $m_D\ll M_M$, the eigenvalues of the mass matrix, which represent the masses of the two physical states, read
\begin{align}
    m_\nu&=-\frac{m_D^2}{M_M}+\mathcal{O}\left(\frac{m_D^4}{M_M^3}\right)\,,
    \\
    M_N&=M_M+\mathcal{O}\left(\frac{m_D^2}{M_M}\right)\,.
\end{align}
There is clearly a hierarchy in the masses of the two states, with $m_\nu\ll M_N$. In fact, the larger the Majorana mass (and thus the heavier the second state), the lighter the first state: hence the naming of this mechanism. 

The physical states can be written in terms of the interaction basis by means of the rotation angle $\theta$:
\begin{align}
    \nu=\cos{\theta}\nu_L-\sin{\theta}N_R^c\,,
    \\
    N=\sin{\theta}\nu_L+\cos{\theta}N_R^c\,,
\end{align}
where the mixing angle, expected to be small if $m_D\ll M_M$, is given by
\begin{equation}
    \theta=\frac{m_D}{M_M}+\mathcal{O}\left(\frac{m_D^2}{M_M^2}\right)\simeq\sqrt{\frac{m_\nu}{M_N}}\,.    
\end{equation}
The spectrum thus contains a light neutrino, basically composed of the LH state (usually dubbed \say{active}), and a heavy state, mostly RH (commonly referred to as \say{sterile}, as they do not partake in the gauge interactions). When it comes to interactions, light neutrinos inherit practically the same terms as the LH states, whereas the heavy neutrinos hardly interact. They are only able to do so via insertions of the small mixing; although not completely sterile, their interactions are \say{weaker-than-weak}. 

The approximation $m_D\ll M_M$ is well motivated by the observation of such light neutrino masses. If these two quantities were of the same order, there would be no large scale suppressing the light masses, and a strong fine-tuning would be required to satisfy the experimental data (similarly to the case with only a Dirac mass). Besides, in this case there would be a large mixing between light and RH states, which would directly affect how weak interactions of light neutrinos behave. As no significant deviations from the SM predictions have been observed, such a scenario is disfavoured by most experiments. 

This mechanism is motivated by the fact that a very large Majorana mass generates very small light neutrino masses in a natural way. In fact, for a Majorana mass of order $10^{15}$ GeV, the observed masses could be explained with a Yukawa of order 1. This possibility is intriguing, especially because such a large Majorana mass could lie close to the scale of Grand Unification Theories (GUTs)~\cite{Georgi:1974sy,Pati:1974yy}. However, it would lead to untestable experimental signals, as the heavy neutrinos would be completely beyond the energy reach of any machine. The measured light masses could still be reproduced with a Majorana mass at the TeV scale and a Yukawa of order $10^{-6}$, which is somewhat fine-tuned, although at the same level as the electron Yukawa. There is clearly a balance between how natural the theory is allowed to be and how large the Majorana mass scale is, with no reason a priori to favour any specific regime. In fact, the Majorana mass would be the first energy scale to appear in particle physics not related to the electroweak one, and thus there is no intuition concerning its value.

This simplified scenario only includes one light neutrino; a more realistic one needs to describe the three of them. Assuming that only one RH neutrino is added to the SM field content,
\begin{equation}
        \mathcal{L}_{\rm seesaw}=-\sum_\alpha y_{\nu_\alpha}\overline{L_L}_{\alpha}\widetilde{H}N_R-\frac{1}{2}\overline{N_R^c}M_MN_R+\rm h.c.\,,
\end{equation}
and the mass matrix now reads
\begin{equation}
    \mathcal{M}=\begin{pmatrix}
        &&&y_{\nu_e}\\
        &\Large{\mathbf{0_{3\times 3}}}&&y_{\nu_\mu}\\
        &&&y_{\nu_\tau}\\
        y_{\nu_e}&y_{\nu_\mu}&y_{\nu_\tau}&M_M
    \end{pmatrix}\,.
\end{equation}
The rank of this matrix is 2, so it is clear that only two physical states will be massive: the very large, almost sterile, heavy neutrino, and a light one, mostly active. The other two eigenvalues of the matrix vanish, showing that the addition of only one RH field is not enough to explain the experimental data. It is thus necessary to include at least a second RH neutrino, enlarging the mass matrix.

Similarly, the matrix that relates the interaction and mass bases (a generalization of Eq.~\ref{eq:rot_matrix}) will now have a dimension $(3+n)\times(3+n)$, where $n$ is the number of RH neutrinos. The upper-left $3\times3$ block of that matrix will correspond to the PMNS matrix, which relates the LH interaction eigenstates ($\nu_e$, $\nu_\mu$ and $\nu_\tau$) with the light mass states ($\nu_1$, $\nu_2$ and $\nu_3$). The off-diagonal elements will correspond to the overlap of the RH and LH states, and they are expected to be small, following the seesaw approximation. The fact that the mixing matrix is enlarged has the important physical consequence of the $3\times3$ PMNS matrix not being unitary anymore; only the whole $(3+n)\times(3+n)$ matrix is. This could cause a loss of probability in certain processes; as such a phenomenon has not been observed so far, this argument allows to set constraints on the mixing between active and heavy neutrinos~\cite{Petcov:1976ff,Bilenky:1977du,Cheng:1977vn,Marciano:1977wx,Lee:1977qz,Lee:1977tib,Shrock:1980vy,Shrock:1980ct,Shrock:1981wq,Schechter:1981bd,Langacker:1988ur,Blennow:2023mqx}. 

Aside from the extremely large scale it involves (or, in turn, the fine-tuning it introduces), the seesaw mechanism has the serious problem of worsening the Higgs hierarchy problem~\cite{Vissani:1997ys,Casas:2004gh}. The addition of RH neutrinos, which would couple to the Higgs boson, would imply that the mass of this scalar would get new quantum corrections, proportional to the mass of the heavy states. If these lay at, say, the GUT scale, the fine-tuning needed to obtain such a low physical mass for the Higgs boson would be huge, on top of that already needed due to the SM contributions. 

\section{Majorana neutrinos, lepton number and low-scale seesaws}
\label{sec:LN_LSS}
The possibility of neutrinos being Majorana particles constitutes one of the main
unknowns in particle physics, that is inherently linked to the mechanism behind neutrino masses. For instance, the seesaw mechanism predicts the neutrino mass eigenstates to be Majorana particles, but other setups grant them a Dirac nature, as is the case for the rest of the SM fermions. Whether neutrinos are Dirac or Majorana particles has very relevant implications.

One of them regards the PMNS matrix. When parametrizing a general matrix that rotates 3 Dirac fermions, 5 out of the 6 complex phases can be rotated away by redefinitions of the fields. The remaining one is the physical CP phase in Eq.~\ref{eq:pmns}, analogous to the one of the CKM. However, if neutrinos are Majorana fermions, not all those phases can be reabsorbed in the fields, due to the presence of the Majorana mass. This term is not invariant under certain field redefinitions, rendering three complex phases as physical parameters in the PMNS matrix. These two extra parameters are introduced as
\begin{equation}
       U_{\rm PMNS}= \mathcal{R}(\theta_{23})\mathcal{R}(\theta_{13},\delta)\mathcal{R}(\theta_{12})P_M\,,
\label{eq:pmns_majorana}
\end{equation}
where 
\begin{align}
    P_M=\begin{pmatrix}
        e^{i\alpha_1}&0&0\\
        0&e^{i\alpha_2}&0\\
        0&0&1
    \end{pmatrix}\,.
\end{align}
$\alpha_1$ and $\alpha_2$ are the so-called Majorana phases, as they only appear if neutrinos are Majorana particles. Oscillations are unable to provide information on these parameters, as they cancel in the oscillation probability expression (Eq.~\ref{eq:osc_prob}). Thus, rarer processes are required to determine the Majorana phases. 

It has been argued above that a Majorana mass for RH neutrinos is allowed because they are not charged under any symmetry. This is true for the gauge symmetries of the SM, and thus their Majorana mass does not violate gauge invariance. However, neutrinos have a lepton number, so they are affected by $U(1)_{\rm L}$, which is a global, anomalous and accidental symmetry of the SM. As lepton number is not imposed while building the SM, it does not forbid a Majorana mass for neutrinos, which in fact breaks explicitly this symmetry and gives rise to a violation of lepton number by two units, $\Delta L=2$. 

Processes that involve a violation of lepton number would constitute a smoking gun of the Majorana nature of neutrinos. They are also the only ones that could be potentially sensitive to Majorana phases; as charge conjugate fields are involved, the corresponding rates depend on squared elements of the PMNS (instead of their squared moduli), thus depending on the complex phases. However, these are extremely rare phenomena. According to the confusion theorem, Dirac and Majorana neutrinos cannot be distinguished in the massless limit. Hence, it is clear that the rate of any lepton number violating process must be proportional to neutrino masses, and thus very suppressed with respect to the energy scale of any possible experiment. 

The huge suppression that light neutrino masses imply hinders the detection of any such scenario, so no evidence for Majorana neutrinos has been found so far. Thus, many experimental efforts are still ongoing in the search for lepton number violation. Chief among them are neutrinoless double-beta ($0\nu\beta\beta$) experiments.

\begin{figure}[t!]
\centering
\includegraphics[width=0.7\textwidth]{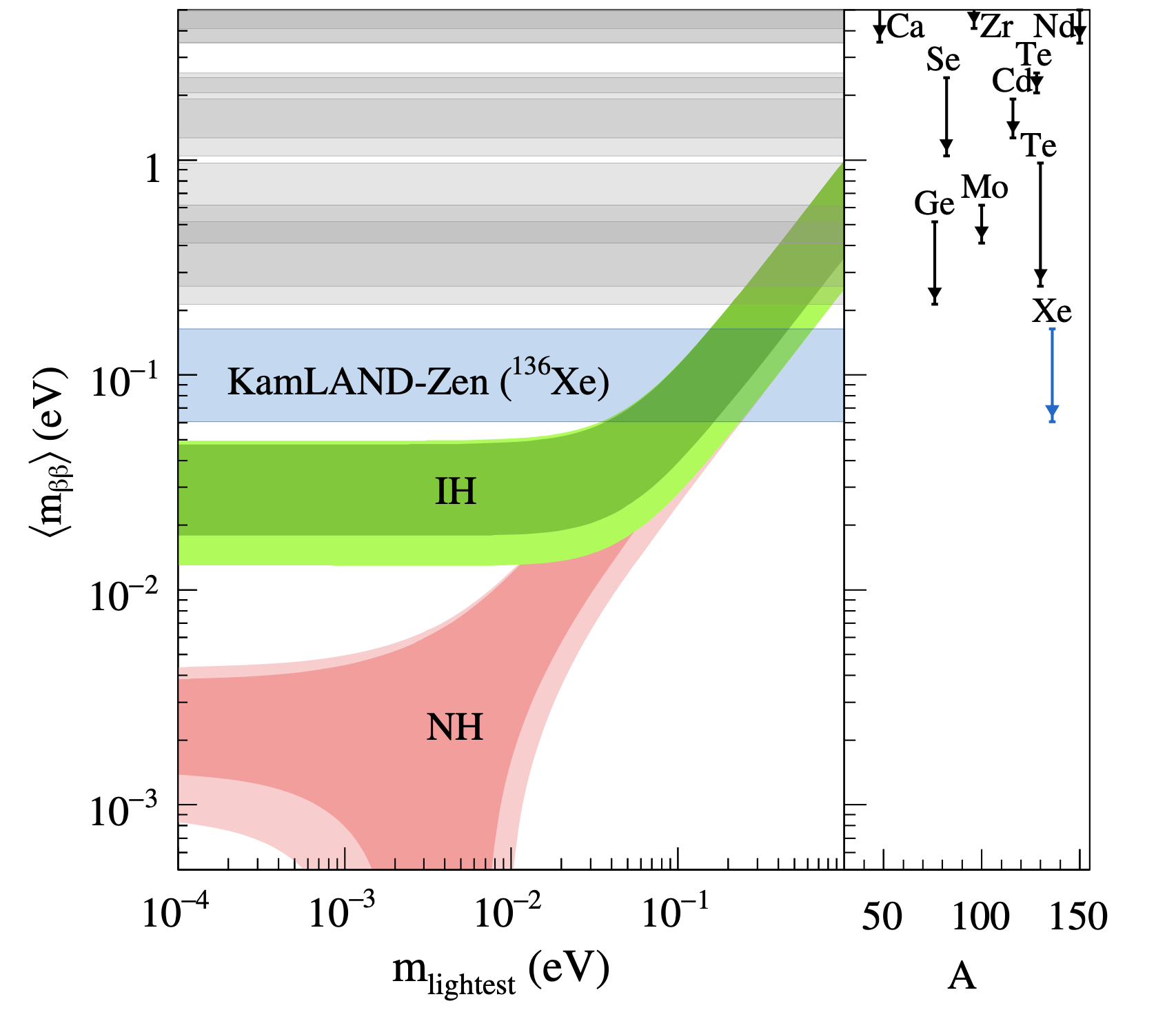}
\caption{Effective Majorana neutrino mass as a function of the lightest neutrino mass, both for normal and inverted orderings~\cite{KamLAND-Zen:2016pfg}. The most stringent limits are displayed, as well as the isotopes employed to obtain them (right panel).}
\label{fig:neutrinoless}
\end{figure}

Double beta decays are processes in which two neutrons decay into two protons, two electrons and two electron antineutrinos. These are rare phenomena, which have only been observed in a few isotopes, which endure the process $(A,Z)$ $\to (A,Z+2)+2e^-+2\overline{\nu_e}$. If neutrinos were Majorana particles, the external legs associated to neutrinos could become a single internal line, yielding a process with only electrons in the final state, $(A,Z)\to (A,Z+2)+2e^-$. In contrast to the usual double beta decay, its neutrinoless version would exhibit a monochromatic spectrum with respect to the energy of the electron pair. Such a novel experimental signature has not been found so far, with upper limits on the corresponding rate currently at the order of $10^{26}$ years~\cite{KamLAND-Zen:2022tow,GERDA:2020xhi}. However, this is not enough to rule out the Majorana nature of neutrinos; in fact, if neutrinos are normally ordered, $0\nu\beta\beta$ decays could still be suppressed. The rate of this process is controlled by the so-called effective Majorana mass, given by
\begin{equation}
    m_{\beta\beta}=\left|\sum_iU_{ei}^2m_{\nu_i}\right|\,.
\end{equation}
This parameter is sensitive to cancellations between different entries of the PMNS, which, in normal ordering, may lead to extremely small (or even vanishing) rates even if neutrinos were Majorana. This can be seen in Fig.~\ref{fig:neutrinoless}, which shows the dependence of $m_{\beta\beta}$ with the lightest neutrino mass.

Other searches for lepton number violation have been performed, for instance, at colliders~\cite{Antusch:2017ebe,Drewes:2019byd}. Final states composed of same-sign leptons would directly point to a different number of leptons in initial and final states. These searches are usually motivated by the hypothetical existence of heavy neutrinos, but are yet to produce positive results.

Despite being still unsupported by experimental data, lepton number violation is appealing for several reasons. One of them is rather theoretical. If one tried to build a Majorana mass term for the SM neutrinos without introducing new particles, the only possibility would be the celebrated Weinberg operator, that appears at dimension 5:
\begin{equation}
    \mathcal{O}_{ d=5}=c^{\alpha\beta}_{ d=5}\left(\overline{L_L^c}_\alpha\widetilde{H}^*\right)\left(\widetilde{H}^\dagger L_{L\beta}\right)\,.
\end{equation}
This term obviously breaks lepton number explicitly, by two units, and, upon EWSB, the desired Majorana mass would appear. It has the drawback of being non-renormalizable, as it has an energy dimension of 5. Intriguingly, this is the only dimension-5 operator that can be built out of the SM particle content and respecting its gauge symmetries. It also has the interesting feature that it appears naturally from the seesaw mechanism. If the RH neutrinos are very heavy, they do not play any relevant dynamical role, and can be safely integrated out. Such a procedure yields the Weinberg operator for light neutrinos. In a simplified scenario with one light and one heavy state (Eq.~\ref{eq:seesaw_lag}), its coefficient is given by~\cite{Broncano:2002rw}
\begin{equation}
    c_{ d=5}^{\alpha\beta}=\frac{y_{\nu_\alpha}^*y_{\nu_\beta}^*}{M_M}\,.
    \label{eq:d5_coeff}
\end{equation}
Note that integrating out the heavy degrees of freedom also yields a dimension-6 operator, which is responsible for the non-unitarity of the PMNS matrix:
\begin{equation}
    \mathcal{O}_{ d=6}=c_{ d=6}^{\alpha\beta}\left(\overline{L_L^c}_\alpha\widetilde{H}\right)i\slashed{\partial}\left(\widetilde{H}^\dagger L_{L\beta}\right)\,,
\end{equation}
with
\begin{equation}
    c_{ d=6}^{\alpha\beta}=\frac{y_{\nu_\alpha}y_{\nu_\beta}^*}{M_M^2}\,.
\end{equation}
As it could be expected, the Weinberg operator inherits the suppression given by the large Majorana mass, equivalent to the suppression of light neutrino masses that arises in the seesaw mechanism. Note that the operator discussed above only includes one family of leptons, but it can be trivally generalized, promoting Eq.~\ref{eq:d5_coeff} to a matrix relation. Variations of the seesaw mechanism, that will be discussed in the following, lead to very similar results, only differing in the particular expression of the coefficient of the dimension-5 operator. It is clear that the Weinberg operator, Majorana masses, lepton number violation and the seesaw mechanism are deeply linked, and constitute a robust framework.

Also, lepton number violation could be the explanation to the baryon asymmetry of the Universe. If an excess of antileptons was produced in the early Universe, sphaleron transitions~\cite{Kuzmin:1985mm} could in turn generate an excess of baryons over antibaryons. These non-perturbative processes, predicted by the SM, respect $B-L$, so they could be able to generate a baryon-antibaryon asymmetry from a violation of lepton number. This mechanism, known as leptogenesis~\cite{Fukugita:1986hr}, is still a possible answer to this problem, and relies on neutrinos being Majorana fermions.

However, there are symmetry arguments also in favour of neutrinos being Dirac particles. In fact, from the naturalness point of view, it would make sense for Majorana masses to be small, as the Lagrangian would gain an extra symmetry (lepton number) if this parameter vanished. This reasoning could explain why neutrino masses are so small, as their mass would be protected by lepton number~\cite{Branco:1988ex,Kersten:2007vk,Abada:2007ux}, which could be considered an approximate global symmetry.

This is the point of view of the so-called low-scale seesaw mechanisms, such as the inverse seesaw (ISS)~\cite{Mohapatra:1986aw,Mohapatra:1986bd,Bernabeu:1987gr}. This setup extends the SM particle content with two different types of RH neutrinos, usually denoted as $N_R$ and $S_R$, which are of course singlets of the SM gauge group. Assuming a single lepton flavour and only one species of each type of RH neutrinos, the Lagrangian of this model reads
\begin{equation}
    \mathcal{L}_{\rm ISS}=-y\overline{L_L}\widetilde{H}N_R-\overline{S_R^c}MN_R-\frac{1}{2}\overline{S_R^c}\mu S_R+\rm h.c. 
    \label{eq:ISS_Lag}
\end{equation}
In the basis $(\nu_L, N_R^c,S_R^c)$, the Majorana mass matrix takes the form
\begin{equation}
    \mathcal{M}=\begin{pmatrix}
        0&m_D&0\\
        m_D&0&M\\
        0&M&\mu
    \end{pmatrix}\,,
\end{equation}
with $m_D=yv/\sqrt{2}$. If the $N_R$ field is assigned a lepton number of 1, and $S_R$ a lepton number of -1, it is clear that the Lagrangian in Eq.~\ref{eq:ISS_Lag} conserves lepton number, except for the Majorana mass term of $S_R$. This can be considered a soft breaking, controlled by $\mu$. Thus, if lepton number is an approximate symmetry, this parameter is expected to be small, and may give rise naturally to the measured light neutrino masses. In fact, the usual seesaw approximation, $m_D\ll M$, is generalized by assuming that $\mu\ll m_D\ll M$. Diagonalizing this matrix yields a mass for light neutrinos given by
\begin{equation}
    m_\nu\simeq\mu\frac{m_D^2}{M^2}\,,
\end{equation}
and a light-active mixing $\theta\simeq m_D/M$. The two heavy states will be almost degenerate, with a mass approximately equal to $M$, and a mass splitting given by $\mu$. Light neutrino masses are now directly proportional to the Majorana mass, instead of inversely proportional; hence the name of this new model. 

Similar setups, such as the linear seesaw~\cite{Malinsky:2005bi}, exhibit very similar features, although with different textures for the mass matrix. Their common characteristic is that light neutrino masses are protected by lepton number, being naturally small without the need to introduce a new, large energy scale; thus the notion of low-scale seesaws. 

These models may give rise to a rich phenomenology, much more testable than the predictions of the original seesaws. The suppression given by lepton number allows the new states to be somewhat light, lying even close to the electroweak scale.  In fact, for Yukawas of order 1, and $\mu\sim\mathcal{O}$(keV), the observed light neutrino masses can be reproduced for $M\sim\mathcal{O}$(TeV), yielding a mixing of order $10^{-2}$. This means that the new states could be even reachable by current experiments, and their interactions would be controlled by a sizeable mixing. Note that these are just very crude estimations; besides, such a large mixing is already ruled out by indirect bounds due to electroweak and flavour precision observables that constrain the non-unitarity of the PMNS matrix~\cite{Blennow:2023mqx}. 

It is clear that low-scale seesaw mechanisms are well motivated, as the argument of small neutrino masses protected by lepton number symmetry is quite appealing. The fact that the new states are not extremely heavy is also satisfactory from a theoretical point of view, as it would not worsen the Higgs hierarchy problem: the existence of new particles close to the electroweak scale would not require an extra fine-tuning to explain the Higgs boson mass. Besides, these setups offer much better prospects to be experimentally probed than the original seesaw, being also interesting from the phenomenological point of view.

\section{Other seesaws}
The setup we have so far named seesaw mechanism is actually the \say{canonical type I} seesaw, as it was the first to be proposed and is arguably the simplest way to generate neutrino masses. Its low-scale versions are just particular cases of this generic setup, in which a certain texture of the mass matrix is chosen. However, instead of adding fermionic singlets (RH neutrinos) to the SM particle content, other new states can play a similar role; the induced neutrino masses will still be suppressed by the masses of these new particles.

The type II seesaw~\cite{Magg:1980ut,Schechter:1980gr,Wetterich:1981bx,Lazarides:1980nt,Mohapatra:1980yp} introduces a new scalar, $\Delta$, which transforms as a triplet under $SU(2)_L$, thus containing three degrees of freedom. Apart from its associated potential, this new field couples both to the Higgs and to the lepton doublets of the SM:
\begin{equation}
    \mathcal{L}_{\rm seesaw}^{\rm II}= -y_\Delta\left( \overline{L_L^c}\tau^aL_L\right)\Delta_a-\mu\left(H^\dagger\tau^aH\right)\Delta_a\,.
\end{equation}
The new scalar will take a non-zero vev, generating a Majorana mass for neutrinos. As the vev is also directly related to the mass of the scalar, $M_\Delta$, the light neutrino masses will end up being suppressed by this new scale:
\begin{equation}
    m_\nu\simeq\frac{y_\Delta\mu v^2}{2M_\Delta^2}\,.
\end{equation}
Although the pattern of generation of light neutrino masses seems similar to the type I, an extra suppression is controlled by the parameter $\mu$. If this quantity is small, very light neutrino masses can be achieved without the need for a very heavy new particle. This philoshophy is related to the protection of neutrino masses by lepton number, in a similar spirit to low-scale seesaws. 

Neutrino masses could also be generated by introducing a fermionic triplet, $\Sigma$, as is the case of the type III seesaw~\cite{Foot:1988aq}:
\begin{equation}
    \mathcal{L}_{\rm seesaw}^{\rm III}=-\frac{1}{2}\overline{\Sigma_a^c}M_\Sigma{\Sigma^a}-y_\Sigma\overline{\Sigma_{a}}\widetilde{H}\tau^aL_L+\rm h.c.
\end{equation}
After EWSB, Dirac masses for neutrinos will be generated. The neutral component of the new triplet will play a very similar role to that of the RH neutrinos in the type I, mixing with active neutrinos and acquiring a Majorana mass that will suppress light neutrino masses:
\begin{equation}
    m_\nu\simeq\frac{\vert y_\Sigma\vert^2v^2}{2M_\Sigma}\,.
\end{equation}
As a unique feature of the type III seesaw, the charged components of $\Sigma$ will combine into Dirac fermions, that will mix with the SM charged leptons.

\part[Heavy neutrinos as solutions to open problems]
     {\cinzel The chosen ones\\[\bigskipamount] 
      \normalfont\Large Heavy neutrinos as solutions to open problems}
\fancyhead[LE]{\scshape \color{lightgray}II. Heavy neutrinos as solutions to open problems}
\fancyhead[RO]{}
\PB
\lettrine[depth=1, loversize=0.55,lraise=-0.4]
    {A}{}huge number of models have been proposed in order to account for light neutrino masses and mixings, reconciling the SM with the observation of neutrino oscillations. These Beyond the Standard Model (BSM) scenarios usually introduce new particles or fundamental interactions, generating neutrino masses through a variety of mechanisms. Arguably the simplest of them consists in introducing the RH counterpart of the SM neutrino fields, yielding the presence of heavy neutrinos in the spectrum, as discussed previously.

Although motivated by neutrino masses, most of these setups offer solutions to several open problems of the SM. This has been the case since the origin of neutrino mass models, as the canonical type I seesaw  contains the necessary ingredients to generate the baryon asymmetry of the Universe via leptogenesis. 

Also regarding the structure of the Universe, a great amount of neutrino mass models provide suitable candidates for dark matter. It is common for these scenarios to include new fundamental forces, usually dubbed \say{dark} or \say{secret}, which hardly affect the SM particles. New particles could partake in these interactions, potentially constituting DM candidates. Interestingly, if they exhibit some particular characteristics, heavy neutrinos themselves could add up to a large fraction of the DM energy budget~\cite{Dodelson:1993je,Shi:1998km}.

It is clear that neutrinos play a relevant role in cosmology, so it is interesting to study whether they can pose solutions to some open problems in this field. One of the most striking among them is the so-called \say{Hubble tension}. This is an inconsistency in the different determinations of the Hubble constant, $H_0$, that controls the expansion rate of the Universe. This disagreement may well be due to systematics in the involved measurements or to a misunderstanding of the cosmological model, but could also be explained by BSM particle physics, such as new interactions that affected neutrinos.

On the other hand, the properties of neutrinos clearly make the flavour puzzle of the SM yet more confusing. The fact that neutrinos are at least a million times lighter than the electron implies that the fermion masses span an even wider range than in the SM. The observed anatomy of the PMNS matrix is also quite puzzling, as it is opposite to that of the CKM. While the latter exhibits a strong hierarchy, suppressing flavour change, the
PMNS is rather anarchic, with entries of the same order and large mixing angle. A complete theory of flavour should also contain an explanation to the lightness of neutrinos, as well as to their mixing pattern. Thus, it is appealing to explore setups which combine solutions to neutrino and flavour physics into a common framework.

Following up on fine-tuning issues, the strong CP problem has also led to the appearance of many BSM extensions, which aim at providing an explanation to the smallness (or even disappearance) of the QCD $\theta$-term (Eq.~\ref{eq:theta_term}). These scenarios also involve new symmetries and fields; although traditionally not that related to neutrino mass models, they may have some characteristics in common, possibly pointing to interesting links. 

In this episode we will explore some heavy neutrino models that may provide explanations to different open problems of the SM, other than the massive nature of neutrinos. We will first summarize briefly the status of these puzzles and some proposed solutions. Then, we will present a mechanism that links neutrino physics to a possible explanation of the Hubble tension via the introduction of a Majoron. Its phenomenological consequences and possible relation to the strong CP problem and flavour physics will be addressed. Finally, we will show a second model, which dives deeper into such a relation, providing a dynamically arising inverse seesaw setup. Its impact on experimental observables will also be discussed.

\chapter{Several open problems}
\fancyhead[LE]{\scshape \color{lightgray}II. Heavy neutrinos as solutions to open problems}
\fancyhead[RO]{\scshape \color{lightgray}6. Several open problems}
\section{The Hubble tension}
\label{sec:hub_tens}

The $\Lambda$CDM model defines the Hubble parameter as the logarithmic derivative of the scale factor, $a(t)$:
\begin{equation}
    H(t)\equiv\frac{\dot{a}(t)}{a(t)}\,,
\end{equation}
where $\dot{a}$ denotes the time derivative of the scale factor. By definition, this parameter quantifies the expansion rate of the Universe. The Hubble constant, $H_0$, is the value of the Hubble parameter at our current cosmological time, determining the expansion rate of the Universe that can be observed nowadays.

Measurements of the CMB allow to determine a value of the Hubble constant, provided a certain cosmological model. The Planck collaboration set its value to $H_0=67.4\pm0.5$ km s$^{-1}$Mpc$^{-1}$~\cite{Planck:2018vyg}, assuming $\Lambda$CDM. An alternative prediction of this model for the Hubble constant can be obtained by using the abundance of light elements to callibrate Baryon Acoustic Oscillations (BAO) and Big Bang Nucleosynthesis (BBN); this method provides very similar figures to those given by Planck. All these are the so-called \say{early-Universe} observations. 

An independent determination of the Hubble constant can be performed by means of local, or \say{late-Universe} measurements, arising, from instance, from type Ia supernovae or strong lensing. The former provide a result of $H_0=73.4\pm1.04$ km s$^{-1}$Mpc$^{-1}$, given by the SH0ES collaboration~\cite{Riess:2021jrx}. Other local observations have come up with results in a similar ballpark. 

Both sets of measurements exhibit a tension of roughly $5\sigma$. The exact quantification of the disagreement depends on the observables taken into account (see Refs.~\cite{Verde:2019ivm,Riess:2019qba} for some reviews). This discrepancy is statiscally robust, and is present in several independent, uncorrelated datasets. This so-called \say{Hubble tension} possibly points to a systematic bias in the local measurements (such as the callibration of the supernovae distances) or, interestingly, to a flaw in the cosmological model. A plethora of solutions have been proposed, such as an scalar field that acts as an early component of dark energy, decaying or interacting dark matter or a breakdown of general relativity (see Refs.~\cite{DiValentino:2021izs,Schoneberg:2021qvd} for some reviews). The most successful attempts modify the Hubble parameter in the early Universe, but are hard to reconcile with other CMB measurements. 

Some of the proposed solutions to the Hubble tension are rather motivated by particle physics (see Refs.~\cite{Archidiacono:2016kkh,Ko:2016uft,DiValentino:2017oaw,DEramo:2018vss,Agrawal:2019lmo,Agrawal:2019dlm,Alexander:2019rsc,Ghosh:2019tab,Escudero:2019gzq,Escudero:2019gvw,Gelmini:2019deq,Park:2019ibn,Kreisch:2019yzn,Smith:2019ihp} for an incomplete list), in particular, by neutrinos. The introduction of new neutrino species, likely partaking in new, BSM interactions, can potentially raise the $\Lambda$CDM predictions for $H_0$, narrowing the gap with late-Universe determinations. In particular, Ref.~\cite{Escudero:2019gvw} suggests that a Majoron that couples to light neutrinos could reduce the tension, provided this new particle exhibits some specific characteristics. This scenario will be further studied in the following.

\section{The flavour puzzle}

Arguably, one of the least appealing aspects of the SM is its inability to predict the masses and mixings of the fermions. Their Yukawa couplings, which eventually control the masses after EWSB, can only be fixed thanks to experimental data (either by measuring the masses themselves or the couplings of the fermions to the Higgs boson). 

The masses of the SM fermions have been measured to be wildly different, ranging from the 173 GeV of the top quark to the 0.511 MeV of the electron. As the vev of the Higgs sets an overall energy scale, this means that the Yukawa couplings range from $\mathcal{O}(1)$ to $\mathcal{O}(10^{-6})$. Although this wide range does not directly pose any problem for the theory, it is hardly natural; the fact that several dimensionless parameters that control the same type of physics take such different values constitutes a large fine-tuning problem. It is also confusing why the fermions are organized in three progressively heavier families. 

Similarly, the theory does not offer any explanation to the mixing pattern in the quark sector; the entries of the CKM matrix cannot be predicted from first principles, needing experimental input. Data has shown that the CKM is quite close to the identity: small mixing angles strongly suppress flavour-violating processes with respect to flavour-conserving ones, especially if they connect the first and third generations.

The observations of non-vanishing neutrino masses and mixings made the flavour puzzle even more complex. Experimental data showed that these particles are extremely light; although a measurement of the exact masses of neutrinos has still not occurred, current limits place their scale below the eV: the range of masses of the fermions now spans across twelve orders of magnitude. However, the way in which neutrino masses are generated clearly depends on the assumed BSM setup, as well as the amount of fine-tuning needed to reproduce such small masses. 

Oscillations have shown that the structure of the PMNS matrix contains no particular hierarchy among its entries. Flavour change in the lepton sector is not suppressed at all, and the corresponding mixing angles are large (in fact, $\theta_{23}$ is currently compatible with the maximal mixing value of 45$^\circ$). This pattern is strikingly opposite to that of the quarks, clearly lacking a rationale.

Let us stress again that the flavour puzzle does not constitute a contradiction between the predictions of the SM and experimental observations; it is rather a naturalness problem. Nevertheless, many efforts have been performed pursuing a theoretical explanation of the masses and mixings of the fermions, shedding some light on their apparently chaotic pattern. Although many of these proposals only focus on the quark sector, dealing with their masses and the CKM matrix, it is clear that a complete theory should also explain the smallness of neutrino masses and their anarchical mixing pattern, as they play a key role in the flavour puzzle. On the other hand, the most basic neutrino mass models do not offer insights on flavour, as they usually do not link the lepton and quark sectors. It is then necessary to develop more complex scenarios, which provide an explanation for the masses and mixings of all the fermions, including the SM ones and massive neutrinos. Thus, it is especially intriguing to explore whether solutions to both problems may stem from a common theory. 

The flavour puzzle is usually tackled by means of new symmetry groups. Most of these are \say{horizontal}, in the sense that they connect different flavour copies across the three generations, in contrast to the gauge symmetries of the SM, which rather link particles belonging to the same family (they are \say{vertical} in this sense). The philosophy of these mechanisms relies on explaining the masses and mixings of the fermions according to their corresponding transformation properties under the flavour symmetry, providing predictions for them from first principles, rather than leaving these quantities as free parameters to be fixed by experimental inputs.

There exists a huge variety of proposals in the literature. Although some of them promote these new \say{flavour symmetries} to gauge groups~\cite{Grinstein:2010ve,Feldmann:2010yp,Buras:2011wi,Buras:2011zb,Guadagnoli:2011id,Alonso:2016onw}, global symmetries are usually the chosen option. Many efforts have appeared employing discrete flavour symmetries (examples include Refs.~\cite{Ma:2001dn,Babu:2002dz,Altarelli:2005yp,Ishimori:2010au,Altarelli:2012ss,Hernandez:2012ra,Grimus:2011fk,King:2013eh}), which usually also contain explanations for the measured neutrino masses and mixings. This kind of approach was behind the so-called tri-bimaximal mixing patterns for neutrinos~\cite{Harrison:2002er,Xing:2002sw}, that predicted a maximal $\theta_{23}$ and a vanishing $\theta_{13}$. This scenario ended up being disfavoured by observations after reactor experiments measured a sizable $\theta_{13}$, which pushed the discrete symmetry approach to more complex scenarios.

On the other hand, the flavour symmetry can be associated to a continuous group. This point of view is motivated by the fact that the SM gauge interactions are flavour-blind, as they treat equally the three generations of fermions. These scenarios need to deal with the Yukawa interactions of fermions, as they explicitly break the flavour symmetry. A common way to do so is to promote the Yukawas to spurions, matrices that transform non-trivially under the flavour symmetry. These spurions can be thought of as vevs of scalars that are charged under the new symmetry, so they share their transformation properties, making the Yukawa terms now invariant under flavour. This is the rationale behind the Minimal Flavour Violation (MFV) prescription~\cite{Chivukula:1987py,DAmbrosio:2002vsn}. Other options promote the Yukawa interactions to effective, $d>4$ terms, adding extra fields to the coupling between fermions and the Higgs doublet. These new particles would also exhibit the right transformation properties to make the whole Yukawa Lagrangian invariant under the flavour symmetry. The BSM scenarios presented in this episode contain this kind of features, so let us describe them is some more detail.

\subsection{The Froggatt-Nielsen mechanism}
Froggatt-Nielsen (FN) models~\cite{Froggatt:1978nt,Buchmuller:2011tm,Altarelli:2012ia,Bergstrom:2014owa} constitute the simplest way to address the flavour puzzle by means of a continuous global symmetry, an Abelian $U(1)_{\rm FN}$ group. All the fermion fields are charged under this symmetry, with different assignments for the LH and RH components. The particle content is extended with the inclusion of a scalar, $\Phi$, usually named \say{flavon}. This particle is added to the Yukawa couplings of the fermions, which now schematically read
\begin{equation}
    \overline{\psi_L} H\psi_R\to \left(\frac{\Phi}{\Lambda_\Phi}\right)^{x_{\psi_R}-x_{\psi_L}}\overline{\psi_L} H\psi_R\,.
\label{eq:Frog-Niel}
\end{equation}
Here, $\Lambda_\Phi$ is the cut-off scale that characterizes the effective field theory, and $x_{\psi_{L(R)}}$ is the FN charge of $\psi_{L(R)}$. Note that the charge of $\Phi$ is fixed to -1, but this is somewhat arbitrary, as it only sets an overall scale for the rest of the FN charges. 

The FN symmetry will be spontaneously broken once the flavon takes a non-zero vev, $v_\Phi$. This means that the fermion masses will not only be controlled by their Yukawas and the vev of the Higgs, but also by the vev of the new scalar, weighted by the corresponding FN charges. Explicitly,
\begin{equation}
    m_\psi=\varepsilon^{x_{\psi_R}-x_{\psi_L}}\frac{y_\psi v}{\sqrt{2}}\,,
\end{equation}
where $\varepsilon=\frac{v_\Phi}{\sqrt{2}\Lambda_\Phi}$.

The interesting feature of this mechanism is the exponential dependence of the fermion masses on their FN charges. This makes it possible to span the huge range exhibited by the fermion masses with charge assignments that lie roughly in the same order of magnitude. In fact, there is enough room to fix the charges of the fermions and the $\varepsilon$ parameter to reproduce the SM spectrum. However, it is likely to end up in situations that could exhibit a larger amount of flavour violation than that of the SM, which is only given by the CKM matrix. For instance, the Goldstone boson associated to the breaking of FN symmetry could couple to quarks of different flavours. These processes are FCNCs, that induce flavour-violating meson decays, neutral meson-antimeson oscillation or similar phenomena. These transitions are present in the SM, but with rates very strongly suppressed due to the GIM mechanism~\cite{Glashow:1970gm}. Experiments have set extremely constraining limits on new sources of flavour violation, strongly disfavouring many BSM flavour scenarios. This issue is the motivation behind the MFV prescription, that will be described in the following.

\subsection{Minimal Flavour Violation}
Commonly, BSM setups that attempt to provide an explanation for the flavour puzzle induce new couplings involving the SM fermions. These new interactions may introduce new sources of flavour violation, different from that present in the SM. In the latter, the Yukawa interactions are the only source of flavour violation, which ends up appearing via the CKM matrix in charged currents that involve quarks. Any violation of flavour that does not follow this structure is strongly disfavoured by experiments; in particular, FCNCs are very strongly constrained, constituting the most stringent limits when it comes to building a flavour model. 

Minimal Flavour Violation is a prescription that aims at protecting any model from unwanted flavour violation, stating that it may only be due to the SM Yukawa couplings. If this principle is imposed, any BSM setup will inherit a suppression of flavour violation similar to the one present in the SM.

This framework is motivated by the flavour symmetry exhibited by the SM Lagrangian, save for the Yukawa terms,
\begin{equation}
    \mathcal{G}_F=U(3)^6\,,
\end{equation}
once three RH neutrinos are added to the SM particle content\footnote{Note that RH neutrinos require a more delicate treatment when it comes to flavour symmetry, as their potential Majorana mass terms are not invariant under such a group.}. If this symmetry group is to be respected also by the Yukawa interactions, the Yukawa couplings need to transform non-trivially. 

This treatment is no more than a low-energy point of view. If there existed a high-energy theory responsible for the generation of the Yukawas, there could possibly exist scalars that acquired vevs at energies much higher than those we can probe. These fields would be charged under $\mathcal{G}_F$, coupling to the Yukawa terms and making them invariant. Their transformation properties are inherited by their vevs, which are encoded into the spurions that we perceive as the Yukawa couplings. Such parameters must then take some particular values in order to satisfy the observed masses of the fermions. Thus, MFV is agnostic about the full theory that generates the Yukawa interactions; in that sense, it is not a model, it is rather a prescription that sets some principles in order to build a theory of flavour. 

The first flavour models attempted to reproduce the quark masses and the anatomy of the CKM, so MFV was originally developed to protect the quark sector from unwanted flavour violation. Neutrino oscillations constitute a new source of flavour change, unrelated to the CKM. Thus, an extended flavour model that predicts the masses and mixings in the lepton sector would also require a wider MFV approach. However, Minimal Lepton Flavour Violation (MLFV) presents some difficulties, not present in the quark sector. This is partly due to the possible presence of Majorana masses for RH neutrinos, which incorporate their own spurion. This extra component makes neutrinos conceptually different to the rest of fermions, and introduces new degrees of freedom that may lead to a loss of predictivity. In the following we will discuss some ways to tackle this issue, studying how neutrino masses may arise naturally in models built from an MLFV point of view.

\section{The strong CP problem}

The strong CP problem is one of the main fine-tuning issues of the SM, and refers to the smallness of the so-called $\theta$-term of QCD:
\begin{equation}
    \mathcal{L}_{\cancel{\rm CP}}=\frac{g_s^2}{16\pi}\theta G_{\mu\nu}^a\widetilde{G}^{\mu\nu a}\,.
\end{equation}
This part of the Lagrangian explicitly breaks CP invariance. This discrete symmetry, which is not imposed when writing the SM Lagrangian, is mostly conserved by the fundamental interactions. In fact, the only other source of CP violation in the SM is the phase of the CKM matrix, which only generates a small breaking of the symmetry.

This term can be rewritten as a total derivative, which, in general, can be removed from the Lagrangian, as gauge fields are expected to vanish at infinity. However, non-perturbative effects allow the vacuum of QCD to present non-trivial solutions that lead to a non-vanishing $\theta$-term.

The parameter $\theta$ itself is not physical. This is due to the fact that the $\theta$-term is directly related to the chiral anomaly, as it can be seen by computing the divergence of the chiral current, defined as $j^\mu_5=\overline{q}\gamma^\mu\gamma_5q$:
\begin{equation}
    \partial_\mu j^\mu_5=2im\overline{q}\gamma_5 q+\frac{g_s^2}{16\pi}G_{\mu\nu}^a\widetilde{G}^{\mu\nu a}\,.
\end{equation}
The first term is due to the fact that quark masses\footnote{For simplicity, only one quark with mass $m$ is considered here.} explicitly break chiral symmetry; although the SM kinetic terms are invariant under a rotation of the form $q\to e^{i\alpha(x)\gamma_5}q$, mass terms are not. 

The second term in the equation above is proportional to the $\theta$-term, showing that this part of the Lagrangian is directly connected to the quark mass matrix through the chiral anomaly. This means that the $\theta$ parameter is not observable; the actual physical quantity is 
\begin{equation}
    \overline{\theta}=\theta + \text{arg} (\text{det} (M_uM_d))\,,
    \label{eq:thetabar}
\end{equation}
where the six quark flavours have now been considered. $M_{u(d)}=Y_{u(d)}v/\sqrt{2}$ is the mass matrix of up (down)-type quarks in the flavour basis. 

The parameter $\overline{\theta}$ controls some observables that explicitly break CP symmetry, such as the neutron electric dipole moment (nEDM)~\cite{Baluni:1978rf,Crewther:1979pi}. So far there has been no experimental evidence of such effects, pointing to a very small, even vanishing $\overline{\theta}$. The strongest limits arise from measurements on the mentioned nEDM, that result in~\cite{Abel:2020gbr}
\begin{equation}
    \overline{\theta}<10^{-10}\,.
\end{equation}
From a naturalness point of view, it is quite striking that a dimensionless parameter of the Lagrangian takes such a tiny value. No symmetry argument can be invoked either, as CP is already broken in the SM: no extra symmetry would be gained by setting $\overline{\theta}$ to 0. Thus, it seems that it is necessary to remove by hand a term that is perfectly allowed in the Lagrangian. It is even more worrisome that this very small parameter is actually a combination of others, as seen in Eq.~\ref{eq:thetabar}. While $\theta$ is linked to the strong interactions, the quark masses stem from the weak sector. It seems that parameters arising from different physics conspire to cancel up to the level of $10^{-10}$, which is a rather unlikely coincidence.

This fine-tuning issue is usually attacked by arguing that some extra symmetry, apart from the SM ones, is present, and yields a vanishing $\theta$-term. Some of these BSM extensions actually promote CP to an exact symmetry that is spontaneously broken~\cite{Nelson:1983zb,Barr:1984qx,Bento:1991ez,Hiller:2001qg}. Others instead rely on the fact that a massless quark would solve the strong CP problem. In fact, it can be seen from Eq.~\ref{eq:thetabar} that if a single quark mass vanished, $\theta$ and $\overline{\theta}$ would coincide. As $\theta$ is unphysical, it can be safely set to 0, yielding no strong CP problem. This situation is obviously incompatible with the measured quark masses, and would require an extended QCD sector~\cite{Hook:2014cda}.

Arguably, the most elegant and accepted solution to the strong CP problem is the axion. This mechanism relies on the invariance under a new global, Abelian, anomalous symmetry, $U(1)_{\rm PQ}$, dubbed Peccei-Quinn~\cite{Peccei:1977hh,Peccei:1977ur}. This setup also introduces a new scalar field, that transforms non-trivially under $U(1)_{\rm PQ}$; in fact, the potential associated to this scalar spontaneously breaks the Peccei-Quinn symmetry. This breaking gives rise to the appearance of a Goldstone boson, corresponding to the angular degree of freedom of the new scalar field, called axion, $a$~\cite{Weinberg:1977ma,Wilczek:1977pj}. The fact that $U(1)_{\rm PQ}$ is anomalous will translate into a coupling of axions to gluons via the chiral anomaly,
\begin{equation}
    \mathcal{L}_{aGG}=\frac{a}{f_a}\frac{g_s^2}{32\pi}\theta G_{\mu\nu}^a\widetilde{G}^{\mu\nu a}\,,
\end{equation}
where $f_a$ is the PQ breaking scale (also dubbed axion decay constant). This coupling is clearly very similar to the $\theta$-term, thus promoting $\theta$ to a dynamical quantity and yielding the effective parameter
\begin{equation}
    \theta_{\rm eff}=\overline{\theta}+\frac{a}{f_a}\,.
\end{equation}
The key to this mechanism is the fact that the axion potential will be minimized in such a way that its vev will cancel the whole $\theta$-term. This is due to non-perturbative, low-energy QCD effects, that explicitly break $U(1)_{\rm PQ}$. Thus, the axion acquires a mass, and its potentital is tilted, such that the corresponding vev exactly cancels $\theta_{\rm eff}$.

Many different axion models have been proposed (see Ref.~\cite{DiLuzio:2020wdo} for a review), realizing the PQ breaking and the axion coupling to gluons in different ways. They all share the philosophy of promoting the $\theta$-term to a field, explaining its cancellation dynamically. These BSM scenarios come with an unexpected boon, as axions may constitute suitable DM candidates. They could even compose the whole energy budget of DM if their mass and decay constant lie within certain ranges. 
\begin{figure}[b!]
\centering
\includegraphics[width=0.73\textwidth]{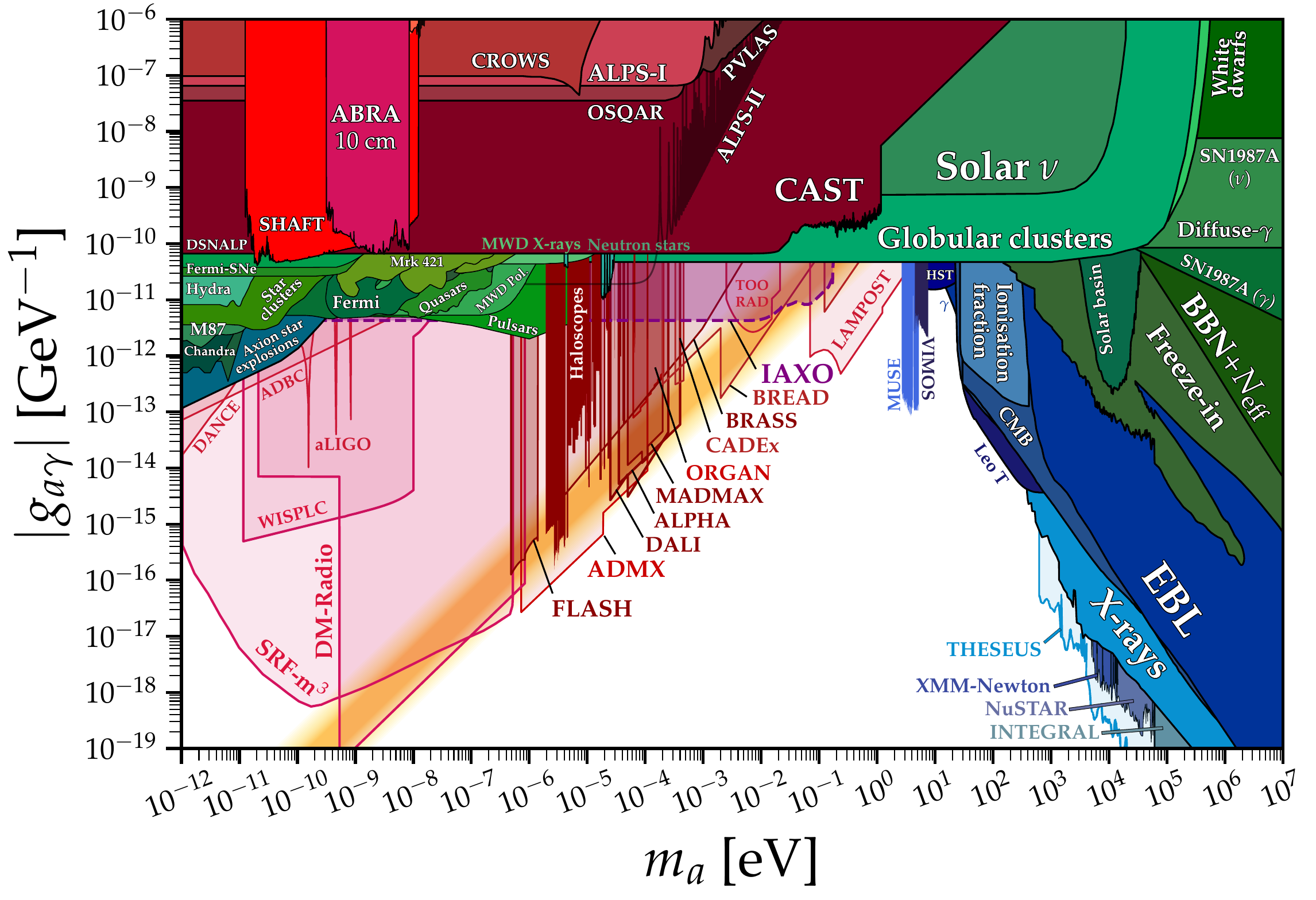}
\includegraphics[width=0.73\textwidth]{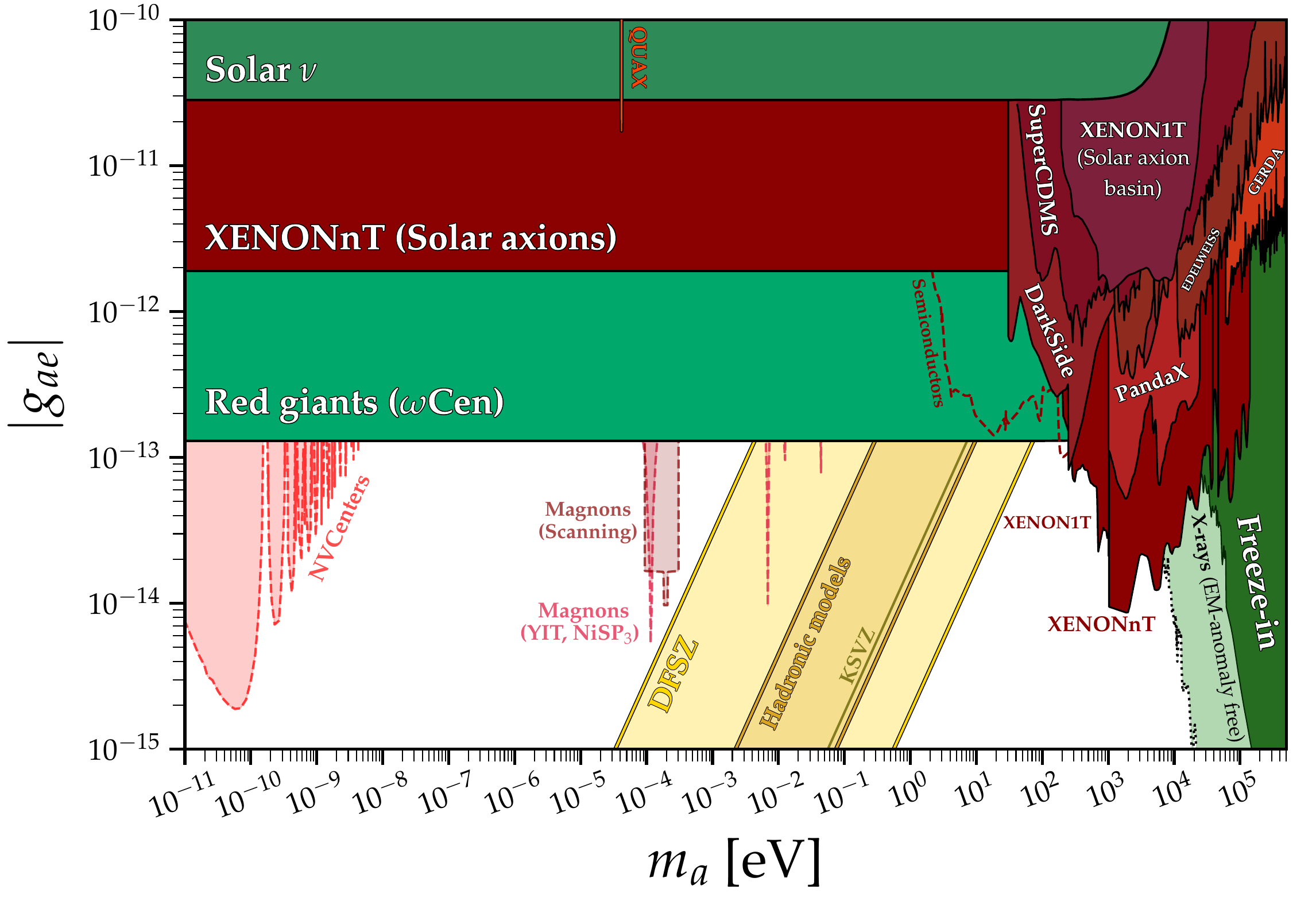}
\caption{Current bounds and future prospects on axion couplings to photons (upper panel) and electrons (lower panel), as a function of the axion mass~\cite{AxionLimits}. The strong CP problem is solved along the diagonal yellow bands.}
\label{fig:axion_bounds}
\end{figure}

A great amount of experimental efforts have been dedicated to search for axions employing different techniques. The theory has been unable to predict values for the mass and couplings of these particles, so a large parameter space must be covered. No positive signal has been found so far, but future experiments expect to reach unprecedented sensitivities. For instance, Fig.~\ref{fig:axion_bounds} shows the current bounds on axion couplings to photons (upper panel) and electrons (lower panel), as well as prospects for upcoming efforts. These are displayed as a function of the axion mass. Note that the strong CP problem is in principle only solved along the diagonal yellow bands, while the rest of the parameter space corresponds to axion-like particles (ALPs). These exhibit the same phenomenology as axions, but typically do not arise from the same models, and do not solve the strong CP problem.

\chapter{A Majoron mechanism}
\fancyhead[RO]{\scshape \color{lightgray}7. A Majoron mechanism}
As it has been described in Sec.~\ref{sec:hub_tens}, the Hubble tension constitutes one of the main problems in current cosmology, potentially pointing to a flaw in the $\Lambda$CDM model. Some of the many proposed solutions suggest that particle physics may be key to solve this anomaly. In particular, Ref.~\cite{Escudero:2019gvw} showed that a Majoron, $\omega$ (the Goldstone boson associated to lepton number breaking), that coupled to neutrinos could considerably alleviate the tension, provided its mass and coupling lie in the ranges
\begin{align}
    m_\omega&\in[0.1,\,1]\eV\,,
\label{MajoronMassWindow}\\
\lambda_{\omega\nu\nu}
&\in[5\cdot 10^{-14},\,10^{-12}]\,.
\label{MajoronNeutrinoMixingWindow}
\end{align}
For such small couplings and masses, Majorons only partially thermalize after Big Bang Nucleosynthesis, or either never thermalize~\cite{Chacko:2003dt}, enhancing the effective number of neutrino species, $N_\text{eff}$, by at least $0.03$ and at most $0.11$\footnote{These values may be tested with future CMB-S4 experiments~\cite{Abazajian:2016yjj}.}. Moreover, a non-vanishing $\lambda_{\omega\nu\nu}$ would reduce neutrino free-streaming, modifying the neutrino anisotropic stress-energy tensor~\cite{Bashinsky:2003tk}. This has an impact on the CMB, that results in modifying the posterior for the Hubble constant: the inclusion of Majoron-neutrino interactions slightly shifts the central value of $H_0$, but largely broadens its profile, reducing the tension from roughly $5\sigma$ to $2.5\sigma$. For larger couplings, $\lambda_{\omega\nu\nu}>10^{-12}$, these effects are too large, being already excluded by Planck data.

Interestingly, Ref.~\cite{Escudero:2019gvw} found that the best $\chi^2$ in a Markov Chain Monte Carlo corresponds to Majoron masses and couplings as in Eqs.~\ref{MajoronMassWindow} and~\ref{MajoronNeutrinoMixingWindow}, and to $\Delta N_\text{eff}=0.52\pm0.19$. The uncertainty in this last observable is very large, with $\Delta N_\text{eff}=0$ being compatible within $3\sigma$. However, such a central value can be achieved if a thermal population of Majorons is produced in the early Universe and is not diluted during inflation. This may occur if the reheating temperature is larger than the RH neutrino masses~\cite{Escudero:2019gvw}. Alternatively, other relativistic species, such as axions, may contribute to $\Delta N_\text{eff}$~\cite{Masso:2002np,Salvio:2013iaa,Ferreira:2018vjj,Arias-Aragon:2020qtn}, and their presence may justify such a large value.

As Majorons arise from the breaking of lepton number, it would seem natural that they were connected to the mechanism behind the generation of neutrino masses. In the following, and relying on Ref.~\cite{Arias-Aragon:2020qip}, we will explore such a possiblity, provided the Majoron exhibits a mass and coupling that allow to alleviate the Hubble tension.

\section{The model}
\label{sec:Mechanism}

In the type I seesaw framework, the Majorana mass term for RH neutrinos can be promoted to a Yukawa term, in which the fermion bilinear couples to a new scalar field, $\chi$, singlet under the SM gauge group. Similarly, the Yukawa coupling of the LH leptons, RH neutrinos and Higgs doublet will now contain insertions of $\chi$. If this particle has the right charge under $U(1)_{\rm L}$, LN will be an exact global symmetry at the Lagrangian level. However, it will be spontaneously broken once $\chi$ takes a non-zero vev, generating a Majorana mass term for RH neutrinos and, following the type I philosophy, small masses for the active neutrinos. Such a breaking will yield the appearance of a Goldstone boson, the Majoron, corresponding to the angular degree of freedom of $\chi$. 

The features of such a setup strongly depend on the $U(1)_{\rm L}$ charges of the LH lepton doublet, the RH neutrinos and the new scalar. Although it seems natural to fix the LN of the former to unity, the assignments of the RH neutrinos and $\chi$ are in principle free. However, in order to preserve the invariance of the Lagrangian under LN, these fields need to have charges with opposite signs. These can be read in Tab.~\ref{tab:LN}. We will assume that three RH neutrinos are present, although this is not a key requirement of this model.

\begin{table}[b!]
\centering
\begin{tabular}{cc} \toprule\vspace{1mm}
    &$U(1)_{\rm L}$ charge \\ \toprule
    $L_L$ & $1$ \\
    $\ell_R$ & $1$ \\
    $N_R$ & $-L_N$  \\
    $\chi$ & $L_\chi$ \\\bottomrule
\end{tabular}
\caption{Lepton number assignments. Fields that are not listed here do not transform under $U(1)_{\rm L}$.}
\label{tab:LN}
\end{table}

The most general Lagrangian that can be written assuming LN invariance reads
\begin{equation}
\LL_{\nu} = -\left(\frac{\chi}{\Lambda_\chi}\right)^{\frac{1+L_N}{L_\chi}}\overline{L_L}\widetilde{H}Y_\nu N_R - \frac{1}{2}\left(\frac{\chi}{\Lambda_\chi}\right)^{\frac{2L_N-L_\chi}{L_\chi}} \chi \overline{N^c_R}Y_NN_R+\hc\,,
\label{Lag}
\end{equation}
where $Y_\nu$ and $Y_N$ are the corresponding Yukawa matrices and $\Lambda_\chi$ is the cut-off scale up to which the effective operator approach holds. 

Note that other terms can be added to this Lagrangian by inserting $\chi^\dag$ instead of $\chi$. However, if the terms in Eq.~\ref{Lag} are local, then their siblings with $\chi^\dag$ are not. The only exception is the term
\be
\frac{1}{2}\left(\frac{\chi}{\Lambda_\chi}\right)^{\frac{2L_N+L_\chi}{L_\chi}} \chi^\dag \overline{N^c_R}Y_NN_R\,.
\ee
However, this operator provides a correction to the Majorana mass term that is suppressed with respect to the one written in Eq.~\ref{Lag}, and it can be safely neglected.

Although the LN charges of $N_R$ and $\chi$ seem to be in principle free, random choices could render non-local the terms in Eq.~\ref{Lag}. This argument poses a first condition on $L_{N,\chi}$.  Requiring that all the terms are local implies that the exponents of $\chi$ must be natural numbers:
\be
\dfrac{1+L_N}{L_\chi}\,,
\dfrac{2L_N-L_\chi}{L_\chi} \in \mathbb{N}\,.
\label{FirstConditionxchi}
\ee
In the LN broken phase, the field $\chi$ can be parametrized as
\begin{equation}
\chi = \frac{\sigma+v_\chi}{\sqrt{2}}e^{i\frac{\omega}{v_\chi}},
\label{chiVEV}
\end{equation}
where $v_\chi$ is its vev, $\sigma$ is the radial component and the angular part, $\omega$, is the Goldstone boson identified as a Majoron. Notice that the scale appearing in the denominator of the exponent is also $v_\chi$, in order to obtain canonically normalized kinetic terms for the Majoron. A useful notation that will be employed in the following is the ratio
\be
\varepsilon_\chi=\frac{v_\chi}{\sqrt{2}\Lambda_\chi}.
\ee
This parameter is expected to be smaller than 1, in order to guarantee a good expansion in terms of $1/\Lambda_\chi$. Consequently, the vev of $\chi$, which represents the overall scale of LN breaking, is expected to be smaller than the scale $\Lambda_\chi$.

Once electroweak symmetry is also spontaneously broken and the SM Higgs also develops its vev, masses for the light neutrinos are generated, according to the type I seesaw mechanism:
\begin{equation}
\LL_{\rm mass}=-\dfrac{1}{2}\overline{\nu_L^c}\mathcal{M}_\nu\nu_L +\hc\,,
\end{equation}
where the mass matrix reads
\begin{equation}
\mathcal{M}_\nu=\frac{\varepsilon_\chi^{\frac{2+L_\chi}{L_\chi}} v^2}{\sqrt{2} v_\chi}Y_\nu Y_N^{-1}Y_\nu^T\,.
\label{eq:mass_matrix}
\end{equation}
In the basis where the charged lepton mass matrix is already diagonal, $\mathcal{M}_\nu$ can be diagonalized by the PMNS matrix, $U$:
\be
\widehat{\mathcal{M}}_\nu\equiv\mathrm{diag}\left(m_1,\,m_2,\,m_3\right)= U^\dagger \mathcal{M}_\nu\,U^*\,.
\ee
The overall scale of active neutrino masses, roughly given by the atmospheric mass splitting (see Tab.~\ref{tab:osc_params}), can be written in terms of the parameter $\varepsilon_\chi$, the ratio of the vevs and a product of Yukawa matrices\footnote{Note that in the following equations, matrices will be generically treated as numbers; these approximate expressions denote the scale of the entries of such matrices, as no strong hierarchies are expected among them.}:
\be
\frac{\varepsilon_\chi^{\frac{2+L_\chi}{L_\chi}} v^2}{\sqrt{2} v_\chi}Y_\nu Y_N^{-1}Y_\nu^T\simeq \sqrt{|\Delta m^2_\text{atm}|}\,.
\label{SecondConditionxchi}
\ee
The heavy neutrinos, that mostly coincide with the RH states, have a mass matrix that in first approximation can be directly read from the Lagrangian in Eq.~\ref{Lag},  
\be
\mathcal{M}_N\simeq\varepsilon_\chi^{\frac{2L_N-L_\chi}{L_\chi}} \frac{v_\chi}{\sqrt{2}}Y_N\,.
\label{MN}
\ee
Obviously, in order for the seesaw approximation to hold, the heavy neutrinos must be much heavier than the light ones, a condition that translates into:
\be
\varepsilon_\chi^{\frac{2L_N-L_\chi}{L_\chi}} \frac{v_\chi}{\sqrt{2}}Y_N\gg\sqrt{|\Delta m^2_\text{atm}|}\,.
\label{ThirdConditionxchi}
\ee
These conditions will be employed in the following to determine possible values for the parameters of the model. 

The Lagrangian that describes the Majoron appears after the electroweak and LN breakings, and can be written as follows:
\begin{align}
\LL_\omega =& \,\frac{1}{2}\partial_\mu\omega\partial^\mu\omega -\frac{1}{2}m_\omega^2 \omega^2-
\left(i\frac{1+L_N}{L_\chi}\varepsilon_\chi^{\frac{1+L_N}{L_\chi}}\right)\frac{v}{\sqrt{2}v_\chi}\overline{\nu_L}Y_\nu N_R\, \omega \,+
\nn\\
&\left(-i\frac{L_N}{L_\chi}\frac{\varepsilon_\chi^{\frac{2L_N-L_\chi}{L_\chi}}}{\sqrt{2}}\right)\overline{N^c_R}Y_NN_R\, \omega +\hc\ ,
\label{MajoronLag}
\end{align}

where the $m_\omega^2$ term parametrizes the Majoron mass, and is introduced here as an explicit breaking of the corresponding shift symmetry. The Majoron directly couples to the active neutrinos:
\be
\LL_\omega\supset \frac{i}{2}\lambda_{\omega\nu\nu}\omega\overline{\nu_L^c}\nu_L+\hc\footnote{This expression coincides with the one in Ref.~\cite{Escudero:2019gvw} identifying $\omega$ with $\phi$ and $\lambda_{\omega\nu\nu}$ with $\lambda$.}\,,
\ee
with
\be
\lambda_{\omega\nu\nu}=2\dfrac{\mathcal{M}_\nu}{L_\chi v_\chi}\,.
\label{lambdaomeganunu}
\ee
Once again, this is strictly a matrix relation, but it suffices to identify $\mathcal{M}_\nu$ with the atmospheric mass splitting. From the results in Ref.~\cite{Escudero:2019gvw}, shown in Eq.~\ref{MajoronNeutrinoMixingWindow}, it is then possible to infer a bound on the product $L_\chi v_\chi$,
\be
|L_\chi| v_\chi\simeq\dfrac{2\sqrt{|\Delta m_\text{atm}^2|}}{\lambda_{\omega\nu\nu}}\in[0.1,\,2]\TeV\,.
\label{xchivchiRelation}
\ee
Substituting this relation in Eqs.~\ref{SecondConditionxchi} and~\ref{ThirdConditionxchi}, new conditions can be found:
\begin{equation}
|L_\chi| \varepsilon_\chi^{\frac{2+L_\chi}{L_\chi}}Y_\nu Y_N^{-1}Y_\nu^T
\simeq\dfrac{2\sqrt2}{\lambda_{\omega\nu\nu}}\dfrac{|\Delta m_\text{atm}^2|}{v^2}\in[0.12\,,2.4]\cdot 10^{-12}\,,
\label{SecondConditionxchiNew}
\end{equation}
and
\begin{equation}
\dfrac{\varepsilon_\chi^{\frac{2L_N-L_\chi}{L_\chi}}}{|L_\chi|}Y_N\gg\dfrac{\lambda_{\omega\nu\nu}}{\sqrt2}\simeq3.5\cdot 10^{-14}\,.
\label{ThirdConditionxchiNew}
\end{equation}
These relations clearly show that the parameter space of the model is greatly affected by the choices of the $U(1)_{\rm L}$ charges. Interestingly, it is possible to choose a set of values that leads to a renormalizable Lagrangian, that we will dub \say{Case R} in the following:
\be
L_N=-1\,,L_\chi=-2\,.
\ee
With these assignments, the powers of $\chi/\Lambda_\chi$ in Eq.~\ref{Lag} are not present in neither the Dirac nor the Majorana mass terms, and the numerical relations found above simplify considerably. The one in Eq.~\ref{xchivchiRelation}, given by the preferred range of values for the Majoron coupling, translates into
\begin{equation}
    v_\chi\simeq\dfrac{\sqrt{|\Delta m_\text{atm}^2|}}{\lambda_{\omega\nu\nu}}\in[0.05,\,1]\TeV\,.
\end{equation}
Eq.~\ref{SecondConditionxchiNew}, needed to satisfy the lightness of neutrino masses, now becomes
\begin{equation}
Y_\nu Y_N^{-1}Y_\nu^T
\simeq\dfrac{\sqrt2}{\lambda_{\omega\nu\nu}}\dfrac{|\Delta m_\text{atm}^2|}{v^2}\in[1.2\cdot 10^{-13},\,2.4\cdot 10^{-12}]\,.
\end{equation}
Finally, the requirement posed by the seesaw approximation, given by Eq.~\ref{ThirdConditionxchiNew}, now reads:
\begin{align}
Y_N\gg\dfrac{2\lambda_{\omega\nu\nu}}{\sqrt2}\simeq7\cdot 10^{-14}\,.
\end{align}
All in all, these conditions set a range of values for $v_\chi$, as well as a lower bound on the overall scale of $Y_N$. Reproducing the light neutrino masses implies that the product $Y_\nu Y_N^{-1}Y_\nu^T$ should be tuned to be very small. 

For values of $L_{N,\chi}$ different from the ones in Case R, the Lagrangian is necessarily non-renormalizable. An interesting question is whether the extremely small values of the product $Y_\nu Y_N^{-1}Y_\nu^T$ can be avoided exploiting the suppression given by $\varepsilon_\chi$, in a similar fashion to the Froggatt-Nielsen approach. 

Considering first the case in which $L_{N,\chi}>0$, it can be seen from Eq.~\ref{SecondConditionxchiNew} that $\varepsilon_\chi$ gets smaller for larger values of $L_\chi$ (unless tuning the product $Y_\nu Y_N^{-1}Y_\nu^T$ to a rather unnatural value, as in Case R). Although this is not a problem by itself, it hardens the constraint in Eq.~\ref{ThirdConditionxchiNew}, forcing $Y_N$ to be also quite small. From Eq.~\ref{MN}, it follows that the heavy neutrinos would be roughly as light as the active, and therefore the expansion in the type I seesaw would break down. Thus, only two appealing possiblities are left, which we will refer to as \say{Case NR1} and \say{Case NR2}:
\begin{align}
\text{NR1:}\,\,&L_N=1\,, L_\chi=1\,,\\
\text{NR2:}\,\,&L_N=1\,, L_\chi=2\,.
\end{align}
Once again, imposing that the light neutrino masses are reproduced and that the Majoron coupling lies in the preferred range, the vev and scale of the new scalar (and consequently the parameter $\varepsilon_\chi$) are constrained to particular intervals. In turn, the masses of the heavy neutrinos are also determined (Eq.~\ref{MN}). All these values are reported in Tab.~\ref{tab:param} for both Cases NR1 and NR2.
\begin{table}[b!]
\begin{center}
\begin{tabular}{ccccccc} \toprule\vspace{1mm}
   & $L_N$ & $L_\chi$ & $v_\chi$ (TeV) & $\varepsilon_\chi\,(10^{-6})$ & $ M_N$ (GeV)& $\Lambda_\chi$ (PeV)\\ \toprule\vspace{1.5mm}
  Case NR1 	& $1$ & $1$ & $[0.1,2]$ & $[0.5,1.4]$ & $[0.003,0.2]$ & $[1.4,11]$\\
Case NR2 	& $1$ & $2$ & $[0.05,1]$ & $[0.24,1.1] $ & $[35.4,707]$ & $[140,650]$ \\\bottomrule
\end{tabular}
\caption{Parameter ranges in the two phenomenologically interesting scenarios. $M_N$ denotes the overall scale of the heavy neutrino masses.}
\label{tab:param}
\end{center}
\end{table}

If $L_N$ were positive but $L_\chi$ were negative, it would be possible to obtain the same results discussed above by substituting $\chi$ by $\chi^\dagger$ in the Lagrangian. In this case, the signs in the denominators of the exponents would be flipped, compensating the negative sign of $L_\chi$. The opposite situation, $L_N<0$ and $L_\chi>0$, is not allowed by the locality conditions.

For $L_{N,\chi}<0$, besides Case R, only one other choice is allowed by the locality conditions:  $L_N=L_\chi=-1$. However, this case would require $\varepsilon_\chi \gg1$, leading to an even more extreme fine-tuning than in Case R, but without the appeal of renormalizability.

In summary, out of the apparently many possibilities of LN charges, only two choices are phenomenologically interesting. Many others are possible, but would lead to fine-tuned scenarios, to Majoron couplings that would not alleviate the Hubble tension, or to light neutrino masses incompatible with observations. Thus, in the following we will focus on Cases NR1 and NR2. 

So far we have dealt with a Majoron that exhibits a coupling to neutrinos in the range given by Eq.~\ref{MajoronNeutrinoMixingWindow}, which translates into the condition in Eq.~\ref{xchivchiRelation}. However, this is only one of the ingredients necessary to lower the $H_0$ tension. A second relevant requirement is Eq.~\ref{MajoronMassWindow}, regarding the Majoron mass, which, for the sake of simplicity, we have introduced by hand in the Majoron Lagrangian (Eq.~\ref{MajoronLag}). In our setup, $U(1)_{\rm L}$ is an exact symmetry at the Lagrangian level, so its spontaneous breaking would yield a pure, massless Goldstone boson. However, there are plenty of arguments to believe that the Majoron would very unlikely be exactly massless. In fact, the origin of its mass has been widely discussed in the literature, and constitutes in itself an interesting research topic. Any violation of the global LN symmetry would induce a mass for the Majoron. An obvious example are gravitational effects, which are expected to break all accidental global symmetries. Estimations of their size from non-perturbative arguments via wormhole effects~\cite{Alonso:2017avz} fall too short of the required values for the Majoron mass. On the other hand,  Planck-suppressed effective operators~\cite{Akhmedov:1992hi} would render it too large, although several possibilities have been discussed that would prevent the lower-dimension operators from being generated~\cite{Dobrescu:1996jp,Lillard:2018fdt,Hook:2019qoh,Dienes:1999gw,Choi:2003wr,Cox:2019rro,Fukuda:2017ylt,Carone:2020nlx}. A simpler possibility, given the singlet nature of the $N_R$, is an explicit breaking of LN via a direct Majorana mass term, not generated by means of the vev of a scalar field. The Majoron would thus develop a mass, slightly below this breaking scale, from its coupling to the $N_R$ through self-energy diagrams. Here we will remain agnostic to the origin of the Majoron mass, and a value consistent with Eq.~\ref{MajoronMassWindow} will be assumed.

The mechanism illustrated in this section allows to soften the $H_0$ tension, explaining at the same time the lightness of the active neutrinos. In the next section, this setup will be embedded into a specific model that allows to account for the flavour puzzle, without violating any bounds from flavour observables, and also contains a QCD axion that solves the strong CP Problem.

\section{The Majoron and axion from an MFV setup}
\label{sec:Model}

Out of the many possible frameworks for a theory of flavour, we will focus here on the MFV setup, that will be shown to naturally suggest the presence of the Majoron and of a QCD axion. Considering the SM spectrum augmented by three RH neutrinos, as in the model described above, the flavour symmetry of the fermion kinetic terms is a product of a $U(3)$ term for each fermion species,
\be
\cG_F=U(3)^6\,.
\ee
The non-Abelian components of this group will be responsible for the intergenerational fermion mass hierarchies and the mixing matrices, while the Abelian ones will be associated to the hierarchies among the masses of the different third family fermions. The latter can be written as follows:
\begin{equation}
\cG_F^{\rm A}=U(1)_{Q_L}\times U(1)_{u_R}\times U(1)_{d_R}\times U(1)_{L_L}\times U(1)_{e_R} \times U(1)_{N_R}\,.
\end{equation}
It is possible to rearrange this product, provided that the new combinations are still linearly independent, identifying among them baryon and lepton number, weak hypercharge and the Peccei-Quinn symmetry:
\be
\cG_F^{\rm A}=U(1)_{\rm B}\times U(1)_{\rm L} \times U(1)_Y \times U(1)_{\rm PQ} \times U(1)_{e_R} \times U(1)_{N_R}\,.
\label{U1symmetries}
\ee
We will assign the charges of the fermions under baryon number and hypercharge as in the SM, while the LN assignments are given in Tab.~\ref{tab:LN}. Moreover, PQ charges are chosen so as to explain the suppression of the bottom and tau masses with respect to the top mass, as will be discussed below. The last two symmetries in Eq.~\ref{U1symmetries} do not play any role in this model, and are explicitly broken. 

In an analogous way to LN, the PQ symmetry is exact at the Lagrangian level after introducing a second scalar field, $\Phi$, which is a singlet of the SM gauge group, but transforms non-trivially under PQ with a charge of $-1$\footnote{In the same fashion as the $U(1)_{\rm L}$ charge of $\chi$, the PQ charge of $\Phi$ can be set to -1 with no loss of generality, as it only sets an overall scale for the rest of the charges.}:
\begin{equation}
\Phi=\frac{\rho+v_\Phi}{\sqrt{2}}e^{i\frac{a}{v_\Phi}},
\end{equation}
where $\rho$ and $a$ are its radial and angular components and $v_\Phi$ its vev. This scalar solves the strong CP problem in the usual way: after the PQ symmetry is spontaneously broken, the axion arises as the associated Goldstone boson, coupling to the $\theta$-term via the anomaly. The axion potential will yield a vev that will cancel the $\overline{\theta}$ parameter, yielding no CP violation in the strong sector. 

The Yukawa couplings of the SM fermions will turn into non-renormalizable terms, just as in the Froggatt-Nielsen model (Eq.~\ref{eq:Frog-Niel}). This way, the masses of the fermions will depend on their PQ charges (generally denoted as $x$), as well as on the parameter $\vep=v_\Phi/\sqrt{2}\Lambda_\Phi$. As the top quark Yukawa coupling is close to $1$, then the whole Yukawa term for the up-type quarks should arise at the renormalizable level. The only PQ charge choice consistent with this requirement is
\be
x_u-x_Q=0\,.
\label{cau=0}
\ee
From the comparison of the bottom and tau masses with that of the top, two additional requirements follow:
\begin{align}
x_d-x_u\simeq\log_\vep(m_b/m_t)\,,\\
x_e-x_\ell\simeq\log_\vep(m_\tau/m_t)\,,
\label{xchargesdefinition}
\end{align}
where $x_d$, $x_u$ and $x_e$ are the PQ charges of RH down-type quarks, up-type quarks and charged leptons, and $x_\ell$ that of the lepton doublet. The specific ultraviolet theory that gives rise to the effective Yukawa operators would determine the value of $\vep$, that in turn fixes the PQ charges in order to reproduce the observed masses. Two possible benchmarks are:
\be
x_d-x_u=x_e-x_\ell=
\begin{cases}
1 \quad\text{for}\quad \vep=0.01\,,\\[2mm]
3 \quad\text{for}\quad \vep=0.23\,,
\end{cases}
\label{TwoCasesEpsilon}
\ee
where the first equality follows from the closeness of the bottom and tau masses. The choice $\vep=0.23$ corresponds to the Cabibbo angle, as traditionally considered in the Froggatt-Nielsen framework, while setting $\vep=0.01$ allows for smaller PQ charges with no need for fine-tuning. It is then possible to set $x_u=x_\ell=0$, so that only RH charged leptons and down-type quarks transform under PQ. 

According to the MFV philosophy, the Yukawas are not simple matrices anymore, but are promoted to spurions, that transform under the non-Abelian part of the flavour symmetry group, $\cG_F$. These can be thought of as vevs of scalar fields that are integrated out, inheriting their transformation properties. In the MFV approach, in order to correctly reproduce quark masses and mixings, as well as charged lepton masses, the background values of $Y_{u,d,e}$ should read
\be
\begin{aligned}
\left\langle Y_u \right\rangle & = c_t\,V^\dagger\, \mathrm{diag}\left(\frac{m_u}{m_t},\frac{m_c}{m_t},1 \right) \,,\\
\left\langle Y_d \right\rangle & = c_b\,\mathrm{diag}\left(\frac{m_d}{m_b},\frac{m_s}{m_b},1 \right) \,,\\
\left\langle Y_e \right\rangle & = c_\tau\, \mathrm{diag}\left(\frac{m_e}{m_\tau},\frac{m_\mu}{m_\tau},1 \right) \,,
\end{aligned}
\ee
where $V$ is the CKM mixing matrix and $c_i$ are $\mathcal{O}(1)$ parameters. The possible origin of these values is under study~\cite{Alonso:2011yg,Alonso:2012fy,Alonso:2013mca,Alonso:2013nca}. 

Any non-renormalizable operator that describes flavour-violating processes should be invariant under the flavour symmetry. This is accomplished by inserting proper powers of the spurions, whose background values suppress the operator under consideration. As a consequence, the scale that can be probed considering flavour observables is at the level of 1-10 TeV~\cite{DAmbrosio:2002vsn,Grinstein:2006cg,Grinstein:2010ve,Feldmann:2010yp,Guadagnoli:2011id,Redi:2011zi,Buras:2011zb,Buras:2011wi,Alonso:2012jc,Alonso:2012pz,Lopez-Honorez:2013wla,Merlo:2018rin}, instead of $100\TeV$ in a generic case~\cite{Isidori:2010kg}, opening up the possibility of discovering new physics at colliders.

The discussion is slightly more complicated when it comes to the neutrino sector, as it involves two spurions, $Y_\nu$ and $Y_N$, which both enter in the definition of the active neutrino masses (Eq.~\ref{eq:mass_matrix}). It follows that it is not possible to identify univocally either $Y_\nu$ or $Y_N$ in terms of neutrino masses and entries of the PMNS matrix. Thus, the suppression in the non-renormalizable flavour violating operators cannot be directly linked to neutrino masses and mixings, losing the predictivity that characterizes the MFV approach in the quark sector. In order to simplify the picture, it is necessary to assume some properties of these Yukawa matrices. The solutions that have been proposed consider either $Y_N\propto\unity$~\cite{Cirigliano:2005ck,Davidson:2006bd} or treating $Y_\nu$ as a unitary matrix~\cite{Alonso:2011jd}. In both cases, the constraints on new physics, considering the present available data on flavour changing processes in the lepton sector, are as low as a few TeV~\cite{Cirigliano:2005ck,Cirigliano:2006su,Davidson:2006bd,Gavela:2009cd,Alonso:2011jd,Alonso:2016onw,Dinh:2017smk}.

If $Y_N$ is assumed to be proportional to the identity matrix, the three RH neutrinos become degenerate in mass, and their associated non-Abelian symmetry, $SU(3)_{N_R}$, is broken down to $SO(3)_{N_R}$. The non-Abelian flavour symmetry in the lepton sector is reduced to $\cG^{\rm NA}_{\rm L}\to SU(3)_{L_L}\times SU(3)_{e_R}\times SO(3)_{N_R}\times \rm CP$. The additional assumption of no CP violation in the lepton sector is meant to force $Y_\nu$ to be real\footnote{Strictly speaking, the condition of CP conservation forces the Dirac CP phase to be equal to 0 or $\pi$, and the Majorana CP phases to be 0, $\pi$ or $2\pi$. However, $Y_\nu$ is real only if the latter are equal to 0 or $2\pi$, so the value of $\pi$ for the Majorana phases needs to be disregarded in order to guarantee predictivity. The CP conservation condition assumed in Refs.~\cite{Cirigliano:2005ck,Davidson:2006bd} is then stronger than the strict definition.}. With these simplifications, the expression for the active neutrino mass in Eq.~\ref{eq:mass_matrix} simplifies to
\be
\mathcal{M}_\nu=\frac{\varepsilon_\chi^{\frac{2+L_\chi}{L_\chi}} v^2}{\sqrt{2} v_\chi}Y_\nu Y_\nu^T\,.
\label{massMLFVI}
\ee
Thus, all flavour changing effects in the lepton sector can be written in terms of $Y_\nu Y_\nu^T$ and $Y_e$, which are the only relevant combinations entering any non-renormalizable operator. It follows that any flavour-changing process can be predicted in terms of lepton masses and mixings.

In this scenario, diagonalizing $\mathcal{M}_\nu$ corresponds to diagonalizing the product $Y_\nu Y_\nu^T$ and, given the fact that the the lepton mixing angles are relatively large, no strong hierarchies should be expected among the entries of $Y_\nu Y_\nu^T$ (in contrast with the quark sector). Besides, the overall scale of this product should be $\OO(1)$, in order to avoid any fine-tuning explanation to neutrino masses. Note that some setups, such as the so-called sequential dominance scenarios, lead to the obtention of large mixing angles even if there exists a strong hierarchy among the Yukawa couplings~\cite{King:1999mb}. However, this possibility is disfavoured by the general philosophy of MLFV. 

The second usual choice to simplify the picture in the lepton sector is to assume that the three RH neutrinos transform as a triplet under the same symmetry group of the lepton doublets, 
\begin{align}
L_L, N_R\sim ({\bf 3},\,\bf 1)_{\cG^{\rm NA}_L}\,,
\\
e_R\sim (\bf 1,\,{\bf 3})_{\cG^{\rm NA}_L}\,,
\end{align}
so the non-Abelian flavour symmetry is now $\cG^{\rm NA}_{\rm L}\to SU(3)_{L_L+N_R}\times SU(3)_{e_R}$. In this case, Schur\textquotesingle s Lemma guarantees that $Y_\nu$ transforms as a singlet. Therefore, $Y_\nu$ is a unitary matrix~\cite{Bertuzzo:2009im,AristizabalSierra:2009ex}, which can always be rotated to the identity by a suitable unitary transformation acting only on the RH neutrinos. The only meaningful quantities in this context are $Y_e$ and $Y_N$, so neutrino masses and mixings are encoded uniquely into $Y_N$,
\be
\mathcal{M}_\nu=\frac{\varepsilon_\chi^{\frac{2+L_\chi}{L_\chi}} v^2}{\sqrt{2} v_\chi}Y_N^{-1}\,.
\label{massMLFVII}
\ee
In this scenario, any flavour change in the lepton sector can be parametrized solely in terms of $Y_e$ and $Y_N$, allowing to predict any such process in terms of lepton masses and mixings. 

The diagonalization of the active neutrino mass now coincides with the diagonalization of $Y_N^{-1}$, that therefore does not present any strong hierarchy among its entries, which should generally be $\OO(1)$ according to the MLFV construction approach.

In summary, the Majoron and the axion constitute the natural Abelian completion of MFV scenarios. Note that both sectors are somehow independent, as the Majoron does not affect (at tree level) the axion and flavour physics. In the following section, we will analyze some of the phenomenological signatures of this model, concentrating on the charge assignments given by Cases NR1 and NR2.

\section{Phenomenological signatures}
\label{sec:Signatures}

\subsection{Scalar potential and Higgs physics}
\label{sec:Majoron_Higgs}
So far, we have only discussed the role of the angular degree of freedom of the scalar responsible for LN breaking, $\chi$. However, the addition of such a particle has other consequences, possibly affecting Higgs observables. The most general potential that contains both this new field and the Higgs doublet reads
\be
V(H,\chi)=-\mu^2H^\dag H + \lambda\left(H^\dag H\right)^2-\mu_\chi^2 \chi^*\chi+\lambda_\chi \left( \chi^*\chi\right)^2 + gH^\dag H\chi^*\chi\,.
\ee
The minimization of this potential provides expressions for the vevs of both fields:
\begin{align}
v^2&=\dfrac{4\lambda_\chi\mu^2-2g\mu_\chi^2}{4\lambda \lambda_\chi-g^2}\,,\\
v_\chi^2&=\dfrac{4\lambda\mu_\chi^2-2g\mu^2}{4\lambda \lambda_\chi-g^2}\,.
\end{align}
The parameters of this scalar potential need to be such that $v$ takes the value given by the electroweak scale and $v_\chi$ acquires the values in Tab.~\ref{tab:param}. Due to the mixed quartic term, the two physical scalars $h$ and $\sigma$ mix in the broken phase, with a mass matrix given by
\be
\cM^2=
\left(
\begin{array}{cc}
2\,\lambda\, v^2& g\,v\,v_\chi \\
g\,v\,v_\chi& 2\,\lambda_\chi\, v_\chi^2 \\
\end{array}\right)\,.
\ee
The two eigenvalues that arise after diagonalizing this mass matrix are the following:
\be
M_{h,\sigma}^2=\lambda v^2+\lambda_\chi v^2_\chi\pm\left(\lambda v^2-\lambda_\chi v_\chi^2\right)\sqrt{1+
\tan^22\vartheta}\,,
\label{ScalarMasses}
\ee
where
\be
\tan2\vartheta=\dfrac{g\,v\,v_\chi}{\lambda_\chi v_\chi^2-\lambda v^2}\,.
\ee
The lightest mass in Eq.~\ref{ScalarMasses} corresponds to the eigenstate mainly aligned with the SM Higgs, while the heaviest state is mostly composed of the radial component of $\chi$. From the relation between the mixed quartic coupling, $g$, and the physical parameters, 
\be
g=\dfrac{M_\sigma^2-M_h^2}{2\,v\,v_\chi}\sin2\vartheta\,,
\ee
it is possible to straightforwardly study the dependence of $M_\sigma$ with the other parameters of the model. 

The mixing of the Higgs with the new scalar also yields a coupling of the former to two Majorons. Expanding the kinetic term of $\chi$ yields a $\sigma\omega\omega$ term, that in turn produces a coupling $h\omega\omega$ after an insertion of the scalar mixing. Thus, a new decay channel for the Higgs opens up, contributing to its invisible width. The rate of this process is given by 
\begin{equation}
\Gamma_{h\to\omega\omega}=\dfrac{\sin^2{\vartheta}M_h^3}{32\pi v_\chi^2}\,.
\end{equation}
These effects on Higgs phenomenology have an impact on the signal strength $\mu_h$, defined as the ratio of observed Higgs events with respect to the SM expectation. This quantity has been measured to be in perfect agreement with the SM by both ATLAS and CMS~\cite{ATLAS:2022vkf, CMS:2022dwd}. In our setup, both the production cross section and the visible decay rates are modified. The former is suppressed by a factor $\cos^2{\vartheta}$, while the latter diminishes due to the appearance of the invisible channel. Explicitly,
\begin{equation}
    \mu_h= \cos^2{\vartheta}\left(1-\frac{\Gamma_{h\to\omega\omega}}{\Gamma_{h}}\right)\,,
\end{equation}
where $\Gamma_{h}$ is the total decay width of the Higgs boson. The combination of the ATLAS and CMS measurements yields a lower limit of $\mu_h>0.94$~\cite{Fernandez-Martinez:2022stj} at the 95\% CL, which translates into an upper bound on $\sin{\vartheta}$ that is proportional to $v_\chi$. 

\begin{figure}[b!]
\centering
\includegraphics[width=0.6\textwidth,keepaspectratio]{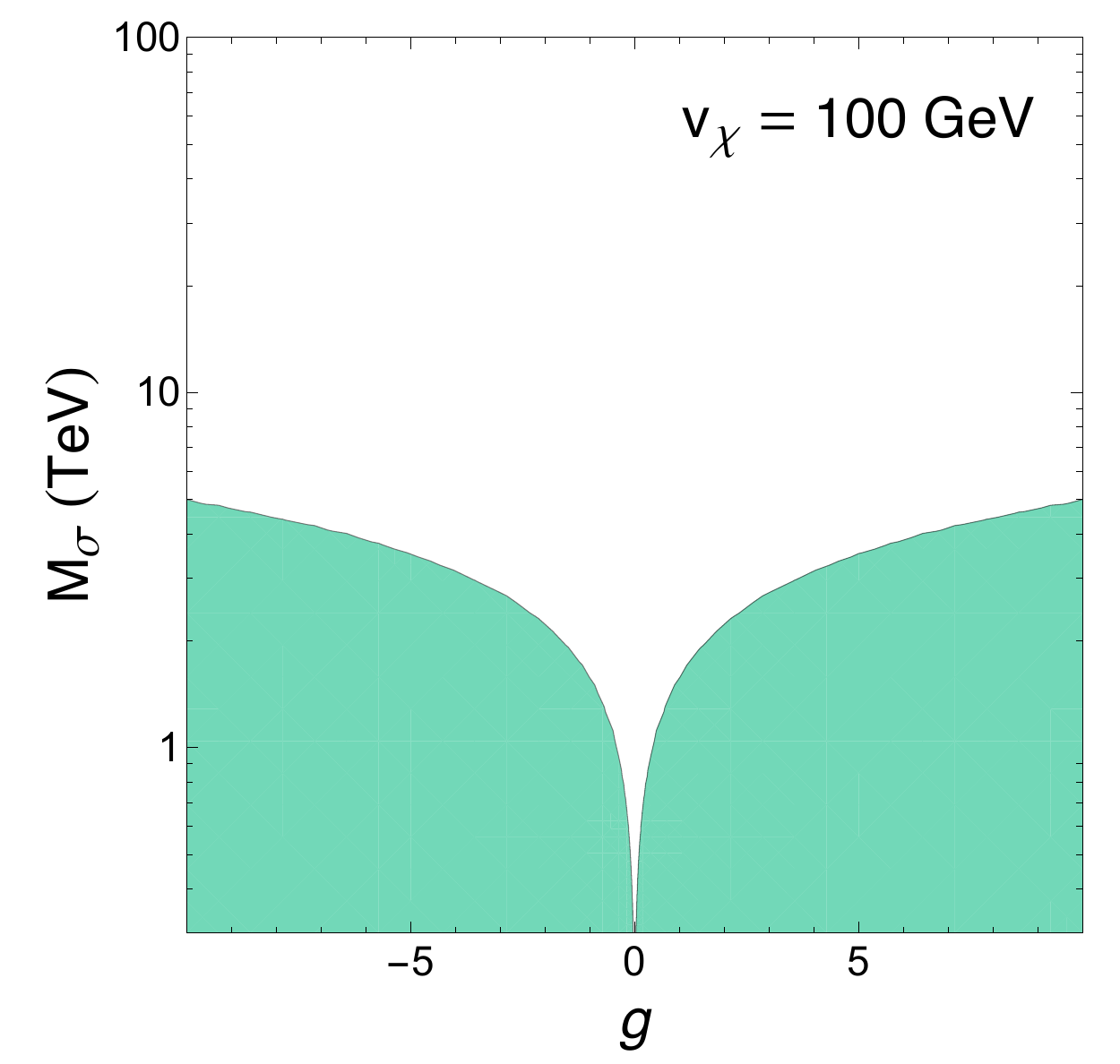}
\caption{Parameter space of the mass of the radial component of the new scalar as a function of its quartic coupling with the Higgs doublet. The green region is excluded by bounds on Higgs signal strength. We choose $v_\chi=100$ GeV, although the bounds are basically insensitive to this quantity.}
\label{Fig:SigmaMassPlots}
\end{figure}
These constraints can be portrayed in the parameter space composed of the mass of $\sigma$ and the quartic coupling $g$, as seen in Fig.~\ref{Fig:SigmaMassPlots}, in which the green areas are ruled out. Although we choose a particular value for $v_\chi$, we find that the bounds are basically independent on this quantity. Clearly, $M_\sigma$ can reach very large values without requiring any fine-tuning on $g$; conversely, $\sigma$ can only be light if $g$ is close to zero. Note that the upper bound on the Higgs invisible branching ratio (that currently stands at a 13\%~\cite{ParticleDataGroup:2022pth}) constitutes an independent source of limits, which exhibit a very similar dependence with the parameters of the model. However, these bounds are always looser than those given by signal strengths, so we do not include them in our results.

\subsection{Majoron coupling to photons and electrons}

In our setup, the Majoron only exhibits tree-level couplings to active neutrinos. However, at the quantum level, couplings to gauge bosons or other SM fermions  are originated. In particular, the probes for photons and electrons coupling to very light bosons pose the strongest constraints.

The searches for very light pseudoscalars, usually addressed to axions, can also apply to Majorons. In the range of masses in Eq.~\ref{MajoronMassWindow}, the strongest constraints on the effective coupling to photons are set by CAST~\cite{Anastassopoulos:2017ftl}, which establishes the upper bound 
\be
\lambda_{\omega\gamma\gamma}\lesssim 10^{-10}\GeV^{-1}\,,
\ee
where $\lambda_{\omega\gamma\gamma}$ is defined as
\be
\LL_{\omega\gamma\gamma} =- \frac{1}{4}\,\lambda_{\omega\gamma\gamma}\,\omega\, F^{\mu\nu} \widetilde{F}_{\mu\nu}\,,
\label{EffectiveOmega2Photons}
\ee
with $\widetilde{F}_{\mu\nu}\equiv \frac{1}{2}\varepsilon_{\mu\nu\rho\sigma} F^{\rho\sigma}$.

As the Majoron does not couple at tree-level to charged particles, the process $\omega\to\gamma\gamma$ occurs only at two loops. Ref.~\cite{Garcia-Cely:2017oco} provides an estimate for its decay width: under the assumption $m_\omega \ll m_e$,
\begin{equation}
\Gamma_{\omega\to \gamma\gamma}=\frac{\alpha^2}{1536^2\pi^7}\dfrac{m_\omega^7}{v^2m_e^4}\left(\Tr\left[\dfrac{m_Dm_D^\dagger}{vv_\chi}\right]\right)^2\,,
\end{equation}
where $\alpha\equiv e^2/4\pi$ and $m_D$ is the Dirac neutrino mass matrix, that can be directly read from Eq.~\ref{Lag}:
\begin{equation}
    m_D=\frac{v}{\sqrt{2}}\varepsilon_\chi^{\frac{1+L_N}{L_\chi}}Y_\nu\,.
\end{equation}
Computing the same process by means of the effective coupling in Eq.~\ref{EffectiveOmega2Photons} yields
\be
\Gamma_{\omega\to \gamma\gamma}=\dfrac{\left|\lambda_{\omega\gamma\gamma}\right|^2m_\omega^3}{32\pi}\,.
\ee
Matching these two decay rates provides an expression for the $\lambda_{\omega\gamma\gamma}$ coupling in terms of the parameters of our model:
\begin{equation}
\lambda_{\omega\gamma\gamma}=\dfrac{\alpha m_\omega^2}{384\sqrt{2}\pi^3m_e^2v_\chi}\varepsilon_\chi^{\frac{2+2L_N}{L_\chi}}\Tr\left[Y_\nu Y_\nu^\dagger\right]\,.
\end{equation}
Assuming the Yukawas are $\OO(1)$, as argued in the previous sections, the values for this coupling turn out to be much smaller than the current experimental limits, due to the two-loop suppression. The figures for Cases NR1 and Cases NR2, as well as the upper bounds, can be found in Tab.~\ref{tab:exp-bounds}.

\begin{table}[t!]
\centering
\begin{tabular}{cccc} \toprule\vspace{1mm}
    &$\lambda_{\omega\gamma\gamma}$(GeV$^{-1}$)&$\lambda_{\omega ee}$&$\lambda_{\omega\nu\nu}$\\ \toprule\vspace{1.5mm}
    Case NR1 &$[10^{-39},10^{-36}]$&$[10^{-25},10^{-24}]$&\multirow{2}{*}{$[10^{-14},10^{-12}]$}\\
    Case NR2&$[10^{-34},10^{-32}]$&$10^{-20}$&\\\midrule
    Exp. upper bounds&$10^{-10}$&$10^{-13}$&$10^{-5}$\\\bottomrule
\end{tabular}

\caption{Predictions for the Majoron effective couplings to electrons, photons and neutrinos for the relevant LN assignments. The corresponding experimental upper bounds are shown for comparison.}
\label{tab:exp-bounds}
\end{table}

Astrophysical measurements can also constrain Majoron couplings. In particular, Ref.~\cite{Viaux:2013lha} provides an upper bound on the Majoron coupling to electrons, based on observations of Red Giants, of $\lambda_{\omega ee}<4.3\cdot 10^{-13}$. Given the effective Lagrangian
\be
\LL_{\omega ee} = -i\, \lambda_{\omega ee}\,\omega\,\bar{e}\,e\,,
\ee
the decay width of the Majoron to two electrons reads~\cite{Garcia-Cely:2017oco}
\be
\Gamma_{\omega ee}\simeq\dfrac{1}{8\pi}|\lambda_{\omega ee}|^2\, m_\omega\,,
\ee
where
\be
\lambda_{\omega ee}\simeq\dfrac{1}{8\pi^2}\frac{m_e}{v}\left(\left[\dfrac{m_Dm_D^\dagger}{vv_\chi}\right]_{11}-\dfrac12\Tr\left[\dfrac{m_Dm_D^\dagger}{vv_\chi}\right]\right)\,,
\ee
with $[\ldots]_{11}$ standing for the $(1,1)$ entry of the matrix in the brackets. Writing this coupling in terms of the parameters of our model, we find
\begin{equation}
\lambda_{\omega ee}=\dfrac{1}{16\pi^2}\dfrac{m_e}{v_\chi}\varepsilon_\chi^{\frac{2+2L_N}{L_\chi}}\left(\left[Y_\nu Y_\nu^\dagger\right]_{11}-\Tr\left[\mathcal{Y}_\nu\mathcal{Y}_\nu^\dagger\right]\right)\,.
\end{equation}
Once again, provided that the elements of the product $Y_\nu Y_\nu^\dagger$ are $\OO(1)$, the loop suppression yields couplings much smaller than the experimental limit. Tab.~\ref{tab:exp-bounds} displays the values for Cases NR1 and NR2.

\subsection{Majoron emission in \boldmath{$0\nu\beta\beta$} decays}

\begin{figure}[t!]
\centering
\includegraphics[width=0.6\textwidth,keepaspectratio]{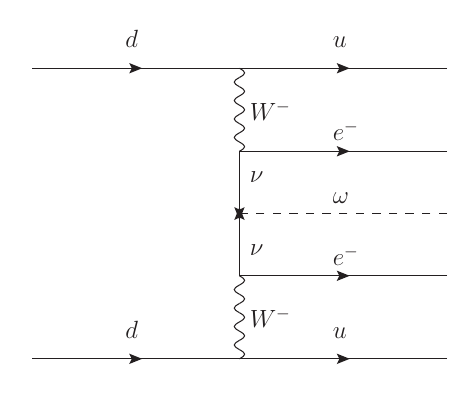}
\caption{Feynman diagram depicting the emission of a Majoron in a $0\nu\beta\beta$ decay.}
\label{Fig:NeutrinolessMajoron}
\end{figure}
The tree level coupling of the Majoron to neutrinos does not have an impact only on cosmology, but may also be relevant for low-energy terrestrial experiments. In particular, searches for neutrinoless double-beta decay could also be sensitive to processes in which Majorons are produced in the annihilation of the two neutrinos (see Fig.~\ref{Fig:NeutrinolessMajoron}).

As mentioned earlier, the lack of such a signal has set very strong bounds on the lifetime of $0\nu\beta\beta$ decay. However, these cannot be directly employed to constrain a process in which a Majoron is also produced, due to the different energy distribution of the emitted electrons. The NEMO-3 collaboration performed a dedicated search~\cite{Arnold:2018tmo}, setting a lower bound of the order of $10^{22}$ years on Majoron emission in $0\nu\beta\beta$ decay. This corresponds to a limit on the Majoron-neutrino coupling that reads
\be
\lambda_{\omega\nu\nu}\lesssim 10^{-5}\,,
\ee
where $\lambda_{\omega\nu\nu}$ is defined in Eq.~\ref{lambdaomeganunu}. Note that the exact constraint depends on the choice of nuclear matrix elements.

This value is much larger than the values required to alleviate the Hubble tension. As Cases NR1 and NR2 were chosen to reproduce such ranges, they are perfectly safe from this bound. The corresponding values are also included in Tab.~\ref{tab:exp-bounds} for completeness.
\subsection{Heavy neutrinos}
\label{sec:Majoron_pheno_HNL}

In both Cases NR1 and NR2, the heavy neutrino masses lie in ranges that may lead to interesting signatures in various experimental facilities. Neutrinos with masses ranging from tens to hundreds of MeV can be probed for, or even potentially detected, at beam dumps or at near detectors of oscillation experiments, such as DUNE or SHiP~\cite{Gorbunov:2007ak,Atre:2009rg,Bondarenko:2018ptm,SHiP:2018xqw,Bondarenko:2019yob,Ballett:2019bgd,Berryman:2019dme,Coloma:2020lgy}. If the masses are rather in the range of tens to hundreds of GeV, there are interesting prospects of production at the LHC or future colliders~\cite{delAguila:2008cj,Atre:2009rg,Antusch:2015mia,Deppisch:2015qwa,Antusch:2016ejd,Cai:2017mow,Dev:2018kpa,Pascoli:2018heg}. 

On the other hand, strong constraints arise from cosmology. Given their extremely small couplings, the heavy neutrinos produced in the early Universe would not be Boltzmann suppressed when decoupled from the thermal bath, possibly leading to an unacceptably large contribution to the relativistic degrees of freedom~\cite{Sato:1977ye,Gunn:1978gr,Hernandez:2013lza,Hernandez:2014fha,Vincent:2014rja}. This is the case if the heavy neutrinos are too long-lived and survive BBN. However, if their decay rate is large enough, they may disentegrate before the onset of BBN; the decay products would then quickly thermalize and BBN would then proceed as in the standard $\Lambda$CDM scenario. In an intermediate situation, in which the decay takes place during BBN, the production of primordial helium would be altered, and strong constraints also apply~\cite{Dolgov:2000pj,Ruchayskiy:2012si,Gelmini:2020ekg,Boyarsky:2020dzc}. All these arguments pose lower bounds on the mixing of the heavy neutrinos, which depend on their mass.

In Case NR2, the larger neutrino masses render them somewhat short-lived: they would decay before BBN and elude these cosmological constraints. In contrast, case NR1 is disfavoured, as it predicts lifetimes comparable to or larger than the onset of BBN. In fact, for heavy neutrino masses in the range $[3.5,200] \MeV$, predicted by Case NR1, the bound on the mixing is~\cite{Vincent:2014rja}:
\be
\sin^2\theta= \dfrac{ m_\nu}{ M_N}\lesssim10^{-15}-10^{-17}\,,
\ee
where $m_\nu$ and $M_N$ represent the overall scale expected of the light and heavy neutrino masses. Tab.~\ref{tab:HeavyNetrinos} shows the masses and mixings predicted by both Cases NR1 and NR2. Case NR1 is clearly quite disfavoured, whereas Case NR2 is safe from these cosmological constraints; although it predicts mixings larger than the limit shown above, this does not apply due to the larger masses of the heavy neutrinos.

\begin{table}[t!]
\centering
\begin{tabular}{ccc} \toprule\vspace{1mm}
    &$ M_N$ (GeV) & $\sin^2\theta$\\ \toprule\vspace{1.5mm}
  Case NR1 	& $[0.003,0.2]$ & $[2.5\cdot 10^{-10},\,1.4\cdot 10^{-8}]$\\
Case NR2 	& $[35.4,707]$ & $[7.1\cdot 10^{-14},\,1.4\cdot 10^{-12}]$ \\\bottomrule
\end{tabular}
\caption{Predictions for heavy neutrino masses and mixings in the two most relevant cases.}
\label{tab:HeavyNetrinos}
\end{table}

\section{Summary}
We have presented a setup characterized by promoting lepton number to an exact global symmetry, which is spontaneously broken by a new scalar field. Once three RH neutrinos are also introduced, light neutrino masses are generated via a type I seesaw. If the LN charges of these fermions and the new scalar are properly chosen, the Majoron associated to the lepton number breaking may exhibit a coupling to active neutrinos that allows to alleviate the Hubble tension. Those charge assignments provide extra suppressions to light neutrino masses, that can satisfy experimental observations without the need for extremely heavy sterile neutrinos. 

This mechanism can be embedded in a more general theory that accounts for the flavour puzzle. Lepton number is part of the Abelian flavour symmetry, which also contains a Peccei-Quinn $U(1)$ factor. The addition of a second extra scalar produces the breaking of the PQ symmetry, which, at the same time, solves the strong CP problem through the usual axion solution. Provided some particular PQ charges, the masses of the SM fermions are explained too.

The phenomenology of this setup is quite safe from most experimental bounds. The couplings of the Majoron to SM particles are very feeble, escaping the strongest constraints. Besides, the radial mode of the new scalar can be quite heavy with no need for fine-tuning, yielding negligible scalar mixing and Higgs invisible decay into Majorons. More interesting signatures could arise in the neutrino sector, as this model predicts new states at scales reachable in different experiments. However, cosmological arguments must be taken into account, as they may disfavour the parameter space predicted by some lepton number assignments.

\chapter{An inverse seesaw Majoraxion mechanism}
\fancyhead[RO]{\scshape \color{lightgray}8. An inverse seesaw Majoraxion mechanism}
As described in Sec.~\ref{sec:LN_LSS}, the inverse seesaw is a specific realization of the so-called low-scale seesaw mechanisms. These models are based on an approximate global lepton number symmetry, which is softly broken. As active neutrino masses are protected by this symmetry, such a small breaking naturally generates very light neutrinos, with no need for an extra suppression given by very large Majorana masses. In fact, the new states predicted by these models can be somewhat light while still mixing considerably, being potentially testable at current or future experiments.

Low-scale seesaws extend the SM particle content with two different sets of heavy neutral leptons, $N_R$ and $S_R$, which exhibit different transformation properties under LN. Generically, their mass terms are given by 
\be
\sL_{\rm mass}=-\dfrac{1}{2}\ov{N_L^c}\cM N_L+\hc\,,
\label{GenericNeutralMassLag}
\ee
where $N_L\equiv(\nu_L,\, N_R^c,\, S_R^c)^T$ contains all the neutrino fields. The mass matrix reads
\be
\cM=
\begin{pmatrix}
0 & \dfrac{v}{\sqrt2} Y_\nu & \epsilon\dfrac{v}{\sqrt2} Y^\prime_\nu\\
\dfrac{v}{\sqrt2} Y_\nu^T & \mu' & \Lambda\\[4mm]
\epsilon\dfrac{v}{\sqrt2} Y^{\prime T}_\nu & \Lambda^T & \mu \\
\end{pmatrix}\,,
\label{genericISSmatrix}
\ee
where $\epsilon$ is a real parameter, $Y_\nu$ and $Y'_\nu$ are the Dirac Yukawa matrices that couple the SM lepton doublet to $N_R$ and $S_R$ respectively, and $\Lambda$, $\mu$ and $\mu'$ are matrices in the flavour space of $N_R$ and $S_R$.  Particular models exhibit different textures in this matrix. For instance, in the inverse seesaw, the (1,3) block vanishes, and the violation of lepton number is given by $\mu$ and $\mu^\prime$; the linear seesaw focuses on the opposite case, where $\mu$ and $\mu^\prime$ are negligible and lepton number breaking is controlled by $\epsilon$. Note that the shape of the matrix shown above applies generally for any number of $N_R$ and $S_R$, which could even be different.

The Majorana masses of $N_R$ and $S_R$, as well as the coupling of the latter to the lepton doublet, explicitly violate lepton number. Thus, $\mu$, $\mu^\prime$ and $\epsilon$ (or their entries) should be small parameters, and $\Lambda$ should dominate, as it conserves lepton number. This pattern yields a suppressed mass for active neutrinos, which has two tree level contributions,
\be
\mathcal{M}_\nu\simeq\dfrac{v^2}{2}\left[\left(Y_\nu\dfrac{1}{\Lambda^T}\mu\dfrac{1}{\Lambda}Y_\nu^T\right)-\epsilon\left(Y'_{\nu}\dfrac{1}{\Lambda}Y_\nu^T+Y_{\nu}\dfrac{1}{\Lambda^T}Y^{\prime T}_\nu\right)\right]\,.
\label{genericISSmnu}
\ee
A rough estimate shows the low-scale nature of this mechanism: assuming that $\Lambda$ is at the TeV scale, the observed neutrino masses can be reproduced with $\mu\sim\OO$(keV). This result arises with $\OO(1)$ Yukawas, showing that no fine-tuning is required to reproduce the small neutrino masses. Besides, this can be achieved with no need for small mixing between active and heavy neutrinos. Upon integrating out the heavy fields, such mixing is controlled by the coefficient of the only dimension-6 operator generated at tree level, which reads
\be
c_{d=6}=Y_\nu\dfrac{1}{\Lambda\Lambda^\dag}Y^\dag_\nu\,.
\label{Genericd6Operator}
\ee
This coefficient does not depend on $\mu$, $\mu^{\prime}$ or $\epsilon$, so the mixing is not suppressed by the small breaking of lepton number. This may lead to possibly interesting phenomenological effects in direct and indirect searches.

Several proposals~\cite{Ma:2009gu,Bazzocchi:2010dt,DeRomeri:2017oxa,Mandal:2021acg,Fernandez-Martinez:2021ypo} have been suggested in order to provide a dynamical origin for the smallness of the parameters $\mu$, $\mu^\prime$ and $\epsilon$. In a similar fashion to the mechanism described in the previous chapter, this can be realized by imposing invariance under a global $U(1)$ group, whose breaking generates these small values. This $U(1)$ can be embedded in a larger flavour symmetry, composing its Abelian part, and, interestingly, can be identified with the Peccei-Quinn symmetry, responsible of solving the strong CP problem via the introduction of an axion. Identifying the same $U(1)$ group with both LN and PQ implies the presence of a single Goldstone boson, responsible for both breakings, and thus baptized as Majoraxion. 

In order to avoid unwanted flavour violation, strongly constrained by experiments, we will once again work in the framework of Minimal Flavour Violation, which has already been discussed in an inverse seesaw setup~\cite{Gavela:2009cd,Dolan:2018yqy}. As mentioned in the previous chapter, the inclusion of neutrinos in MFV poses problems, as the presence of two new spurions hinders predictivity. This can be avoided in a minimal case: Ref.~\cite{Gavela:2009cd} showed that if only one $N_R$ and one $S_R$ are introduced, it is possible to completely determine $Y_\nu$ and $Y_\nu'$, and therefore the coefficient of the dimension-6 operator can be uniquely linked to the active neutrino masses. This represents a very predictive scenario: the radiative rare charged lepton decay rates can be expressed in terms of active neutrino masses, lepton mixing angles and Majorana phases. However, this is in general not true if more RH neutrinos are present in the spectrum, and further assumptions on the Yukawas are needed. We will study three scenarios, characterized by the different transformation properties of the corresponding spurions, and discuss how their phenomenological impact can be quite distinct. The results and figures of this chapter stem from those found in Ref.~\cite{Arias-Aragon:2022ats}.

\section{Dynamical inverse seesaw within MFV}
\label{sec:MFV}

We will extend the SM spectrum with the addition of six heavy neutral leptons, divided into two groups, three $N_R$ and three $S_R$. The main difference between these two types of fermions will reside in their transformations properties under $\UPQ$. With such a particle content, the largest possible symmetry of the kinetic terms of the Lagrangian in the lepton sector is 
\be
\cG_F=U(3)_{L_L}\times U(3)_{e_R}\times U(3)_{N_R}\times U(3)_{S_R}\,,
\ee
where each term corresponds to a lepton species. $\cG_F$ can be rewritten as a product of the non-Abelian terms,
\be
\cG_F^\text{NA}=SU(3)_{L_L}\times SU(3)_{e_R}\times SU(3)_{N_R}\times SU(3)_{S_R}\,,
\ee
and of four Abelian factors,
\be
\cG_F^\text{A}=U(1)_{L_L}\times U(1)_{e_R}\times U(1)_{N_R}\times U(1)_{S_R}\,,
\ee
that can be rearranged in order to make explicit some interesting properties of the theory. Once the flavour symmetry of the quark kinetic terms is also taken into account, the product of the seven $U(1)$ symmetries can be rewritten as the global version of the hypercharge, $U(1)_Y$, baryon and lepton number, the PQ symmetry $\UPQ$, and three rotations on $e_R$, $N_R$ and $S_R$.

The PQ charge assignments can account for the appearance of a low-scale seesaw structure in the mass matrix. The condition $x_{N_R}=-x_{S_R}$ suppresses the entries proportional to $\mu$ and $\mu^{\prime}$ with respect to those proportional to $\Lambda$. If the relation $x_\ell=x_{N_R}$ is also enforced, the elements controlled by $Y'_\nu$ will be much smaller than those given by $Y_\nu$. Interestingly, with this choice of charges, PQ is directly identified with lepton number. Aside from these theoretical considerations, several observables set experimental constraints on the values of the PQ charges. The most important are given by radiative rare charged lepton decays, $\mu\to e$ conversion in nuclei, the effective number of neutrinos and the $W$ boson mass.

Notice that this setup shares several features with the model presented in the previous chapter. The spontaneous breaking of the PQ symmetry solves the strong CP problem via the presence of an axion; besides, the pattern of the SM fermion masses can be reproduced if the PQ charges are properly chosen. The main difference is that now the PQ and LN symmetries coincide, instead of being independent groups. 

In order to guarantee the invariance of the entire Lagrangian under the flavour symmetry, the Yukawa and mass matrices are promoted to spurion fields, transforming only under the flavour symmetry. The generic Lagrangian for the lepton sector reads as follows:
\begin{align}
\sL_{\rm mass}=\,&-\ov{L_L}\,H\,Y_e\, e_R-
\ov{L_L}\,\tH\, Y_\nu\, N_R-
\epsilon\,\ov{L_L}\,\tH\, Y'_\nu\, S_R\,\nonumber\\
&-\dfrac12\,\ov{N_R^c}\,\mu'\,N_R-
\dfrac12\,\ov{S_R^c}\,\mu\, S_R-
\dfrac12\left(\ov{N_R^c}\,\Lambda\,S_R+\ov{S_R^c}\,\Lambda^T\,N_R\right)
+\hc\,,
\label{GenericLagISS}
\end{align}
matching Eq.~\ref{genericISSmatrix}. These operators are invariant under $\cG_F^\text{NA}$ only if the six spurions transform as

\begin{align}
Y_e\sim({\bf3},\ov{\bf3},\bf1,\bf1)\,,\\
Y_\nu\sim({\bf3},\bf1,\ov{\bf3},\bf1)\,,\\
Y'_\nu\sim({\bf3},\bf1,\bf1,\ov{\bf3})\,,\\
\mu'\sim(\bf1,\bf1,\bf1,\ov{\bf6})\,,\\
\mu\sim(\bf1,\bf1,\ov{\bf6},\bf1)\,,\\
\Lambda\sim(\bf1,\bf1,\ov{\bf3},\ov{\bf3})\,.
\end{align}
Eq.~\ref{genericISSmnu} shows that not all these spurions are equally relevant for the generation of light neutrino masses; for instance, $\mu'$ does not contribute at tree level. Moreover, if one of the two contributions dominates, only three spurions are required: $\Lambda$, $Y_\nu$ and either $\mu$ or $Y'_\nu$. However, this is not sufficient to achieve the predictive power desirable within the MFV context. The first simplification that can be adopted is to identify the $SU(3)_{N_R}$ and $SU(3)_{S_R}$ groups into a single one, $SU(3)_{N_R+S_R}$. This way, there is no need to introduce neither two different Dirac spurions, $Y_\nu$ and $Y'_\nu$, nor two distinct Majorana mass terms, $\mu$ and $\mu'$. Furthermore, the Majorana mixed mass term, $\Lambda$, would have the same transformation properties as $\mu$ and $\mu'$:
\begin{align}
Y_\nu\sim Y'_\nu\sim({\bf3},\bf1,\ov{\bf3})\,,\\
\mu\sim\mu'\sim\Lambda\sim(\bf1,\bf1,\ov{\bf6})\,.
\end{align}
This pattern resembles the setup of the type I seesaw mechanism, characterized by two neutrino spurions, as shown in the previous chapter. Once again, this scenario, which is already simplified, is not predictive enough. To overcome this problem, we will identify three possible frameworks, dubbed \say{Case A}, \say{Case B} and \say{Case C}. Despite the differences in the flavour symmetries and in the field transformations, all three realizations correctly describe the lepton masses and mixings, once the spurions acquire specific background values. The latter depend on the particular setup, giving rise to different associated phenomenologies, as will be discussed in the following. 

\subsection{Case A}
\label{sec:CASEA}

In this scenario, the $SU(3)_{N_R+S_R}$ group is further reduced to its orthogonal version, and the flavour group results in
\be
\cG_F^{SO(3)}\equiv SU(3)_{L_L}\times SU(3)_{e_R}\times SO(3)_{N_R+S_R}\times \UPQ\,.
\label{FsymCaseA}
\ee
This way, the mass terms are proportional to the identity matrix, 
\be
\mu,\,\mu'\,,\Lambda\propto\unity\,,
\ee
while the only remaining neutrino spurion with non-trivial flavour structure transforms as
\be
Y_\nu\sim Y'_\nu\sim({\bf3},\bf1,{\bf3})\,.
\ee
The relevant transformation properties can be read in Tab.~\ref{Tab.CaseA}. Interestingly, in this scenario, the number of heavy neutral leptons can be free, although Ref.~\cite{Gavela:2009cd} showed that the minimal scenario compatible with data from neutrino oscillations contains only one $N_R$ and one $S_R$. We will only consider the $3+3$ scenario, but many of the results can be easily extended to the $n+n$ case.

\begin{table}[t!]
\centering

\begin{tabular}{ccccc} \toprule\vspace{1mm}
    &  $SU(3)_{L_L} $ & $SU(3)_{e_R}$ & $SO(3)_{N_R+S_R}$ & $U(1)_{\rm PQ}$ \\ \toprule
    $L_L$&$\bf 3$ & \bf{1} &  \bf{1} & $x_\ell$\\
    $e_R$   	& \bf{1}  			& $\bf 3$	& \bf{1}    & $x_e$\\\midrule
    $N_R$  		& \bf{1}	    & \bf{1}	& $\bf 3$   & $x_\ell$\\
    $S_R$    	& \bf{1}  			& \bf{1}	& $\bf 3$	& $-x_\ell$\\
    $\Phi$  	& \bf{1}		& \bf{1}	& \bf{1}		& $-1$ \\\midrule
  $Y_e$   	& $\bf3$  		& $\bf\ov3$	& \bf{1}   & 0\\
  $Y_\nu$	    & $\bf3$ 		& \bf{1}	& $\bf3$	& 0\\ \bottomrule
\end{tabular}
\caption{Transformation properties of the SM leptons, new fields and spurions under the global symmetries $\cG_F^{SO(3)}$.}
\label{Tab.CaseA}
\end{table}

The mass Lagrangian invariant under this symmetry can be written as 
\begin{align}
\sL^\text{A}_{\rm mass}=&-\ov{L_L}HY_ee_R\left(\dfrac{\Phi}{\Lambda_\Phi}\right)^{x_e-x_\ell}-
\ov{L_L}\tH Y_\nu N_R-
c_\nu\ov{L_L}\tH Y_\nu S_R\left(\dfrac{\Phi}{\Lambda_\Phi}\right)^{2x_\ell}\\
&-\dfrac12c_N\ov{N_R^c}N_R\Phi\left(\dfrac{\Phi}{\Lambda_\Phi}\right)^{2x_\ell-1}-
\dfrac12c_S\ov{S_R^c}S_R\Phi^\dag\left(\dfrac{\Phi^\dag}{\Lambda_\Phi}\right)^{2x_\ell-1}-
\Lambda\ov{N_R^c}S_R+\hc\,,\nn
\end{align}
where $c_\nu$, $c_N$, and $c_S$ are free real parameters. The terms that contain them present trivial flavour structure; the only flavour information is contained in $Y_\nu$, that appears in both Dirac Yukawa terms. Notice that the powers of $\Phi/\Lambda_\Phi$ play the role of the generic suppressing parameters $\epsilon$, $\mu$ and $\mu^{\prime}$.

After flavour and EW symmetry breaking, the associated neutral lepton mass matrix in the notation of Eq.~\ref{GenericNeutralMassLag} reads
\be
\cM=
\left(
\begin{array}{ccc}
0  
& \dfrac{v}{\sqrt2}Y_\nu  
& c_\nu\dfrac{v}{\sqrt2}\vep^{2x_\ell}Y_\nu  \\[4mm]
\dfrac{v}{\sqrt2}Y^T_\nu  
& c_N\dfrac{v_\Phi}{\sqrt2}\vep^{2x_\ell-1}  
& \Lambda  \\[4mm]
c_\nu\dfrac{v}{\sqrt2}\vep^{2x_\ell}Y^T_\nu  
& \Lambda 
& c_S\dfrac{v_\Phi}{\sqrt2}\vep^{2x_\ell-1}  
\end{array}
\right)\,,
\ee
where the blocks $(2,2)$ and $(3,3)$ are suppressed with respect to the $(2,3)$ one, due to the presence of powers of $\vep$, and similarly for the blocks $(1,3)$ and $(1,2)$. This enforces the low-scale seesaw structure discussed previously. By block diagonalizing the mass matrix, we can identify the mass eigenvalues: at leading order, the light and heavy neutrinos acquire the masses

\begin{align}
\mathcal{M}_\nu&\simeq\dfrac{v^2}{2}\vep^{2x_\ell-1}\left(\dfrac{c_S v_\Phi}{\sqrt2\Lambda^2}-\dfrac{2c_\nu\vep}{\Lambda}\right)Y_\nu Y_\nu^T\,,\\
\mathcal{M}_N&\simeq \Lambda\,.
\label{CaseANeuMassMatrices}
\end{align}
Notice that the heavy states are degenerate at this level of approximation. The new intermediate states, where the mass matrix $\cM$ is block diagonal, are given by $\nu_L'\simeq\nu_L-\theta S_R^c$, where
\be
\theta\simeq\dfrac{v}{\sqrt2\Lambda}Y_\nu\,.
\label{ThetaCaseA}
\ee
The next step consists in writing the spurions in terms of lepton masses and entries of the PMNS matrix. Without any loss of generality, it is possible to perform flavour transformations such that the charged leptons are diagonal in both the mass and flavour bases. As a result, in the charged lepton sector,
\be
Y_e=\dfrac{\sqrt2}{v}\mathrm{diag}(m_e,\,m_\mu,\,m_\tau)\,.
\label{ChargedLeptonSpurion}
\ee
In this basis, the active neutrino mass matrix can be diagonalized by a unitary matrix, $U$, defined as
\be
\widehat{\mathcal{M}}_\nu=U^\dag \mathcal{M}_\nu U^\ast\,,
\label{DiagonalisationOfmnu}
\ee
where $\widehat{\mathcal{M}}_\nu\equiv\mathrm{diag}\left(m_1,\,m_2,\,m_3\right)$. Notice that the matrix that appears in charged current interactions (which we will call PMNS), will be the product of the unitary $U$ times a non-unitary matrix, arising from the previous block-diagonalization through $\theta$. This correction is however bounded to be small~\cite{Antusch:2014woa,Fernandez-Martinez:2016lgt} and can be neglected for neutrino oscillation phenomenology, such that the $U$ matrix is given by the usual mixing angles and Majorana phases (see Eq.~\ref{eq:pmns_majorana}).

By inverting Eq.~\ref{CaseANeuMassMatrices}, we obtain a constraint on the neutrino Yukawa spurion, such that
\be
Y_\nu\,Y_\nu^T=\dfrac{1}{f}\,U\,\widehat{\mathcal{M}}_\nu\,U^T\,,
\ee
where
\be
f\equiv\dfrac{v^2\,\vep^{2x_\ell-1}}{2\,\sqrt2\,\Lambda^2}\left(c_S\,v_\Phi-2\,\sqrt2\,c_\nu\,\vep\,\Lambda\right)\,.
\label{fCASEA}
\ee
This condition translates into the following expression for $Y_\nu$:
\be
Y_\nu=\dfrac{1}{f^{1/2}} U \widehat{\mathcal{M}}_\nu^{1/2} \cH^T\,,
\ee
where $\cH$\footnote{In the Casas-Ibarra parametrization~\cite{Casas:2001sr}, a generic complex orthogonal matrix, dubbed $R$, appears instead of $\cH$. However, $R$ can be decomposed as a product of $\cH$ and a real and orthogonal matrix, absorbing the latter by a transformation of the heavy neutrinos~\cite{Cirigliano:2006nu}, given the fact that all the Majorana terms are  proportional to the identity matrix in the present case.} is a complex orthogonal and Hermitian matrix, that can be parametrized as
\be
\cH\equiv e^{i\phi}=\unity-\dfrac{\cosh r-1}{r^2}\phi^2+i\dfrac{\sinh r}{r}\phi\,,
\ee
with $\phi$ being a matrix that depends on three additional real parameters,
\be
\phi=\left(
\begin{array}{ccc}
0
& \phi_1
& \phi_2\\
-\phi_1
& 0
& \phi_3\\
-\phi_2
& -\phi_3
& 0
\end{array}
\right)\,,
\ee
and $r\equiv\sqrt{\phi_1^2+\phi_2^2+\phi_3^2}$.
The presence of the $\cH$ matrix, that leads to a more cumbersome phenomenological discussion, is a characteristic feature of this choice of flavour symmetry, and is absent in the other two frameworks that we will consider in the following. 

The only dimension-6 operator generated at tree level by integrating out the heavy neutral leptons is the one introduced in Eq.~\ref{Genericd6Operator}, with the Wilson coefficient given by
\be
c_{d=6}=\dfrac{Y_\nu Y_\nu^\dag}{\Lambda^2}=\frac{1}{f\Lambda^2}
U\,\widehat{\mathcal{M}}_\nu^{1/2}\,\cH^T\,\cH^*\,\widehat{\mathcal{M}}_\nu^{1/2}\,U^\dagger \,.
\ee
 The dependence on $\cH$ is therefore non-trivial, and prevents the prediction of low-energy observables exclusively in terms of neutrino masses and PMNS matrix elements. 

\boldmath
\subsection{Case B}
\unboldmath
\label{sec:CASEB}
\begin{table}[b!]
\centering
\begin{tabular}{cccc} \toprule\vspace{1mm}
    &  $SU(3)_{\rm V} $ & $SU(3)_{e_R}$ & $U(1)_{\rm PQ}$ \\ \toprule
    $L_L$&$\bf 3$&\bf{1}& $x_\ell$\\
    $e_R$   	& \bf{1} 			& $\bf 3$		    & $x_e$\\\midrule
    $N_R$  		& \bf{3} 		    &\bf{1}		   & $x_\ell$\\
    $S_R$    	& \bf{3}	& \bf{1}		& $-x_\ell$\\
    $\Phi$  	& \bf{1} 			& \bf{1} 			& $-1$ \\\midrule
  $Y_e$   	& $\bf3$  		& $\bf\ov3$		    & 0\\
  $Y_\nu$	    & $\bf6$ 		& \bf{1}  			& 0\\ \bottomrule
\end{tabular}
\caption{Transformation properties of the SM leptons, new fields and spurions under the global symmetries $\cG_F^{SU(3)_{\rm V}}$ in Case B.}
\label{Tab.CaseB}
\end{table}
A second choice, Case B, consists of identifying the symmetries associated to the lepton doublets and to the heavy neutral leptons into a single vectorial unitary group. Thus, the flavour symmetry in the lepton sector reduces to
\be
\cG_F^{SU(3)_{\rm V}}=SU(3)_{\rm V}\times SU(3)_{e_R}\times U(1)_{\rm PQ}\,.
\ee
Both $N_R$ and $S_R$ transform as triplets under the vectorial group (see Tab.~\ref{Tab.CaseB} for all the relevant properties). In this case, Schur\textquotesingle s Lemma guarantees that the neutrino Dirac Yukawa matrices are singlets of the symmetry group~\cite{Bertuzzo:2009im,AristizabalSierra:2009ex}, so $Y_\nu$ and $Y'_\nu$ are proportional to the identity matrix. The mass terms keep non-trivial transformation properties:
\be
\mu\sim\mu'\sim\Lambda\sim(\ov{\bf6},\bf1)\,.
\ee
The mass Lagrangian invariant under this symmetry can be written as
\begin{align}
\sL^\text{B}_{\rm mass}=&-\ov{L_L}HY_ee_R\left(\dfrac{\Phi}{\Lambda_\Phi}\right)^{x_e-x_\ell}-
c_{\nu N}\ov{L_L}\tH N_R-
c_{\nu S}\ov{L_L}\tH S_R\left(\dfrac{\Phi}{\Lambda_\Phi}\right)^{2x_\ell}\\
&-\dfrac12c_N\ov{N_R^c}Y_N N_R\Phi\left(\dfrac{\Phi}{\Lambda_\Phi}\right)^{2x_\ell-1}-
\hspace{-1mm}\dfrac12\ov{S_R^c}Y_NS_R\Phi^\dag\left(\dfrac{\Phi^\dag}{\Lambda_\Phi}\right)^{2x_\ell-1}
\hspace{-3mm}-\Lambda\,\ov{N_R^c}Y_NS_R+\hc\,,\nn
\end{align}
where $c_i$ are once again free real parameters. The only neutrino spurion is now $Y_N$, which is symmetric, $Y_N=Y_N^T$. The mass matrix in the broken phase reads
\be
\cM=
\left(
\begin{array}{ccc}
0  
& c_{\nu N} \dfrac{v}{\sqrt2}  
& c_{\nu S} \dfrac{v}{\sqrt2}\vep^{2x_\ell}  \\[4mm]
c_{\nu N} \dfrac{v}{\sqrt2}  
& c_N\dfrac{v_\Phi}{\sqrt2}\vep^{2x_\ell-1}  \,Y_N 
& \Lambda\,Y_N  \\[4mm]
c_{\nu S}\dfrac{v}{\sqrt2}\vep^{2x_\ell}  
& \Lambda\, Y_N 
& \dfrac{v_\Phi}{\sqrt2}\vep^{2x_\ell-1}  \,Y_N 
\end{array}
\right)\,,
\ee
with the light and heavy mass matrices given by 
\begin{align}
\mathcal{M}_\nu&\simeq\dfrac{v^2}{2}\vep^{2x_\ell-1}\left(\dfrac{c^2_{\nu N} v_\Phi}{\sqrt2\Lambda^2}-\dfrac{2c_{\nu N}c_{\nu S}\vep}{\Lambda}\right)Y_N^{-1}\,,\\
\mathcal{M}_N&\simeq \Lambda\,Y_N\,.
\label{CaseBNeuMassMatrices}
\end{align}
In this case, the new neutrinos are not all degenerate, as $\mathcal{M}_N$ contains flavour information through the dependence on $Y_N$.
The mixing between the active and heavy neutral leptons, $\theta$, is given in this case by
\be
\theta\simeq\dfrac{c_{\nu N}v}{\sqrt2\Lambda}Y_N^{-1}\,.
\label{ThetaCaseB}
\ee
Once again, it is possible to invert Eq.~\ref{CaseBNeuMassMatrices} to obtain an expression for the relevant spurion: 
\be
Y_N=f\, U^\ast \widehat{\mathcal{M}}_\nu^{-1} U^\dag\,,
\ee
where
\be
f\equiv\dfrac{v^2}{2}\vep^{2x_\ell-1}\left(\dfrac{c_{\nu_N}^2v_\Phi}{\sqrt2\Lambda^2}-\dfrac{2c_{\nu_N}c_{\nu_S}\vep}{\Lambda}\right)\,.
\label{fDefinitionCaseB}
\ee
Differently from the previous case, the spurion is now uniquely determined in terms of neutrino masses and mixings and the scale $f$, guaranteeing a strong predictive power of this scenario. The Wilson coefficient of the dimension-6 operator now reads
\be
c_{d=6}=\dfrac{c_{\nu N}^2}{\Lambda^2}Y_N^{-1}\left(Y_N^{-1}\right)^\dag=\dfrac{c_{\nu N}^2}{f^2\Lambda^2}
U\,\widehat{\mathcal{M}}_\nu^2\,U^\dagger \,.
\ee

\subsection{Case C}
\label{sec:CASEC}

Finally, Case C consists of the same flavour symmetry as in the previous scenario, but $N_R$ and $S_R$ transform differently: $N_R\sim({\bf3},\bf1)$ and $S_R\sim(\ov{\bf3},\bf1)$ (see Tab.~\ref{Tab.CaseC}). Once again, Schur\textquotesingle s Lemma reduces the total number of spurions that transform under flavour, but with a different structure:
\begin{align}
&Y_\nu,\,\Lambda\propto\unity\,\\
&\mu'\sim Y^{\prime\dagger}_\nu\sim\mu^\dag\sim(\ov{\bf6},\bf1)\,.
\end{align}
\begin{table}[t!]
\centering
\begin{tabular}{cccc} \toprule\vspace{1mm}
    &  $SU(3)_{\rm V} $ & $SU(3)_{e_R}$ & $U(1)_{\rm PQ}$ \\ \toprule
    $L_L$&$\bf 3$&\bf{1}& $x_\ell$\\
    $e_R$   	& \bf{1} 			& $\bf 3$		    & $x_e$\\\midrule
    $N_R$  		& \bf{3} 		    &\bf{1}		   & $x_\ell$\\
    $S_R$    	& $\bf\ov3$	& \bf{1}		& $-x_\ell$\\
    $\Phi$  	& \bf{1} 			& \bf{1} 			& $-1$ \\\midrule
  $Y_e$   	& $\bf3$  		& $\bf\ov3$		    & 0\\
  $Y_\nu$	    & $\bf6$ 		& \bf{1}  			& 0\\ \bottomrule
\end{tabular}
\caption{Transformation properties of the SM leptons, new fields and spurions under the global symmetries $\cG_F^{SU(3)_{\rm V}}$ in Case C.}
\label{Tab.CaseC}
\end{table}
The mass Lagrangian now reads
\begin{align}
\sL^\text{C}_{\rm mass}=&-\ov{L_L}HY_ee_R\left(\dfrac{\Phi}{\Lambda_\Phi}\right)^{x_e-x_\ell}-
c_{\nu N}\ov{L_L}\tH N_R-
c_{\nu S}\ov{L_L}\tH\,Y_N^\dag S_R\left(\dfrac{\Phi}{\Lambda_\Phi}\right)^{2x_\ell}\\
&-\dfrac12c_N\ov{N_R^c}Y_N N_R\Phi\left(\dfrac{\Phi}{\Lambda_\Phi}\right)^{2x_\ell-1}-
\hspace{-1mm}\dfrac12\ov{S_R^c}Y_N^\dag S_R\Phi^\dag\left(\dfrac{\Phi^\dag}{\Lambda_\Phi}\right)^{2x_\ell-1}\hspace{-1mm}-\Lambda\,\ov{N_R^c}S_R+\hc\,\nn
\end{align}

As in Case B, the only neutrino spurion is the symmetric $Y_N$. Notice the differences with respect to $\sL^\text{B}_{\rm mass}$: the flavour information now appears both in the Majorana and Dirac mass terms. The corresponding mass matrix is

\be
\cM=
\left(
\begin{array}{ccc}
0  
& c_{\nu N} \dfrac{v}{\sqrt2}  
& c_{\nu S} \dfrac{v}{\sqrt2}\vep^{2x_\ell} \,Y_N^\dag \\[4mm]
c_{\nu N} \dfrac{v}{\sqrt2} 
& c_N\dfrac{v_\Phi}{\sqrt2}\vep^{2x_\ell-1}  \,Y_N 
& \Lambda  \\[4mm]
c_{\nu S}\dfrac{v}{\sqrt2}\vep^{2x_\ell}\,Y_N^\dag
& \Lambda 
& \dfrac{v_\Phi}{\sqrt2}\vep^{2x_\ell-1}  \,Y_N^\dag 
\end{array}
\right)\,.
\ee 

The difference with respect to the previous case in the position of the spurion $Y_N$ leads to slightly different expressions for the neutrino masses:

\begin{align}
\mathcal{M}_\nu&\simeq\dfrac{v^2}{2}\vep^{2x_\ell-1}\left(\dfrac{c^2_{\nu N} v_\Phi}{\sqrt2\Lambda^2}-\dfrac{2c_{\nu N}c_{\nu S}\vep}{\Lambda}\right)Y_N^\dag\,,\\
\mathcal{M}_N&\simeq \Lambda\,.
\label{CaseCNeuMassMatrices}
\end{align}
In this case, the six heavy neutral leptons are degenerate at this level of approximation. The mixing between the active and heavy neutral leptons is now given by
\be
\theta\simeq\dfrac{c_{\nu N}v}{\sqrt2\Lambda}\,.
\label{ThetaCaseC}
\ee
In contrast to the two previous scenarios, this mixing is independent of any flavour information, leading to sensibly different phenomenological results. 

The neutrino spurion now reads
\be
Y_N=\dfrac1f\, U^\ast \widehat{\mathcal{M}}_\nu\,U^\dag\,,
\ee
where $f$ has the same expression as in Case B (Eq.~\ref{fDefinitionCaseB}). Although the spurions are uniquely determined in terms of lepton masses and mixing and the scale $f$, the low-energy phenomenology does not contain any flavour information, as the coefficient of the dimension-6 operator simply reads
\be
c_{d=6}=\dfrac{c_{\nu N}^2}{\Lambda^2}\,.
\ee

\section{Phenomenological signatures}
\label{sec:Majoraxion_pheno}
In the three scenarios we have presented, low-energy observables are affected by the non-unitarity of the PMNS matrix, giving rise to possible deviations from SM predictions, or even to new processes, such as lepton flavour violation. In all generality, these non-unitarity effects can be parametrized as 
\be
\cN=(\unity-\eta)\,U\,,
\ee
where $\cN$ is the actual matrix that relates the mass and interaction bases, $U$ is unitary and $\eta$ is expected to contain small entries. The latter is given by the mixing between active and heavy neutrinos, $\theta$~\cite{Fernandez-Martinez:2007iaa}:
\be
\eta\equiv\dfrac12\theta\theta^\dag.
\ee
Depending on the flavour symmetry under study, $\eta$ will exhibit different dependences on neutrino masses, PMNS entries, LN charges and other inputs, leading to quite distinct phenomenologies.

We will focus on a set of precision electroweak and flavour observables, that are particularly sensitive to this kind of new physics. In particular, we will discuss the radiative rare decay of the muon, muon conversion in nuclei, the effective number of neutrinos, $N_\nu$, (as determined by the invisible width of the $Z$) and the mass of the $W$ boson. The analytical expressions for these observables in terms of $\eta$ are well known in the literature (see Refs.~\cite{Langacker:1988ur,Antusch:2006vwa,Antusch:2014woa,Fernandez-Martinez:2016lgt})\footnote{For simplicity, we write all SM expressions for the different observables at tree level. However, the SM loop corrections are numerically very relevant for these electroweak precision observables, and have been taken into account in our numerical analysis. Conversely, loop corrections involving the new heavy neutral leptons can be safely neglected~\cite{Fernandez-Martinez:2015hxa}.}. We will employ as input parameters the fine structure constant, the $Z$ boson mass, and Fermi\textquotesingle constant, which is extracted from the muon decay. The corresponding values are~\cite{ParticleDataGroup:2020ssz}
\begin{align}
\alpha&=\,7.2973525693(11)\cdot 10^{-3},\\
G_\mu&=\,1.1663787(6)\cdot 10^{-5}\,\mathrm{GeV}^{-2},\\
M_Z&=\,91.1876(21)\,\mathrm{GeV}\,.
\end{align}
The non-unitarity of the mixing matrix $\cN$ implies a modification of the dominant muon decay, whose decay rate reads
\be
\Gamma_\mu\simeq\dfrac{m_\mu^5G_F^2}{192\pi^3}\left(1-2\eta_{ee}-2\eta_{\mu\mu}\right)\equiv\dfrac{m_\mu^5G_\mu^2}{192\pi^3}\,.
\ee
This implies the following relation between the Fermi constant parameter, $G_F$, that enters the Fermi Lagrangian, and its experimental determination via muon decay, $G_\mu$:
\be
G_F=G_\mu\left(1+\eta_{ee}+\eta_{\mu\mu}\right)\,.
\ee
Once the non-unitarity corrections are included, the relation between the $W$ boson mass and $G_\mu$ now reads
\be
    M_W = M_Z \sqrt{\frac{1}{2}+\sqrt{\frac{1}{4}-\frac{\pi\alpha(1-\eta_{\mu\mu}-\eta_{ee})}{\sqrt{2}G_\mu M_Z^2}}}\,,
\ee
while its experimental determination is $M_W=80.379(12)\GeV$~\cite{ParticleDataGroup:2020ssz}. Imposing a match between both quantities will lead to limits on the entries of $\eta$.

Another consequence of the non-unitarity of $\cN$ is the modification of the $Z$ boson invisible decay into neutrinos:
\be
\Gamma_Z^{\rm inv}=\frac{G_\mu M_Z^3}{12\sqrt{2}\pi}(3-4\eta_{\tau\tau}-\eta_{ee}-\eta_{\mu\mu})\equiv \dfrac{G_\mu M_Z^3N_\nu}{12\sqrt{2}\pi}\,,
\ee
where $N_\nu$ is the number of active neutrinos. By comparing with the experimental determination, $N_\nu=2.9963(74)$~\cite{Janot:2019oyi}, it is possible to further constrain $\eta$.

Lepton flavour violating processes may be generated if $\eta$ contains non-zero off-diagonal entries. An instance of such are radiative rare charged lepton decays. If the heavy neutrino masses are much larger than the EW scale, the corresponding branching ratios are given by
\be
\BR(\ell_i\to \ell_j\gamma)\equiv\dfrac{\Gamma(\ell_i\to \ell_j\gamma)}{\Gamma(\ell_i\to \ell_j\nu\ov\nu)}=\frac{3\alpha}{2\pi}\vert\eta_{ij}\vert^2\,.
\label{BRmutoegamma}
\ee
Similarly, $\mu \to e$ conversion in nuclei may be allowed. The corresponding relative rate reads~\cite{Alonso:2012ji}
\begin{equation}
R_{\mu\to e}={\scriptsize\frac{\sigma\left(\mu^- X\rightarrow e^- X\right)}{\sigma\left(\mu^- X\rightarrow \text{Capture}\right)}}\normalsize
\simeq\frac{G_\mu^2 \alpha^5m_\mu^5}{2s_w^4\pi^4\Gamma_{\text{capt}}}
\frac{Z^4_{\text{eff}}}{Z}\vert\eta_{e\mu}\vert^2F_p^2\Big[\left(A+Z\right)F_u+\left(2A-Z\right)F_d \Big]^2\,,
\label{MutoeConversion}
\end{equation}
where $X$ stands for a nucleus with atomic mass number $A$ and (effective) atomic number $Z$ ($Z_\text{eff}$). $F_p$ is a nuclear form factor~\cite{Kitano:2002mt,Suzuki:1987jf}, whose values are summarized in Tab.~\ref{Tab.MutoEConversionParameters} for some relevant elements. On the other hand, $F_u$ and $F_d$ are form factors associated to the neutrino physics parameters, defined in Refs.~\cite{Alonso:2012ji,Fernandez-Martinez:2016lgt}:
\begin{align}
F_u=&\dfrac23s^2_w\dfrac{16\log\left(x_N^2\right)-31}{12}-\dfrac{3+3\log\left(x_N^2\right)}{8}\,,\\
F_d=&-\dfrac13s^2_w\dfrac{16\log\left(x_N^2\right)-31}{12}-\dfrac{3-3\log\left(x_N^2\right)}{8}\,,
\end{align}
where $x_N$ is the ratio between the masses of the heavy neutrinos and the $W$ boson. $s^2_w$ can be taken at its SM predicted value, $s^2_w=0.22377(10)$ (on-shell scheme)~\cite{ParticleDataGroup:2020ssz}, as the non-unitarity  deviations would correspond to higher order corrections. Imposing the current upper bounds on the rates of these flavour violating processes, that are summarized in Tab.~\ref{Tab.ExperimentalBounds}, yields constraints on the off-diagonal elements of $\eta$.
\begin{table}[t!]
\centering

\begin{tabular}{cccc} \toprule\vspace{1mm}
    &  $Z_\text{eff} $ & $F_p$ & $\Gamma_\text{capt} \,(10^6\,\text{s}^{-1})$ \\ \toprule\vspace{1.5mm}
    ${}^{27}_{13}$ Al& $11.5$ & $0.64$ & $0.7054$ \\\vspace{1.5mm}
    ${}^{48}_{22}$ Ti& $17.6$ & $0.54$ & $2.59$\\
    ${}^{197}_{\,\,\,79}$ Au& $33.5$ & $0.16$ & $13.07$ \\\bottomrule
\end{tabular}
\caption{Atomic mass number $A$, (effective) atomic number $Z$ ($Z_\text{eff}$), form factor $F_p$ and capture rate $\Gamma_\text{capt}$ of the aluminum, titanium and gold nuclei. Values from Refs.~\cite{Suzuki:1987jf,Kitano:2002mt}.}
\label{Tab.MutoEConversionParameters}
\end{table}

\begin{table}[t!]
\centering
\begin{tabular}{ccc} \toprule\vspace{1mm}
      & Experimental Bound& Future Sensitivity \\ \toprule\vspace{1.5mm}
    $\BR(\mu\to e \gamma)$ &$4.2\cdot 10^{-13}$~\cite{MEG:2016leq}& $6\cdot 10^{-14}$~\cite{Baldini:2021kfb} \\\vspace{1.5mm}
    $\BR(\tau\to e \gamma)$ & $3.3\cdot 10^{-8}$~\cite{BaBar:2009hkt}& $9\cdot 10^{-9}$~\cite{Banerjee:2022xuw}  \\
    $\BR(\tau\to \mu \gamma)$ & $4.2\cdot 10^{-8}$~\cite{Belle:2021ysv}& $6.9\cdot 10^{-9}$~\cite{Banerjee:2022xuw} \\\midrule\vspace{1.5mm}
    $R_{\mu\to e}\,(\text{Al})$& -- &$6\cdot 10^{-17}$~\cite{Kutschke:2011ux}\\ \vspace{1.5mm}
    $R_{\mu\to e}\,(\text{Ti})$&$4.3\cdot 10^{-12}$~\cite{ParticleDataGroup:2020ssz} &$10^{-18}$ ~\cite{Barlow:2011zza} \\
    $R_{\mu\to e}\,(\text{Au})$&$7\cdot 10^{-13}$~\cite{ParticleDataGroup:2020ssz}& --\\ \bottomrule
\end{tabular}
\caption{Experimental determinations and future sensitivities, at $90\%$ CL, for a selected list of lepton flavour violating processes.}
\label{Tab.ExperimentalBounds}
\end{table}

We will employ all these experimental inputs to study the parameter space in terms of $x_\ell$ and the mass of the heavy neutral leptons. We will require that the neutrino oscillation parameters, summarized in Tab.~\ref{tab:osc_params}, are satisfied. We will assume that the neutrino mass hierarchy is normal, and that the lightest neutrino is massless. A wider study was performed in Ref.~\cite{Arias-Aragon:2022ats}, showing that no relevant differences appear if inverted hierarchy is assumed, or if the mass of the lightest neutrino is left as a free parameter.

Finally, an additional constraint that strongly affects the parameter space is associated to the consistency of the theory, as we will impose that the spurions always remain in the perturbative regime, requiring that the absolute values of their entries are never larger than 1.

\subsection{Case A}
\label{sec:pheno_caseA}
In this scenario, the mixing is proportional to the spurion $Y_\nu$ (Eq.~\ref{ThetaCaseA}), so the matrix $\eta$ reads
\begin{equation}
\eta=\frac{v^2}{4f\Lambda^2}
U\,\widehat{\mathcal{M}}_\nu^{1/2}\,\cH^T\,\cH^*\,\widehat{\mathcal{M}}_\nu^{1/2}\,U^\dagger \,.
\label{eq:eta_caseA}
\end{equation}
The presence of the matrix $\cH$ complicates the analysis, as it introduces extra dependences on the three parameters $\phi_i$. Also, this dependence is exponential, so perturbativity will set strong constraints, as it may be compromised even for moderate values of these quantities. To simplify the analysis, we will assume the $c_i$ coefficients to be equal to unity, $c_\nu=c_S=1$, as they are expected to be of order 1. Notice that the parameter $c_N$ does not affect the analysis at the level of approximation considered. 

\begin{figure}[t!]
    \centering
    \includegraphics[width=\textwidth]{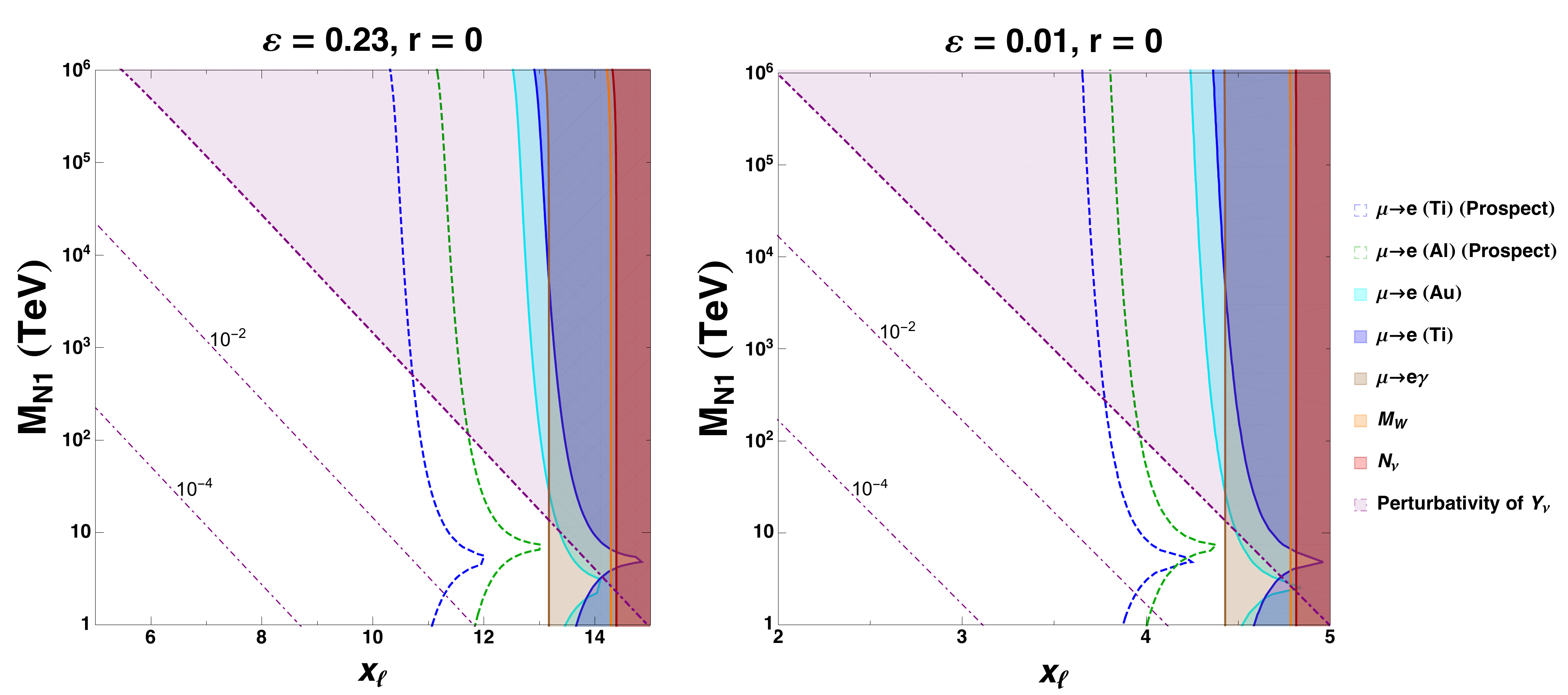}
    \caption{Parameter space of the mass of the lightest heavy neutral lepton as a function of the PQ charge, assuming $r=0$. The plot in the left (right) holds for $\vep=0.23$ ($\vep=0.01$).}
    \label{fig:CaseA_MFV}
\end{figure}

Fig.~\ref{fig:CaseA_MFV} provides a first intuition on the parameter space of this scenario, composed of the PQ charge $x_\ell$ and the mass of the lightest heavy neutral lepton, $M_{N1}$, in the specific case where $\phi_i=0$. The PQ-expanding parameter is set to $\vep=0.23$ ($\vep=0.01$) on the left (right) panel. In this simplified case, as $\cH=\unity$, the dependence on the Majorana phases disappears, according to Eq.~\ref{eq:eta_caseA}.

The shaded areas are excluded at $90\%$ CL by current experimental data, namely by the number of active neutrinos, the $W$ boson mass, $\mu\to e\gamma$ decay and $\mu\to e$ conversion in titanium and gold nuclei. The future prospects on the latter observable are represented by the dashed green and blue lines. We do not show the future prospect for $\mu\to e\gamma$, as it is not competitive with $\mu\to e$ conversion. The region in light purple is disfavoured by perturbativity arguments, as it leads to entries of $Y_\nu$ larger than 1. The dot-dashed purple lines indicate the regions where the absolute value of its largest entry is equal to $10^{-2}$ and $10^{-4}$.

These plots show that the bounds on $\mu\to e\gamma$ set upper limits on $x_\ell$, in particular, $x_\ell\lesssim13$ ($x_\ell\lesssim 4$) for $\vep=0.23$ ($\vep=0.01$). On the other side, the perturbativity of $Y_\nu$ sets upper bounds on the mass of the lightest heavy neutral lepton. Furthermore, if a natural value for this parameter is required, such a mass is constrained to a particular range, given by the chosen $x_\ell$. For instance, for $\vep=0.23$ and $x_\ell=13$, requiring $Y_\nu\in[0.01,1]$ imposes $M_{N1}\in[0.2,17]\TeV$. Conversely, lower values of $x_\ell$ imply heavier states. A comparison between both plots shows that changing the value of $\vep$ does not affect the shape of the parameter space, as it basically amounts to a rescaling of $x_\ell$. This is due to the fact that the dominant term depends on the combination $\vep^{2 x_\ell -1}$ (see the left term within the brackets in Eq.~\ref{fCASEA}).

\begin{figure}[t!]
    \centering
    \includegraphics[width=\textwidth]{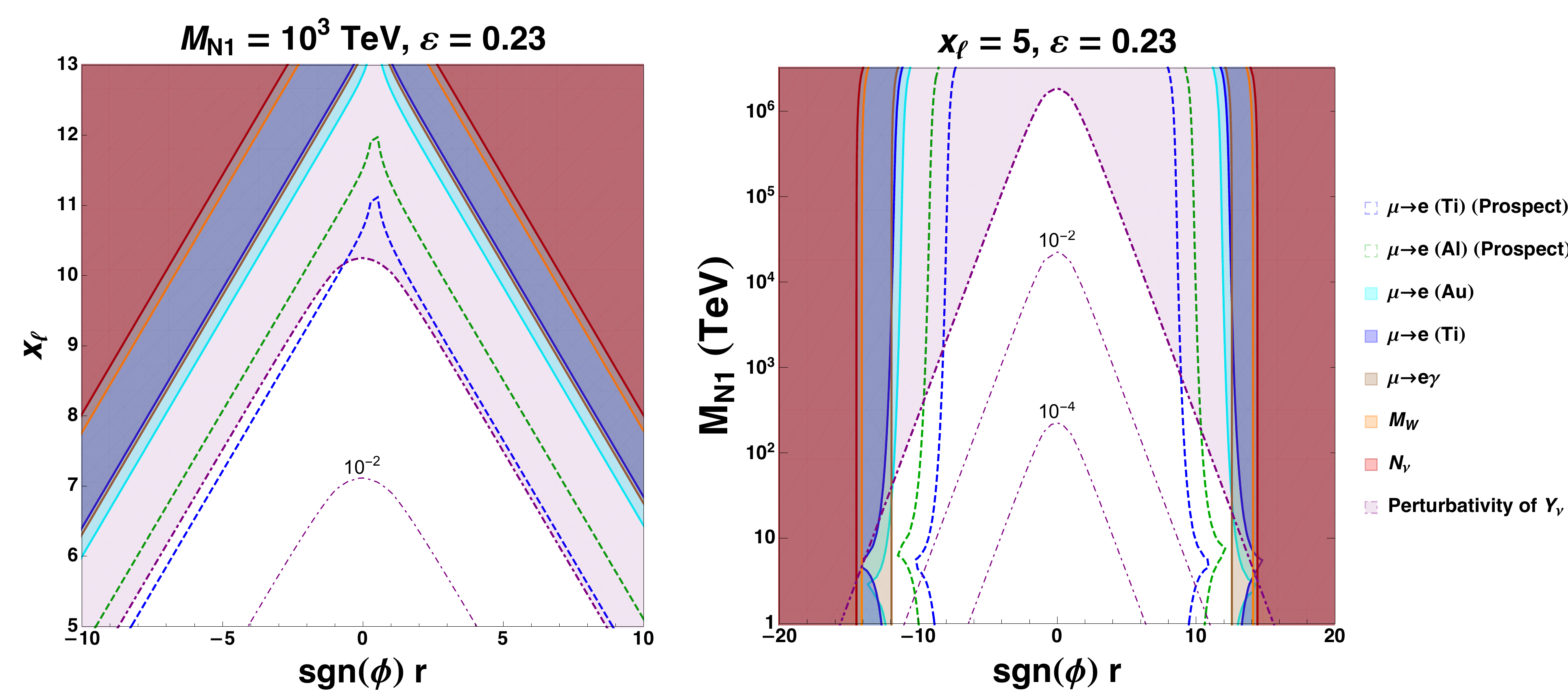}
    \caption{Parameter space of $x_\ell$ having fixed $M_{N1}=10^3\TeV$ (on the left) and of $M_{N1}$ having fixed $x_\ell=5$ (on the right), both as a function of the parameter $\sign{\phi}r$.}
    \label{fig:CaseA_Benchmarks}
\end{figure}

We will generalize this simplified analysis by considering a non-trivial $\cH$, but assuming the three $\phi_i$ phases are equal: $\phi_1=\phi_2=\phi_3\equiv\phi$. In this case, the Majorana phases do not exactly vanish in the entries of $\eta$; however, their impact is negligible, so they can be set to 0 for simplicity. In the left (right) panel of Fig.~\ref{fig:CaseA_Benchmarks} we show the dependence of $x_\ell$ ($M_{N1}$) on $r$, for $\vep=0.23$. We also consider positive and negative values of $\phi$, represented in the horizontal axis by the $\sign{\phi}$. 

Increasing the value of $r$ quickly leads to a very large neutrino Yukawa spurion, so perturbativity strongly constrains this parameter. In fact, the left panel of Fig.~\ref{fig:CaseA_Benchmarks} shows that these bounds are even stronger than the present experimental bounds. For the bechmark chosen in this plot, $M_{N1}=10^3$ TeV, $r$ has to be smaller than $\sim8$. Interestingly,  when imposing a \say{natural} value of $Y_\nu$, that is $\left|\max\left(Y_\nu\right)\right|\in[10^{-2},\,1]$, small values of $x_\ell$ require non-vanishing values of $r$. Nevertheless, values of $r>1$ imply, by construction, that $Y_\nu Y_\nu^\dagger$ is significantly larger than $Y_\nu Y_\nu^T$. This would mean that the terms generating the mixing are much larger than those generating neutrino masses, even more than what could be expected from the typical low-scale seesaw suppression, given by $\vep^{2 x_\ell -1}$. This scenario is arguably fine-tuned, as it would require a very particular texture for $Y_\nu$.

The right panel of Fig.~\ref{fig:CaseA_Benchmarks} is similar to the left one, although in this case the experimental bounds are more competitive than perturbativity ones in some regions of the parameter space. Having fixed $x_\ell=5$, the upper limit is $r<13$, which is achieved at low masses; however, the allowed range shrinks for heavier neutrinos. In this plot it can also be appreciated how non-vanishing values of $r$ are necessary in order to keep $Y_\nu$ natural.

In order to explore the parameter space without imposing any relation between the three $\phi_i$, we perform a numerical scan, generating 2000 points that satisfy $M_{N1}\in[1,10^3]\TeV$, $r\in[0.1,\,10]$, $\phi_1/\phi_2\in\pm[0.1,\,10]$ and $\phi_2/\phi_3\in\pm[0.1,\,10]$. We require that $x_\ell$ is an integer in the range $[5,13]$; larger values are disfavoured by experimental and perturbativity bounds, while the seesaw approximation does not hold for $x_\ell<5$. 

\begin{figure}[h!]
    \centering
    \includegraphics[width=\textwidth]{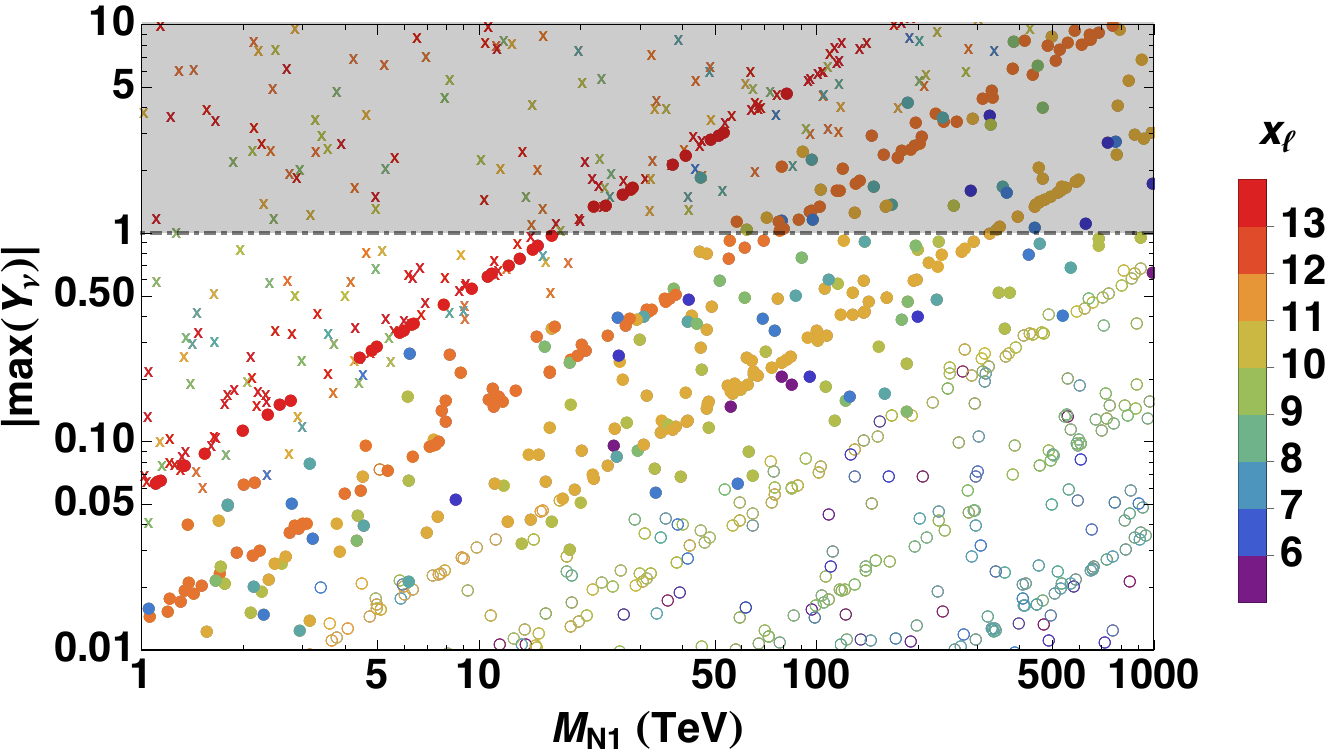}
    \caption{Scatter plot in the plane $\left|\max\left(Y_\nu\right)\right|$ vs. $M_{N1}$, assuming $\vep=0.23$. See text for details.}
    \label{fig:CaseA_Scatter_Parameters}
\end{figure}

The results of this scan are shown in Fig.~\ref{fig:CaseA_Scatter_Parameters}, which shows the parameter space composed by $\left|\max\left(Y_\nu\right)\right|$ and $M_{N1}$. The shape of the points denotes their experimental status, as crosses are ruled out and circles are allowed. Full circles correspond to points that could be probed by future experiments; if they are empty, their corresponding predictions lie beyond the prospects of future sensitivities. The color code depicts the value of $x_\ell$; the fact that this quantity is discrete explains why the points tend to cluster in diagonal lines. The region shaded in gray is disfavoured by perturbativity arguments.

This figure shows a clear pattern, as the top-left corner is populated by crosses, the center is composed of full circles and the bottom-right corner contains mainly empty circles. This tendency is followed by the color code, as the value of $x_\ell$ diminishes from top-left to bottom-right. These conclusions are similar to those obtained from Fig.~\ref{fig:CaseA_MFV}, which showed that large values of the PQ charge are disfavoured, especially for low masses. If $x_\ell$ is chosen to be somewhat small, the new neutrinos need to be quite heavy in order to avoid the bounds. However, in contrast to the case of that figure, now the allowed regions do not require very small Yukawas. This is due to the presence of a non-trivial $\cH$, which enhances them without spoiling other predictions. 

\begin{figure}[h!]
    \centering
     \includegraphics[width=\textwidth]{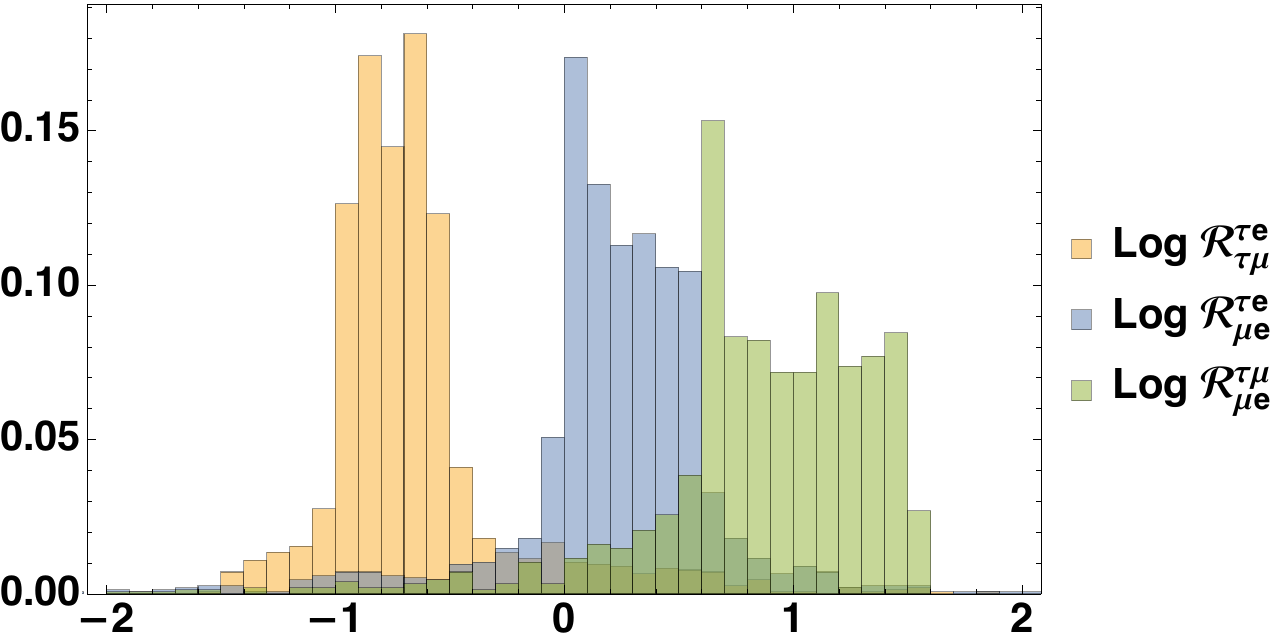}\qquad
    \caption{Density histograms for the ratios of branching ratios $\cR^{ij}_{ks}$. See text for details.}
    \label{fig:CaseA_Scatter_RatiosHisto}
\end{figure}

For each of the points generated in the numerical scan, we have computed the ratio of branching ratios of radiative rare charged lepton decays, 
\be
\cR^{ij}_{ks}\equiv \dfrac{\BR(\ell_i\to\ell_j\gamma)}{\BR(\ell_k\to\ell_s\gamma)}=\dfrac{\left|\eta_{ij}\right|^2}{\left|\eta_{sk}\right|^2}\,.
\ee
The three ratios are not independent, as $\cR^{\tau e}_{\mu e} = \cR^{\tau e}_{\tau \mu} \cR^{\tau \mu}_{\mu e}$. The corresponding distributions are shown in the histogram in Fig.~\ref{fig:CaseA_Scatter_RatiosHisto}. Although they are not too narrow, some specific values are clearly preferred, pointing to a well determined hierarchy among the branching ratios of the different processes:
\be
\BR(\mu\to e\gamma)\lesssim\BR(\tau\to e\gamma)<\BR(\tau\to\mu\gamma)\,.
\ee
Assuming the observation of $\mu\to e \gamma $ with a BR close to the present bound, the predicted values for the tau decays are $\BR(\tau\to e\gamma)\sim 1.8\cdot 10^{-13}$ and $\BR(\tau\to \mu\gamma)\sim 6\cdot 10^{-13}$, that are far from the present experimental bounds and hardly observable in the future.

\subsection{Case B}
In this case, the active-heavy mixing is inversely proportional to $Y_N$, which is the only relevant spurion. Thus, the $\eta$ matrix reads
\begin{equation}
    \eta=\dfrac{c_{\nu N}^2v^2}{4\Lambda^2}Y_N^{-1}\left(Y_N^{-1}\right)^\dag=\dfrac{c_{\nu N}^2v^2}{4f^2\Lambda^2}U\,\widehat{\mathcal{M}}_\nu^2\,U^\dagger\,.
\end{equation}
This scenario is much more predictive than Case A, due to the absence of the matrix $\cH$. This also implies that the Majorana phases do not affect the observables, further reducing the parameter space in this scenario. As in the previous case, we set $c_{\nu N}=c_{\nu S}=1$. 

\begin{figure}[t!]
    \centering
    \includegraphics[width=\textwidth]{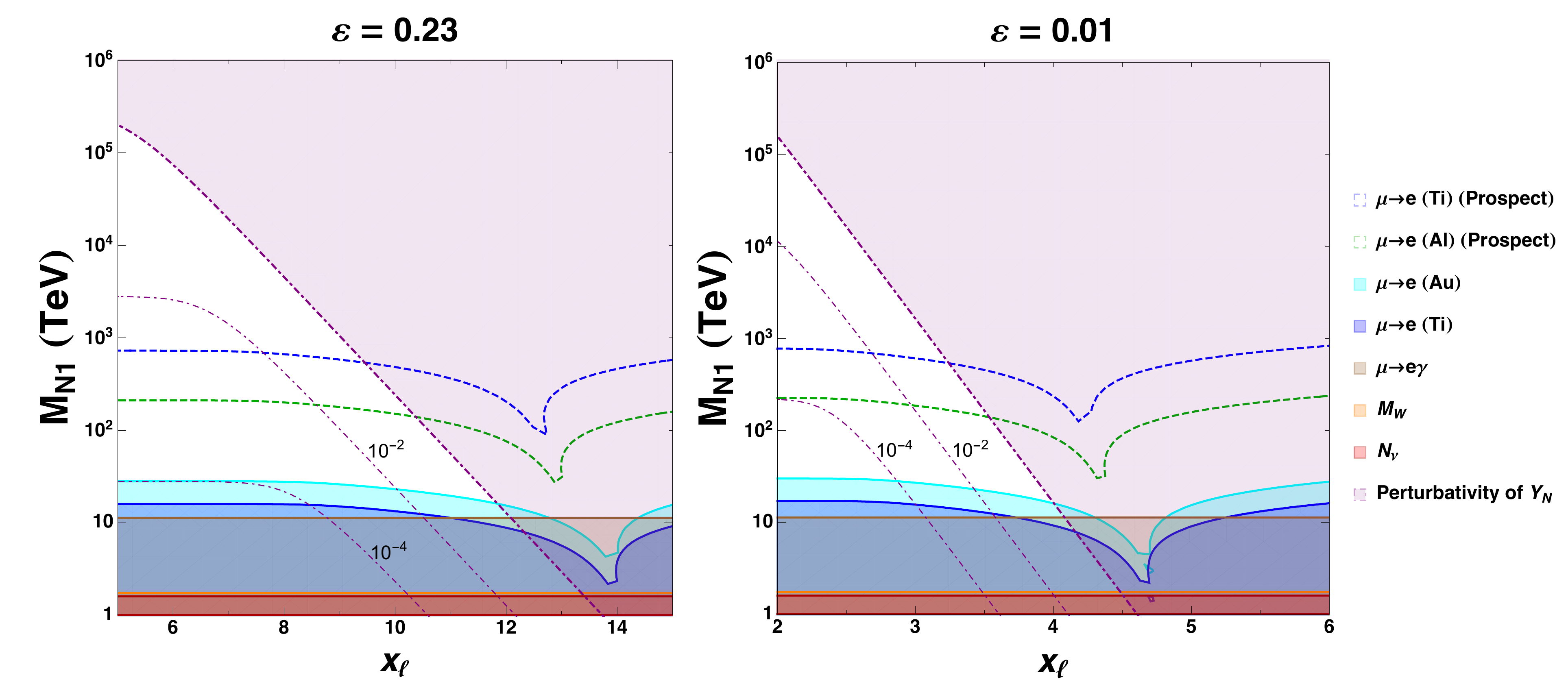}
    \caption{Parameter space of the mass of the lightest heavy neutrino as a function of the PQ charge $x_\ell$, assuming $\vep=0.23$ (0.01) in the left (right) panel.}
    \label{fig:Case2_Observables}
\end{figure}

Fig.~\ref{fig:Case2_Observables} is analogous to Fig.~\ref{fig:CaseA_Benchmarks}, showing the experimental and perturbativity bounds in the parameter space composed of the lightest heavy neutrino mass and the Peccei-Quinn charge $x_\ell$. Both plots only differ in the value of $\vep$, choosing $\vep=0.23\,(0.01)$ in the left (right) panel. As in Case A, the main effect of this variable is just a rescaling in the values of $x_\ell$, as well as slightly stronger perturbativity constraints for $\vep=0.01$.

The experimental bounds provide a lower bound for the lighest heavy neutrino mass, which cannot be smaller than, roughly, 10 TeV. In contrast to Case A, the new neutrinos now remain heavy. The perturbativity bounds provide upper bounds on both this mass, that cannot exceed $\sim10^5$ TeV, and the PQ charge, constrained to $x_\ell\lesssim 13(4)$ for $\vep=0.23\,(0.01)$. This interplay can be understood from Eq.~\ref{CaseBNeuMassMatrices}, that shows that the dominant contribution to light neutrino masses reads
\begin{equation}
    \mathcal{M}_\nu\propto \dfrac{\vep^{2x_\ell-1}}{\mathcal{M}_N^2}Y_N\,.
\end{equation}
Clearly, if $Y_N$ is to remain natural, larger values of $x_\ell$ push the heavy neutrino to be lighter, in order to reproduce the observed neutrino masses. Otherwise, the suppression given by $\vep$ would be too strong, requiring a Yukawa larger than 1. 

The predictive power of this scenario allows to obtain particular values for the ratio of branching ratios of radiative rare charged lepton decays, which roughly read\footnote{As mentioned previously, here we restrict ourselves to normal ordering. In the IO case, these values are slightly different, but they remain in the same order of magnitude and exhibit the same hierarchies. See Ref.~\cite{Arias-Aragon:2022ats} for more details on IO.}
\begin{align}
\cR^{\tau e}_{\mu e}&=1.5\,,\\
\cR^{\tau e}_{\tau \mu}&=0.06\,,\\
\cR^{\tau \mu}_{\mu e}&=27\,.
\end{align}
Once again, the $\tau\to e \gamma$ channel dominates, but the predicted rates are too small to be probed in the future, assuming the BR of the $\mu\to e\gamma$ channel lies close to the current limits. Notice that this hierarchy is in apparent contradiction to the one obtained in Ref.~\cite{Dinh:2017smk} (case EFCII). This is due to the different spurion analysis: Ref.~\cite{Dinh:2017smk} obtained that $\eta\propto Y_N^\dagger Y_N$, differently to our results. This yields a strong dependence on the lightest neutrino mass in NO, explaining the different hierarchies in this kind of decays. 

\subsection{Case C}
This scenario is by far the simplest from a phenomenological point of view, as the mixing does not depend on any spurion, and thus does not contain any flavour information. The $\eta$ matrix now reads
\be
\eta=\dfrac{c_{\nu N}^2v^2}{4\Lambda^2}\,\unity\,,
\label{CASECeta}
\ee
which, furthermore, does not depend on light neutrino masses or the PQ charge $x_\ell$. After setting $c_{\nu N}=1$, the only quantity that affects the observable is $\Lambda$, which at first approximation coincides with the heavy neutrino masses. Besides, as $\eta$ is diagonal, no lepton flavour violation is induced. Thus, the most relevant observables are the number of active neutrinos, $N_\nu$, determined in the invisible decay of the $Z$ boson, and the mass of the $W$ boson, extracted from muon decay via $G_F$. 

\begin{figure}[t!]
    \centering
    \includegraphics[width=0.485\textwidth]{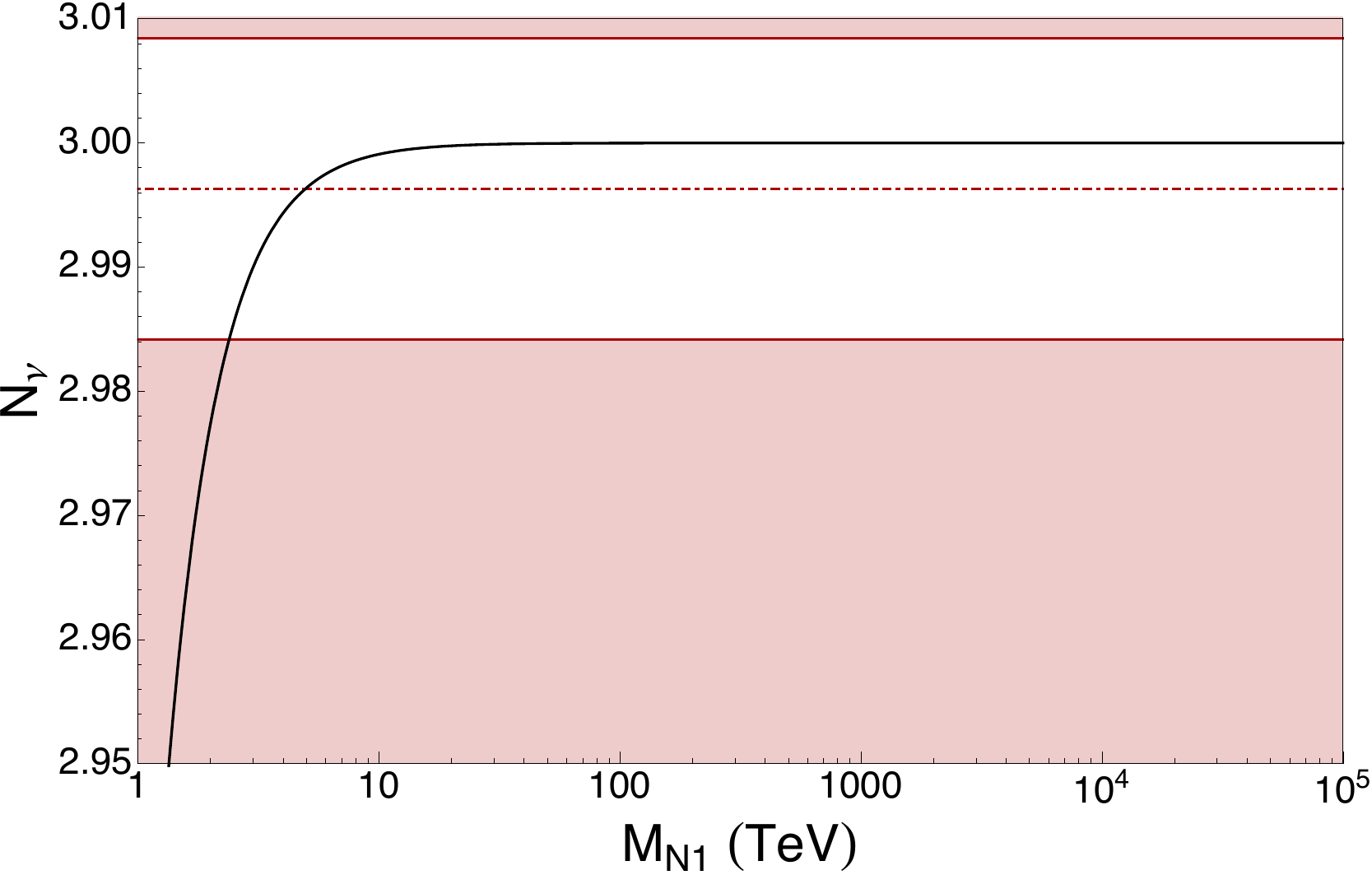}
    \includegraphics[width=0.505\textwidth]{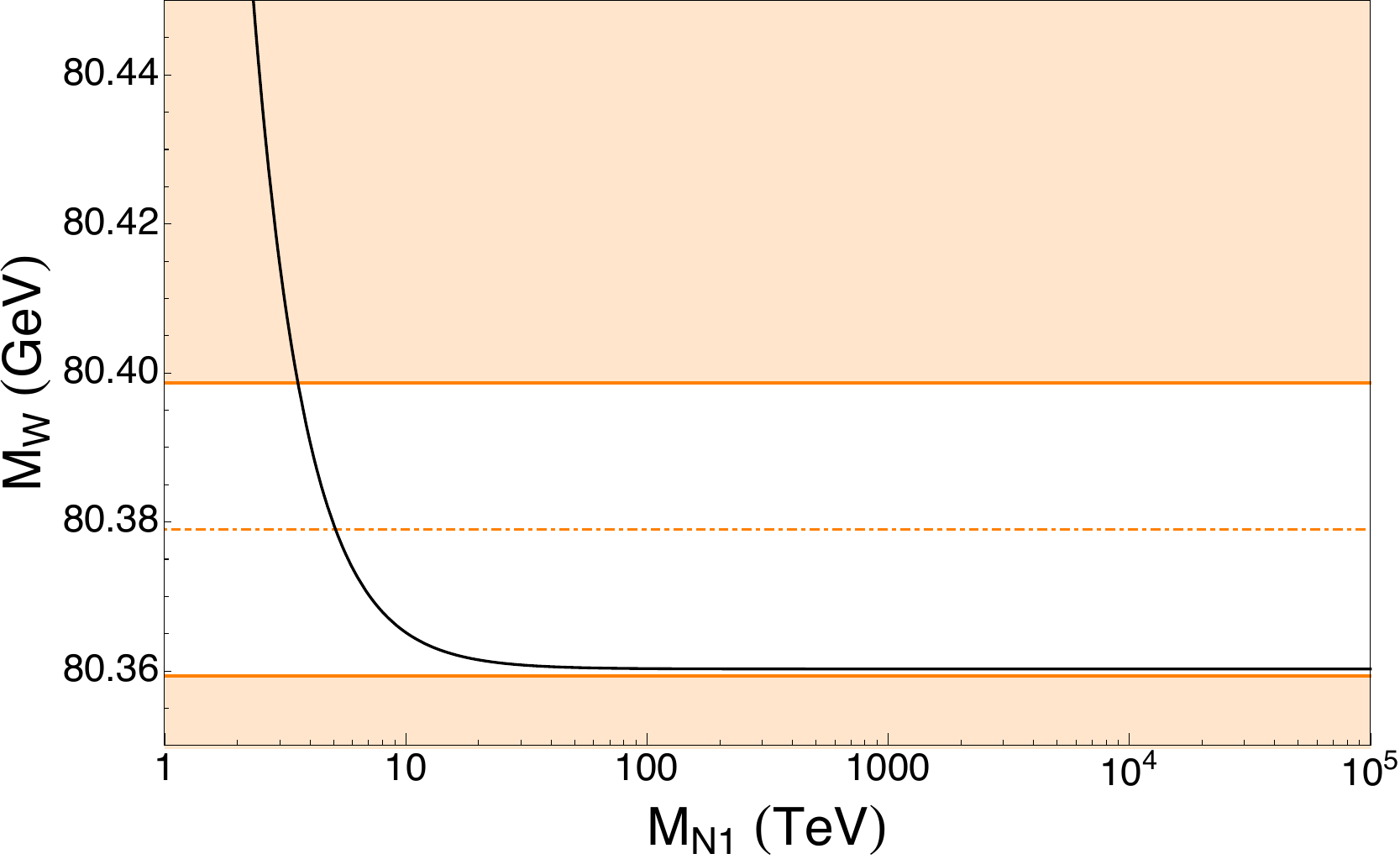}
    \caption{Predictions for the effective number of active neutrinos (left) and the $W$ gauge boson mass (right) as a function of the mass of the lightest heavy neutrino. The white areas are experimentally allowed at 90\% CL, while the dotted-dashed lines represent the best fit values.}
    \label{fig:Case3_Observables}
\end{figure}

The black lines in Fig.~\ref{fig:Case3_Observables} show the predictions of this framework for those two observables. The shaded regions are outside the 90\% CL, and are thus disfavoured. Clearly, in order to respect the experimental determinations, the heavy neutrinos cannot be too light, with lower limits at the order of the TeV. 

According to Eq.~\ref{CaseCNeuMassMatrices}, lower limits on heavy neutrino masses imply upper limits on $x_\ell$, once $\vep$ is fixed. For $\vep=0.23$, and assuming a natural Yukawa ($Y_N\in[0.01,1]$), the largest value compatible with the determination of $M_W$ is $x_\ell=13$. This would yield heavy neutrinos at roughly 4 TeV. If $\vep=0.01$ is chosen instead, the values of $x_\ell$ are once again rescaled, with $x_\ell=5$ being the largest possible value. 

\section{Summary}
We have explored a possible dynamical origin of an inverse seesaw mechanism, that generates small neutrino masses via an approximate lepton number symmetry. We have shown how this symmetry can be identified with a Peccei-Quinn one, whose spontaneous breaking both solves the strong CP problem through the usual axion solution and generates small lepton number violating parameters. Thus, the same scalar field is responsible for both breakings, and a single Goldstone boson, the Majoraxion, appears in the spectrum. The SM particle content is also extended with six heavy neutrinos, separated in two categories that exhibit different transformation properties.

This new symmetry can be embedded in a Minimal Flavour Violation setup, being part of the Abelian sector of the whole flavour symmetry. If the charges of the SM fermions are properly chosen, a Froggatt-Nielsen mechanism can explain the strong hierarchies between their masses. 

However, it is well known that the neutrino sector complicates the MFV approach, as the larger number of spurions diminishes its predictive power. In order to tackle this issue, we have identified three simplified scenarios, that differ in the flavour symmetry of the lepton sector and in the transformation properties of the new neutrinos. All three of them are able to reproduce the measured neutrino masses and mixings, but also predict different phenomenologies, that allow to constrain the corresponding parameter spaces.

The first scenario, characterized by an orthogonal symmetry group for the heavy neutrinos, predicts flavour violating processes, that pose strong constraints. However, these can be satisfied with heavy neutrinos not much beyond the TeV scale. In the second framework, where the heavy neutrinos and the charged leptons transform under the same group, the limits arising from flavour violation can only be avoided if the new states are quite heavy, around tens of TeV at least. Finally, the third setup, which only differs from the second in the properties of the heavy neutrinos, does not predict flavour violation. The mass of the $W$ boson yields the strongest limits, that still allow for heavy neutrinos at the TeV scale.

\part[Analyzing and constraining heavy neutrinos]
     {\cinzel The dark side\\[\bigskipamount] 
      \normalfont\Large Analyzing and constraining heavy neutrinos}
\fancyhead[LE]{}
\fancyhead[RO]{}
\PB
\lettrine[depth=1, loversize=0.55,lraise=-0.4]
    {T}{}he great majority of SM extensions that explain the neutrino masses and mixings observed in the oscillation phenomena include new neutrino species. These particles receive many different names, such as right-handed neutrinos, sterile neutrinos, heavy neutrinos or heavy neutral leptons (HNLs), that we will use equally. Finding these states would constitute the first discovery of a particle not included in the SM, and would very likely point towards the mechanism behind neutrino masses.

Such a firm theoretical motivation has sparked a plethora of experimental efforts in search for these particles. Either in dedicated experiments or through specific strategies in already existing facilities, heavy neutrinos have been 
studied for decades in experiments with increasing sensitivities, but no compelling evidence of their existence has been found so far. This implies that, if they exist, HNLs interact in an extremely feeble way with the particles of the SM, or are too heavy to be produced in current machines.

One of the main difficulties when it comes to probing heavy neutrinos is the lack of predictivity from the theoretical point of view, especially when it comes to their mass. For instance, the type I seesaw provides a 1-to-1 correspondence between the mass of the HNLs and their interactions with the SM, but does not provide a rationale on the scale at which these particles lie. This means that an extremely wide range has to be probed, requiring many different techniques, from oscillations (which are sensitive to eV-scale sterile neutrinos) to colliders (that could test neutrinos as heavy as the TeV). 

The experimental efforts usually opt for a model-independent point of view, as it greatly simplifies the analyses. Empirical searches aim at constraining the strength of the mixing between the active and heavy states, for a range of masses that depends on the nature of the performed search. Usually, those bounds can be directly translated into limits on the parameters of each model. 

In this episode we will study some ways to probe heavy neutrinos in a model-independent way. We will first focus on the simplest scenario, in which the HNLs only interact with the SM via their mixing with light neutrinos. We will review the basic framework and the main existing constraints on the mixing parameters. As we will see, the interactions of heavy neutrinos with mesons play a very relevant role in these searches, so we will derive a low-energy theory that includes such couplings. Employing this formalism, we will estimate the sensitivity of the future Deep Underground Neutrino Experiment to heavy neutrino mixing. Later, we will study an effective field theory approach, that encodes possible new HNL interactions, but being agnostic regarding the full model that generates them. We will collect the most relevant operators and analyze the processes they can mediate, deriving bounds on the intensity of the interactions given by the most relevant observables.

\chapter{Heavy neutrino mixing}
\fancyhead[LE]{\scshape \color{lightgray}III. Analyzing and constraining heavy neutrinos}
\fancyhead[RO]{\scshape \color{lightgray}9. Heavy neutrino mixing}

Any model that introduces RH neutrinos predicts a mixing of these particles with the light mass eigenstates (provided the former contribute to neutrino masses). This is independent of any possible new interactions that affect the HNLs, or of their Majorana/Dirac nature. If $s$ sterile neutrinos are introduced, the flavour eigenstates now read
\begin{equation}
    \nu_\alpha=\sum_{i=1}^3U_{\alpha i}\nu_i+\sum_{i=4}^{3+s}U_{\alpha i}N_i\equiv U_{\alpha i}n_i\,,
\end{equation}
where we have introduced the mass eigenbasis $n = (\nu , N)$, with an index $i$ that runs over the light and heavy mass eigenstates. The matrix $U$ relates the interaction and mass bases. Its upper-left $3\times3$ block, that rotates $\nu_{1,2,3}$ into $\nu_{e,\mu,\tau}$, is the usual PMNS, whereas the other off-diagonal entries control the mixing between the flavour eigenstates and the HNLs. These elements are expected to be small, so the heavy states will almost coincide with the RH, sterile neutrinos, which are interaction eigenstates. 

The information regarding the theory that includes heavy neutrinos is encoded in the particular expression for the mixing, as this quantity depends on the parameters of the concrete model. For instance, in the type I seesaw, the mixing is given by the Majorana mass of RH neutrinos and the Yukawa coupling that links them to the lepton doublets. However, in the following we will treat the mixing angles as free parameters, in order to be agnostic on the model responsible for generating neutrino masses. Notice that inducing neutrino masses without a heavy-active mixing requires quite contrived scenarios, usually requiring extra symmetry arguments or some fine-tuning, as a vanishing mixing would require to set some parameters to 0 or unexpected cancellations. 

As mentioned in Sec.~\ref{sec:seesaw}, the presence of mixing angles induces weaker-than-weak interactions for heavy neutrinos. As neutrino flavour eigenstates have a small component of these new particles, any gauge interaction that affects neutrinos also involves HNLs. This way, even if no new fundamental forces are assumed, heavy neutrinos can be searched for in any SM process that would produce light neutrinos, simply replacing these by the new heavy states. However, these processes are suppressed by insertions of the small mixing parameters. These interactions are described by the following Lagrangian:
\begin{equation}
\mathcal{L}^{\nu}_{\mathrm{EW}} = \frac{g}{\sqrt{2}}W^+_\mu \sum_{\alpha} \sum_{i} U_{\alpha i}^* \bar n_{i}\gamma^\mu P_L \ell_\alpha +\frac{g}{4c_w} Z_\mu \sum_{i, j} C_{ij}\bar n_{i}\gamma^\mu P_L n_j + \mathrm{h.c.} \,,
\label{eq:neutrino-currents} 
\end{equation}
where $C_{ij} \equiv \sum_\alpha U^*_{\alpha i} U_{\alpha j}$. 

A general study of heavy neutrino mixing is usually difficult to handle. The angles may possibly span very wide ranges, and the dimension of the parameter space may be quite large, especially if many heavy neutrinos are introduced. Thus, most phenomenological and experimental studies deal with simplified situations, in which only one HNL is considered. This assumption can be a good approximation if the other heavy neutrinos are too heavy or their mixing too small, in a way that they would hardly play any relevant role in the phenomenology. Furthermore, out of the three mixings that control the interactions of the HNL, it is common to assume that one dominates over the other two. This \say{single flavour dominance} implies that the heavy neutrino would only mix with one lepton flavour, appearing only in processes involving either electrons, muons or taus. This assumption is very powerful from the experimental point of view, as these charged particles usually determine the relevant processes. Thus, most searches assume that only one of the three mixings is active, setting bounds on either $\vert U_{e4}\vert^2$, $\vert U_{\mu 4}\vert^2$ or $\vert U_{\tau 4}\vert^2$, and assuming that the other two vanish. Note that the relevant quantities are the moduli squared of the mixings, as the $U_{\alpha 4}$ entries themselves are not observable. The single flavour dominance assumption often provides conservative limits, as this situation allows for less interactions of the HNL. In contrast, several non-vanishing elements would open more channels and, in general, set stronger bounds. However, let us stress that this simplified setup does not reflect the flavour structure typically found in realistic neutrino mass models. As a better approximation to their associated phenomenology, some additional benchmarks have been recently proposed in the context of the CERN’s Physics Beyond Colliders initiative~\cite{Drewes:2022akb}.

\section{Searches for heavy neutrinos}
\label{sec:mixing_bounds}
As mentioned before, there is no definitive theoretical rationale on how heavy the new neutrinos should be. This means that the experimental searches for these particles differ greatly depending on the mass of the hypothetical HNL. In the following we will summarize the main strategies employed to probe these particles.
\begin{itemize}[topsep=-1pt,itemsep=0pt]
    \item[$\bullet$]{Oscillations.} If the new neutrinos are light enough, roughly below the MeV scale, they would have an impact on the oscillation of the light states. These may oscillate into the steriles (and vice versa), altering the predictions of the 3 neutrino oscillation paradigm. Those new oscillations would be controlled by the active-sterile mixing, and the absence of their observation allows to set constraints on those parameters. 
    \item[$\bullet$]{Peak searches.} Many SM particles, including most mesons, exhibit dominant decay channels with neutrinos in the final state. Via mixing, heavy neutrinos could also be produced this way. If such a decay is 2-body, the energy spectrum of the decay channels is monochromatic. Any peak different from the SM expectation could indicate the presence  of a heavy neutrino in the final state. This strategy is employed mainly in the decays of charged mesons, that decay into a neutrino and a charged lepton. The energy of the latter, that depends on the mass of the neutrino, can be precisely determined. So far, all the experiments that employ this technique have observed energies of charged leptons consistent with massless neutrinos. This type of search allows for high luminosities, as light mesons, such as pions and kaons, can be produced in huge amounts. In fact, some of the strongest bounds on neutrino mixing have been obtained by this type of experiments, for heavy neutrinos with masses around tens or hundreds of MeV. 
    \item[$\bullet$]{Beam dumps.} This sort of experiment also exploits the decays of mesons into heavy neutrinos, but with a different strategy. A primary beam, usually composed of protons, is directed into a target, producing a great number of secondary mesons, that possibly decay in flight into HNLs. The difference with peak searches is that these experiments look for the decays of the heavy neutrinos inside a detector. If that took place, unexpected SM particles, mainly charged leptons and lighter mesons, would be produced. This type of searches can be performed in the near detectors of long-baseline neutrino oscillation facilities. In fact, these searches provide quite strong limits if the HNL lies around the GeV range.
    \item[$\bullet$]{Colliders.} If HNLs are heavier than a few GeV, they are more efficiently probed at higher energies. Colliders offer the best possibilities, generating heavy neutrinos in interactions with $W$ and $Z$ bosons. Such HNLs could then decay back into SM particles, producing rather exotic signatures, such as displaced vertices or same-sign lepton pairs (if they are Majorana). This strategy suffers from a lack of intensity, as the current luminosities are not high enough to produce a large number of neutrinos. However, the corresponding bounds remain the most competitive up to even the TeV scale.
\end{itemize}
Apart from these direct laboratory searches, there are other sources of information on heavy neutrino mixing. One of them is cosmology, as briefly mentioned in Sec.~\ref{sec:Majoron_pheno_HNL}. Even if the mixing was not large enough to thermalize HNLs with the SM plasma, these states would still be produced, and, upon becoming non-relativistic, their contribution to the energy density of the Universe could be sizable. For small masses and mixings, the lifetime of the HNLs could be larger than the age of the Universe, and their predicted population could exceed the measured dark matter component. Conversely, for heavier masses, the HNL decays could lead to an unacceptably large contribution to the energy density in radiation, $N_{\rm eff}$. Their decay products to charged particles could also raise the ionization floor after recombination and alter the CMB temperature and polarization power spectra. The combination of these effects sets very stringent bounds, stronger than the laboratory constraints in some parts of the parameter space, as shown in Refs.~\cite{Vincent:2014rja,Langhoff:2022bij}. Similarly, if HNLs decay during BBN, they might alter its predictions for the primordial abundance of light elements, leading to stringent constraints~\cite{Dolgov:2000pj,Ruchayskiy:2012si,Gelmini:2020ekg,Sabti:2020yrt,Boyarsky:2020dzc}. For larger mixing angles, however, HNLs decay before the BBN epoch, and cosmological constraints are no longer effective. In that sense, they provide an upper as well as a lower limit for the mixing of HNLs, with a direct complementarity to laboratory-based bounds.

As mentioned in Sec.~\ref{sec:seesaw}, a further way to constrain heavy neutrino mixing is given by the unitarity of the PMNS matrix. If $s$ extra neutrinos are added to the SM particle content, the matrix that rotates the flavour basis into the mass one is not $3\times3$ anymore, but rather $(3+s)\times(3+s)$. The upper-left $3\times3$ block of the whole matrix corresponds to the usual PMNS, with the difference that, in contrast to the three neutrino paradigm, it is no longer unitary, since the whole matrix is. As discussed in Sec.~\ref{sec:Majoraxion_pheno}, this has an impact on flavour and electroweak precision observables, possibly yielding discrepancies with respect to the SM predictions. These effects can be parametrized in different ways, and the corresponding parameters are directly related to the heavy neutrino mixing, as the whole matrix needs to be unitary. Non-unitarity arguments only need the experimental input of SM processes, not requiring the observation of an HNL. As such, they provide ways to bound their mixing independently of their mass.  Although dedicated searches provide stronger limits, these arguments are the only ones that exist for extremely heavy neutrinos (above the TeV scale).

From now on, we will focus on HNLs with masses in the MeV-GeV range. This is quite below the scale that appeared naturally in the first formulations of the type I seesaw. In those proposals, neutrino masses were linked to new symmetries, which broke spontaneously at very high energies, thus yielding very heavy new states. In fact, if that were not the case, very small, arguably fine-tuned, Yukawa couplings were needed in order to generate light neutrino masses. This naturalness issue is absent in low-scale seesaws, in which these light-ish particles are motivated by an approximate lepton number symmetry. HNLs in that range of masses have other appealing characteristics: they could explain the baryon asymmetry of the Universe via Akhmedov-Rubakov-Smirnov (ARS) leptogenesis~\cite{Akhmedov:1998qx,Asaka:2005pn,Shaposhnikov:2008pf}, and do not worsen the Higgs hierarchy problem, as they do not introduce an extremely high scale that corrects the Higgs mass at loop level. Also, these masses are quite interesting experimentally, as they can be probed at different facilities. As mentioned above, the constraints on heavy neutrino mixing exhibit the best sensitivities for HNLs in this range (hundreds of MeVs), thanks to the luminosity achievable at beam dumps and peak searches when producing charged pions and kaons. In fact, this is the only region of parameter space where current experiments are (almost) sensitive to mixings as small as those predicted by the type I seesaw. 

\begin{figure}[t!]
\centering
\includegraphics[width=\columnwidth]{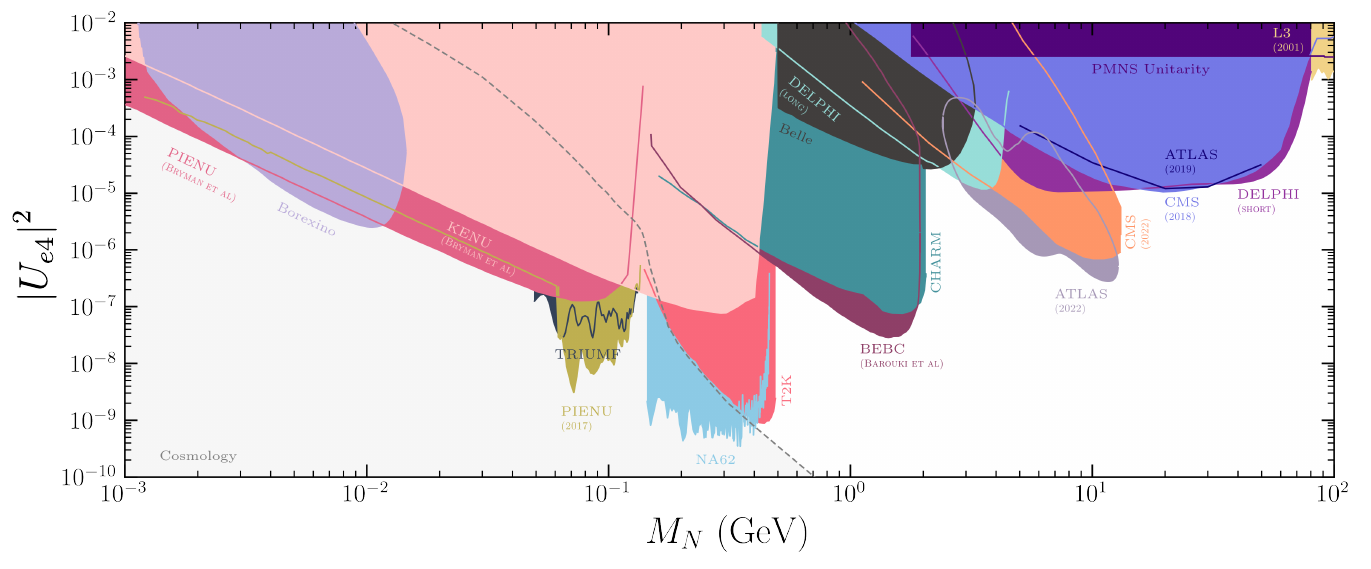}
\includegraphics[width=\columnwidth]{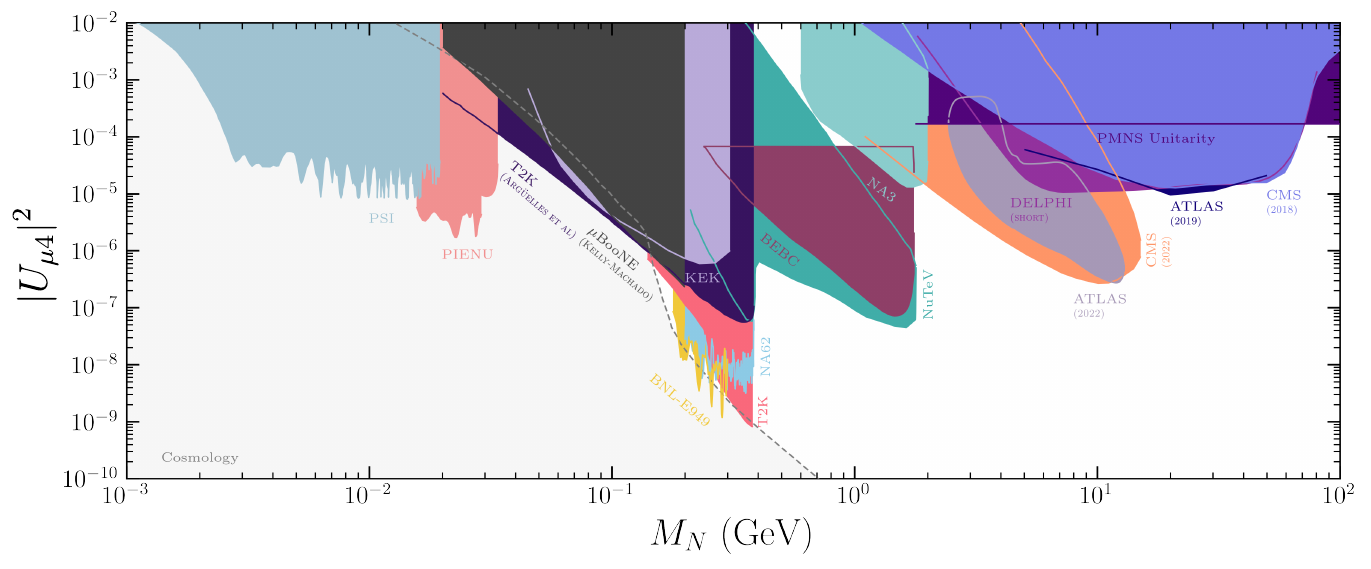}
\includegraphics[width=\columnwidth]{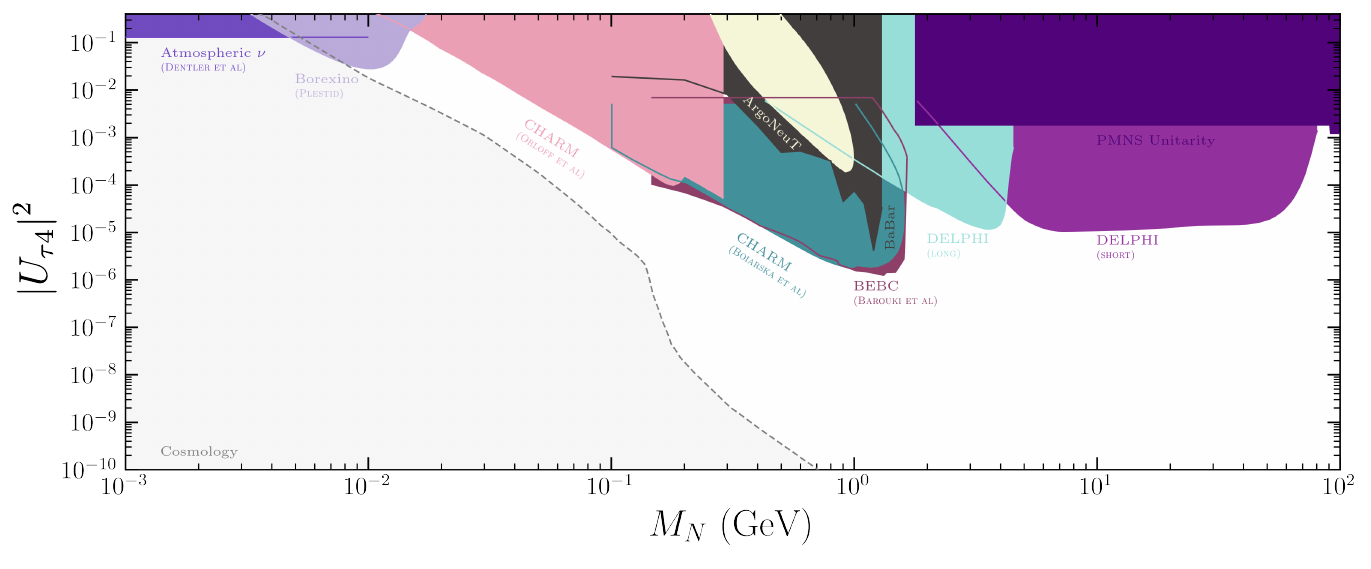}
\caption{Present constraints on heavy neutrino mixing at 90\% CL, as a function of the HNL mass~\cite{Fernandez-Martinez:2023phj}. The bounds set by the different experiments are displayed separately, but we only show those more relevant in each mass window. Single flavour dominance is assumed, with the limits on $|U_{e 4}|^2$, $|U_{\mu 4}|^2$, and $|U_{\tau 4}|^2$ depicted on the upper, middle, and lower panels, respectively.}
\label{fig:mixing_bounds_dom}
\end{figure}

A comprehensive collection of experimental limits on HNLs in this mass range, under the single flavour dominance assumption, is available on \gitlink. This database adds to the existing collection of limits on other light particles, such as axions and dark photons~\cite{Ilten:2018crw,AxionLimits}, as well as complementary efforts for HNLs~\cite{Bolton:2019pcu}. For previous studies summarizing constraints on HNLs, see Refs.~\cite{Atre:2009rg,Ruchayskiy:2011aa,deGouvea:2015euy,Drewes:2015iva,Antusch:2015mia,Fernandez-Martinez:2016lgt,Bolton:2019pcu,Chrzaszcz:2019inj} as well as recent reviews~\cite{Agrawal:2021dbo,Abdullahi:2022jlv}. Let us briefly point out the experiments that provide the strongest limits, which are displayed in Fig.~\ref{fig:mixing_bounds_dom}.

\begin{itemize}[topsep=-1pt,itemsep=0pt]
\item[$\bullet$]{Electron flavour dominance.} 
At the lowest masses, the most stringent bounds are given by the TRIUMF~\cite{Britton:1992xv} and PIENU~\cite{Aguilar-Arevalo:2017vlf} collaborations (as well as phenomenological uses of their data~\cite{Bryman:2019bjg}), employing peak searches in $\pi^+\to e^+ N$ decays. Below $16$~MeV, Borexino has searched for the $N\to \nu e^+e^-$ decays of HNLs produced in $^8$B decays in the Sun, also providing competitive limits~\cite{Borexino:2013bot}. Above the pion mass, peak searches in the $K^+\to e^+ N$ channel, as those performed by the NA62 collaboration~\cite{NA62:2020mcv}, dominate. In a similar range, competitive limits are also provided by the T2K collaboration, that employed kaons as the possible source of HNLs and searched for their decay in their near detector~\cite{Abe:2019kgx}. Above the kaon mass, the beam dump experiments CHARM~\cite{Bergsma:1985is} and BEBC~\cite{WA66:1985mfx,Barouki:2022bkt} set the most relevant limits. Finally, in the collider regime, strong bounds were placed by the DELPHI collaboration~\cite{Abreu:1996pa} studying $Z$ decays into HNLs at the LEP accelerator, searching for the prompt or displaced decays of the heavy neutrinos. As this production channel occurs via neutral currents, this bound is largely insensitive to the flavour structure of the HNL mixing. At the LHC, ATLAS~\cite{ATLAS:2019kpx, ATLAS:2022atq} and CMS~\cite{CMS:2022fut} have set some of the strongest limits by using $W$ and $Z$ decays into HNLs, that subsequently disintegrate into dilepton final states.
\item[$\bullet$]{Muon flavour dominance.}
This case follows a very similar pattern to the one above. Peak searches in pion decays, performed at the PSI~\cite{Daum:1987bg} and PIENU experiments~\cite{Aguilar-Arevalo:2019owf}, provide the best limits at low masses. Phenomenological studies have recasted limits from MicroBooNE~\cite{Kelly:2021xbv} and T2K~\cite{Arguelles:2021dqn}, yielding competitive bounds for masses at tens of MeV. Above the pion mass, the dominant studies are the decay-in-flight searches of PS191~\cite{Bernardi:1985ny,Bernardi:1987ek}\footnote{From now on, we will omit the PS191 limits, as they have been shown to have been overestimated and are subdominant to the constraints shown here~\cite{Arguelles:2021dqn,Gorbunov:2021wua}.} and T2K~\cite{Abe:2019kgx} and the peak searches (in kaon decays) performed at KEK~\cite{Hayano:1982wu,Yamazaki:1984sj}, E949~\cite{Artamonov:2009sz} and NA62~\cite{NA62:2021bji}. For HNLs heavier than the kaon, the strongest bounds are set by beam dumps, mainly BEBC~\cite{WA66:1985mfx,Barouki:2022bkt} and NuTeV~\cite{Vaitaitis:1999wq}. Lastly, as in the electron case, the leading limits on heavy neutrinos above the GeV scale are given by collider searches, namely those carried out by ATLAS~\cite{ATLAS:2019kpx, ATLAS:2022atq}, CMS~\cite{CMS:2022fut} and DELPHI~\cite{Abreu:1996pa} (with the same strategies as in the electron flavour). 
\item[$\bullet$]{Tau flavour dominance.} This case is the less probed, as the tau lepton is difficult to handle experimentally. At low masses, the only constraints are given by atmospheric neutrino data~\cite{Dentler:2018sju,Arguelles:2022tki,Blennow:2016jkn}, followed by Borexino~\cite{Plestid:2020ssy}, that searched for the decays of HNLs produced by solar neutrinos. For larger masses, heavy neutrinos can only be generated in beam dumps, via $D$ and tau decays, so the best limits are those obtained in phenomenological reanalyses of CHARM~\cite{Orloff:2002de,Boiarska:2021yho} and BEBC~\cite{WA66:1985mfx,Barouki:2022bkt} data. Note that dedicated searches were performed at the ArgoNeuT~\cite{ArgoNeuT:2021clc} and BaBar~\cite{BaBar:2022cqj} experiments, finding looser bounds. For HNLs above the GeV scale, the LHC is not able to pose constraints in this case, and the dominant limits are those set by DELPHI~\cite{Abreu:1996pa}, as they are mostly flavour blind.
\end{itemize}
Clearly, if HNLs lie close to the GeV scale, their interactions with mesons play a key role in their phenomenology. This topic has been widely studied in the literature (see Refs.~\cite{Atre:2009rg,Gorbunov:2007ak,Ballett:2019bgd,Bondarenko:2018ptm} for some examples), although some discrepancies are present in these studies. In the following, we will systematically derive the relevant interactions between mesons and heavy neutrinos.

\setuplargerchapters

\chapter{Heavy neutrino interactions with mesons}
\label{sec:mesons}
\fancyhead[RO]{\scshape \color{lightgray}10. Heavy neutrino interactions with mesons}
As it has just been explained, the interactions between mesons and heavy neutrinos are very relevant to understand the phenomenology of the latter, if their mass lies at, roughly, the GeV scale. In fact, the majority of the most stringent limits on heavy neutrino mixing are provided by experiments in which HNLs are produced in the decay of (charged) mesons. Heavy neutrino decays into final states that include mesons may also be key to detect these hypothetical particles. 

In order to describe these interactions, a low-energy effective theory is a very useful tool. On one hand, the quark degrees of freedom are not suitable in this energy regime, due to the non-perturbativity of QCD. It is necessary to introduce hadronic matrix elements, parametrized by decay constants or form factors, that mediate the transition between the initial and final meson states. From the electroweak point of view, these processes take place at energies much below the electroweak scale, so Fermi\textquotesingle s theory can be perfectly employed. These procedures yield effective operators involving neutrinos, mesons, and possibly charged leptons, that allow to compute all the processes relevant for HNL phenomenology.

All the interactions studied in this chapter were derived in Ref.~\cite{Coloma:2020lgy}, and are collected in a public FeynRules~\cite{Alloul:2013bka} model. This tool can be found in Ref.~\cite{webfeynrules}, and allows for full event generation at the differential level. The effective theory obtained here applies for any number of HNLs and for any BSM theory that generates neutrino mixing. However, in order to write a concrete FeynRules model, some simplifications are needed. For instance, this model only includes one species of heavy neutrino. Two versions of the model are available, one for Majorana neutrinos, based in the type I seesaw, and another for Dirac neutrinos, inspired by low-scale seesaws. Each of these setups exhibits distinct mixing structures (see Appendix B of Ref.~\cite{Coloma:2020lgy}), but with no deep qualitative differences. Finally, this tool contains mesons as heavy as the $B_c$; however, in the following we will only consider the $D_s$ and lighter states. These results can be easily generalized for the heavier mesons.

For convenience, let us first review the weak interactions of quarks and the corresponding currents, as they are necessary to compute the meson interactions. The weak Largangian involving quarks reads
\begin{equation}
{\mathcal{L}}^q_{\mathrm{EW}} = \frac{g}{\sqrt{2}} W^{+ \mu} j_{W, \mu}  + \frac{g}{4 c_w} Z^\mu j_{Z, \mu}  + \mathrm{h.c.} \,,
\label{eq:quark_interactions}
\end{equation}
with
\begin{equation}
j_{Z,\mu}  =  \sum_q \overline q \gamma_\mu (T_3^q - 2 Q^q s_w^2) q + \sum_q \overline q \gamma_\mu \gamma_5 (-T_3^q) q \,,
\label{eq:Z-current}
\end{equation}
and 
\begin{equation}
j_{W,\mu} =   \sum_{q=u,c,t} \sum_{q'=d,s,b} V_{q q' } \overline{q} \gamma_\mu P_L q' \,.
\label{eq:W-current}
\end{equation}
Here, $Q^q$ and $T_3^q$ stand for the electric charge and the isospin of the quark $q$ (from now on the index $q$ will be dropped for simplicity), and $V_{q q'}$ is the corresponding element of the CKM mixing matrix. It is useful to separate both currents in their vector and axial parts. This way, the $Z$ current can be decomposed as
\begin{equation}
j_{Z,\mu}=j_{Z,\mu}^V+j_{Z,\mu}^A \, ,
\label{eq:Z_VA_current}
\end{equation}
with 
\begin{align}
j_{Z,\mu}^{V} & =  \sum_q \overline{q}   (T_3 - 2Q s_w^2) \gamma_\mu q  \, ,\label{eq:Z-V} \\
j_{Z,\mu}^{A} & =  - \sum_q \overline q \gamma_\mu \gamma_5 T_3 q \,. \label{eq:Z-A}
\end{align}
Analogously, the $W$ current may be written as
\begin{equation}
j_{W,\mu}=j_{W,\mu}^{V}+j_{W,\mu}^{A}\,,
\label{eq:W_VA_current}
\end{equation}
where its vector and axial parts are given by 
\begin{align}
 j_{W,\mu}^{V} & =  \frac{1}{2} \sum_{q=u,c,t} \sum_{q'=d,s,b} V_{q q'}\overline{q} \gamma_\mu  q'  \, ,  \label{eq:W-V}\\
 j_{W,\mu}^{A} & =  - \frac{1}{2} \sum_{q=u,c,t} \sum_{q'=d,s,b} V_{q q'} \overline{q} \gamma_\mu \gamma_5 q' \,. \label{eq:W-A}
\end{align}
 We adopt a definition of the meson decay constants such that $f_\pi =130$~MeV, namely:
\begin{align}
\langle 0 | j_{a, \mu}^{A} | P_b \rangle & =  i \delta_{ab} \frac{f_P}{\sqrt{2}} p_\mu \, , \label{eq:fP}   \\
\langle 0 | j_{a, \mu}^{V} | V_b \rangle & =  \delta_{ab} \frac{f_V}{\sqrt{2}} \epsilon_\mu \, , \label{eq:fV}
\end{align}
for the pseudoscalar ($P$) and vector ($V$) mesons respectively. Here, $p_\mu$ stands for the momentum of the pseudoscalar meson and $\epsilon_\mu$ for the polarization of the vector meson (note that, with this definition, the decay constants $f_V$ have units of $[E]^2$). The values of the decay constants of the mesons considered here are summarized in Tab.~\ref{tab:Fpi}. The corresponding currents are defined as:
\begin{align}
j_{a, \mu}^{A} & =   \overline{q} \lambda_a \gamma_\mu \gamma_5 q , \label{eq:axial-currents}\\
j_{a, \mu}^{V} & =   \overline{q} \lambda_a \gamma_\mu q , \label{eq:vector-currents}
\end{align}
where
\begin{equation}
q \equiv \left( \begin{array}{c} u \\ d \\ s \end{array} \right)\,.
\end{equation}
In this notation, the set $\{\lambda_a \}$ corresponds to linear combinations of the eight Gell-Mann matrices (generators of $SU(3)$) plus the identity, normalized such that ${\rm Tr}\left\lbrace \lambda_a \lambda_b \right\rbrace = \delta_{ab}/2$ (see Appendix A of Ref.~\cite{Coloma:2020lgy} for the explicit matrices).
\begin{table}[t!]
\begin{center}
\begin{tabular}{cccc} \toprule
    \multicolumn{2}{c}{Pseudoscalars} & \multicolumn{2}{c}{Vectors} \\ \toprule\vspace{1.5mm}
    $f_{\pi}$ & 0.130 GeV & $f_{\rho}$ & 0.171 GeV$^2$ \\\vspace{1.5mm}
    $f_{K}$ &  0.156 GeV     & $f_{\omega}$ & 0.155 GeV$^2$  \\\vspace{1.5mm}
    $f_{D}$ &   0.212 GeV   & $f_{\phi}$ &0.232 GeV$^2$ \\\vspace{1.5mm}
    $f_{D_s}$ &  0.249 GeV & $f_{K^*}$ & 0.178 GeV$^2$\\\vspace{1.5mm}
    $f_\eta$ & 0.082 GeV &&\\
    $f_{\eta^\prime}$ & -0.096 GeV&&\\
    \bottomrule
\end{tabular}
\end{center}
\caption{Decay constants for pseudoscalar and vector mesons, defined as in Eqs.~\ref{eq:fP} and~\ref{eq:fV}. Most pseudoscalar decay constants are directly taken from Ref.~\cite{ParticleDataGroup:2018ovx}, except those of the $\eta$ and $\eta^\prime$, which are defined in Eqs.~\ref{eq:eta_decay_constant} and~\ref{eq:etap_decay_constant}. The decay constants of the vector mesons have been computed as described in Sec.~\ref{sec:vect_meson_decay_const}.}
\label{tab:Fpi}
\end{table}

\section{Pseudoscalar mesons}

\subsection{Neutral mesons: \boldmath{$\pi^0, \eta, \eta^\prime$}}
\label{sec:neutral-pseudo}

The quark content of the neutral pseudoscalar mesons will correspond to linear combinations of the diagonal generators $\lambda_0, \lambda_3$ and $\lambda_8$. Substituting the explicit expressions for the generators into Eq.~\ref{eq:axial-currents} we obtain:
\begin{align}
j_{3, \mu}^{A } & =  \frac{1}{2} \left[ \overline u \gamma_\mu \gamma_5 u - \overline d \gamma_\mu \gamma_5 d \right]  , \nonumber \\
j_{8, \mu}^{A } & =  \frac{1}{2\sqrt{3}} \left[ \overline u \gamma_\mu \gamma_5 u + \overline d \gamma_\mu \gamma_5 d - 2 \overline s \gamma_\mu \gamma_5 s \right]  , \label{eq:pseudoscalar-mesons} \nonumber \\
j_{0, \mu}^{A } & =  \frac{1}{\sqrt{6}} \left[ \overline u \gamma_\mu \gamma_5 u + \overline d \gamma_\mu \gamma_5 d 
+\overline s \gamma_\mu \gamma_5 s \right]  . 
\end{align}
The neutral pion can be directly identified with the current $j_{3,\mu}^{A}$, being the neutral member of the $SU(2)$ triplet of pseudo-Goldstone bosons that arises from the flavour symmetry between up- and down-type quarks. Conversely, the $\eta$ and $\eta'$ mainly correspond to the currents $j_{8,\mu}^{A}$ and $j_{0,\mu}^{A}$ respectively, although with significant mixing among them, as discussed in detail below. 

These neutral mesons can be produced or decay through NC interactions mediated by the $Z$ boson. Thus, in order to obtain their effective vertices with neutrinos, we start from Fermi\textquotesingle s theory after integrating out the $Z$, inserting the decay constant of the corresponding meson. The axial current in Eq.~\ref{eq:Z-A} can be expressed as a linear combination of the neutral axial currents as:
\begin{equation}
j_{Z, \mu}^{A} = 
-\frac{1}{2}\left( \overline u \gamma_\mu \gamma_5 u - \overline d \gamma_\mu \gamma_5 d - \overline s \gamma_\mu \gamma_5 s \right) =
-\left(  j_{3, \mu}^A + \frac{1}{\sqrt{3}}j_{8, \mu}^A - \frac{1}{\sqrt{6}}j_{0, \mu}^{A}\right)  \, .
\label{eq:axial-Z}
\end{equation}
At low energies, the amplitude, for example, for $\pi^0 \rightarrow  n \overline{n}$ would read
\begin{equation}
i\mathcal{M}_{\pi^0 n_i \overline{n_j}} = \frac{ig^2}{4c_w^2M_Z^2}C_{ij}\overline{u_i}\gamma^\mu P_L v_j \langle 0 | j_{Z, \mu}^A | \pi^0 \rangle \,,
\end{equation}
where  $\overline{u_i}$ and $v_j$ are the corresponding spinors for the neutrino mass eigenstates. Substituting the $Z$ current from Eq.~\ref{eq:axial-Z} and the corresponding hadronic matrix element from Eq.~\ref{eq:fP}, and introducing Fermi\textquotesingle s constant, 
\begin{equation}
\frac{G_F}{\sqrt{2}} =\frac{ g^2 }{8c_w^2 M_Z^2} \, , 
\end{equation}
the amplitude is given by
\begin{equation}
i\mathcal{M}_{\pi^0 n_i \overline{n_j}} = G_F C_{ij} f_\pi\overline{u_i}\gamma^\mu P_L v_j p_\mu \,,
\label{eq:amppi}
\end{equation}
where $p_\mu$ is the 4-momentum carried by the pion. Translating the momentum into a derivative, we can write down, in configuration space, the effective operator that leads to the amplitude in Eq.~\ref{eq:amppi}:
\begin{equation}
\mathcal{O}_{\pi^0 n_i \overline{n_j}} =\frac{1}{2}G_F C_{ij} f_\pi \partial_\mu  (\overline n_{i}\gamma^\mu P_L n_j) \pi^0 + \textrm{h.c.}\,
\end{equation}
Furthermore, if all particles are on shell, it is possible to apply Dirac\textquotesingle s equation to obtain Yukawa couplings proportional to the neutrino masses: 
\begin{equation}
\mathcal{O}_{\pi^0 n_i \overline{n_j}} =  \frac{i}{2} G_F C_{ij} f_\pi \overline n_i (m_i P_L - m_j P_R ) n_j \pi^0 + \textrm{h.c.}\,
\end{equation}
Since the coupling is proportional to the masses of the neutrinos, the heavy states will dominate the interaction, in complete analogy to the chiral enhancement of the charged pion decay $ \pi \to \mu  \nu_\mu$ compared to $\pi \to e \nu_e$.

Similarly, the operators associated to the other neutral pseudoscalar currents, for on-shell particles, can be obtained as:
\begin{align}
\mathcal{O}_{\eta_0 n_i \overline{n_j}} &= -\frac{i}{2}G_F  C_{ij} \frac{f_0}{\sqrt{6}} \overline n_i (m_i P_L - m_j P_R ) n_j \eta_0 + \textrm{h.c.}\, , \\
\mathcal{O}_{\eta_8 n_i \overline{n_j}}& = \frac{i}{2}G_F  C_{ij} \frac{f_8}{\sqrt{3}} \overline n_i (m_i P_L - m_j P_R ) n_j \eta_8 + \textrm{h.c.} 
\end{align}
However, unlike in the $\pi^0$ case, the $\eta$ and $\eta^\prime$ mesons mix significantly and do not correspond exactly with the quark content of the $\eta_8$ and $\eta_0$, defined through the corresponding currents in Eq.~\ref{eq:pseudoscalar-mesons}. Thus, a change of basis must be performed in order to obtain the effective vertices for the physical states. We adopt the usual parametrization for this change of basis, with two angles, $\theta_0$ and $\theta_8$ (see e.g. Ref.~\cite{Escribano:2015yup}), and define:
\begin{equation}
\left(
\begin{array}{cc}
f_{\eta,8} & f_{\eta,0} \\
f_{\eta^\prime,8} & f_{\eta^\prime,0} 
\end{array} \right) 
=
\left(
\begin{array}{cc}
f_8 \cos\theta_8 & - f_0 \sin\theta_0 \\
f_8 \sin\theta_8 & f_0 \cos\theta_0 
\end{array} \right) \, .
\end{equation}
The values for $f_0, f_8, \theta_0$ and $\theta_8$ have been taken from Ref.~\cite{Escribano:2015yup} and are summarized in Tab.~\ref{tab:eta_mixing} for convenience\footnote{Note that in Ref.~\cite{Escribano:2015yup} the authors use a different normalization for the current definitions from the one adopted here. However, this does not affect our results, since they provide theirs in terms of the ratios $f_8 /f_\pi$ and $f_0/f_\pi$, which remain unaffected by an overall normalization factor.}. Through this change of basis, the currents for the $\eta$ and $\eta'$ can be obtained as combinations of the $j_{0,\mu}^A,j_{8,\mu}^A$ currents as
\begin{align}
j_{\eta, \mu}^A&= \cos\theta_8j_{8, \mu}^A -  \sin\theta_0 j_{0, \mu}^A \, , \\
j_{\eta^\prime, \mu}^A&= \sin\theta_8 j_{8, \mu}^A  + \cos\theta_0 j_{0, \mu}^A \, .
\end{align}
Therefore, the relevant operators in the mass basis will read 
\begin{align}
\mathcal{O}_{\eta n_i \overline{n_j}} & = 
 \frac{i}{2} G_F C_{ij} \left[ \frac{\cos\theta_8 f_8}{\sqrt{3}} + 
\frac{ \sin\theta_0 f_0}{\sqrt{6}} \right] \overline n_i (m_i P_L - m_j P_R ) n_j \eta + \textrm{h.c.} \,  , \label{eq:eta_coupling}\\
\mathcal{O}_{\eta^\prime n_i \overline{n_j}} & = 
 \frac{i}{2} G_F C_{ij} \left[ \frac{\sin\theta_8 f_8}{\sqrt{3}} - 
\frac{ \cos\theta_0 f_0}{\sqrt{6}} \right] \overline n_i (m_i P_L - m_j P_R ) n_j \eta^\prime + \textrm{h.c.} \,, \label{eq:etap_coupling} 
\end{align}
which allow to define the \say{effective} decay constants
\begin{align}
    f_\eta= \frac{\cos\theta_8 f_8}{\sqrt{3}} + 
\frac{ \sin\theta_0 f_0}{\sqrt{6}}\,,
\label{eq:eta_decay_constant}\\
f_{\eta^\prime}= \frac{\sin\theta_8 f_8}{\sqrt{3}} - 
\frac{ \cos\theta_0 f_0}{\sqrt{6}}\,,
\label{eq:etap_decay_constant}
\end{align}
whose values can be found in Tab.~\ref{tab:Fpi}. Note that these figures significantly disagree from those found in Ref.~\cite{Gorbunov:2007ak}.


\begin{table}[t!]
\begin{center}

\begin{tabular}{cccc} \toprule
    \multicolumn{2}{c}{Decay constants}&\multicolumn{2}{c}{Rotation angles} \\ \toprule\vspace{1.5mm}
    $f_{0}$ & 0.148 GeV & $\theta_0$ & -6.9$^\circ$ \\
    $f_8$ &  0.165 GeV  &   $\theta_8$ & -21.2$^\circ$  \\ \bottomrule

\end{tabular}

\end{center}
\caption{Decay constants for the $\eta_0$ and $\eta_8$, and angles that parametrize the rotation to the physical basis, taken from Ref.~\cite{Escribano:2015yup} (see text for details).}
 \label{tab:eta_mixing}
\end{table}

We will not discuss any heavier neutral pseudoscalar mesons, as they are too heavy and unstable to play any relevant role in HNL phenomenology.
\subsection{Charged mesons: \boldmath{$\pi^\pm, K^\pm, D^\pm, D^\pm_s$}}

The normalized combinations of generators that reproduce the quark content of the $\pi^\pm$ and $K^\pm$ are:
\begin{align}
j_{\pi^\pm, \mu}^{A} &= \frac{1}{\sqrt{2}} \overline q \gamma_\mu \gamma_5 (\lambda_1 \mp i \lambda_2) q \, , \\
j_{K^\pm, \mu}^{A} &= \frac{1}{\sqrt{2}} \overline q \gamma_\mu \gamma_5 (\lambda_4 \mp i \lambda_5) q  \, .
\end{align}
Thus, from Eq.~\ref{eq:W-A} we obtain 
\begin{equation}
j_{W, \mu}^{A} = -\frac{1}{\sqrt{2}}\left( V_{ud} \, j_{\pi^-, \mu}^{A} + V_{us} \, j_{K^-, \mu}^{A} \right) \, .
\label{eq:Wcurrent}
\end{equation}
The amplitude for $\pi^- \to \ell^- \overline{n}$ is obtained after integrating out the $W$ boson, following the same procedure used to derive the effective vertex for the $\pi^0 \to \overline{n} n$ decay in the previous section:
\begin{equation}
i\mathcal{M}_{\pi \ell_\alpha \overline{n_i} } = \frac{ig^2}{2M_W^2}U_{\alpha i}\overline{u_\alpha} \gamma^\mu P_L v_i \langle 0 | j_{W, \mu}^A | \pi^- \rangle \, .
\end{equation}
After introducing the $W$ current defined in Eq.~\ref{eq:Wcurrent} and evaluating the hadronic matrix element, the amplitude reads
\begin{equation}
i\mathcal{M}_{\pi \ell_\alpha \overline{n_i} } = \sqrt{2}G_F  U_{\alpha i}V_{ud} f_\pi  \overline{u_\alpha} \gamma^\mu P_L v_i p_\mu\, .
\end{equation}
In the same fashion as before, we translate this amplitude to an effective operator in configuration space, with the 4-momentum $p_\mu$ as a derivative acting on the leptonic current:
\begin{equation}
\mathcal{O}_{\pi \ell_\alpha \overline{n_i} } = \sqrt{2} G_F U_{\alpha i}  V_{ud}  f_\pi \partial_\mu  (\overline {\ell_\alpha}  \gamma^\mu P_L n_i ) \pi^- + \textrm{h.c.}
\end{equation}
Once again, if all the particles involved are on shell, it is possible to obtain Yukawa couplings proportional to the fermion masses via Dirac\textquotesingle s equation:
\begin{equation}
\mathcal{O}_{\pi \ell_\alpha \overline{n_i} } =  i \sqrt{2} G_F U_{\alpha i} V_{ud} f_\pi \overline {\ell_\alpha}  (m_\alpha P_L - m_i P_R) n_i  \pi^- + \textrm{h.c.} \, 
\label{eq:effective-pion}
\end{equation}
This procedure can be repeated for the charged kaons, obtaining the same result once the corresponding decay constant and CKM element are introduced:
\begin{equation}
\mathcal{O}_{K \ell_\alpha \overline{n_i}} =  i \sqrt{2} G_F U_{\alpha i} V_{us} f_K \overline {\ell_\alpha}  (m_\alpha P_L - m_i P_R) n_i  K^- + \textrm{h.c.}\, 
\label{eq:effective-kaon}
\end{equation}
Although this framework does not include the charm quark, these results can be directly generalized to the $D^\pm$ and $D_s^\pm$ mesons, as the corresponding currents and amplitudes are completely analogous to the lighter cases. The corresponding effective operators thus read
\begin{align}
\mathcal{O}_{D \ell_\alpha \overline{n_i}} &=  i \sqrt{2} G_F U_{\alpha i} V_{cd} f_D \overline \ell_\alpha  (m_\alpha P_L - m_i P_R) n_i  D^- + \textrm{h.c.}\, ,\\
\mathcal{O}_{D_s \ell_\alpha \overline{n_i}} &=  i \sqrt{2} G_F U_{\alpha i} V_{cs} f_{D_s} \overline {\ell_\alpha}  (m_\alpha P_L - m_i P_R) n_i  D_s^- + \textrm{h.c.}\, 
\end{align}
\section{Vector mesons}

\subsection{Neutral mesons: \boldmath{$\rho, \omega, \phi $}}
As for the pseudoscalar case, the vector currents associated to the generators can be expressed in terms of the $u$, $d$ and $s$ quarks as
\begin{align}
j_{3, \mu}^{V} & =  \frac{1}{2} \left[ \overline u \gamma_\mu u - \overline d \gamma_\mu d \right]  ,  \\
j_{8, \mu}^{V} & =  \frac{1}{2\sqrt{3}} \left[ \overline u \gamma_\mu u + \overline d \gamma_\mu d - 2 \overline s \gamma_\mu s \right]  ,
\\
j_{0, \mu}^{V} & =  \frac{1}{\sqrt{6}} \left[ \overline u \gamma_\mu u + \overline d \gamma_\mu d 
+\overline s \gamma_\mu s \right]  .  
\end{align}
Considering their respective quark contents, the corresponding normalized currents for the $\rho^0$, $\omega$ and $\phi$ mesons are given by
\begin{align}
j_{\rho^0, \mu}^{V} & =  j_{3, \mu}^{V} \, ,  \\
j_{\omega, \mu}^{V} & =  \sqrt{\frac{1}{3}}j_{8, \mu}^{V} + \sqrt{\frac{2}{3}}j_{0, \mu}^{V} \, , \\
j_{\phi, \mu}^{V} & =  - \sqrt{\frac{2}{3}}j_{8, \mu}^{V} + \sqrt{\frac{1}{3}}j_{0, \mu}^{V}  \, .  
\end{align}
The production and decay of the vector mesons take place via the vector component of the $Z$ current, Eq.~\ref{eq:Z-V}, which can be written as the following linear combination of the vector meson currents:
\begin{equation}
\label{eq:vector-Z}
j_{Z, \mu}^{V} =
\left( 1 - 2 s_w^2\right) j_{\rho^0, \mu}^{V} 
-\frac{2}{3} s_w^2 j_{\omega, \mu}^{V}  
- \sqrt{2}\left( \frac{1}{2} - \frac{2}{3}s_w^2   \right) j_{\phi, \mu}^{V}\, .
\end{equation}
After integrating out the $Z$ boson, the amplitude for the $\rho^0\rightarrow\overline{n}n$ process reads:
\begin{equation}
i\mathcal{M}_{\rho^0 n_i \overline{n_j}}=\frac{ig^2}{4c_w^2M_Z^2}C_{ij}\overline{u_i}\gamma^\mu P_L v_j \left\langle 0\vert j_{Z,\mu}^V\vert\rho^0\right\rangle \, .
\end{equation}
Introducing the $Z$ vector current defined in Eq.~\ref{eq:vector-Z} and evaluating the matrix element according to Eq.~\ref{eq:fV}, we get:
\begin{equation}
i\mathcal{M}_{\rho^0 n_i \overline{n_j}}=iG_FC_{ij}f_\rho\left( 1 - 2 s_w^2\right) \overline{u_i}\gamma^\mu P_L v_j \epsilon_\mu\, ,
\end{equation}
where $\epsilon_\mu$ is the polarization vector of the $\rho^0$ meson. It is then immediate to extract the effective operator in configuration space:
\begin{equation}
\mathcal{O}_{\rho^0 n_i \overline{n_j}} =  -\frac{1}{2} G_F C_{ij} (1 - 2s_w^2) f_\rho \rho^0_\mu  (\overline{n_i} \gamma^\mu P_L n_j) + \textrm{h.c.} \, 
\end{equation}
Analogously, for the other two neutral vector mesons we obtain:
\begin{align}
\mathcal{O}_{\omega n_i \overline{n_j}} & =  \frac{1}{2} G_F C_{ij} \frac{2}{3} s_w^2 f_\omega \omega_\mu (\overline {n_i} \gamma^\mu P_L n_j )  + \textrm{h.c.} \, , \\
\mathcal{O}_{\phi n_i \overline{n_j}} & =   \frac{1}{2}G_F C_{ij} \sqrt{2}\left(\frac{1}{2} - \frac{2}{3} s_w^2\right) f_\phi \phi_\mu (\overline {n_i} \gamma^\mu P_L n_j )  + \textrm{h.c.} \, 
\end{align}
The $\phi$ meson is the only one entirely composed of a $q\overline{q}$ pair. Thus, the operator that describes it can be generalized to heavier neutral vectors with analogous quark contents, such as the $J/\psi$ ($c\overline{c}$) or the $\Upsilon$(1S) ($b\overline{b}$), upon replacing the corresponding decay constant.

\subsection{Charged mesons: \boldmath{$\rho^\pm, K^{*, \pm}$}}

In complete analogy to the charged pseudoscalars, the charged vector meson currents are given by
\begin{align}
j_{\rho^\pm, \mu}^{V} &= \frac{1}{\sqrt{2}} \overline q \gamma_\mu  (\lambda_1 \mp i \lambda_2) q \, , \\
j_{K^{*,\pm}, \mu}^{V}  &= \frac{1}{\sqrt{2}} \overline q \gamma_\mu  (\lambda_4 \mp i \lambda_5) q  \, ,
\end{align}
and the vector component of the $W$ current from Eq.~\ref{eq:W-V} can be written as
\begin{equation}
j_{W, \mu}^{V} = \frac{1}{\sqrt{2}}\left( V_{ud} \, j_{\rho^-, \mu}^{V} + V_{us} \, j_{K^{*,-}, \mu}^{V}\right) \, .
\end{equation}
The computation of the effective operators is done exactly in the same way as for the charged pseudoscalar case. The amplitude for the $\rho^-\rightarrow\overline{n}\ell^-$ process reads
\begin{equation}
i\mathcal{M}_{\rho^- \ell_\alpha \overline{n_i}}=\frac{ig^2}{2M_W^2}U_{\alpha i}\overline{u_\alpha} \gamma^\mu P_L v_i \langle 0 | j_{W, \mu}^V | \rho^- \rangle = i\sqrt{2}G_F U_{\alpha i}V_{ud}f_\rho \epsilon_\mu \overline{u_\alpha} \gamma^\mu P_L v_i \, .
\end{equation}
Thus, we finally obtain
\begin{equation}
\mathcal{O}_{\rho \ell_\alpha \overline{n_i}}  = -\sqrt{2} G_F U_{\alpha i} V_{ud}  f_\rho \rho^-_\mu (\overline {\ell_\alpha}  \gamma^\mu P_L n_i )  + \textrm{h.c.} \, ,
\label{eq:effective-rho}
\end{equation}
and, equivalently, for the $K^{*,\pm}$ meson we get
\begin{equation}
\mathcal{O}_{K^{*} \ell_\alpha \overline{n_i}}  = - \sqrt{2} G_F U_{\alpha i} V_{us}  f_{K^*} K^{*,-}_\mu (\overline {\ell_\alpha}  \gamma^\mu P_L n_i ) + \textrm{h.c.} \, 
\end{equation}
Heavier charged vectors are not too relevant for heavy neutrino production, as their hadronic decay modes are extremely dominant. 

\subsection{Vector meson decay constants}
\label{sec:vect_meson_decay_const}

The interactions involving heavy neutrinos and vector mesons have proved to be somewhat controversial, as several discrepancies are present in the literature. Some of these disagreements stem from the vertices, which have been consistently derived above. In particular, the case of neutral vectors exhibits different dependences on the weak mixing angle, possibly due to an inaccurate treatment of the $Z$ current. 

Another source of discrepancy is the value of the decay constants of the vector mesons\footnote{Note that the normalization of these constants in the literature may differ from ours, as it is common to normalize them to the vector meson mass, in order to obtain decay constants with units of energy. However, we will keep our conventions, defined in Eq.~\ref{eq:fV}, so as to employ the same definitions for pseudoscalars and vectors.}. As these particles are usually unstable under QCD, and thus exhibit large branching ratios into lighter mesons, the lattice determination of their decay constants is usually not trivial, or even not available. It is then useful to identify decay channels mediated by the electroweak interactions, that have been accurately measured by experiments. A matching of the analytical branching ratio and the experimental determination allows to extract the value of the decay constant. 

In the case of neutral vector mesons, the channel $V\to e^+e^-$ is a good choice, as it can be easily computed (it is dominated by photon exchange) and the corresponding branching ratio is usually measured with high precision. The electromagnetic current can be decomposed as 
\[
j_{\mathrm{EM}, \mu}^{V} = i \sum_q e Q^q \overline{q} \gamma_\mu q \, .
\]
Analogously to the case of the $Z$, this expression can be rewritten as a linear combination of the meson currents: 
\begin{equation}
\label{eq:EM-current}
j_{\mathrm{EM}, \mu}^{V} = i e\left[ j_{\rho, \mu}^{V}  + \frac{1}{3} j_{\omega, \mu}^{V} -\frac{\sqrt{2}}{3}  j_{\phi, \mu}^{V}  \right] \, .
\end{equation}
This allows to compute the width for the vector meson decays into $e^+ e^-$ pairs, that read
\begin{align}
\label{eq:rho-ee}
\Gamma(\rho \to e^+ e^-) & =  \frac{2\pi}{3}\frac{\alpha^2 f_\rho^2}{m_\rho^3} \, , \\
\label{eq:omega-ee}
\Gamma(\omega \to e^+ e^-) & =  \frac{2\pi}{27}\frac{\alpha^2 f_\omega^2}{m_\omega^3} \, , \\
\label{eq:phi-ee}
\Gamma(\phi \to e^+ e^-) & =  \frac{4\pi}{27}\frac{\alpha^2 f_\phi^2}{m_\phi^3} \, .
\end{align}
Comparing these results to the corresponding measurements~\cite{ParticleDataGroup:2018ovx}, we find the values for the decay constants $f_V$ listed in Tab.~\ref{tab:Fpi}.

Charged vector mesons do not exhibit sizable decays mediated by the electroweak interactions, so other options have to be considered. In the case of the $\rho^\pm$ mesons, it is a very good approximation to use the $\rho^0$ constant, since the isospin breaking corrections should be negligible. However, the $K^{*,\pm}$ does not have a neutral counterpart, so it is possible to make use of the process $\tau^- \to K^{*,-} \nu_\tau$, that has been observed. The authors of Ref.~\cite{Maris:1999nt} perform such calculation and report the value of the ratio between the $\rho$ and $K^*$ decay constants:
\begin{equation}
\frac{f_{K^*}}{f_{\rho}} = 1.042 \, .
\end{equation}
Therefore, using $f_\rho = 0.171~\mathrm{GeV}^2$, we obtain $f_{K^*} = 0.178~\mathrm{GeV}^2$ as listed in Tab.~\ref{tab:Fpi}.

\section{Semileptonic meson decays}

Some mesons exhibit non-negligible branching ratios for semileptonic decay channels into neutrinos, charged leptons and lighter mesons. These can even dominate over the two-body leptonic decays if the mass of the heavy neutrino is not large enough to sufficiently enhance the latter.

After integrating out the $W$ boson, the amplitude for the $P\rightarrow D\overline{n}\ell$ decay (where now $P$ and $D$ stand for generic parent and daughter mesons, respectively) reads
\begin{equation}
i\mathcal{M}_{P D \ell_\alpha \overline{n_i}}=\frac{ig^2}{2M_W^2}U_{\alpha i}\overline{u_\alpha}\gamma^\mu P_L v_i \left\langle D \vert j^V_{W,\mu}\vert P\right\rangle \, ,
\end{equation}
where $j_{W,\mu}^V$ is defined in Eq.~\ref{eq:W-V}. This hadronic matrix element is usually expressed in terms of two form factors, $f_+$ and $f_-$~\cite{Lubicz:2017syv}:
\begin{equation}
\label{eq:fplus_fminus}
\left\langle D\vert j_{W,\mu}^V \vert P\right\rangle =\frac{1}{2}V_{qq^\prime}\left( p_\mu f_+(q^2) + q_\mu f_-(q^2)\right)  \, ,
\end{equation}
where $V_{qq^\prime}$ is the CKM element corresponding to the interacting quarks, while $p_\mu\equiv p_\mu^D+p_\mu^P$ is the sum of the 4-momenta of the parent and daughter mesons and $q_\mu\equiv p_\mu^D-p_\mu^P$ is the 4-momentum transfer between them. Thus, the amplitude can be written as
\begin{equation}
i\mathcal{M}_{P D \ell_\alpha \overline{n_i}}=i\sqrt{2}G_F V_{qq^\prime}U_{\alpha i}\overline{u_\alpha}\gamma^\mu P_L v_i\left( p_\mu f_+(q^2) + q_\mu f_-(q^2)\right)\, .
\end{equation}
In what follows, it becomes convenient to express this in terms of the 4-momenta of the daughter meson, $p_\mu^D$, and of the leptonic pair, $p_\mu^{n\ell}$:
\begin{equation}
i\mathcal{M}_{P D \ell_\alpha \overline{n_i}}=i\sqrt{2}G_F V_{qq^\prime}U_{\alpha i}\overline{u_\alpha}\gamma^\mu P_L v_i\left[2 p_\mu^Df_+(q^2)+p_\mu^{n\ell}\left(f_+(q^2)-f_-(q^2) \right) \right] \, .
\end{equation}
Note that we have not specified the electric charges of the involved mesons. In fact, this amplitude describes all the processes allowed by charge conservation ($P^-\rightarrow D^0\overline{n}\ell^-$ and $P^0\rightarrow D^+\overline{n}\ell^-$, as well as their CP-conjugates). However,  it should be stressed that, even though electromagnetic contributions to these hadronic form factors are generally small, in some cases the numerical parameters they contain might be slightly different depending on the charge of the mesons, since they come from fits to different datasets.

From this amplitude it is possible to extract the corresponding effective operator in configuration space, writing the 4-momenta as derivatives:
\begin{align}
\mathcal{O}_{P D \ell_\alpha \overline{n_i}} 
& =  -i\sqrt{2}G_FV_{qq^\prime}U_{\alpha i}\left[2f_+(q^2)\overline{\ell_\alpha}\gamma^\mu P_L n_i\left(\partial_\mu \phi_D \right)\phi^\dagger_P  + \right. \nonumber \\ 
& +  \left. \left(f_+(q^2)-f_-(q^2) \right) \partial_\mu(\overline{\ell_\alpha}\gamma^\mu P_L n_i)\phi_D \phi^\dagger_P \right] + \textrm{h.c.} \, ,
\label{eq:semilep}
\end{align}
where $\phi_P$ and $\phi_D$ are the parent and daughter meson fields, respectively. Once more, if the involved fields are on shell, it is possible to apply Dirac\textquotesingle s equation and substitute the derivative acting on the leptonic current by terms proportional to their masses. The resulting operator reads
\begin{align}
\mathcal{O}_{P D \ell_\alpha \overline{n_i}} &= \sqrt{2}G_FV_{qq^\prime}U_{\alpha i}\overline{\ell_\alpha}
  \left[\left(f_+(q^2)-f_-(q^2) \right)  (m_\alpha P_L-m_i P_R) \phi_D  \right. \nonumber \\
 &\left.   - 2if_+ (q^2) (\partial_\mu \phi_D) \gamma^\mu P_L  \right] n_i\phi^\dagger_P + \textrm{h.c.} 
\end{align}
\section{Branching ratios of heavy neutrinos}
\begin{figure}[b!]
\centering
\includegraphics[width=0.48\textwidth]{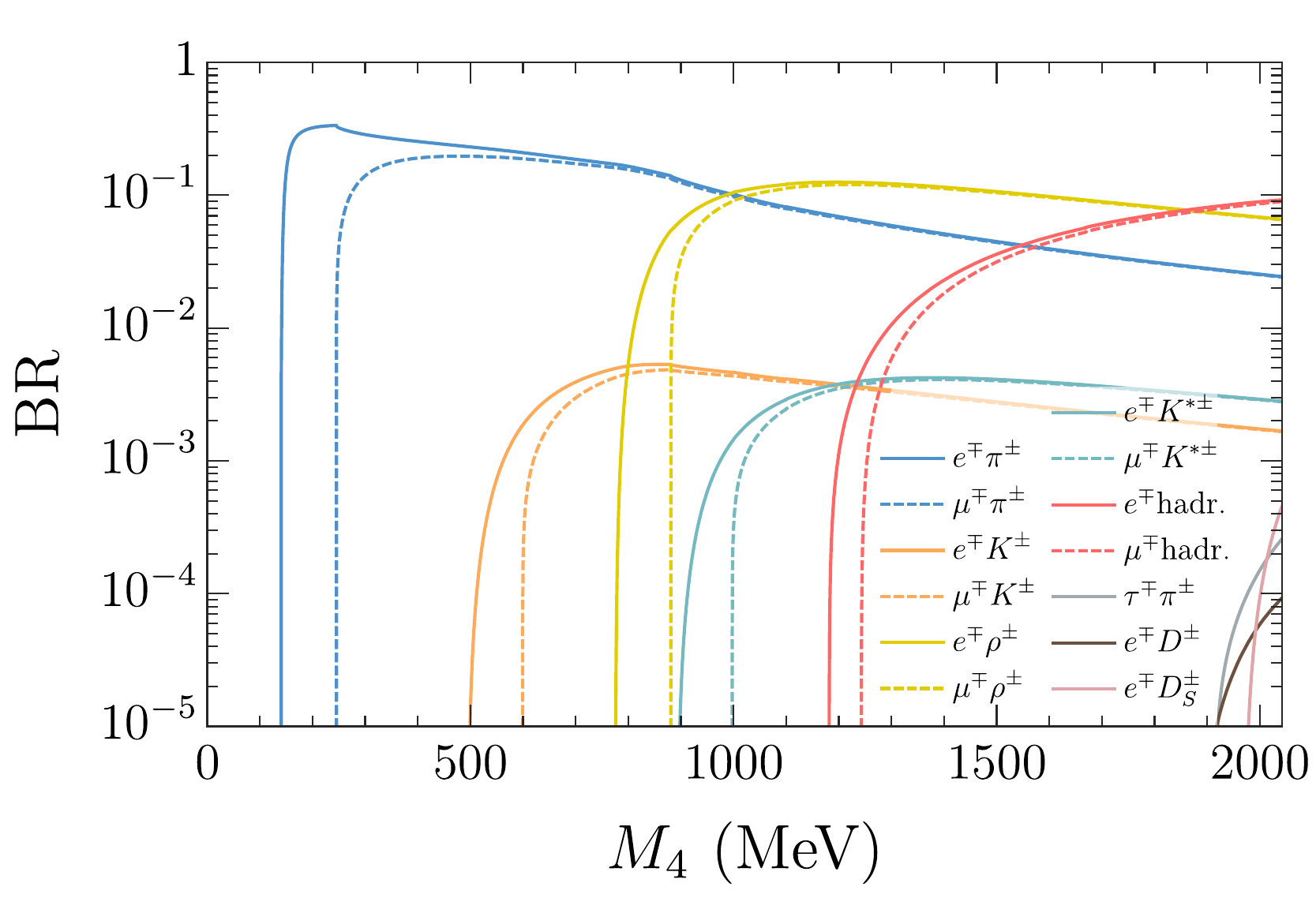}
\includegraphics[width=0.48\textwidth]{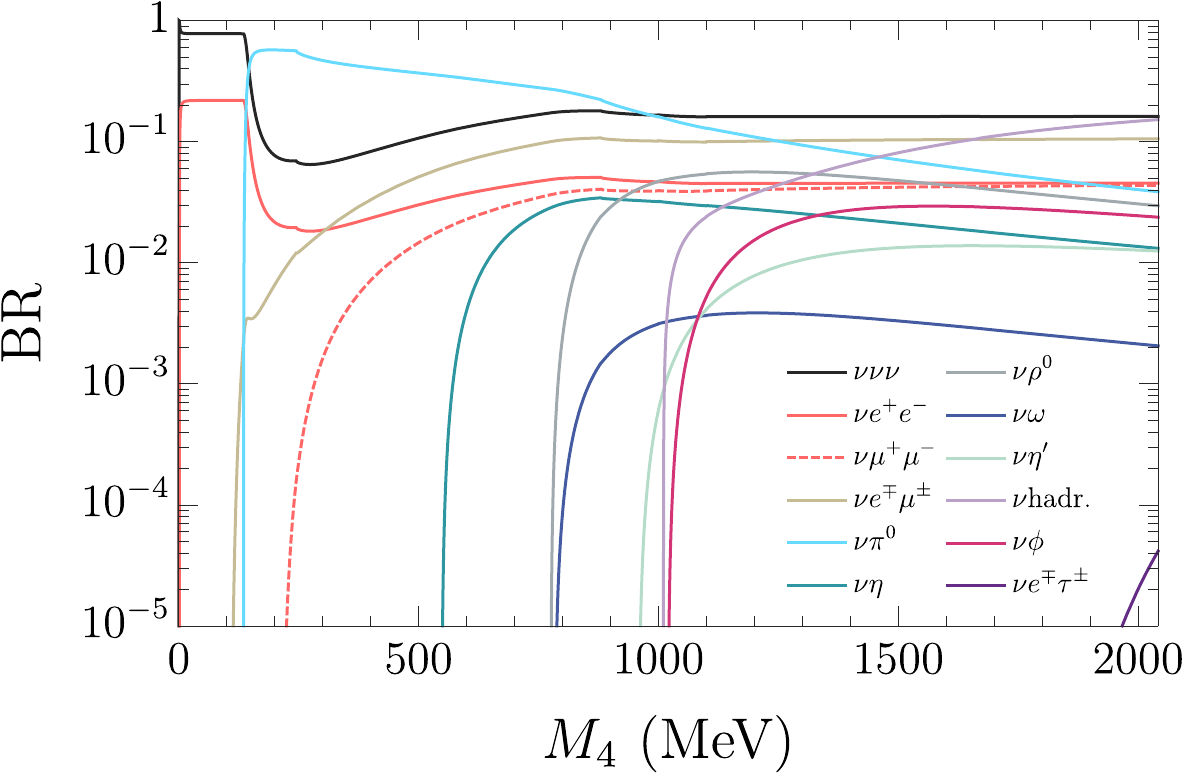}\\\vspace{-0.5cm}\hspace{1.3cm}\includegraphics[height=0.5cm]{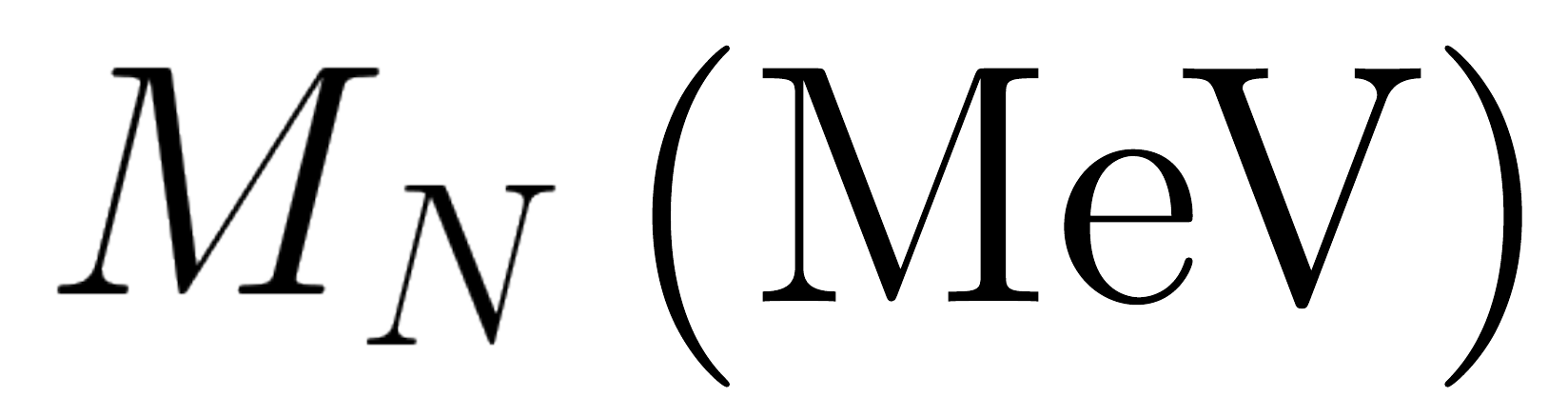}\hspace{5.3cm}\includegraphics[height=0.5cm]{PlotsDuneNuevos/dolorM.pdf}
\caption{Branching ratios of a heavy neutrino as a function of its mass, under the assumption of degenerate mixing ($\vert U_{e 4}\vert^2 = \vert U_{\mu 4}\vert^2 = \vert U_{\tau 4}\vert^2$)~\cite{Coloma:2020lgy}.}
\label{fig:branching_ratios_deg}
\end{figure}
\begin{figure}[t!]
\centering
\includegraphics[width=0.48\textwidth]{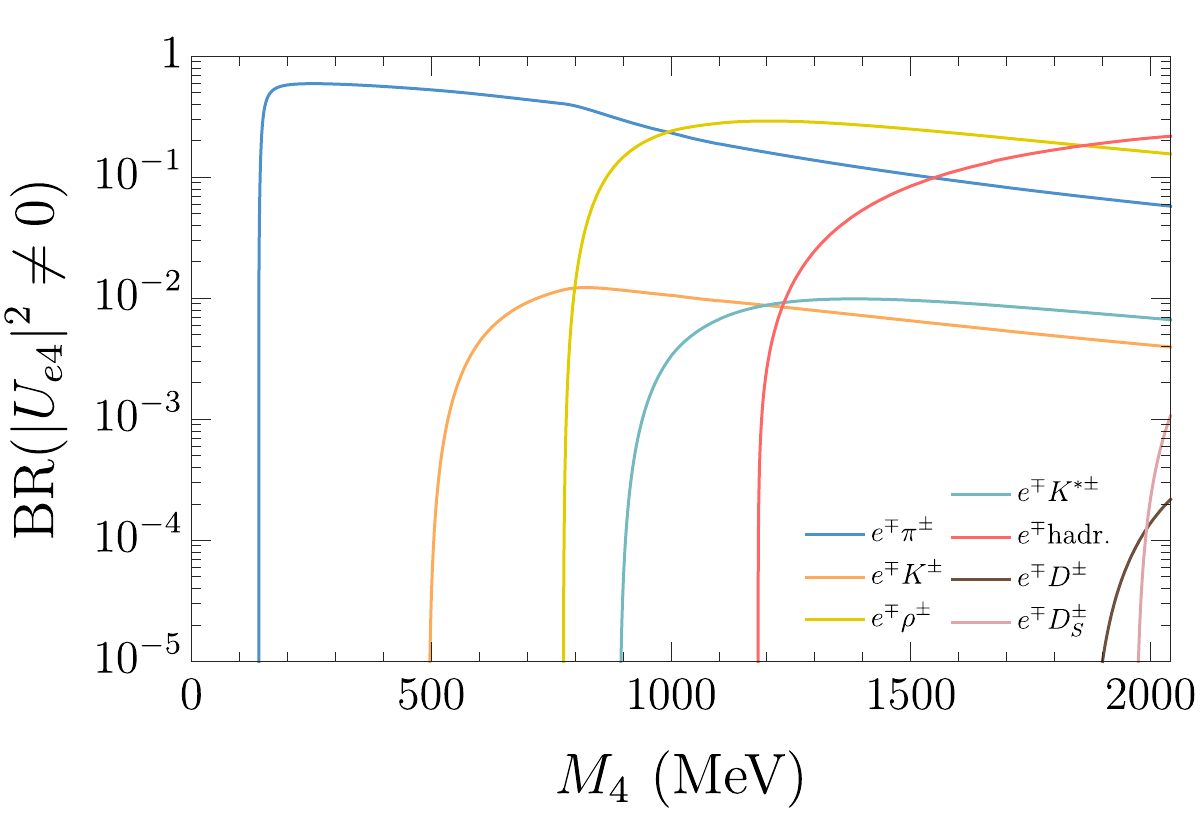}
\includegraphics[width=0.48\textwidth]{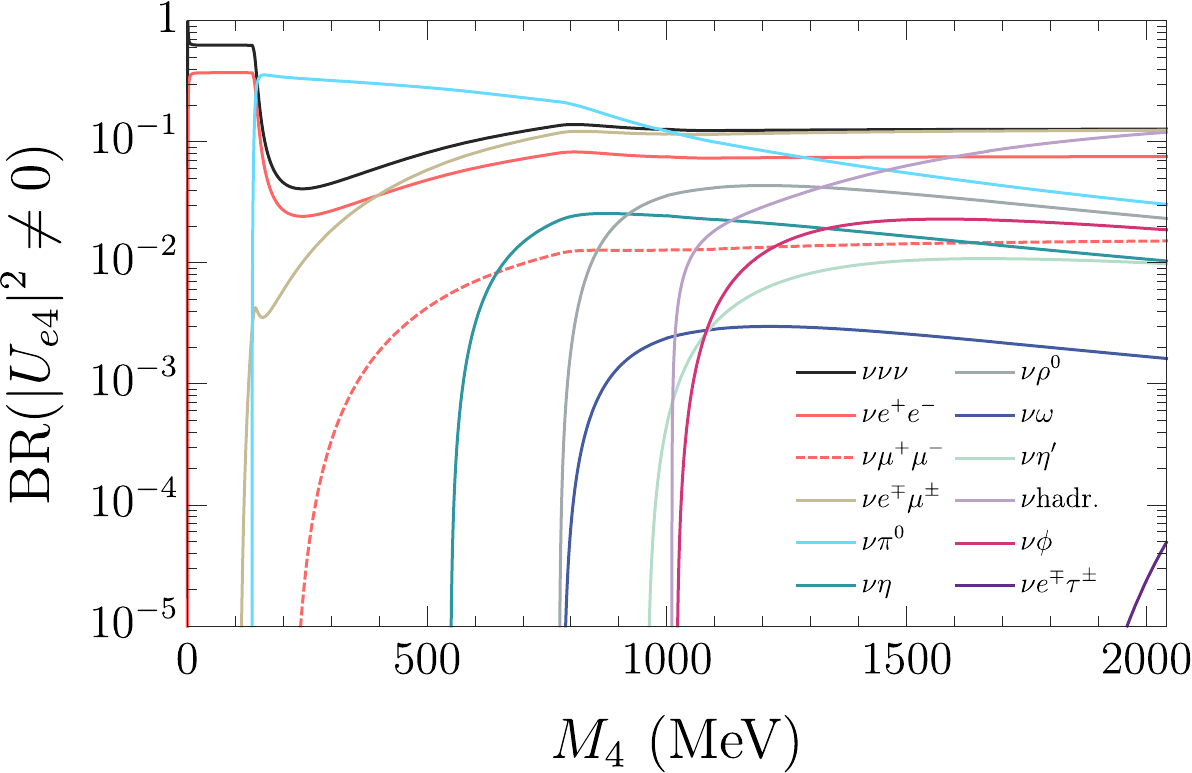}\\\vspace{-0.5cm}\hspace{1.3cm}\includegraphics[height=0.5cm]{PlotsDuneNuevos/dolorM.pdf}\hspace{5.3cm}\includegraphics[height=0.5cm]{PlotsDuneNuevos/dolorM.pdf}
\includegraphics[width=0.48\textwidth]{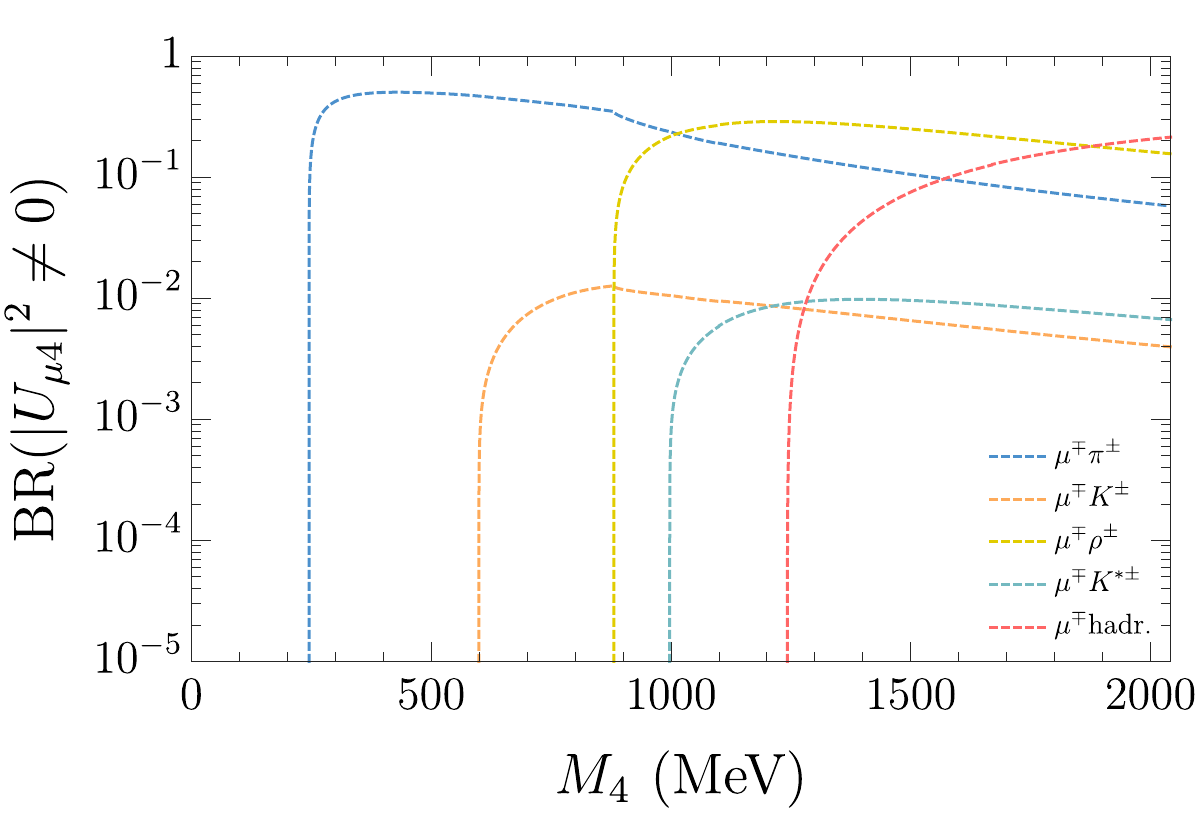}
\includegraphics[width=0.48\textwidth]{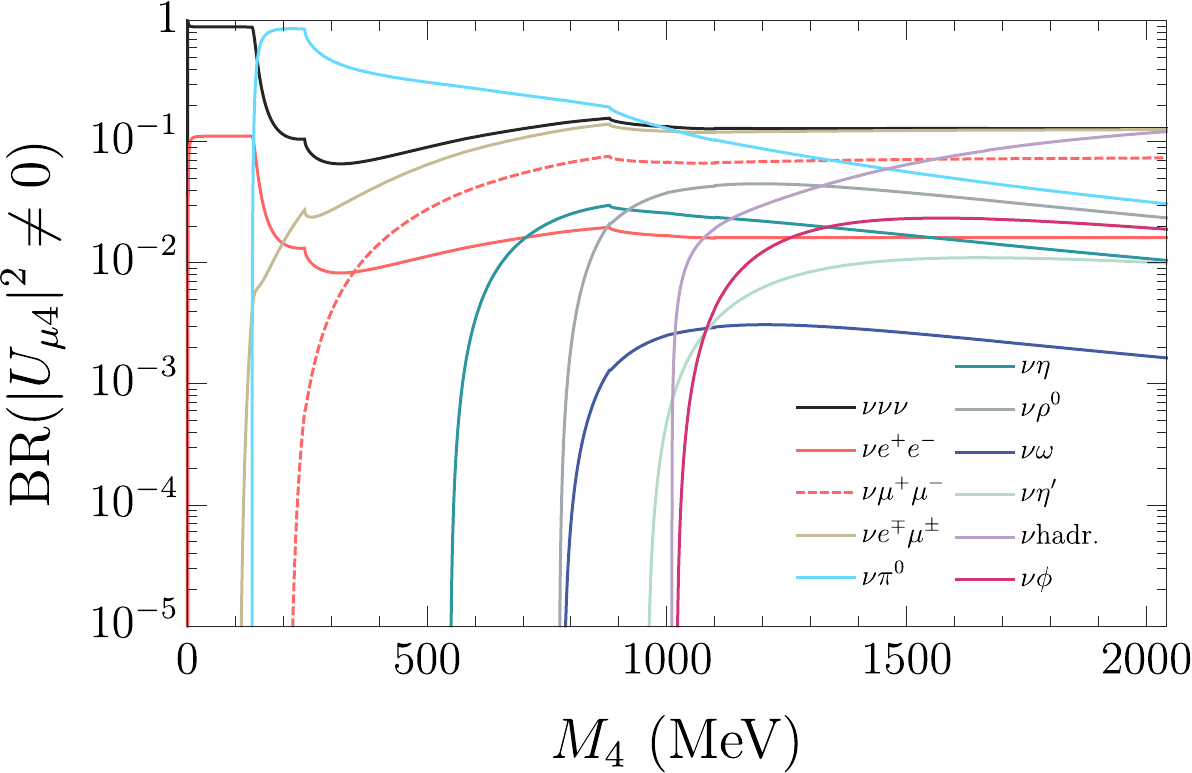}\\\vspace{-0.5cm}\hspace{1.3cm}\includegraphics[height=0.5cm]{PlotsDuneNuevos/dolorM.pdf}\hspace{5.3cm}\includegraphics[height=0.5cm]{PlotsDuneNuevos/dolorM.pdf}
\includegraphics[width=0.48\textwidth]{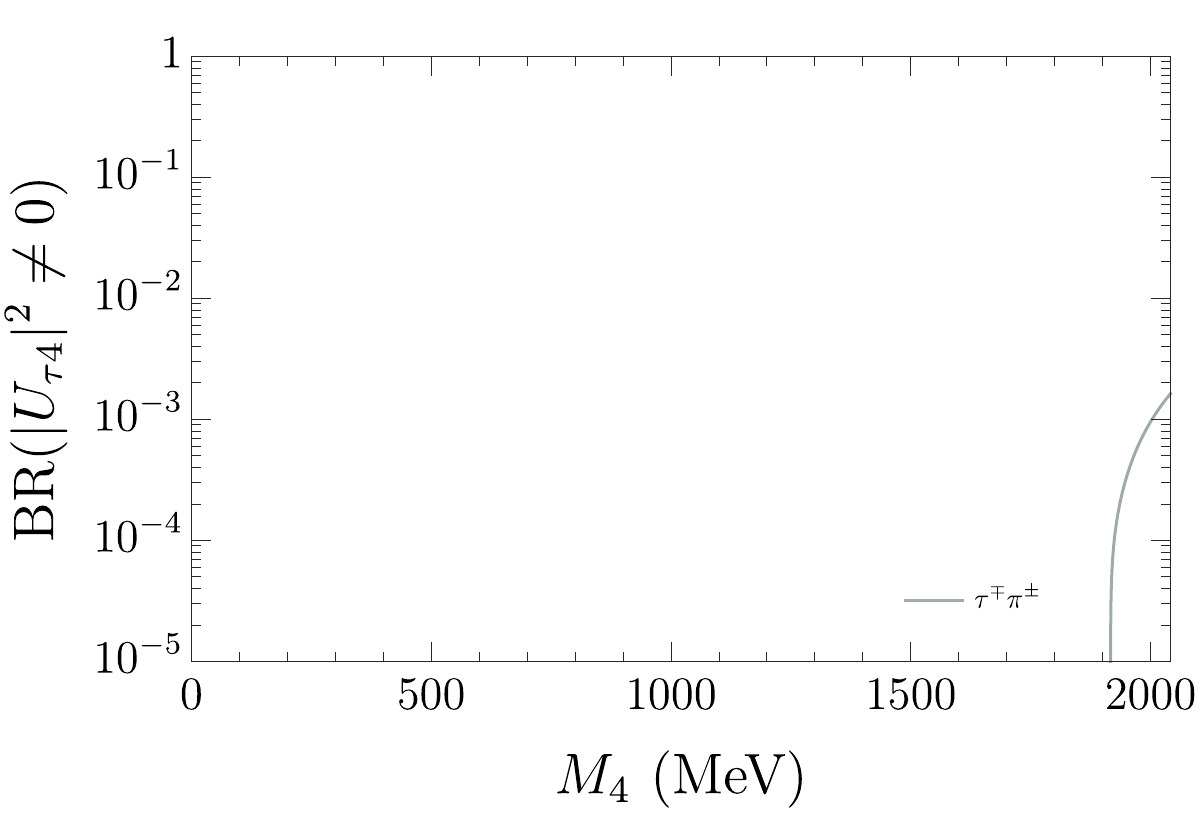}
\includegraphics[width=0.48\textwidth]{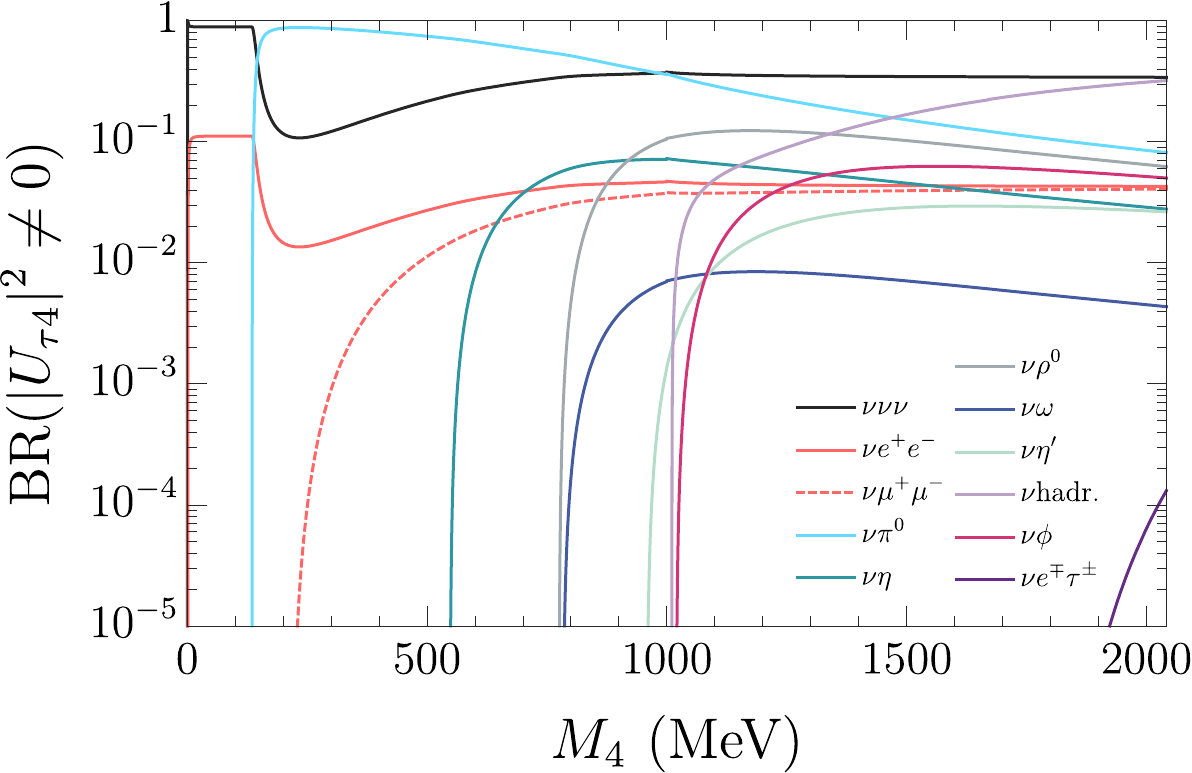}\\\vspace{-0.5cm}\hspace{1.3cm}\includegraphics[height=0.5cm]{PlotsDuneNuevos/dolorM.pdf}\hspace{5.3cm}\includegraphics[height=0.5cm]{PlotsDuneNuevos/dolorM.pdf}
\caption{Branching ratios of a heavy neutrino as a function of its mass, under the single flavour dominance assumption~\cite{Coloma:2020lgy}.}
\label{fig:branching_ratios_flav}
\end{figure}
The effective theory described above allows to compute both HNL production and decay. We summarize in Apps.~\ref{app:hnl_prod_mixing} and~\ref{app:hnl_decay_mixing} the widths for the most relevant channels. Employing those expressions, we have obtained the branching ratio for each HNL decay channel as a function of its mass, as displayed in Figs.~\ref{fig:branching_ratios_deg} and~\ref{fig:branching_ratios_flav}. The first one corresponds to the case where the mixings to all lepton flavours are degenerate ($\vert U_{e4}\vert^2=\vert U_{\mu 4}\vert^2=\vert U_{\tau 4}\vert^2$), whereas the latter represents the single flavour dominance scenario. In both figures, the left (right) panel contains the channels without (with) light neutrinos in the final state, separated just for the sake of clarity. The labels $l^\mp$hadr. and $\nu$hadr. stand for HNL multibody decays, with 3 or more mesons in the final state. It is clear that, although leptonic modes (especially the fully invisible one) are very relevant, the branching ratios into mesons are quite sizable, especially for lower masses.
\section{Summary}
Strong theoretical arguments motivate the existence of heavy neutrinos close to the GeV scale, which are in fact predicted by low-scale seesaw mechanisms, based on an approximate lepton number symmetry. If heavy neutrinos lie at this range, their interactions with mesons play a very relevant role in their phenomenology. The decay of charged mesons, which are very abundantly produced in different kinds of experiments, constitute the main production source of HNLs, whereas the decay into lighter mesons can also be very relevant for heavy neutrinos in this mass range. 

Thus, deriving a low-energy effective theory that accounts for these interactions, integrating out the weak bosons and introducing the hadronic matrix elements, is crucial to study the associated phenomenology. Although the relevant Feynman rules and decay widths were already available in the literature, some sources of discrepancy could be found. We have consistently derived the most important vertices from first principles, carefully treating the corresponding generators and currents. This allows to derive the effective operators that control HNLs interacting with mesons, which were made available in a FeynRules model that allows for full event generation.



\chapter{Heavy Neutral Leptons at DUNE}
\fancyhead[RO]{\scshape \color{lightgray}11. Heavy Neutral Leptons at DUNE}
Long-baseline oscillation experiments often employ a near detector (ND), situated hundreds of meters away from the target, that allows to callibrate the neutrino flux. This ND can also be useful for other purposes, for instance, probing heavy neutrinos. This is the case of the Deep Underground Neutrino Experiment, a facility currently in construction in the United States, whose main target is the measurement of the CP violating phase of the PMNS and the ordering of neutrino masses.

In this experiment, the beam is composed of protons, that, upon impinging on a target, produce a very large number of charged mesons, as heavy as the $D_s$. These mesons are then focused employing very intense magnetic fields, that may be flipped, allowing to alternatively focus positively or negatively charged mesons. These particles may decay into HNLs via mixing, with interactions given by the effective theory derived in the previous chapter. Tau leptons can also play the same role as the mesons in HNL production, although with smaller rates. Depending on their mass and mixings, the heavy neutrinos may travel sizable distances, and possibly decay back into SM particles, upon a second mixing insertion, inside the ND. This strategy is particulary interesting to probe HNLs close to the GeV range, and, as we will see in the following, it may allow to reach excellent sensitivities.

We will use the effective theory derived in the previous chapter to compute the expected heavy neutrino flux at the ND, as well as the expected number of decays for different channels. The results shown in this chapter can also be found in Ref.~\cite{Coloma:2020lgy}. From now on we will assume a Dirac HNL; in the Majorana case, the heavy neutrino decay widths, and thus the number of events, would increase in a factor of 2. We will consider a ND geometry as described in the DUNE Technical Design Report (TDR)~\cite{DUNE:2020lwj}. The ND complex will be located 574~m downstream from the neutrino beam source, and will include three primary detector components: a liquid Argon Time Projection Chamber (LArTPC) called ArgonCube; a high-pressure gaseous TPC surrounded by an electromagnetic calorimeter (ECAL) in a 0.5~T magnetic field, called the Multi-Purpose Detector (MPD); and an on-axis beam monitor called System for on-Axis Neutrino Detection (SAND). We will deal with the detector volume corresponding to the MPD\footnote{To be specific, we consider a cylinder of 5~m diameter and 5~m length, as described in Ref.~\cite{neardet}. We also consider a tilt angle $\alpha = 0.101$ due to the beam inclination with respect to the horizontal~\cite{tilt}. }, for which the beam-induced background is smaller, given its lower density. 

All the computations presented in this chapter have been performed using the nominal beam configuration and luminosity envisioned for DUNE~\cite{DUNE:2020lwj}: 120~GeV protons and $1.1\cdot 10^{21}$ protons on target (PoT) per year, divided equally into positive and negative horn focusing modes, which yields a total of $7.7\cdot 10^{21}$ PoT over 7 years of data taking, which is expected to start in 2027. The simulation of the meson production in the target has been done as follows. For pions and kaons, we use the results of the detailed GEANT4~\cite{Agostinelli:2002hh,Allison:2006ve,Allison:2016lfl} based simulation (G4LBNF) of the LBNF beamline developed by the DUNE collaboration~\cite{DUNE:2020lwj}. The simulation includes a detailed description of the geometry, including the 1.5~m long target, three focusing horns, decay region, and surrounding shielding. The DUNE collaboration provides both neutrino and antineutrino mode predictions, generated for a 120~GeV primary proton beam. For positive horn focusing mode (PHF) we use the results of the full simulation to calculate the event rates at the DUNE ND, while for negative horn focusing (NHF) mode we scale the rates from PHF mode based on the flux ratios between $\pi^-/\pi^+$ and $K^-/K^+$ as predicted by G4LBNF.
\begin{table}[t!]
\begin{center}
\begin{tabular}{cccccc} \toprule
    & $\pi$ & $K$ & $\tau$ & $D$ & $D_s$ \\ \toprule\vspace{1.5mm}
    Positive sign & 6.3 & 0.75 & $2.1 \cdot 10^{-7}$ & $1.2\cdot 10^{-5}$ & $3.3 \cdot 10^{-6}$ \\
    Negative sign & 5.7 & 0.33 & $3.0 \cdot 10^{-7}$ & $1.9\cdot 10^{-5}$ & $4.6 \cdot 10^{-6}$  \\\bottomrule
\end{tabular}
\caption{\label{tab:parent_PoT} Average number of positive and negative parent mesons and tau leptons produced at DUNE in the target per PoT. }
\end{center}
\end{table}

However, G4LBNF does not include the production of $D$ and $D_s$ mesons and tau leptons. Thus, in this case Pythia (v.8.2.44)~\cite{Sjostrand:2014zea} was used to create a pool of events and predict production rates for proton collisions at various momenta, and a GEANT4-based simulation was subsequently used to predict proton inelastic interactions with 120~GeV primary protons impinging on the target. For each inelastic interaction, we randomly pick a Pythia event from the pool of events generated at the corresponding momentum, with a weight proportional to the rate predicted by Pythia. In doing this, we neglect the effect of the magnetic horns, since these heavy particles decay very promptly and, therefore, it is safe to assume that their production will be similar for the PHF and NHF modes. The average number of parent mesons and tau leptons per PoT produced in the target\footnote{The production rates reported in Tab. 3.1 of version 1 of Ref.~\cite{Berryman:2019dme} are a factor 2-3 smaller. The reinteractions that we take into account lead to higher production of low energy pions, which do not have a significant impact in the final sensitivity given their low collimation. For the heavier mesons the discrepancy is due to the use of different Pythia versions. The authors of Pythia looked into the issue and confirmed a bug was introduced in version 8.240 of Pythia that led to the lower rates found in Ref.~\cite{Berryman:2019dme}.} are listed in Tab.~\ref{tab:parent_PoT}.

\begin{table}[t!]
\begin{center}
\begin{tabular}{cccccc}\\\toprule
Parent&$\pi$&$K$&$\tau$&$D$&$D_s$\\\midrule\vspace{1.5mm}
\multirow{3}{*}{2-body decay}&$eN$&$eN$&$\pi N$&$eN$&$eN$\\\vspace{1.5mm}
&$\mu N$&$\mu N$&$\rho N$&$\mu N$&$\mu N$\\
&&&&$\tau N$&$\tau N$\\\midrule\vspace{1.5mm}
\multirow{2}{*}{3-body decay}&\multirow{2}{*}{$-$}&$\pi^0e\,N$&$e\nu N$&$K^0e\,N$&\multirow{2}{*}{$-$}\\\vspace{1.5mm}
&&$K^0\mu\,N$&$\mu\nu N$&$K^0\mu\,N$&
\\\bottomrule
\end{tabular}
\end{center}
\caption{\label{tab:HNL_production}List of 2-body and 3-body decays into HNLs, for the parent particles considered here.}
\end{table}
Tab.~\ref{tab:HNL_production} summarizes the different HNL production channels that have been included in our analysis. This set contains the dominant leptonic and semileptonic decays into heavy neutrinos of the parent mesons ($\pi$, $K$, $D$ and $D_s$) produced in the target. Moreover, since the $D$ and $D_s$ decay very promptly and have sizable branching ratios to tau leptons, a significant tau production rate is expected.
This provides an additional production mechanism for HNL masses below the tau mass, controlled by $\vert U_{\tau 4}\vert^2$, allowing DUNE to significantly improve the sensitivity to this more elusive mixing matrix element. All decay modes of the tau could produce an HNL in the final state, provided it is kinematically allowed. Nevertheless, we have opted to conservatively consider only the tau decay modes $\tau^-\to \rho^- N$, $\tau^-\to \pi^- N$ and $\tau^-\to \ell_\alpha^- N \bar{\nu}$. The decays with 3 or more mesons in the final state have been neglected, since the phase space is reduced for the production of a massive particle and the simulation of the HNL kinematics is more challenging for these channels (see App.~\ref{subsec:multimesons}).

After obtaining the expected flux of HNLs entering the detector, we match it to the 22 different decay modes into SM particles listed in App.~\ref{sec:decays} (and shown in Figs.~\ref{fig:branching_ratios_deg} and~\ref{fig:branching_ratios_flav}), according to their corresponding branching ratios, to obtain the expected signal at the detector. 

In the following we first illustrate the impact of the boost of the HNL on the detector acceptance, and then we compute the expected number of heavy neutrino decays inside the DUNE ND to estimate its sensitivity.

\section{The effect of the HNL mass on the detector acceptance}
\label{subsec:boost}
\begin{figure}[t!]
\centering
\includegraphics[width=0.49\textwidth]{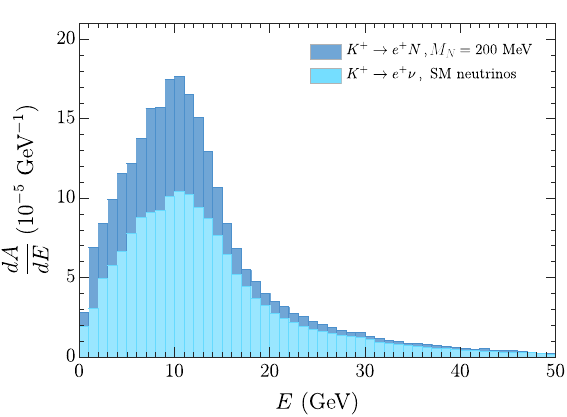}
\includegraphics[width=0.49\textwidth]{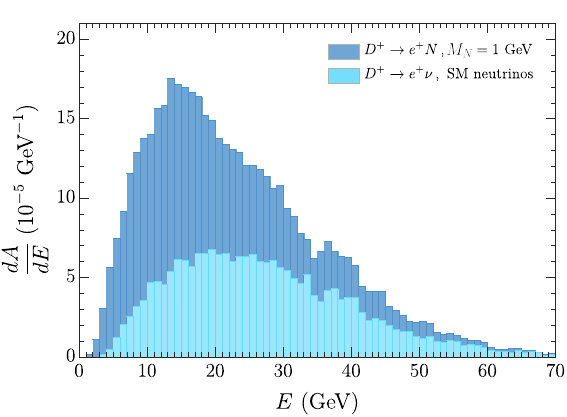}
\caption{Detector acceptance (dark blue in both panels) as a function of the neutrino energy, for neutrinos with a mass of 200~MeV,  produced in $K^+ \to e^+ N$ decays (left panel), and for neutrinos with a mass of 1~GeV, produced in $D^+ \to e^+ N$ decays (right panel). In both panels, the light blue histogram shows the detector acceptance when the neutrino mass is set to zero.}
\label{fig:energy_spectrum}
\end{figure}
The effect of the boost in the beam direction is more efficient for particles with smaller velocities. Therefore, a larger detector acceptance is obtained when the effect of the heavy neutrino mass is properly included in the flux simulations, compared to estimations of the HNL flux based on the massless neutrino distributions. In order to illustrate the impact of the boost on the detector acceptance, we simulate the heavy neutrino flux at the DUNE ND from meson decays, and we compare it to the result obtained for the light neutrino flux. Our results are shown in Fig.~\ref{fig:energy_spectrum}, for neutrinos produced from kaon decays (left panel, where $M_N = 200~\mathrm{MeV}$) and from $D$ decays (right panel, where $M_N = 1~\mathrm{GeV}$). 

As can be seen in this figure, the increase in acceptance is considerable: up to a factor of two for 200~MeV neutrinos from kaon decays, and up to a factor of three for 1~GeV neutrinos coming from $D$ decays. The effect of the boost will also lead to a distortion in the expected spectra, due to the different dependence of the detector acceptance with the neutrino energy, which can be seen from the comparison of the shape of the light and dark histograms in each panel. The net result is a relative increase in the number of neutrinos at low energies that enter the ND, given their smaller velocities and hence stronger collimation.   

\begin{figure}[t!]
\centering
\includegraphics[width=0.6\textwidth]{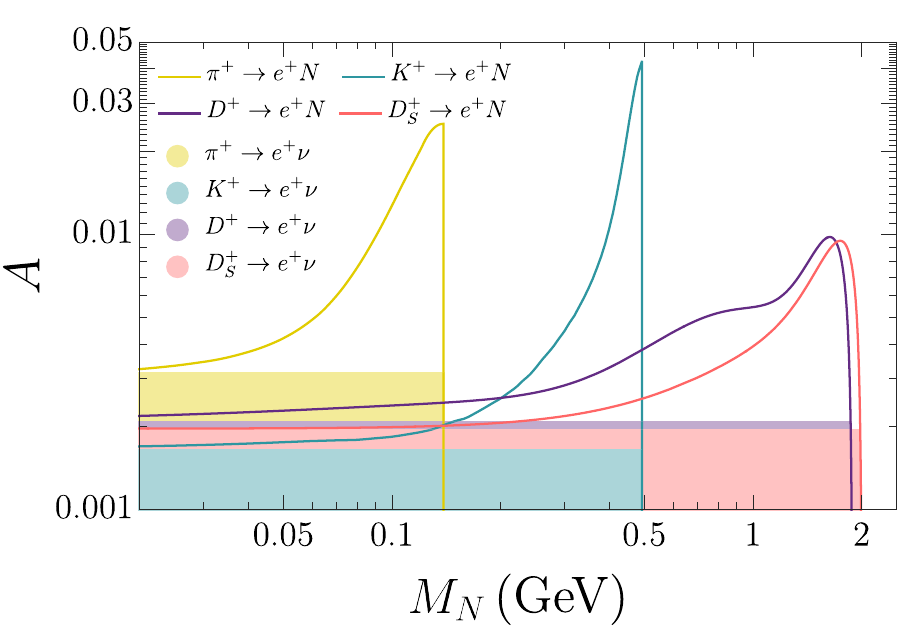}
\caption{Total detector acceptance as a function of the heavy neutrino mass. For reference, the total acceptance of the detector in the light neutrino case is indicated by the shaded regions: $3.2\cdot 10^{-3}$ for neutrinos produced from pion decays, $1.7\cdot 10^{-3}$ for neutrinos from kaon decays, $2.1\cdot 10^{-3}$ for neutrinos from $D$ decays, and $2.0\cdot 10^{-3}$ for neutrinos from $D_s$ decays.}
\label{fig:events}
\end{figure}

Finally, Fig.~\ref{fig:events} shows the total detector acceptance, after integrating over the neutrino energy, as a function of the HNL mass. Different production channels are displayed separately. Note that the acceptance is expected to be different depending on the parent meson that produced the neutrino, due to the effect of the horns: while pions are typically very well-focused at a long-baseline experiment, this is not the case for heavier mesons, which are not only harder to focus due to their larger masses, but also decay much faster. This effect is most significant for $D$ and $D_s$ mesons, which decay very promptly and therefore are practically unaffected by the horn focusing system. As can be seen, as the heavy neutrino mass approaches the production threshold and its velocity decreases accordingly, the acceptance grows very rapidly given the stronger boost in the beam direction. Notice, however, that the phase space is also decreasing, and hence the number of total HNL events will also be reduced. 

\section{Expected sensitivity to HNL decays}

Once the flux of heavy neutrinos that reach the ND, $d\phi_N/dE_N$, has been computed numerically as a function of the neutrino energy $E_N$, the total number of expected neutrino decays into a given decay channel $c$ inside the DUNE ND can be expressed as
\begin{equation}
N_c ({\rm{ND}})={\rm{BR}}_c  \int dE_N P(E_N) \frac{d\phi_N}{dE_N} \, .
\end{equation}
In the above, BR$_c$ is the branching ratio of the corresponding decay channel and $P(E_N)$ stands for the probability of the heavy neutrino decaying inside the ND. This depends on the boost factor and therefore on the neutrino energy:
\begin{equation}
P(E_N)=e^{-\frac{\Gamma  L}{\gamma \beta} } \left(1- e^{-\frac{\Gamma \Delta \ell_{\rm det}}{\gamma \beta }}\right) \, .
\label{eq:prob}
\end{equation}
Here, $\Gamma$ is the total decay width of the heavy neutrino in its rest frame, while $\gamma = E_N/M_N$, $\beta = |\vec{p}_N| / E_N$ and $\vec{p}_N$ stands for the neutrino momentum. $L$ is the distance between the HNL production and the ND, while $\Delta \ell_{\rm det}$ is the length of the HNL trajectory inside the detector. 

From Eq.~\ref{eq:prob} it is easy to see that the neutrino must be sufficiently long-lived to reach the ND, or otherwise the number of decays will be exponentially suppressed. This will be the case for sufficiently large energies and small enough matrix elements $U_{\alpha 4}$, which correspond to the most interesting region of the parameter space. In this limit, $\Gamma L \ll \gamma \beta$, and the decay probability can be further approximated as 
\begin{equation}
P(E_N) \approx \frac{\Gamma \Delta \ell_{det}}{\gamma\beta} \, .
\label{eq:prob-approx}
\end{equation}
Nevertheless, this approximation does not hold anymore for large masses and mixings, and thus we will employ the full expression in Eq.~\ref{eq:prob} to compute the probability of the HNL decaying inside the detector.

Given that the neutrino flux entering the detector will be directly proportional to its aperture, and that the probability for the neutrino to decay inside the ND is proportional to $\Delta \ell_{det}$, it is easy to see that the sensitivity to heavy neutrino decays will scale with the volume of the ND.
\begin{figure}[t!]
\centering
\includegraphics[width=0.49\textwidth]{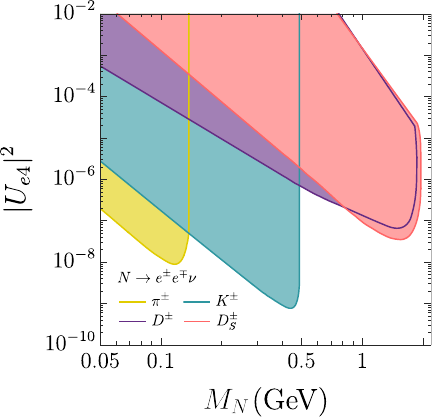}
\caption{90\% CL contours for the expected sensitivity to the mixing $|U_{e4}|^2$ as a function of the heavy neutrino mass, for a total amount of $ 7.7\cdot 10^{21}$ PoT. The different regions show the contributions of heavy neutrinos produced from different parent mesons.  \label{fig:result-per-parent} }
\end{figure}
In order to compute the final sensitivity to HNLs, a detector simulation should be performed, including relevant background contributions from SM neutrino interactions in the ND. Such a fully detailed detector simulation is beyond the scope of this thesis, where we rather show an estimate of the sensitivity of DUNE, that highlights the relevance of the effective interactions derived in the previous chapter. The main source of background for this search comes from light neutrino interactions in the detector volume, and is very significant. In Ref.~\cite{Ballett:2019bgd}, the background rates for argon were estimated at $\sim 3\cdot 10^5$ events/ton/$10^{20}$ PoT. Fortunately, SM neutrino events present a very different topology from that of heavy neutrino decays, and a series of kinematic cuts can heavily reduce the expected background and bring it down to a negligible level. This was the case, for example, in the search for HNLs performed at the T2K near detector~\cite{Abe:2019kgx}, which also used a gas TPC. Therefore, following Refs.~\cite{Ballett:2019bgd,Abe:2019kgx}, hereafter we will assume that this is achievable, showing our expected sensitivity contours to heavy neutrino decays under the assumption of no background. We also assume that the cuts applied to reduce the background will translate in our case into similar signal efficiencies as those obtained in Ref.~\cite{Abe:2019kgx}. Although the efficiency will eventually depend on the mass of the HNL and the considered decay channel (see Fig. 4 in Ref.~\cite{Abe:2019kgx}), here we use 20\% as an educated guess. Finally, for our sensitivity contours we estimate the 90\% CL sensitivity on the signal following the Feldman and Cousins~\cite{Feldman:1997qc} prescription for a Poisson distribution with no background. Under the hypothesis of no events being observed, this corresponds to the expected number of signal events being smaller than 2.44.  

Before showing our sensitivity contours, we display in Fig.~\ref{fig:result-per-parent} an example to illustrate the relative importance of the different HNL production mechanisms. In this example, we show the contours obtained under the assumption of electron flavour dominance, setting $\vert U_{\mu 4}\vert^2=\vert U_{\tau 4}\vert^2=0$. The different regions show the contributions obtained when the heavy neutrinos are produced from the decays of a given parent meson, as indicated by the labels. In this case, the signature in the detector would be electron-positron pairs, corresponding to the decay $N \to \nu e^+ e^-$. As can be seen, for $M_N < m_\pi$ the leading production mechanism is $\pi^\pm$ decay. For masses in the region $m_\pi < M_N < m_K$, $K^\pm$ dominates and, in fact, the sensitivity contour reaches the lowest values of $U_{e4}$ just below the kaon mass. This occurs even though kaons are less frequently produced than pions; this is due to the fact that HNLs lighter than the pion become very long-lived, leading to less decays inside the detector, with the consequent reduction in sensitivity. On the other hand, in the heavy mass region ($M_N > m_K$) the heavy neutrino is predominantly produced from either $D$ or $D_s$ meson decays, with a subdominant contribution from tau decays. While the $D_s$ is heavier than the $D$, and therefore more difficult to produce, its decay to heavy neutrinos is mediated by the CKM element $V_{cs}$ instead of $V_{cd}$. This compensates for its smaller production rate and, as a result, the sensitivity in this region is dominated by $D_s$ decays. Finally, the different slope as a function of $M_N$ for the $D$ contribution is simply due to the fact that, unlike for the $\pi$, $K$ and $D_s$ decays, the $D$ meson production of $N$ is dominated by three-body decays instead of two-body.

\begin{figure}[h!]
\centering
\includegraphics[height=5.686cm,keepaspectratio]{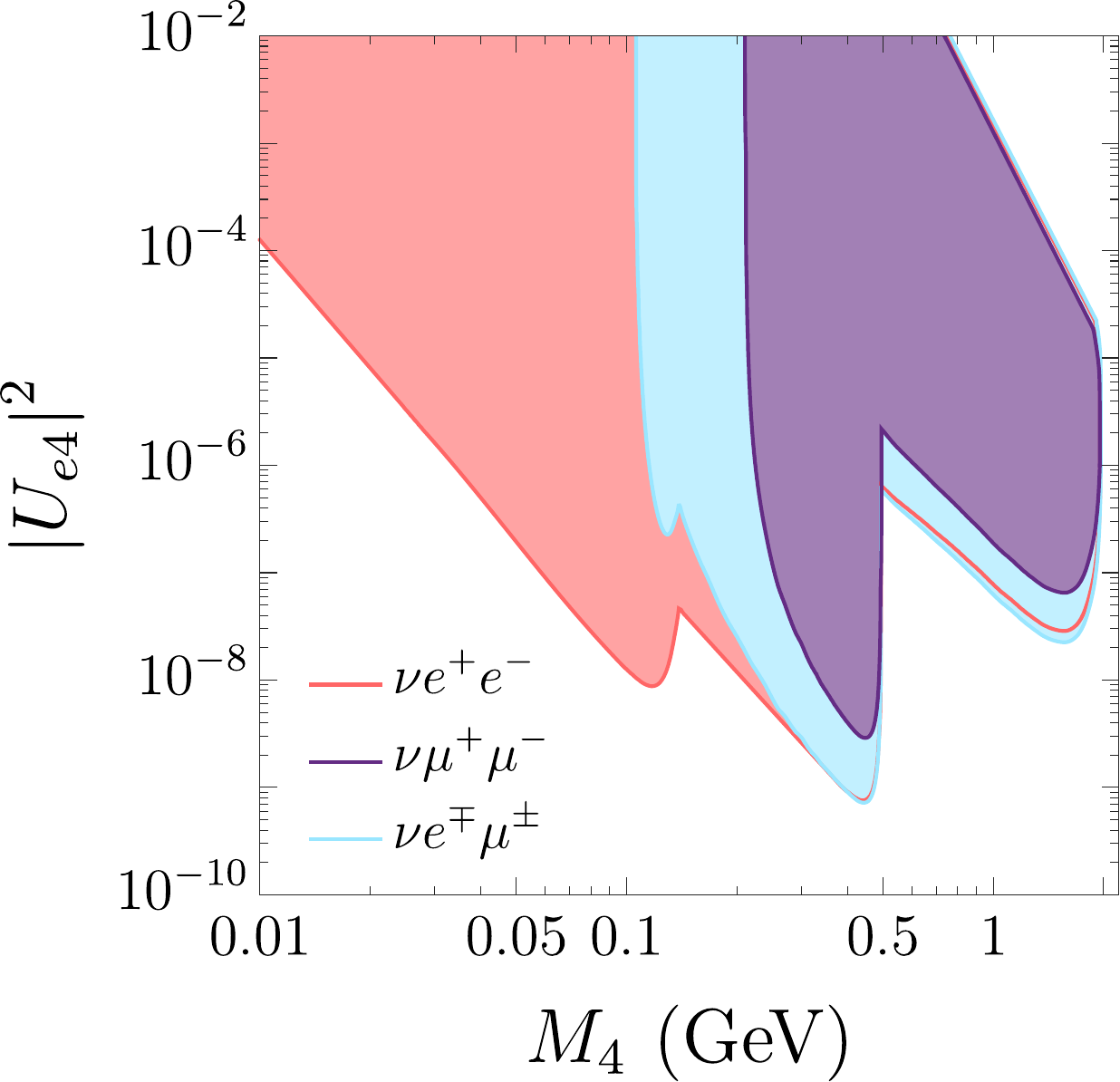} \hspace{-0.47 cm}
\includegraphics[height=5.5cm,keepaspectratio]{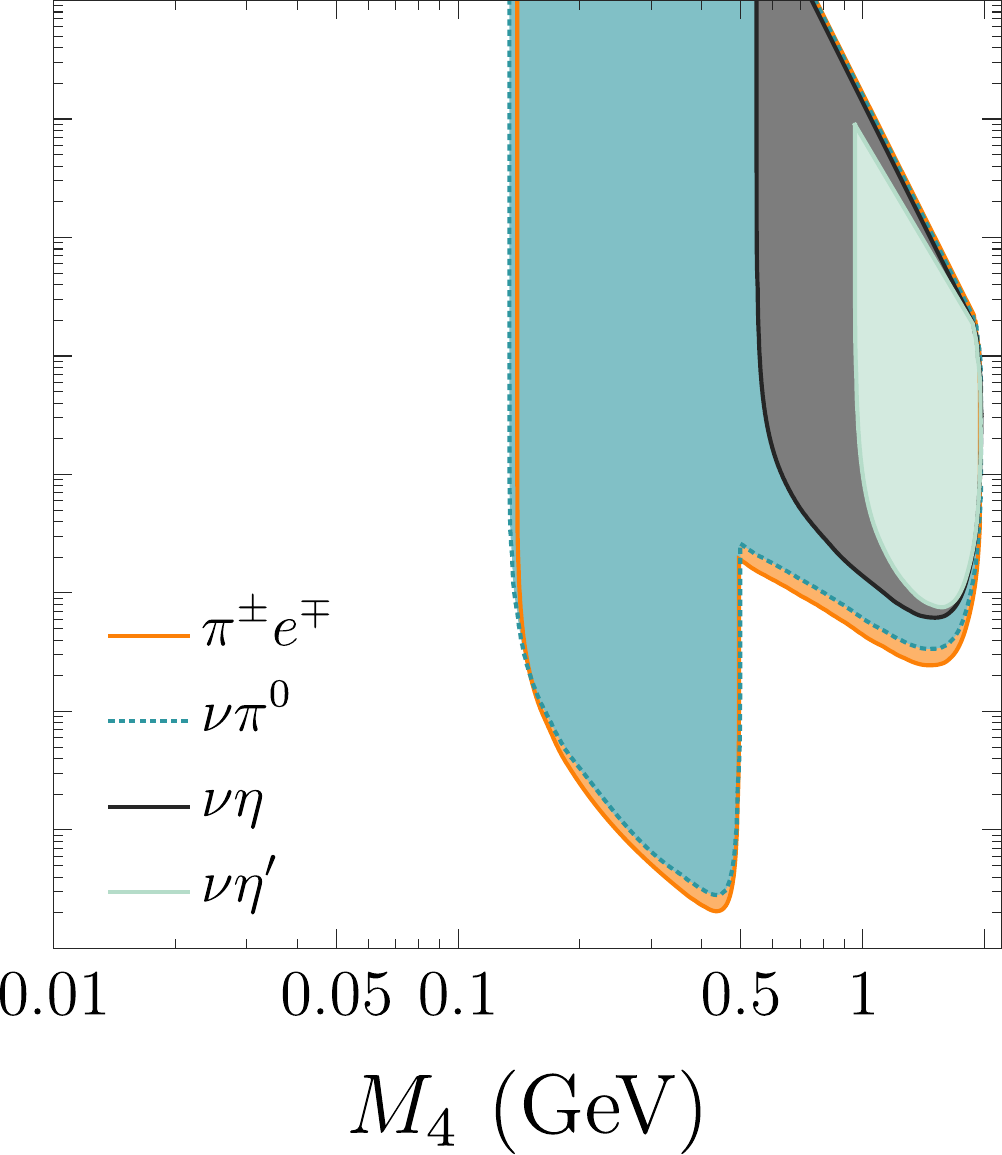} \hspace{-0.47 cm}
\includegraphics[height=5.5cm,keepaspectratio]{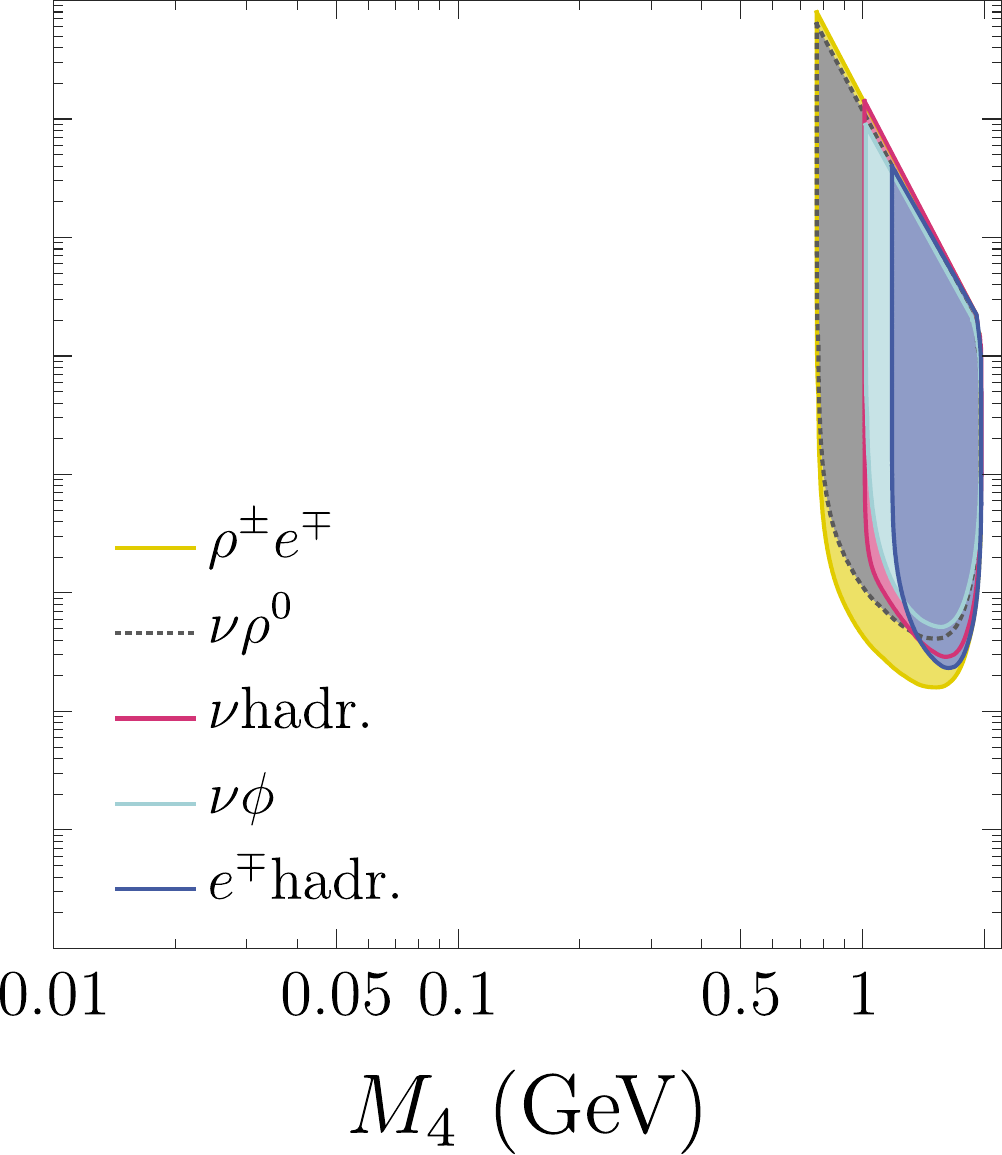}\\\vspace{-0.5cm}\hspace{1cm}
\includegraphics[height=0.55cm,keepaspectratio]{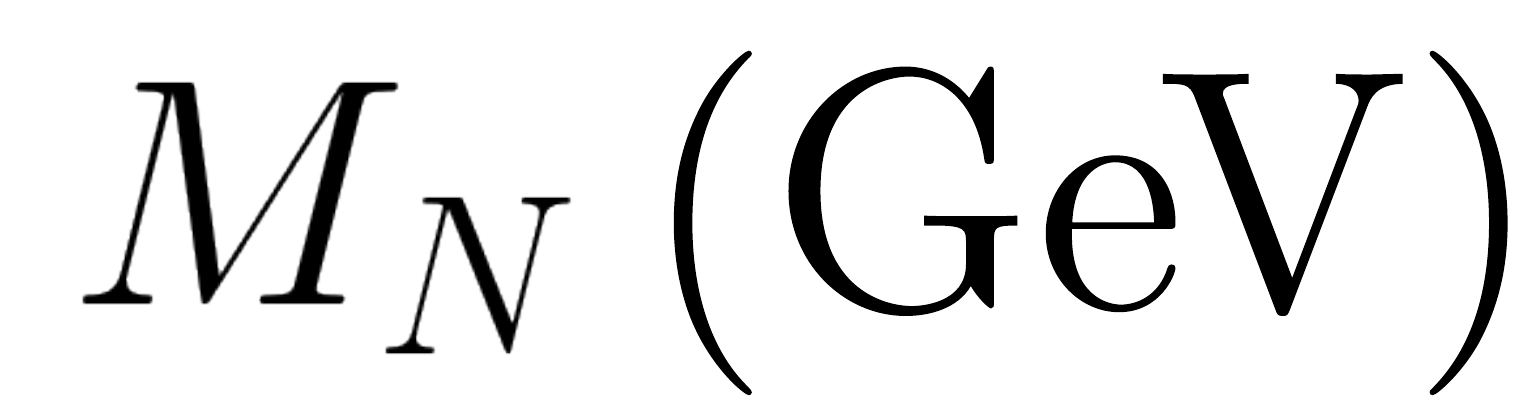}\hspace{2.4cm}
\includegraphics[height=0.55cm,keepaspectratio]{PlotsDuneNuevos/dolor.pdf}\hspace{2.4cm}
\includegraphics[height=0.55cm,keepaspectratio]{PlotsDuneNuevos/dolor.pdf}
\\
\includegraphics[height=5.686cm,keepaspectratio]{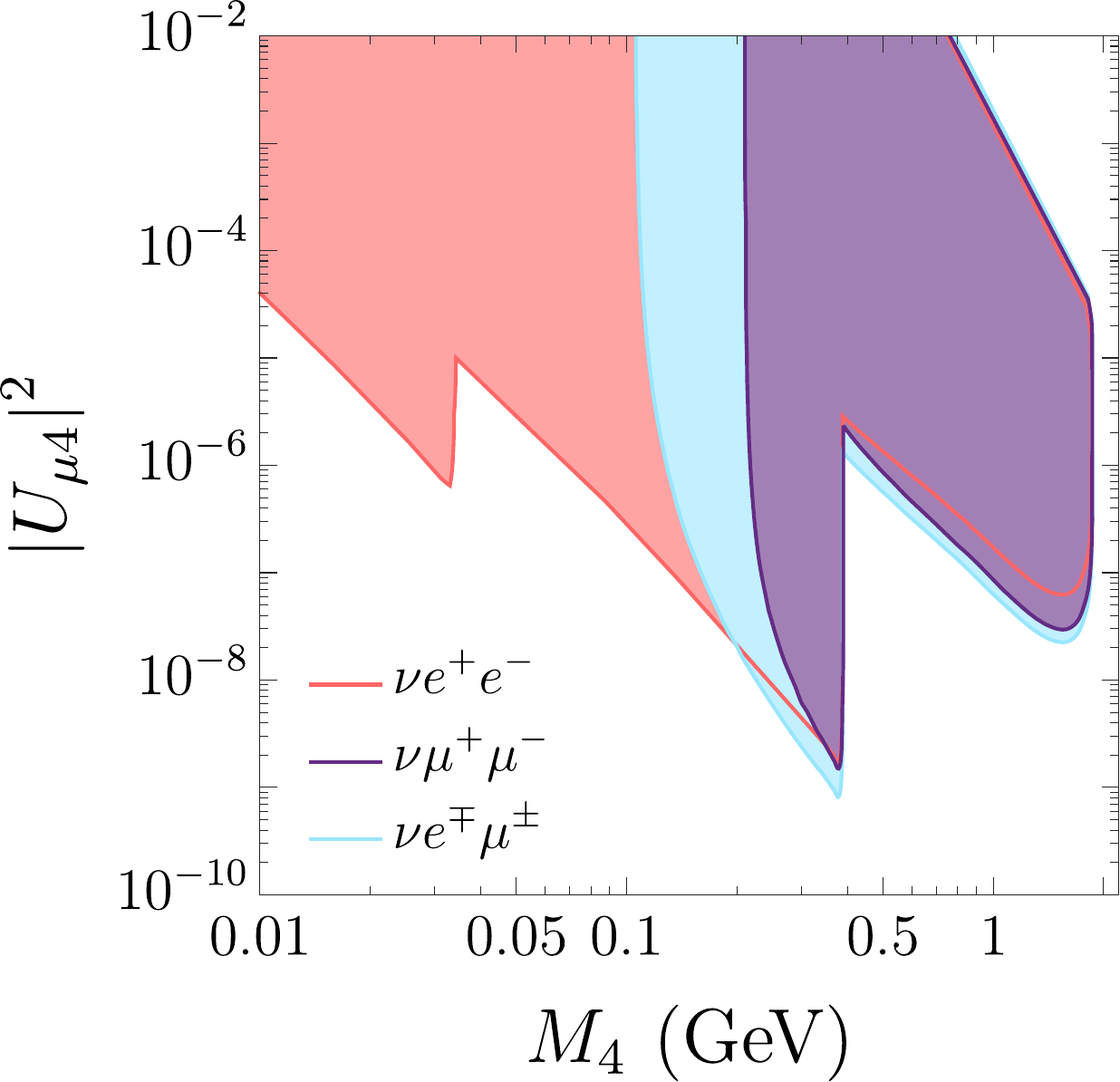} \hspace{-0.47 cm}
\includegraphics[height=5.5cm,keepaspectratio]{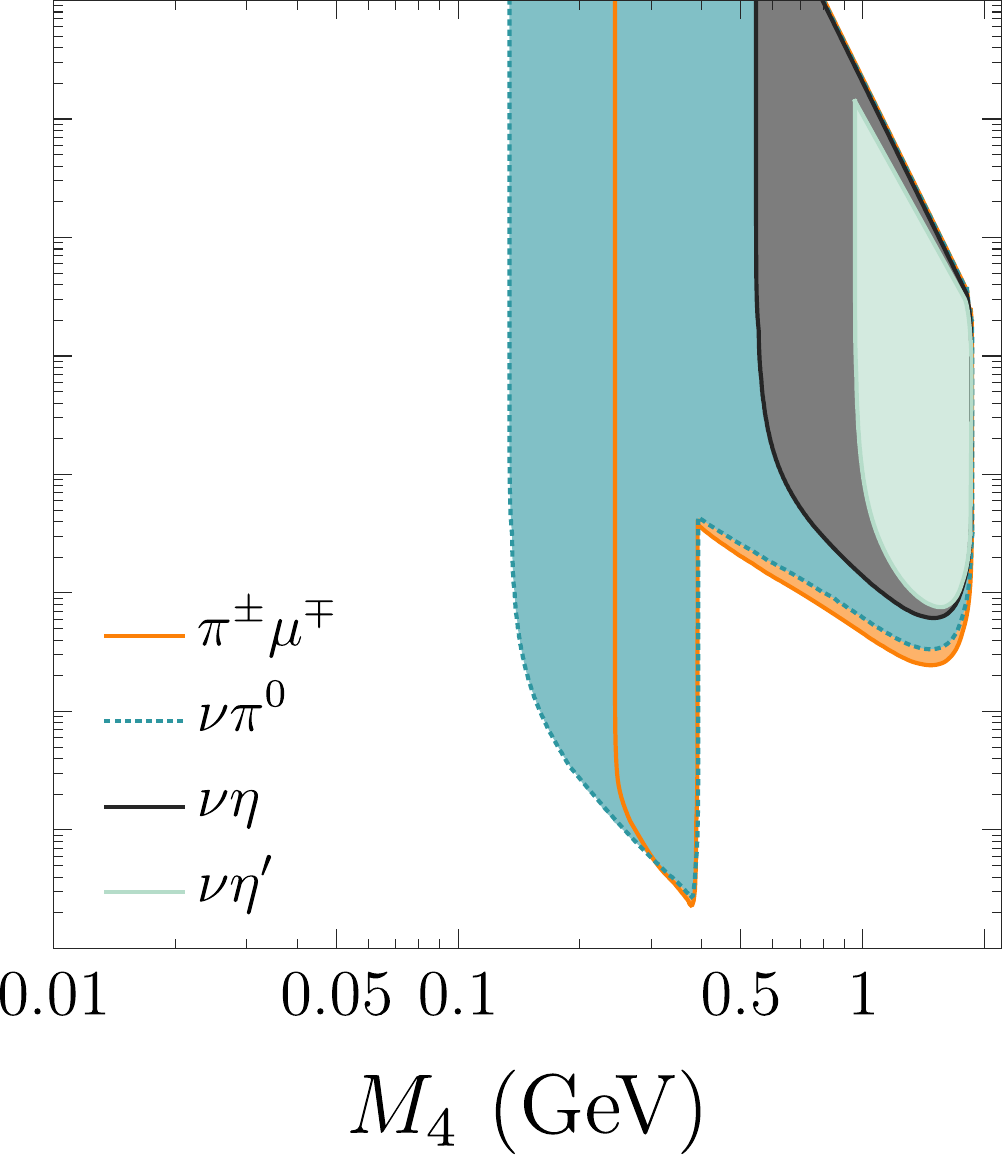} \hspace{-0.47 cm}
\includegraphics[height=5.5cm,keepaspectratio]{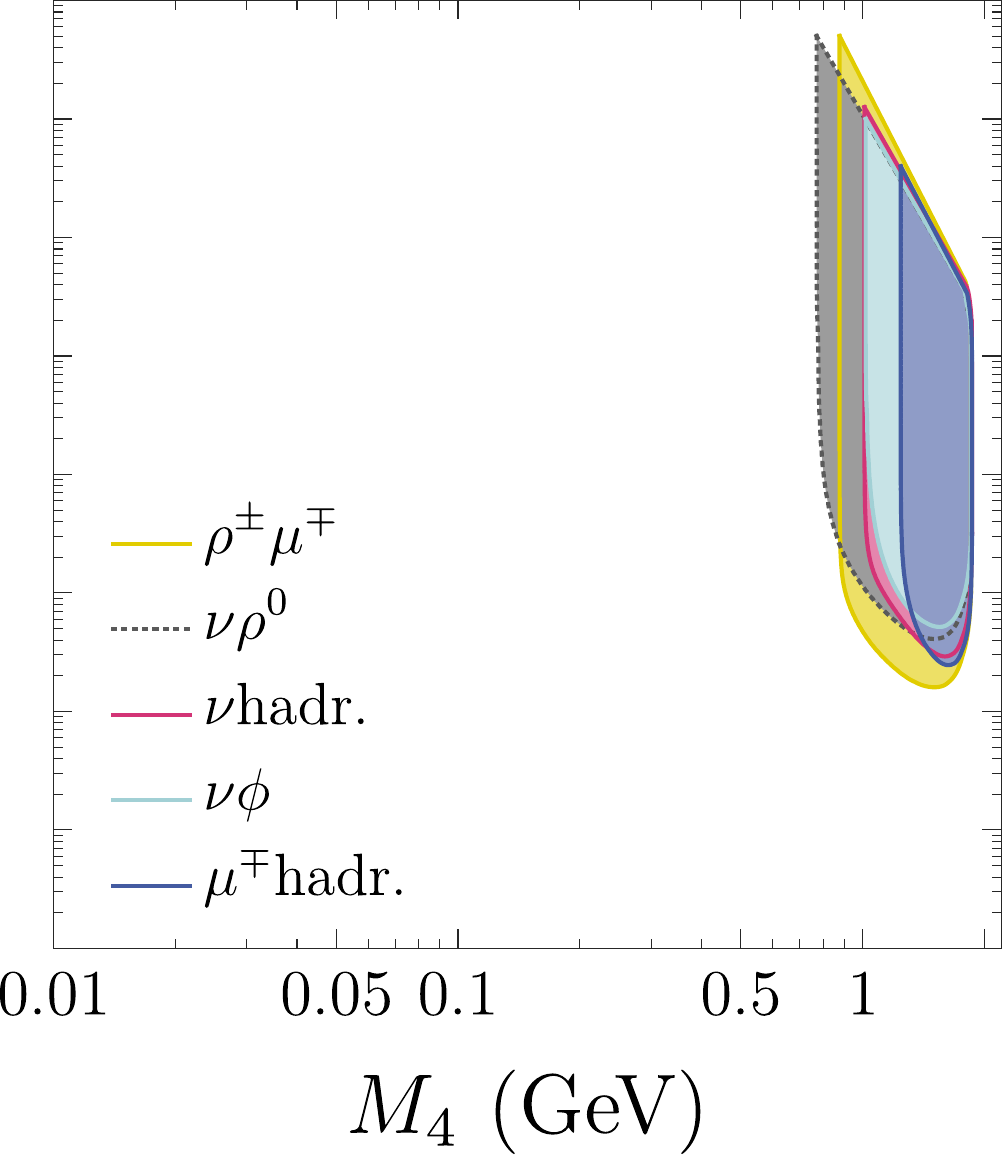}\\\vspace{-0.5cm}\hspace{1cm}
\includegraphics[height=0.55cm,keepaspectratio]{PlotsDuneNuevos/dolor.pdf}\hspace{2.4cm}
\includegraphics[height=0.55cm,keepaspectratio]{PlotsDuneNuevos/dolor.pdf}\hspace{2.4cm}
\includegraphics[height=0.55cm,keepaspectratio]{PlotsDuneNuevos/dolor.pdf}
\\
\includegraphics[height=5.686cm,keepaspectratio]{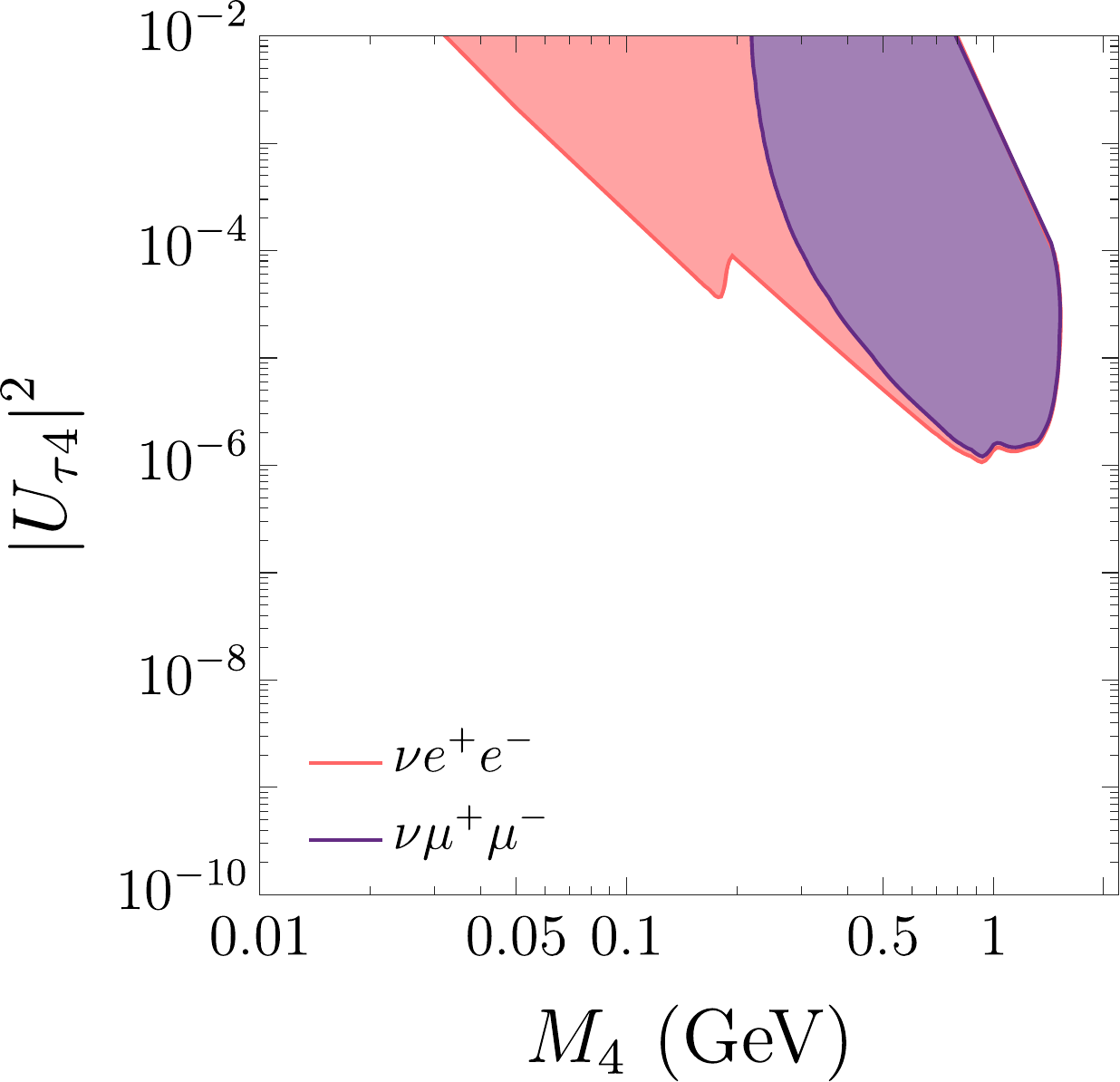} \hspace{-0.47 cm}
\includegraphics[height=5.5cm,keepaspectratio]{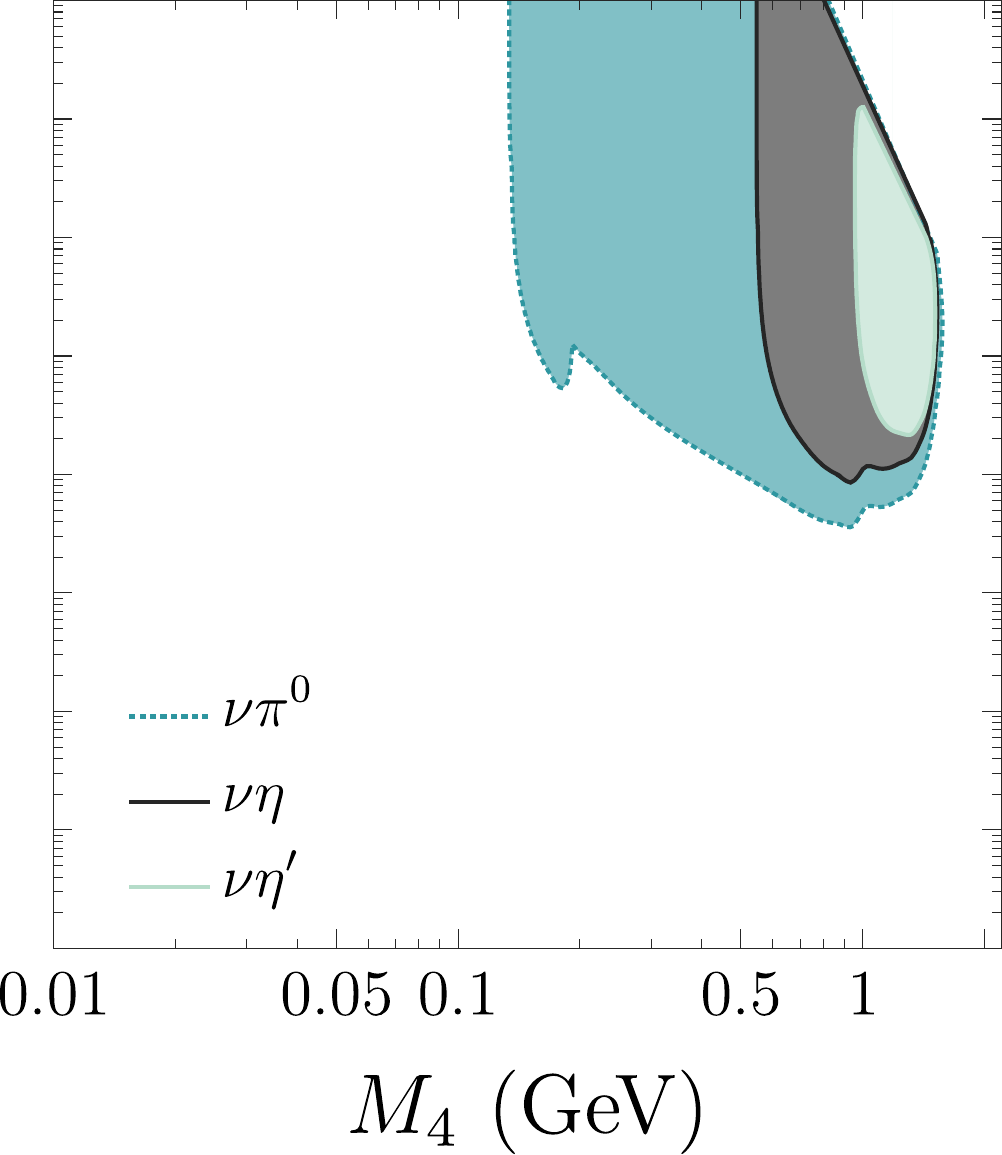} \hspace{-0.47 cm}
\includegraphics[height=5.5cm,keepaspectratio]{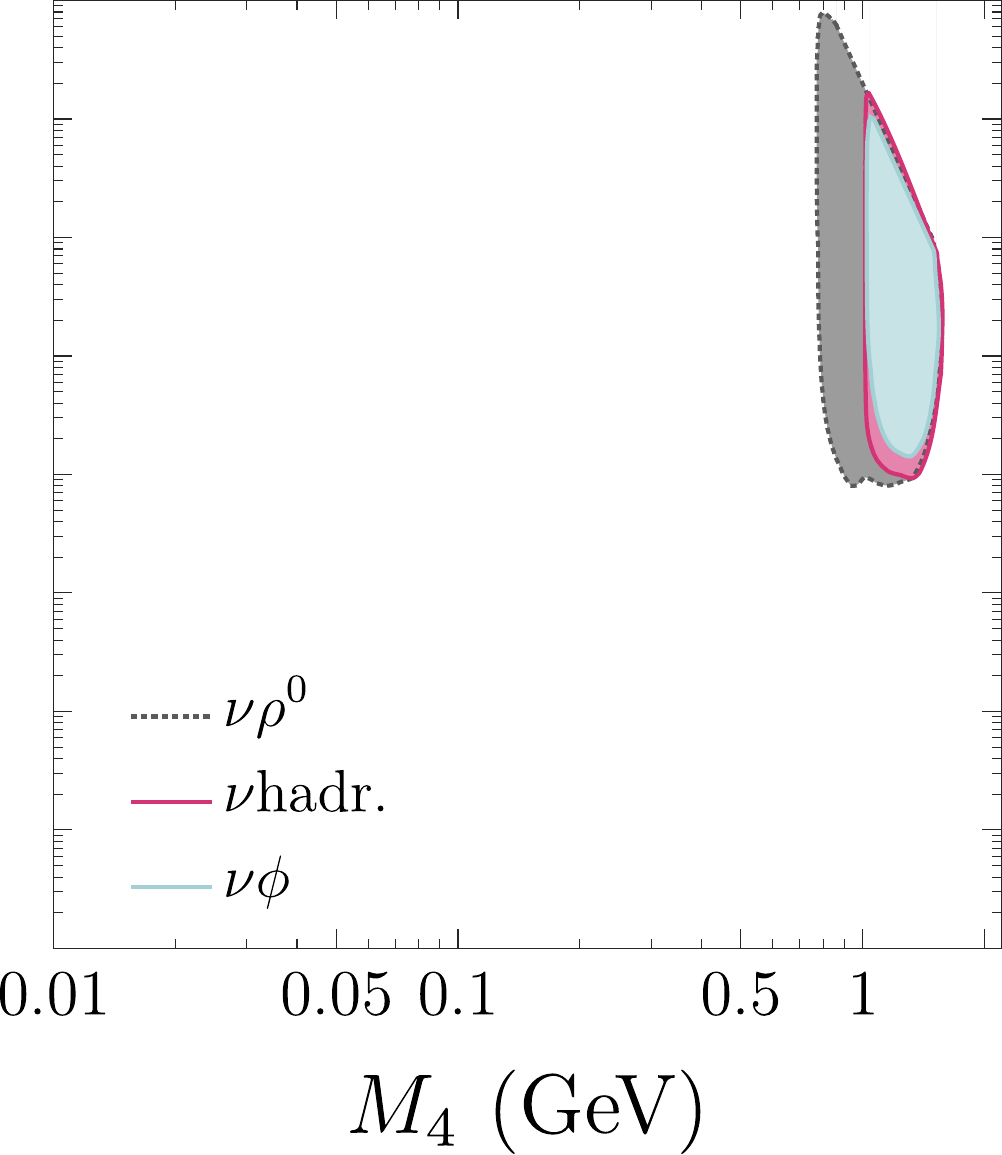}\\\vspace{-0.5cm}\hspace{1cm}
\includegraphics[height=0.55cm,keepaspectratio]{PlotsDuneNuevos/dolor.pdf}\hspace{2.4cm}
\includegraphics[height=0.55cm,keepaspectratio]{PlotsDuneNuevos/dolor.pdf}\hspace{2.4cm}
\includegraphics[height=0.55cm,keepaspectratio]{PlotsDuneNuevos/dolor.pdf}
\\
\caption{Expected DUNE sensitivity (at $90\%$ CL) to the mixing matrix elements $\vert U_{\alpha 4}\vert^2$ as a function of the heavy neutrino mass, for a total of $7.7 \cdot 10^{21}$ PoT. Single flavour dominance is assumed. The different regions correspond to the results for different final states. Left panels correspond to signatures with charged leptons and missing energy, while middle (right) panels correspond to those with pseudoscalar (vector) mesons in the final state.  \label{fig:sens}}
\end{figure}

Fig.~\ref{fig:sens} shows the sensitivity contours in the $(M_N,\vert U_{\alpha 4}\vert^2)$ plane at 90\% CL, for different decay channels. The upper, middle and lower panels in the figure show the results assuming that the heavy neutrino mixes predominantly with the $e$, $\mu$ and $\tau$ sectors respectively. 

\begin{figure}[htb!]
\centering
\includegraphics[height=5.686cm,keepaspectratio]{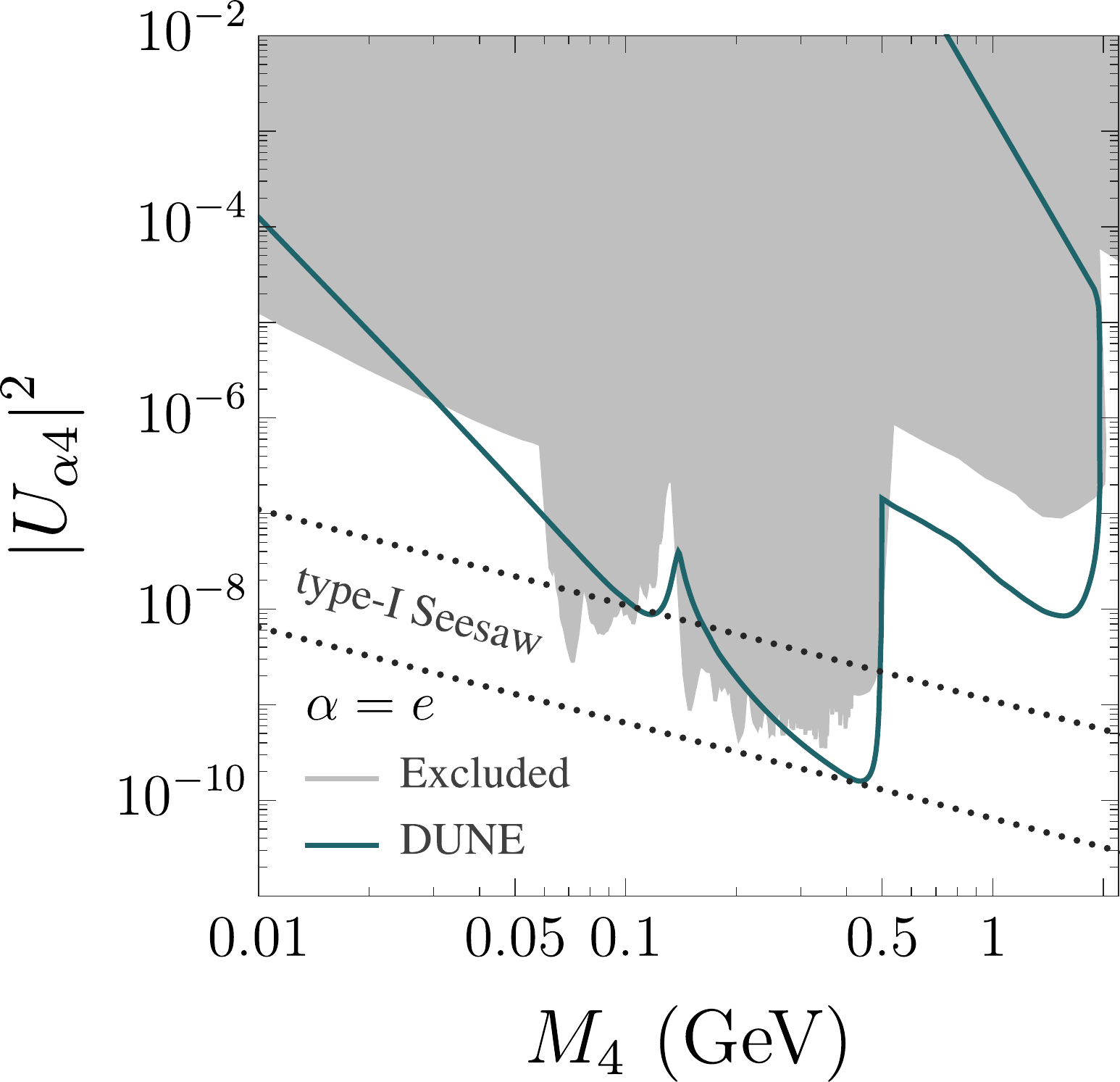} \hspace{-0.47 cm}
\includegraphics[height=5.5cm,keepaspectratio]{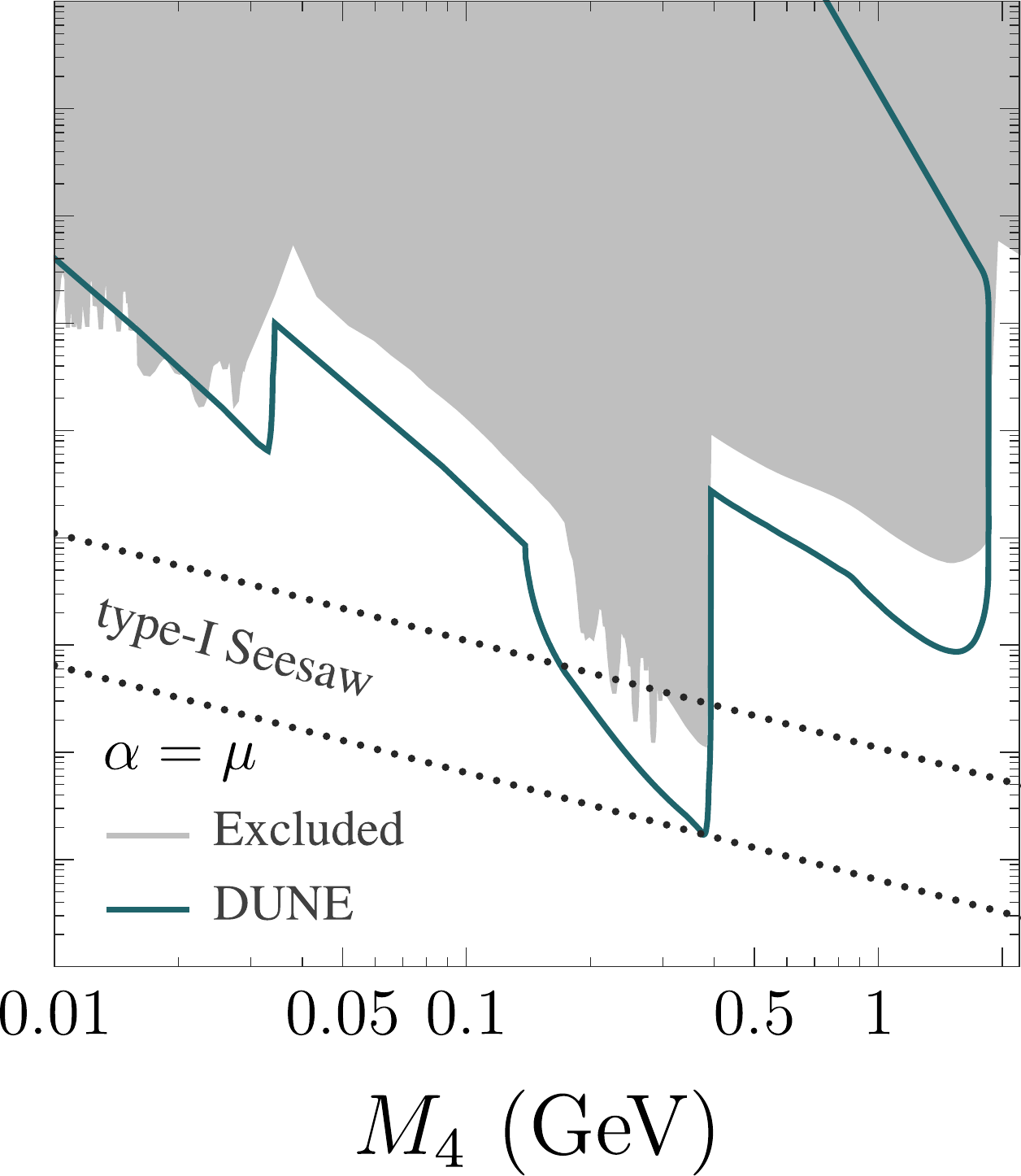} \hspace{-0.47 cm}
\includegraphics[height=5.5cm,keepaspectratio]{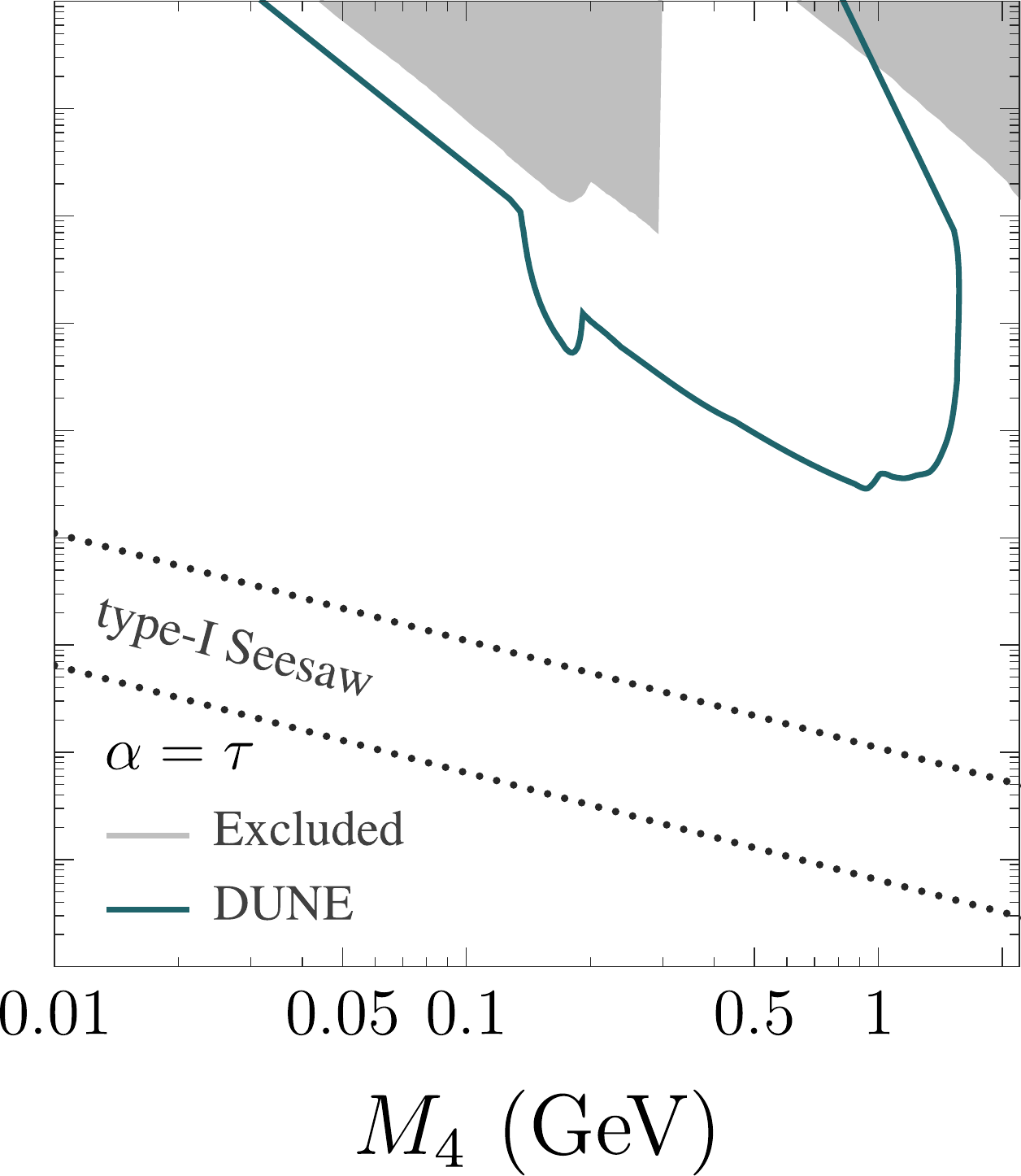}\\\vspace{-0.5cm}\hspace{1cm}
\includegraphics[height=0.55cm,keepaspectratio]{PlotsDuneNuevos/dolor.pdf}\hspace{2.4cm}
\includegraphics[height=0.55cm,keepaspectratio]{PlotsDuneNuevos/dolor.pdf}\hspace{2.4cm}
\includegraphics[height=0.55cm,keepaspectratio]{PlotsDuneNuevos/dolor.pdf}
\\
\caption{Expected DUNE sensitivity (at $90\%$ CL) to the mixing matrix elements $\vert U_{\alpha 4}\vert^2$ as a function of the heavy neutrino mass, for a total of $7.7 \cdot 10^{21}$ PoT, combining all visible decay channels. Single flavour dominance is assumed. The shaded gray areas are disfavored at 90\% CL by present experiments. The dotted gray lines enclose the region predicted by the canonical type I seesaw.  \label{fig:sens_total}}
\end{figure}

Finally, Fig.~\ref{fig:sens_total} summarizes in blue the 90\% CL expected sensitivities at the DUNE near detector to the heavy neutrino mixing $\vert U_{\alpha 4}\vert^2$ as a function of its mass, assuming a Dirac HNL. In the Majorana case, the increase in the number of events would translate into a slightly better sensitivity, although the results would be qualitatively very similar. In this last figure we combine the events from all the channels depicted in Fig.~\ref{fig:sens}, under the same assumption of $20\%$ signal efficiency and negligible background, following Ref.~\cite{Abe:2019kgx}. 

The number of HNL events depends on both their production rate and their decay probability inside the detector. At low masses, the heavy neutrino production is dominated by pion decay, which is roughly proportional to $\vert U_{\alpha 4}\vert^2M_N^2$ (see Eq.~\ref{eq:width_P_N_l}). In this region, the most important HNL decay channel is $N\to \nu e^+e^-$, which is proportional to $\vert U\vert^2M_N^5$ (see Eq.~\ref{eq:width_N_la_la}). Thus, according to Eq.~\ref{eq:prob-approx}, the number of events should scale as $\vert U_{\alpha 4}\vert^4M_N^8$ (an extra $M_N$ power arises due to the $1/\gamma$ factor, proportional to $M_N$). We have indeed verified that the slopes in the low mass regions of Fig.~\ref{fig:sens_total} fit well to $\vert U_{\alpha 4}\vert^2\propto M_N^{-4}$, as expected.

For comparison, the shaded gray areas indicate the parameter space disfavored by current experiments (at $90\%$ CL)\footnote{Note that, since these results were obtained, some works have appeared setting stronger limits. Thus, the gray areas do not represent the full excluded regions, and should only be taken as a reference.}. See Sec.~\ref{sec:mixing_bounds} for a summary of the sources of the most dominant bounds. We do not display here the cosmological bounds, as they are somewhat complementary to those obtained in laboratory experiments, and they should not be directly compared. As a target region, we have also indicated in Fig.~\ref{fig:sens_total} the naive expectation for the mixing matrix elements from the type I seesaw mechanism: $\vert U_{\alpha 4}\vert^2 \sim m_\nu/M_N$, where $m_\nu$ stands for the light neutrino masses. As their absolute scale is not yet determined, we display a band rather than a line. At the lower end we employ the atmospheric mass splitting, $\sqrt{\Delta m^2_\text{atm}}= 0.05$~eV, while the upper line has been set using the latest bound of 0.8~eV from the KATRIN experiment~\cite{KATRIN:2021uub}. Notice, however, that these predictions are somewhat naïve, as single flavour dominance is assumed; besides, the values contained in the band only represent the expected scale of the mixing, but individual elements could either exceed or fall below these ranges.

We find that DUNE is expected to improve over present constraints by several orders of magnitude in a large fraction of the parameter space. In particular, the sensitivity is expected to reach unprecedented levels for HNL masses between the $K$ and $D$ meson thresholds, thanks to the effect of the boost in the detector acceptance, among other factors. Interestingly, in this range of masses, DUNE will be able to explore the region predicted by the type I seesaw mechanism, offering some of the first probes of this model.

We have also compared our results to similar studies in the literature~\cite{Krasnov:2019kdc,Ballett:2019bgd,Berryman:2019dme}. After the number of events have been correspondingly rescaled to account for the different detector volumes, PoT and efficiencies assumed, we find a rather good agreement between the four estimations of the DUNE sensitivity. We find the best overall agreement with Ref.~\cite{Krasnov:2019kdc}. The main difference is a slightly better sensitivity to the $U_{\tau 4}$ mixing in our results, in the small sensitivity peak we find around $M_N \sim 1$~GeV, corresponding to the closure of the $\tau \to N \rho^-$ production channel. This peak also seems absent in the other references. Regarding Ref.~\cite{Ballett:2019bgd}, the main differences we find are at the peaks in sensitivity at the kinematic thresholds of the meson masses, where we find that smaller mixings could be reached. We believe that these differences are due to the effect of the boost factor on the detector acceptance discussed in Sec.~\ref{subsec:boost}, which becomes most relevant close to the kinematic thresholds, as shown in Fig.~\ref{fig:events}. For HNLs heavier than the kaon, Ref.~\cite{Ballett:2019bgd} also finds a lower sensitivity to $U_{\mu 4}$ than the other estimations, which find a similar behavior to that of $U_{e 4}$, as expected from their similar branching ratios. Finally, we also find generally good agreement with Ref.~\cite{Berryman:2019dme}. The main differences are in the areas of parameter space were the HNL decays to $\rho^0$ and especially to $\pi^0$ are most relevant, since these decay modes were not included. The slope of the sensitivity curves is also slightly less steep than the $\vert U_{\alpha 4}\vert^2\propto M_N^{-4}$ found in the other references.

\section{Summary}
\label{sec:conclusions}

Although mainly designed to measure the mass ordering and the CP phase of the PMNS, the future Deep Underground Neutrino Experiment offers great chances to probe other kinds of new physics. Chief among them is the existence of heavy neutrinos. The high luminosity of the primary beam would yield a huge number of charged mesons, that could produce a large amount of HNLs upon their decay. For certain values of their mass and mixing, these new particles could decay back into SM particles (mainly leptons and lighter mesons) inside the DUNE ND.

We have employed the effective theory derived in the previous chapter to simulate the HNL fluxes produced at DUNE, as well as their decays inside the ND, finding its sensitivity to heavy neutrino mixing as a function of its mass. Although our results are just estimations, as proper detector simulations and background analyses are required, we find that DUNE could considerably improve the current limits. In particular, the sensivity is particularly good close to the pion and kaon masses, partially due to the boost experienced by the HNL, that increases the detector acceptance. Interestingly, in that range of masses, very small values of the mixing, as small as those predicted by the type I seesaw mechanism, could be probed.

\chapter{Heavy neutrinos in an effective field theory}
\label{sec:eft}
\fancyhead[RO]{\scshape \color{lightgray}12. Heavy neutrinos in an effective field theory}
Effective field theories (EFTs) have been employed for decades to describe the impact of high-energy theories in low-energy processes. If a given theory contains heavy particles, these will hardly play any dynamical role at lower energies, as they cannot propagate. However, their presence still has an effect on the interactions of the light degrees of freedom. Usually, this occurs by means of propagators of heavy particles, that, upon integration-out, reduce to point interactions involving several light particles. The paradigmatic example of this framework is Fermi\textquotesingle s theory, that describes the weak interactions of SM fermions at the GeV scale. As this energy is much smaller than the electroweak scale, the weak gauge bosons can be integrated out, yielding four-fermion interactions that describe, for instance, $\beta$-decays. This approach is very successful at these energies, but starts to worsen as the scale is raised, completely breaking down when reaching the $W$ mass. The effective theory derived in Ch.~\ref{sec:mesons} is in the end another facet of Fermi\textquotesingle s theory, as the weak gauge bosons are integrated out in order to deal with the interactions of mesons that lie at the GeV scale.

The assumption behind the EFT framework, that implies that the energies of the relevant processes are much lower than the typical scale of the full theory, shows up as an expansion in powers of a certain ratio of energy scales. In the case of Fermi\textquotesingle s theory, this parameter would roughly be $p/M_W$, with $p$ generically denoting the momenta of the involved particles. The usual Fermi Lagrangian is obtained by directly neglecting this quantity, reducing the $W$ propagator to a constant; however, higher-order terms would appear if further terms of the expansion are kept. Thus, the full Fermi Lagrangian, and in general that of any EFT, consists of an infinite tower of operators, increasingly suppressed by the small ratio of scales. In turn, the subsequent pieces of the Lagrangian would exhibit higher energy dimensions, which is also a typical feature of effective theories. Thus, EFTs are only renormalizable order by order.

Fermi\textquotesingle s theory followed a bottom-up approach: it only contained the particles known at the time (protons, neutrons and electrons), being unaware of the weak interactions, which were the underlying full theory. This spirit is still employed today, using EFTs that attempt to describe the low-energy effects of an unknown, high-energy theory. However, a top-bottom point of view is also quite useful, as it allows to perform low-energy computations in a simpler, more intuitive way. 

Following the bottom-up approach, the most general and consistent way to build an EFT is to write any operator containing the SM particles and respecting its fundamental symmetries. The result is the Standard Model Effective Field Theory (SMEFT)~\cite{Buchmuller:1985jz,Grzadkowski:2010es,Brivio:2017vri}. Although this framework contains an infinite tower of operators, the number of terms that can be written at each order is finite, and basically determined by the fundamental ingredients of the SM. As the possible full theory that generates these effective operators is unknown, their particular coefficients, as well as the suppressing energy scale, are a priori free parameters.

The SMEFT Lagrangian can be schematically written as
\begin{equation}
    \mathcal{L} =  \mathcal{L}_{d=4} +\mathcal{L}_{d=5} +  \mathcal{L}_{d=6}+...\,,
\end{equation}
with
\begin{equation}
    \mathcal{L}_{d=N} = \sum_i \mathcal{O}_i = \frac{1}{\Lambda^{N-4}}\sum_i C_{i} \widetilde{\mathcal{O}_i}\,.
\end{equation}
Here, $\Lambda$ is the scale of new physics, which is usually assumed to be the same for all the operators, increasingly suppressing higher-order terms. $\widetilde{\mathcal{O}_i}$ are the different effective operators, whose relative intensity is controlled by the different $C_i$, dimensionless parameters dubbed Wilson coefficients. These are in general unrelated, although some assumptions can be performed. This is arguably one of the main drawbacks of EFTs in general and of the SMEFT in particular, as the number of free, uncorrelated Wilson coefficients is quite large, even at dimension 6. In fact, the higher the dimension, the greater the number of possible operators (although their effect on low-energy observables is more suppressed). This clearly reduces the predictive power of this approach, and some simplifications are needed, such as considering only one effective operator at a time. 

Note that most EFTs contain the SM, extending it with effective operators that encode possible new physics effects. If these were not present, the SM could still be recovered by choosing some values for the parameters that control the effective interactions. For instance, the SMEFT comprises the SM Lagrangian; if all the Wilson coefficients of the higher-dimensional operators are set to 0, or the new physics scale tends to infinity, the SM interactions are recovered. Thus, observing deviations from the SM predictions in a particular process would point to non-vanishing Wilson coefficients. Here lies the appeal of the SMEFT, and of EFTs in general, as they allow to encode BSM effects in low-energy observables in a model-independent way. Although it is not straighforward to obtain information on how the high-energy theory would look like, the EFT approach would at least allow to point at disfavoured or preferred models. 

As it has been lengthly discussed already, neutrino masses require an extension of the SM, but it remains unclear whether that new theory lies at very high energies (as suggested by the original seesaw formulations) or instead contains light degrees of freedom (as predicted by low-scale seesaw mechanisms). In the former case, heavy neutrinos would be the degrees of freedom to integrate out. As mentioned in Sec.~\ref{sec:LN_LSS}, this would directly yield the Weinberg operator, already contained in the SMEFT. Higher-dimensional operators would also be generated, although the strong suppression that yields tiny neutrino masses implies that these terms would be even smaller, explaining why they have not been observed so far. This is the case of the dimension-6 operator, corresponding to active-heavy neutrino mixing, a phenomenon yet to be seen. Other theories that account for neutrino masses would also translate into effective operators. 

However, if heavy neutrinos lie close to the electroweak scale, or even below it, it would be more appropriate to treat these new particles as light degrees of freedom, that appear dynamically in the effective theory together with the SM particles. As argued previously, there are strong arguments to think that heavy neutrinos could lie at the GeV scale, so a scenario in which these states are present in a low-energy EFT is quite sensible. In this situation, the SMEFT needs to be extended, as the presence of gauge singlet fermions allows to write new operators, that couple the heavy neutrinos to SM fields. 

This results in the $\nu$SMEFT. Its operators were first systematically considered in Ref.~\cite{delAguila:2008ir} (see also~\cite{Graesser:2007yj,Graesser:2007pc}), and then in Ref.~\cite{Liao:2016qyd}, where redundant operators were removed, and new dimension-7 operators were included. This effective theory, that has been widely studied in the literature~\cite{Graesser:2007yj,Graesser:2007pc,delAguila:2008ir,Aparici:2009fh,Peressutti:2014lka,Duarte:2014zea,Bhattacharya:2015vja,Duarte:2015iba,Duarte:2016caz,Liao:2016qyd,Duarte:2016miz,Caputo:2017pit,Duarte:2018xst,Alcaide:2019pnf,Chala:2020vqp,Barducci:2020icf,Dekens:2020ttz,Biekotter:2020tbd,Duarte:2020vgj,Barducci:2020ncz,Dekens:2021qch,Cottin:2021lzz,Li:2021tsq,Cirigliano:2021peb,DeVries:2020jbs,Zhou:2021ylt,Zhou:2021lnl,Beltran:2022ast,Delgado:2022fea,Barducci:2022gdv,Talbert:2022unj,Barducci:2022hll,Zapata:2022qwo,Mitra:2022nri,Beltran:2023nli,Dekens:2023iyc}, allows to describe how the behaviour of HNLs would be modified if they were affected by BSM physics, besides their usual mixing with the light states. Thus, new couplings of HNLs to SM particles are introduced, having a potential impact on low-energy observables. 

The $\nu$SMEFT is an interesting tool to probe HNLs beyond the simplest mixing scenario, without relying on any particular model. This framework has been typically explored from the point of view of collider physics, but it is also interesting from the low-energy perspective. In fact, if the HNLs lay at the MeV-GeV range, the effective operators could potentially mediate many different processes, observable at various experimental facilities. Obviously, as heavy neutrinos have not been discovered, these effects must be, at most, very feeble, so the corresponding Wilson coefficients must be very small. 

This will be the focus of this chapter, based on Ref.~\cite{Fernandez-Martinez:2023phj} and devoted to analyzing how low-energy observables can be employed to probe the operators of the $\nu$SMEFT. We will focus on dimension-6 operators, as, beyond that order, the number of operators is hardly manageable, and their effects are also more suppressed. We will collect all the operators at this order that include at least one HNL field, and study the processes they could mediate. We will then employ the current experimental bounds on such observables to set bounds on the Wilson coefficients as a funcion of the heavy neutrino mass\footnote{Apps.~\ref{app:EFT_widths} and~\ref{app:recasting} contain the details of the expressions employed to derive such limits.}. Notice that the constraints on heavy neutrino mixing are relevant for many of the effective operators we will consider: see Sec.~\ref{sec:mixing_bounds} for the main bounds available up to date. 

A generic analysis of all observables and effective operators is quite demanding, and a global fit to all the Wilson coefficients would be the ideal approach. As a first step, we will instead work in a simplified scenario, in which we will only consider one effective operator at a time. In other words, when analyzing the effects of a particular operator, we will assume that all the other Wilson coefficients vanish. Furthermore, we will also neglect the effects of the mixing of the heavy neutrinos with the active states. While these assumptions will not always encompass specific high-energy completions, they are reasonable from a phenomenological point of view, as they greatly simplify the analysis and, in general, provide conservative results. This is due to the fact that activating several operators at the same time allows for more HNL interactions; hence, more observables would be relevant, and strongest contraints would apply. Exceptions to this reasoning can appear in the form of flat directions, as cancellations between different contributions may arise. We will comment on some of these situations when relevant. 

We will work in a framework with a single HNL species, which is a Majorana particle, described by the usual Lagrangian of the type I seesaw (Eq.~\ref{eq:seesaw_lag}). As mentioned before, we will assume that no other terms affect the heavy neutrinos at the renormalizable level. We will treat the mass of these particles as a free parameter, ranging from 1 MeV to 100 GeV, roughly. Reproducing the observed neutrino masses would require a more complex scenario, but this simplified setup is quite useful to probe possible new physics at low energies. Note that our results would be qualitatively the same for Dirac HNLs, with small quantitative differences due to slightly different production or decay rates. 

Before diving on the dimension-6 operators, let us briefly comment on the only two operators that contain HNLs at dimension 5 (see Ref.~\cite{Caputo:2017pit} for a detailed discussion)\footnote{We will denote the HNL fields as just $N$, leaving implicit their RH chirality.}:
\begin{align}
    \mathcal{O}^{d=5}_{\rm Higgs} &=\frac{C^{d=5}_{\rm Higgs}}{\Lambda} \overline{N^c} N |H|^2\,, \label{eq:dim5op_Higgs}\\
    \mathcal{O}^{d=5}_{\rm dipole} &=\frac{C^{d=5}_{\rm dipole}}{\Lambda} \overline{N^c} \sigma_{\mu\nu} N B^{\mu\nu}\,.\label{eq:dim5op_dipole}
\end{align}
The first one contributes to the Majorana mass term and to the invisible decays of the Higgs boson into two HNLs. 
The associated phenomenology was studied in Refs.~\cite{Graesser:2007pc,Caputo:2017pit,Barducci:2020icf,Barducci:2020ncz,Barducci:2022hll}.
The focus of these works was on the sensitivity of collider experiments to
prompt or displaced vertex signals originated by HNL visible decays, which depend on the mixing elements and the HNL mass. Additionally, measurements of the Higgs boson signal strength, $\mu_h$, provide an important constraint on $C^{d=5}_{\rm Higgs}/\Lambda$, since the exotic Higgs decay $h\rightarrow NN$ would affect this quantity. As mentioned in Sec.~\ref{sec:Majoron_Higgs}, the combination of the ATLAS and CMS measurements yields the lower limit $\mu_h>$ 0.94. Taking into account that this operator does not affect the Higgs production cross section, this translates into an upper bound of the 6\% on the invisible branching ratio. For HNLs lighter than roughly 40 GeV, we find 
\begin{equation}\label{eq:dim5bound}
    \frac{C^{d=5}_{\rm Higgs}}{\Lambda}<3 \cdot 10^{-5}~\mathrm{GeV}^{-1}\,.
\end{equation}
This constraint is comparable with the expectation for the LHC sensitivity from direct searches~\cite{Caputo:2017pit} or even stronger. It should be noted that this current bound is independent of the HNL mixing and mass as long as $M_N\lesssim 40$ GeV. The bound becomes increasingly weaker for heavier masses up to $M_h/2$.

The second operator requires at least two different HNL fields to not vanish, and amounts to a transition magnetic moment for the HNLs involved. It has been discussed, for instance, in Refs.~\cite{Aparici:2009fh,Barducci:2022gdv}.

\section{Higgs-dressed mixing}
\label{sec:higgs_mix}
The first and simplest operator we consider at dimension 6 is the addition of a Higgs bilinear to the usual Yukawa coupling (Eq.~\ref{eq:seesaw_lag}),
\begin{align}\label{eq:higgsdressed}
    \mathcal{O}_{\rm LNH}^\alpha = \frac{C_{\rm LNH}^\alpha}{\Lambda^2}(H^\dagger H) \overline{L_\alpha}\widetilde{H}N  \,,
\end{align}
where $\overline{L_\alpha}$ is a left-handed lepton doublet of flavour $\alpha$. Once the extra pair of Higgs doublets acquire vevs, this operator induces extra contributions at tree level to light neutrino masses and their mixing with the HNLs. 
In the limit where $M_N \ll v \ll \Lambda$, the generated mass and mixing read
\begin{align}
    U_{\alpha 4} &= \frac{C_{\rm LHN}^\alpha}{2\sqrt{2}} \frac{v}{M_N}\left(\frac{v}{\Lambda}\right)^2, 
    \\
    m_{\nu} &=
    \frac{(C_{\rm LHN}^\alpha)^2}{8} \frac{v^2}{M_N}\left(\frac{v}{\Lambda}\right)^4\,.
    \label{eq:higgs_mix}
\end{align}
Note that there are three copies of this operator, one for each lepton flavour, whose Wilson coefficients are, in principle, independent.

At low energies, the mixing induced by this operator gives rise to an identical phenomenology to that already studied in the simplest, renormalizable case. Thus, all bounds on mixing from direct constraints apply directly, yielding limits on the Wilson coefficient. 
There could be possible cancellations between this contribution and the tree level dimension-4 Yukawa couplings, leading to possible flat directions; as we work under the assumption of negligible standard mixing, we will not take them into account. In principle, it could be possible to lift this flat direction, as this effective operator also mediates processes involving two HNLs and several Higgs bosons, and the dimension-4 Yukawa does not. 
However, these processes, such as $hh(h)\to NN$, require multi-Higgs production and thus do not offer promising experimental prospects.

The existing limits on the absolute neutrino mass scale would also provide extremely stringent constraints on the Wilson coefficient. However, the expression shown in Eq.~\ref{eq:higgs_mix} corresponds to the inclusion of a single HNL. When several of these new particles are considered, their contributions to  neutrino masses could cancel each other, while this is not a possibility for their mixings. Indeed, this is exactly what would be expected in low-scale realizations of the seesaw mechanism. In those scenarios, the contribution to the neutrino mass cancels, due to its protection from the approximate lepton number symmetry, while the mixing does not. Such a protection mechanism needs to hold in order to have such light HNLs in the first place. Thus, we do not include bounds on the Wilson coefficient arising from the current information on the absolute value of neutrino masses.
\begin{figure}[t!]
\centering
\includegraphics[width=\columnwidth]{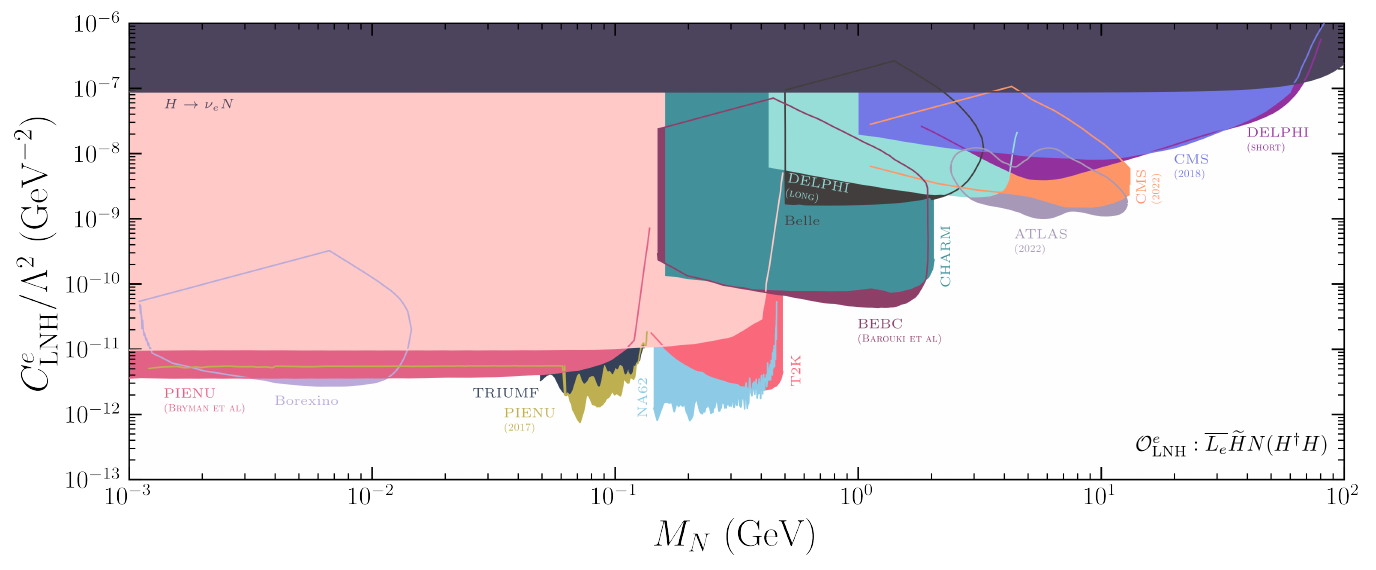}
\includegraphics[width=\columnwidth]{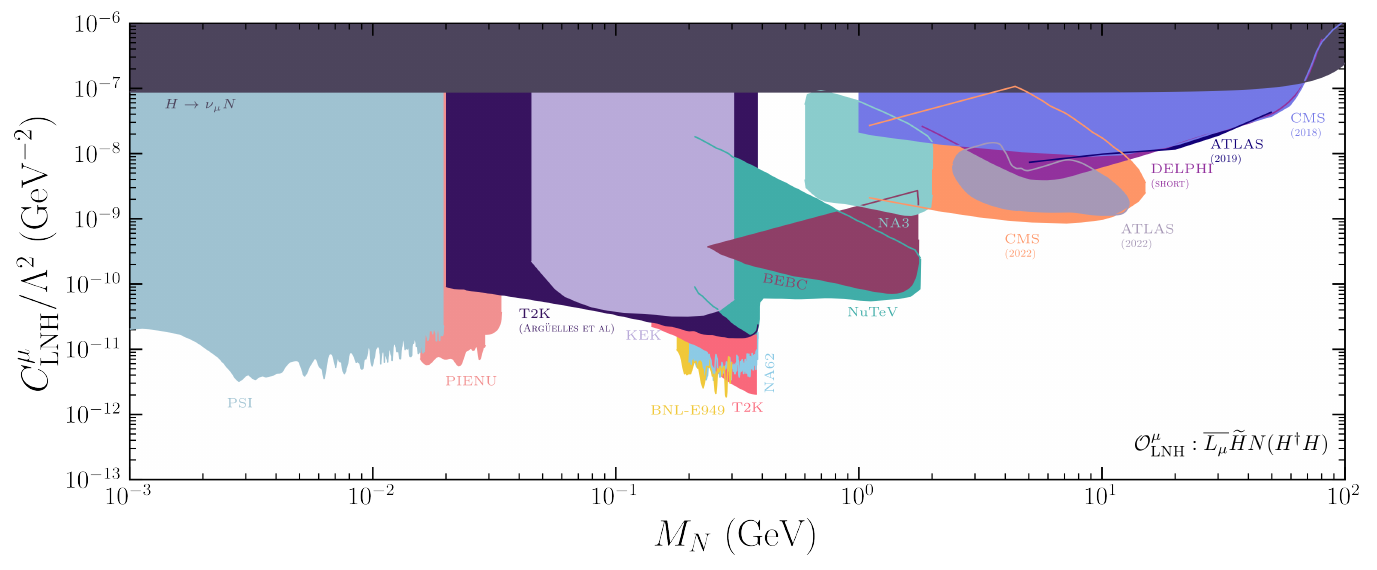}
\includegraphics[width=\columnwidth]{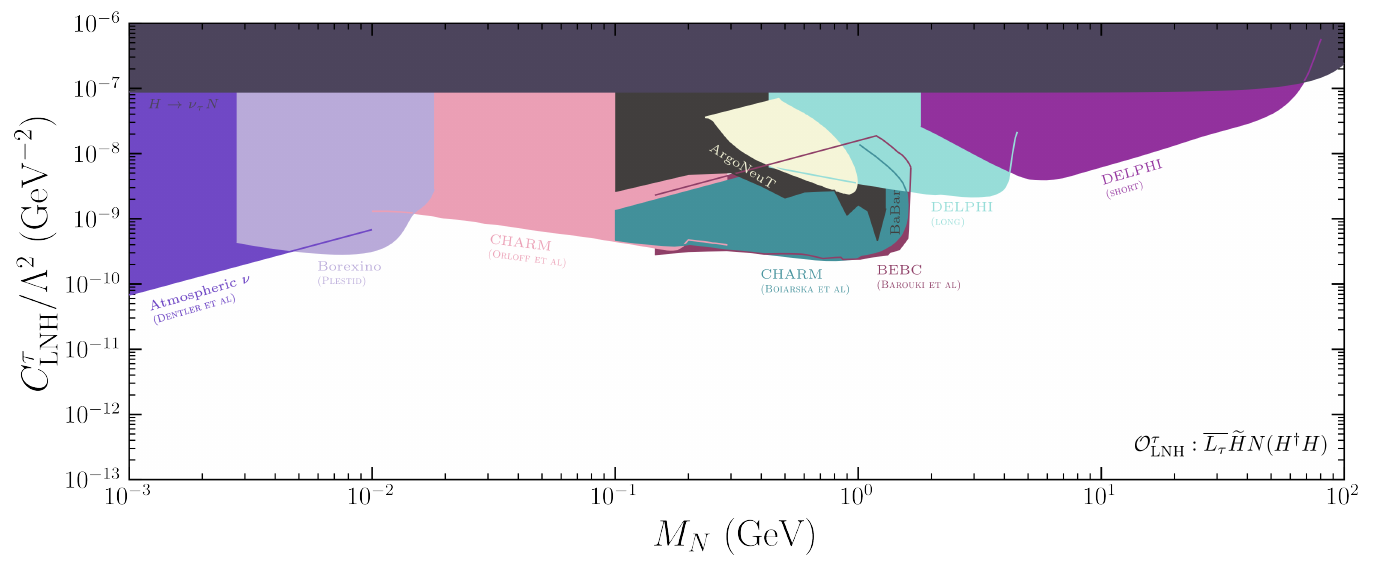}
\caption{90\% CL constraints on the Wilson coefficients of the Higgs-dressed mixing operator (Eq.~\ref{eq:higgsdressed}), as a function of the heavy neutrino mass. We display in separate panels the bounds relevant for each lepton flavour.}
\label{fig:higgsdressed}
\end{figure}

Apart from contributing to the neutrino masses and mixing, this operator induces an invisible Higgs decay channel, into a light and a heavy neutrino. Once again, the corresponding branching ratio is bounded by measurements of Higgs signal strengths~\cite{ATLAS:2022vkf, CMS:2022dwd}, posing alternative constraints on the Wilson coefficients.

In Fig.~\ref{fig:higgsdressed}, we show the bounds on the Wilson coefficients of this operator under the three lepton flavour dominance scenarios, shown in three separate panels. Most shaded regions (except the dark gray bands) correspond to reinterpreted mixing bounds, and, according to the assumption of single flavour dominance, are different for each lepton flavour. The limits from Higgs invisible decay (dark gray region) apply to the dominant Wilson coefficient, independently of its flavour.

\section{Bosonic currents}
\label{sec:boson_curr}

It is possible to couple bosonic currents to the $V+A$ currents constructed out of heavy neutrinos. The resulting operators read
\begin{align}
\label{eq:OHN}
    \mathcal{O}_{\rm HN} &=\frac{C_{\rm HN}}{\Lambda^2} \overline{N}\gamma^\mu N (H^\dagger i \overleftrightarrow{D}_\mu H) \,,
\\
\label{eq:OHNe}
    \mathcal{O}_{\rm HN\ell}^{\alpha} &= \frac{C_{ \rm HN\ell}^{\alpha}}{\Lambda^2} \overline{N}\gamma^\mu \ell_\alpha (\widetilde{H}^\dagger i \overleftrightarrow{D}_\mu H) \,,
\end{align}
where $\ell_\alpha$ is a right-handed charged lepton of flavour $\alpha$. After electroweak symmetry breaking, the first operator will yield a coupling of the $Z$ boson to two HNLs, while the second operator will produce a coupling of the $W$ boson to a charged lepton and an HNL. The latter will exhibit three copies, depending on the flavour of the charged lepton. The phenomenological consequences of these two operators are quite distinct, and so are the constraints placed upon their corresponding Wilson coefficients. 
They are discussed separately below.

\subsection{Neutral bosonic current}
\label{sec:neut_bos_curr}

The operator $\mathcal{O}_{\rm HN}$ yields a vertex between two HNLs and a $Z$, opening a new decay channel for this boson if kinematically allowed by the masses of the HNLs. This decay, controlled by the Wilson coefficient $C_{\rm HN}$, constitutes an extra contribution to the invisible decay of the $Z$ boson, which was measured with high accuracy at LEP~\cite{ParticleDataGroup:2022pth,Janot:2019oyi}. We have performed a $\chi^2$ fit to this observable, obtaining upper limits on $C_{\rm HN}/\Lambda^2$.

Similarly, this neutral bosonic current would mediate decays of neutral mesons ($\pi^0$, $\eta$, $\eta'$, $J/\psi$, $\rho^0$, $\Upsilon$(1S)) into two HNLs, if light enough. Analogous invisible decays are also present in the SM, with light neutrinos in the final state~\cite{Marciano:1996wy,Chang:1997tq}. 
For pseudoscalar mesons, these processes require a chirality flip, so in the SM they are suppressed by the small neutrino masses; for vectors, they are very subdominant with respect to hadronic and electromagnetic channels.
Thus, this kind of decay has not been observed so far, and only upper bounds are available.  
Imposing the decays into HNLs to respect the experimental limits provides another method to constrain the Wilson coefficient. We will only display the upper bounds given by the neutral pion and the $\Upsilon$(1S) invisible decays, as they are the most stringent. In particular, the NA62 collaboration determined the corresponding branching ratio of the pion to be smaller than $4.4 \cdot 10^{-9}$, at 90\% CL~\cite{NA62:2020pwi}. For the $\Upsilon$(1S), we employ the limit of $3 \cdot 10^{-4}$, at 90\% CL~\cite{BaBar:2009gco}.
Stronger limits on the former can be obtained using $\gamma \gamma \to \pi^0 \to N N$ in supernovae (see also the supernova discussion below).
The limit is $3.2 \cdot 10^{-13}$ for the BR~\cite{Natale:1990yx}.

Another probe of this effective operator comes from collider physics, in particular, monophoton searches performed at LEP~\cite{DELPHI:2003dlq}. This channel, intended to look for processes in which one or more invisible particles were produced in the $e^+e^-$ collision, has the photon as the only detectable signal. These searches are blind to the nature of the invisible particle, dark matter being one of the most recurrent candidates. We follow Ref.~\cite{Fox:2011fx}, where an EFT framework was also adopted, with dark matter fermion singlets playing a similar role as our HNLs. We translate this information into constraints on $C_{\rm HN}/\Lambda^2$. Note that Ref.~\cite{Fox:2011fx} employs a basis of effective operators different from the one considered here. To derive our bounds, we compute the monophoton production cross section for the operators of our basis, applying the same cuts in the photon energy and angle, and rescale the constraints accordingly to obtain the same cross section in the signal region as for the operators in Ref.~\cite{Fox:2011fx}. 
Details on this rescaling procedure can be found in App.~\ref{app:monophoton}.

A very similar process could be mediated by this operator replacing electrons with quarks and the photon with a gluon, yielding experimental signals composed of a single jet and missing energy. Thus, monojet searches at the LHC pose an additional constraint on the corresponding Wilson coefficient. 
Limits from these processes are not available for the operator under consideration, and a proper rescaling would involve a detailed simulation, which is out of the scope of this thesis. Nevertheless, a crude estimate, obtained by rescaling parton-level cross sections, yields bounds similar to those from monophoton searches. Monojets will also be relevant for the observables that will be discussed in Sec.~\ref{sec:four-ferm}.

Supernovae also offer competitive probes of particles that couple very feebly with the SM. 
Due to their weakly interacting nature, neutrinos are the dominant cooling mechanism of core-collapse supernovae.
This mechanism is compatible with the observation of neutrinos from SN1987A by the Kamiokande-II~\cite{Hirata:1987hu} and IMB~\cite{Bionta:1987qt} neutrino detectors.
Thus, a BSM particle would be incompatible with observations if it exhibited a mass and couplings such that it extracted energy from the supernova at a faster rate than SM neutrinos~\cite{Raffelt:1987yt}. 
This usually provides upper and lower limits on the couplings: if the new particle couples too strongly with the SM fields, it would be trapped inside the supernova, whereas, if the coupling is too small, the new particle is produced much less often than neutrinos, cooling the supernova in a smaller amount. We follow Ref.~\cite{DeRocco:2019jti}, where these arguments are used to constrain the parameter space of a fermion singlet, also working in an EFT framework. 
We again rescale their results so that the production and scattering cross sections of the HNLs coincide with those for the operators in Ref.~\cite{DeRocco:2019jti} along the lines defining the lower and upper bounds respectively (see App.~\ref{app:supernova}). 
Note that, depending on the HNL mass, the most stringent upper bound may be given by HNL production in $\pi^0$ decay in supernovae, as discussed above.

\begin{figure}[t!]
\includegraphics[width=\columnwidth]{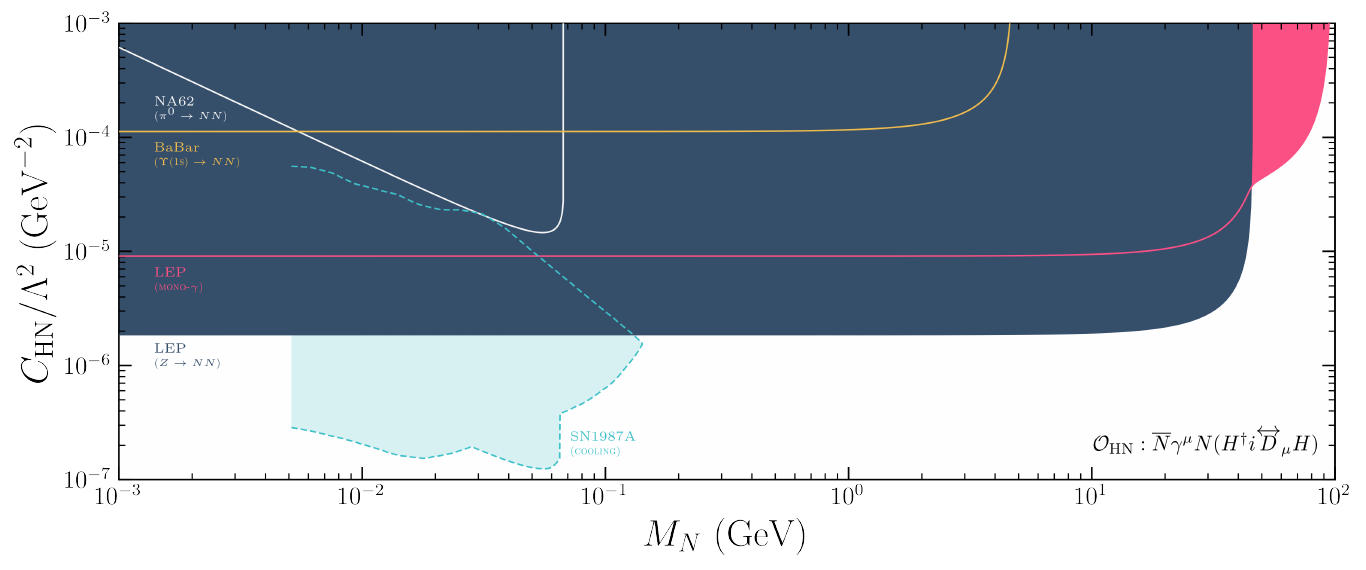}
\caption{90\% CL constraints on the Wilson coefficient of the neutral bosonic current operator (Eq.~\ref{eq:OHN}), as a function of the heavy neutrino mass.}
\label{fig:neutbosoncurr}
\end{figure}

Finally, it should be noted that for this operator, or, more generally, for those involving two HNL fields, and in absence of mixing or other interactions, HNLs would be completely stable. Thus, their production in the early Universe could lead to an overabundant dark matter component and very stringent constraints. As discussed in Sec.~\ref{sec:mixing_bounds}, we regard these important cosmological constraints as complementary to the direct laboratory bounds. Indeed, their comparison provides a test of the necessary assumptions on the cosmological model, which are not present in the laboratory or astrophysical observations. For instance, if the reheating temperature were below the HNL mass, the overproduction constraint would not apply. 

Fig.~\ref{fig:neutbosoncurr} collects the bounds on $C_{\rm HN}/\Lambda^2$ derived from the different observables discussed above: invisible $\pi^0$, $\Upsilon$(1S), and $Z$ decays (white, yellow and dark blue regions respectively), monophoton searches (pink area) and supernova cooling arguments (light blue region). Note that the bounds on the standard HNL mixing do not apply, as they typically involve CC interactions (mostly charged meson decay), which are not mediated by this operator.

\subsection{Charged bosonic current}
After EWSB, the operator $\mathcal{O}_{\rm HN\ell}^{\alpha}$ leads to a vertex between an HNL, a charged lepton, and a $W$. Its chiral structure is similar to the SM charged currents, yielding a coupling of the $W$ to a RH current (analogous to the LH one in the case of standard mixing), controlled by the Wilson coefficient via the \say{effective mixing} 
\begin{equation}
    U_{\alpha 4}^{\rm CC} \equiv \frac{C_{\rm HN\ell}^{\alpha} v^2}{\sqrt{2}\Lambda^2}\,.
    \label{eq:effectiveCCmixing}
\end{equation}
We then reinterpret the current bounds on HNL mixing as limits on $C_{\rm HN\ell}^{\alpha}/\Lambda^2$. 
Most of the bounds only involve processes mediated by the SM charged currents, and thus translate directly by means of $U_{\alpha 4}^{\rm CC} = U_{\alpha 4}$. 
However, some of the experimental searches also consider NC-mediated processes, absent for the effective operator under consideration. 
The limits provided by those searches need to be rescaled, to remove the neutral contributions while keeping constant the expected number of events at the detector.
We describe our rescaling procedure in App.~\ref{app:rescaling}.

Supernova cooling and LEP monophoton searches are also able to probe charged bosonic currents. There are no basic conceptual differences with the neutral case; the main distinctions appear at a diagrammatic level, as the HNLs are now produced via the exchange of a $W$ in a $t$ channel. 
However, the associated bounds are not competitive with those arising from the mixing, as the corresponding processes are NC-like. Thus, we will not display them.

Finally, this effective operator mediates leptonic decays of muon and tau, $\ell_\alpha\to\ell_\beta\nu_\beta N$. We have performed a $\chi^2$ fit of the corresponding decay widths of the muon and the tau to the experimental measurements~\cite{ParticleDataGroup:2022pth}, finding limits on the Wilson coefficient, that dominate at low masses. Note that, in the channel $\tau\to\mu\nu N$, the experimental determination exhibits a slight tension, of roughly 2$\sigma$, with the SM prediction. This means that, at the 90\% CL, the Wilson coefficient in the tau flavour dominance scenario is constrained to a band instead of only an upper bound.
We display the corresponding lines in a dashed style, as they are not completely equivalent to those arising from other observables.

\begin{figure}[h!]
\includegraphics[width=\columnwidth]{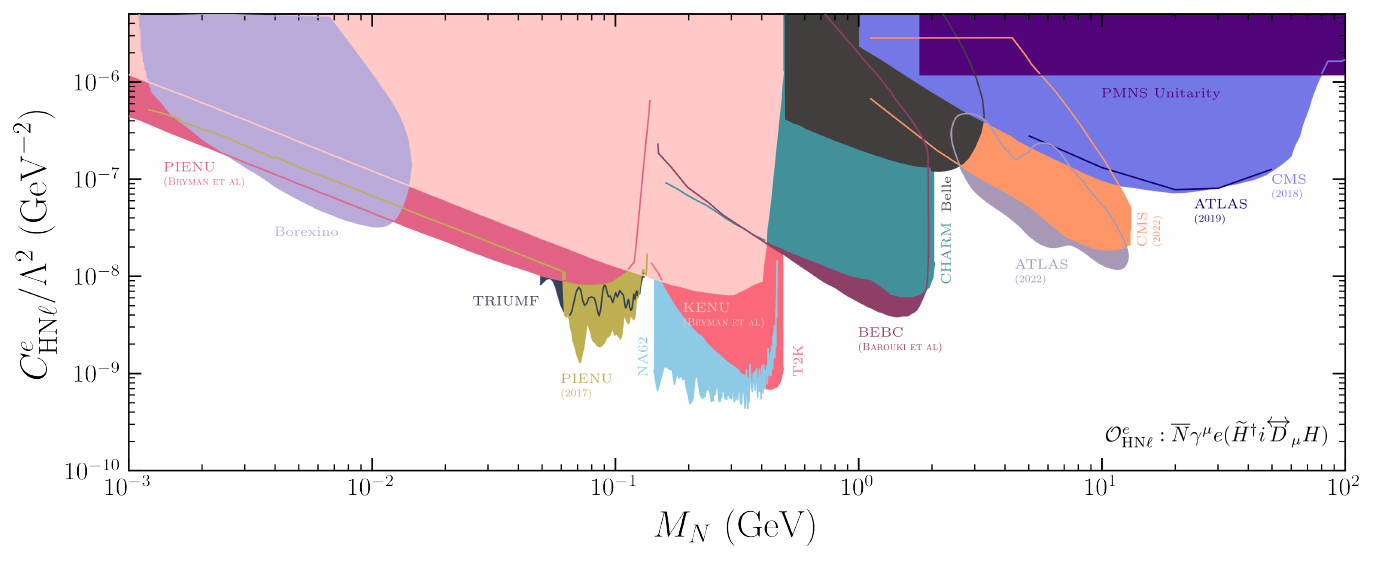}
\includegraphics[width=\columnwidth]{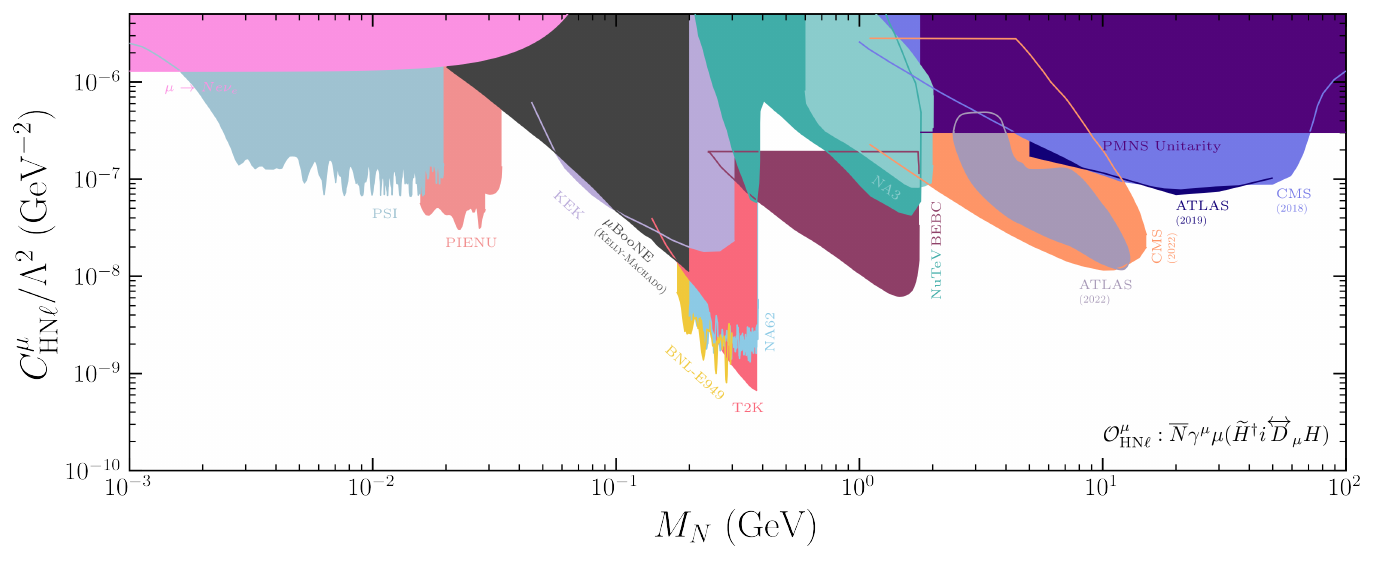}
\includegraphics[width=\columnwidth]{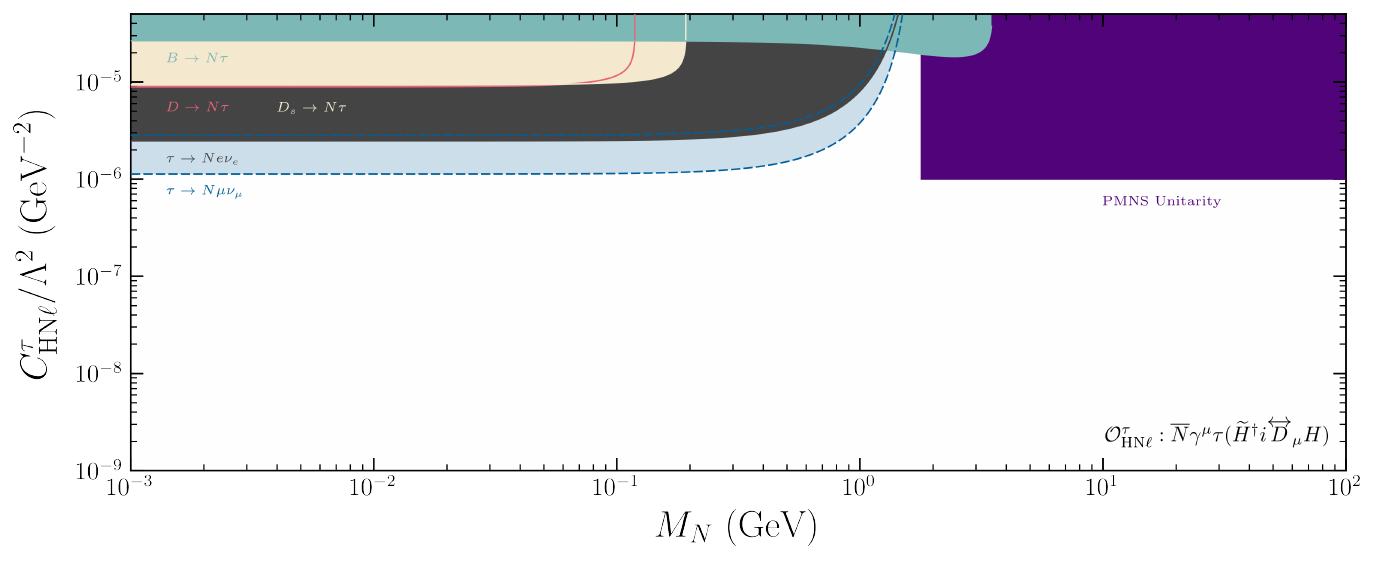}
\caption{90\% CL constraints on the Wilson coefficients of the charged bosonic current operator (Eq.~\ref{eq:OHNe}), as a function of the heavy neutrino mass. We display in separate panels the bounds relevant for each lepton flavour.}
\label{fig:charbosoncurr}
\end{figure}

Fig.~\ref{fig:charbosoncurr} contains the bounds on the Wilson coefficients of this operator, mainly coming from reinterpretations of limits on heavy neutrino mixing. 
Once again, we assume single flavour dominance, displaying on different panels the relevant constraints for each lepton flavour.
\section{Tensor currents}
\label{sec:tensor}

Similar to the dimension-5 operator in Eq.~\ref{eq:dim5op_dipole}, dipole couplings between the HNL and the lepton doublet can be considered. 
Two different operators arise:
\begin{align}\label{eq:neut_tensor}
    \mathcal{O}_{\rm NB}^\alpha &= \dfrac{C_{\rm NB}^\alpha}{\Lambda^2}\left(\overline{L_\alpha} \sigma_{\mu \nu} N\right) \widetilde{H} B^{\mu\nu}+ \rm{h.c.}\,,
    \\\label{eq:char_tensor}
    \mathcal{O}_{\rm NW}^\alpha &= \dfrac{C_{\rm NW}^\alpha}{\Lambda^2}\left(\overline{L }_\alpha\sigma_{\mu \nu} N\right) \tau^a \widetilde{H} W^{\mu\nu a} + \rm{h.c.}\,
\end{align}
Three copies of these two operators exist, one for each lepton flavour. Once again, their Wilson coefficients are, in principle, independent. After the Higgs develops a vacuum expectation value, these operators are translated into three dipoles for the HNL, coupling it to the photon, the $W$, and the $Z$:
\begin{align}
    \label{eq_dipole_coefficient}
    d_\gamma^\alpha &=\dfrac{\mu_{\nu}^{\alpha}}{2}=\dfrac{v \cos{\theta_{w}}}{\sqrt{2} \Lambda^2}\left(C_{\rm NB}^\alpha+\tan{\theta_{w}} C_{\rm NW}^\alpha\right)\,,
    \\
    d_Z^\alpha &=\dfrac{v \cos{\theta_{w}}}{\sqrt{2} \Lambda^2}\left(C_{\rm NW}^\alpha-\tan{\theta_{w}}C_{\rm NB}^\alpha\right)\,,
    \\
    d_W^\alpha &= \dfrac{v}{\Lambda^2} C_{\rm NW}^\alpha\,.
\end{align}
Bounds on these dipoles can be translated into limits on the Wilson coefficients by applying the relations above. Most limits on $C_{\rm NB}$ and $C_{\rm NW}$ are obtained from processes mediated by the magnetic dipole moment $d_\gamma$. Probes of this quantity include searches for exotic electromagnetic interactions of light neutrinos~\cite{Brdar:2020quo,Coloma:2017ppo,Gninenko:1998nn,DONUT:2001zvi} or HNL production and decay~\cite{Coloma:2017ppo,Plestid:2020ssy}. The LEP monophoton searches and supernova cooling bounds, mentioned in the previous sections, also constrain the magnetic dipole moment. Finally, this observable must also be small enough to respect primordial abundances after BBN~\cite{Brdar:2020quo} (see Ref.~\cite{Schwetz:2020xra} for a recent review on all these bounds). Once again, we regard these cosmological bounds as complementary to the laboratory constraints. 

Note that a clear flat direction is present if $C_{\rm NB}/C_{\rm NW} = - \tan{\theta_{w}}$, so that the dipole moment with the photon would vanish even with large Wilson coefficients.
In this case, none of the mentioned limits apply, as they rely either on heavy neutrino production or decay via $d_\gamma$.

The weak dipoles $d_Z$ and $d_W$ are less important for low-energy processes, but can contribute to HNL production, for instance, at colliders. In fact, these observables would be the only way to constrain these operators if they lie along the flat direction mentioned above. We will recast the LHC and LEP limits obtained in Ref.~\cite{Magill:2018jla} to obtain bounds on either $C_{\rm NB}$ or $C_{\rm NW}$, assuming the other vanishes\footnote{For the LHC limits on the $d_W^\alpha$ parameter, we take the $a=0.2$ case of Fig.~9 of Ref.~\cite{Magill:2018jla}.}. In principle, the operator $\mathcal{O}_{\rm NW}$ would also mediate CC interactions, such as the ones involved in the usual bounds on the mixing, but with a different Lorentz structure. Similarly, $\mathcal{O}_{\rm NB}$ would induce couplings between HNLs and neutral mesons. Nevertheless, the couplings with mesons vanish due to their particular Lorentz structure. Thus, all bounds arising from HNL production or decay via meson interactions do not apply to these operators. Finally, the operator $\mathcal{O}_{\rm NB}$ would mediate a new invisible $Z$ decay channel, into an HNL and a light neutrino. Imposing its rate to be consistent with current measurements on invisible $Z$ decay would yield yet another limit on $C_{\rm NB}/\Lambda^2$. However, we find this bound to be an order of magnitude worse than the one arising from monophoton searches, and do not include it in our plots.

\begin{figure}[t!]
\centering
\includegraphics[width=\columnwidth]{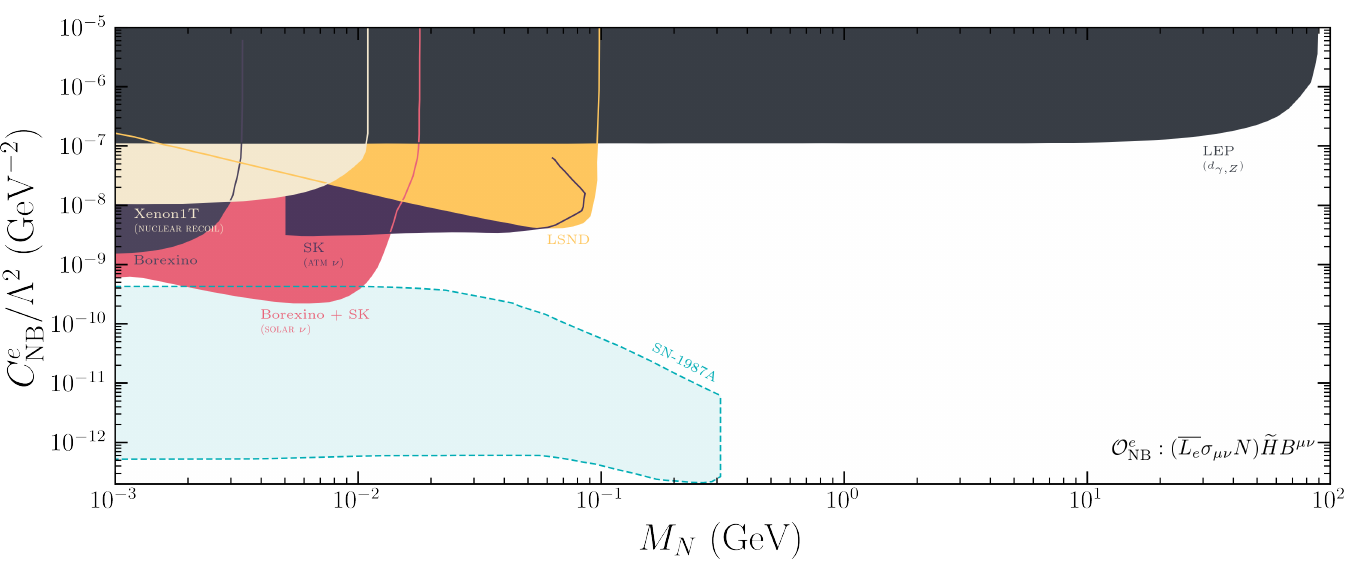}
\includegraphics[width=\columnwidth]{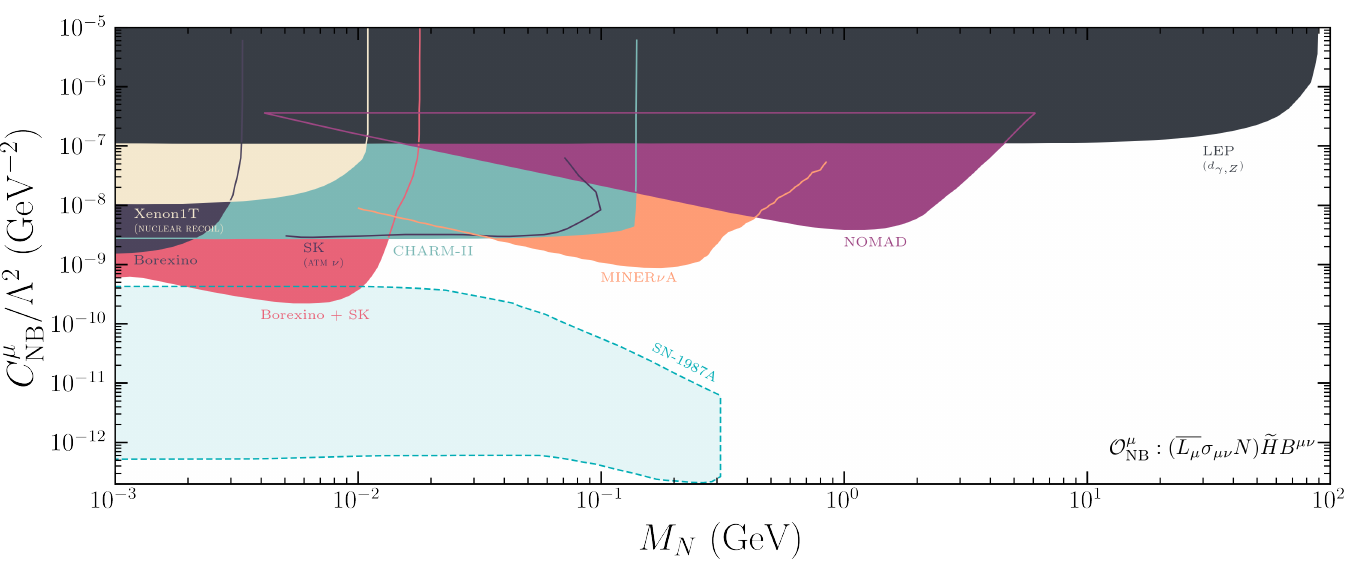}
\includegraphics[width=\columnwidth]{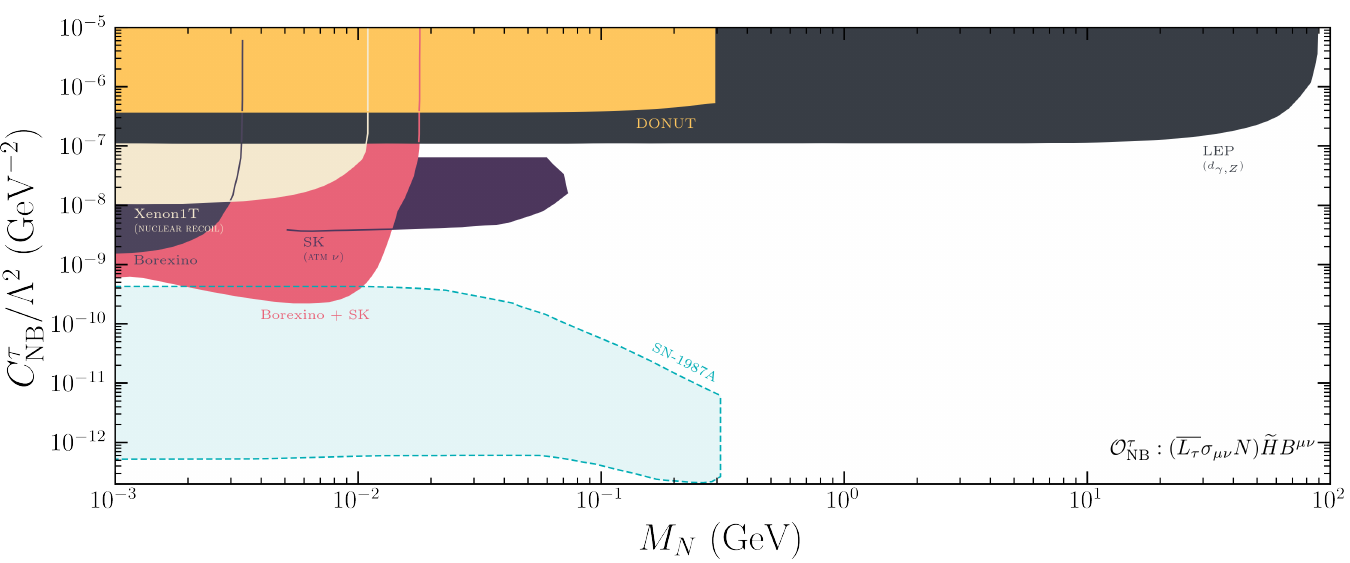}
\caption{90\% CL constraints on the Wilson coefficients of the tensor current operator in Eq.~\ref{eq:neut_tensor}, as a function of the heavy neutrino mass. We display in separate panels the bounds relevant for each lepton flavour.
\label{fig:ONB}
}
\end{figure}

\begin{figure}[ht!]
\centering
\includegraphics[width=\columnwidth]{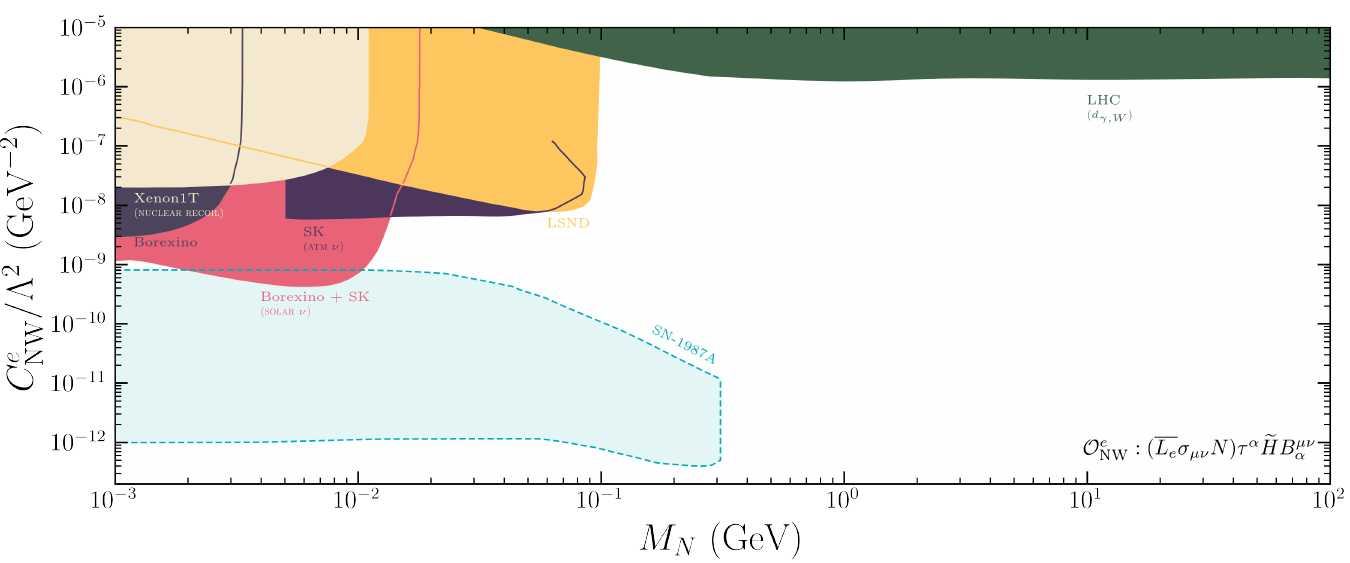}
\includegraphics[width=\columnwidth]{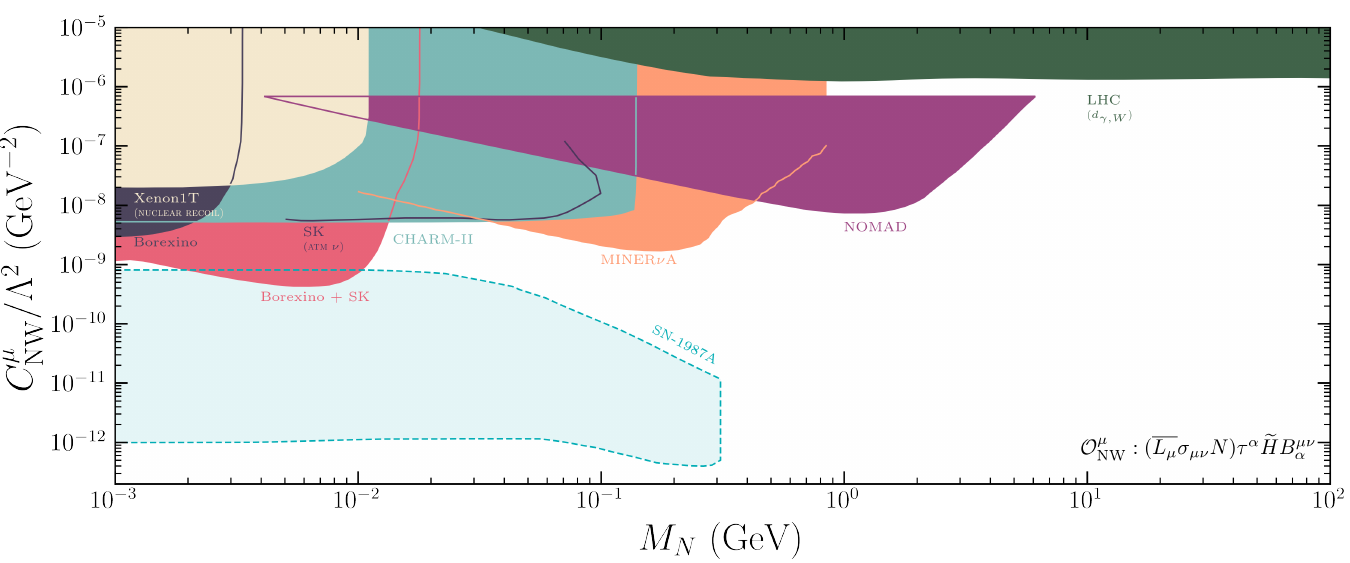}
\includegraphics[width=\columnwidth]{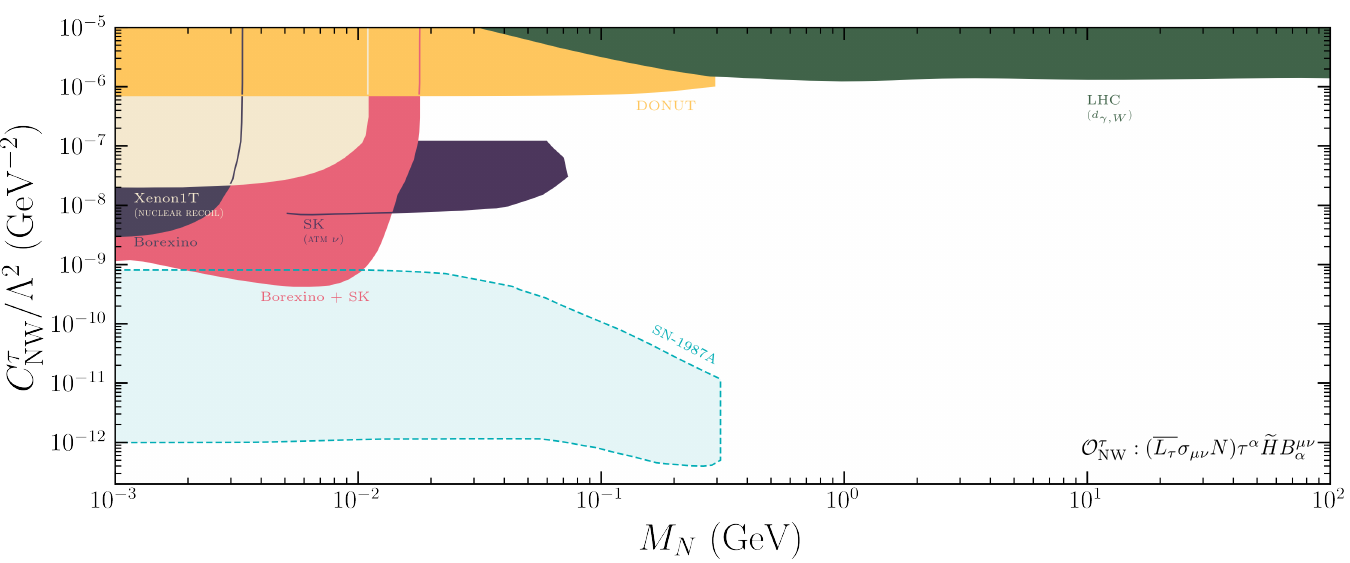}
\caption{90\% CL constraints on the Wilson coefficients of the tensor current operator in Eq.~\ref{eq:char_tensor}, as a function of the heavy neutrino mass. We display in separate panels the bounds relevant for each lepton flavour.
\label{fig:ONW}}
\end{figure}

Figs.~\ref{fig:ONB} and~\ref{fig:ONW} show the present bounds for $C_{\rm NB}/\Lambda^2$ and $C_{\rm NW}/\Lambda^2$, respectively. Both of them contain three panels, one for each lepton flavour, following the single flavour dominance assumption. Most bounds on the dipole moments depend on the flavour of the involved neutrino, so the different panels will, in general, contain limits provided by different experiments. Some sources of constraints are flavour independent, such as those provided by LEP or by supernova cooling arguments.

\section{Four-fermion interactions}
\label{sec:four-ferm}
Several different four-fermion operators involving HNLs can be written at dimension 6. We will separate their study into two main categories, depending on whether the operators mediate CC-like or NC-like processes.

Analyzing their potentially complex flavour structure leads to a large number of operators of each type, a priori independent. As in the previous sections, we will consider separately the operators that affect electrons, muons, and taus, treating the corresponding Wilson coefficients as independent parameters, to be bounded by different observables. 

To derive the constraints on the effective operators involving quark fields, their flavour structure becomes very relevant. Indeed, one of the most advantageous ways of producing HNLs is to exploit decays with a relatively low rate in the SM to compete against, such as $K$ or $B$ meson decays, which must proceed through CKM-suppressed weak processes. Thus, these searches only apply if quark flavour violation is allowed in the effective operators. 

However, some operators contain both charged and neutral couplings and, if allowed to be flavour-violating, could lead to much more strongly constrained flavour-changing neutral processes, such as neutral meson oscillations at the loop level. 
In the spirit of deriving conservative limits on the Wilson coefficients, we will assume that the effective operators are flavour universal and diagonal in the SM flavour basis. Thus, the only source of flavour violation in the quark sector will be the SM CKM matrix, which will appear in all CC-like operators, in an analogous way to the SM CC interactions when going to the mass basis. Indeed, we will also assume that the corresponding rotation for the right-handed fields, which will be physical for some operators, also corresponds to that of the CKM matrix\footnote{Alternatively, more elaborate prescriptions, such as Minimal Flavour Violation, could be considered. However, the additional Yukawa suppression present in these scenarios would be too strong for some operators, so that the mass scales probed would be at odds with EFT validity. 
For the type of operators analyzed here, such a level of protection against quark flavour violation is unnecessary.}.

For comparison, for the flavour-changing CC-like operators, we will show our results both under the assumption of CKM suppression and for the flavour-blind case, where all flavours couple with the same intensity (all flavour copies share a common Wilson coefficient). In this way, it will be apparent which parts of the bounds derive purely from the experimental results, and which rather stem from the flavour alignment prescription. However, notice that, in the flavour-blind case, stronger constraints from FCNC processes could generally apply; we only display this case for the sake of comparison.

\subsection{Neutral currents}
\label{sec:4ferm_neut_curr}
Three sets of four-fermion operators mediate NC-like processes. The first one reads
\begin{equation}
    \mathcal{O}_{\rm ff} =\frac{C_{\rm ff}}{\Lambda^2} (\overline{f} \gamma^\mu f) (\overline{N} \gamma_\mu N)\,,
    \label{eq:Off}
\end{equation}
where $f$ can stand for any right-handed fermion in the SM, either leptons or quarks. 
In the latter case, we will consider couplings to the up- and down-type quarks independently, through the coefficients $C_{\rm uu}$ and $C_{\rm dd}$.
For simplicity, we assume that this coefficient is generation independent.

Analogously, very similar operators can be written out of left-handed SM fields,
\begin{align}
    \label{eq:OLL}
    \mathcal{O}_{\rm LN}^\alpha &= \frac{C_{\rm LN}^\alpha}{\Lambda^2}(\overline{L_\alpha}\gamma^\mu L_\alpha) (\overline{N} \gamma_\mu N)\,,
    \\
    \label{eq:OQN}
    \mathcal{O}_{\rm QN} &= \frac{C_{\rm QN}}{\Lambda^2} \sum_i (\overline{Q_i}\gamma^\mu Q_i) (\overline{N} \gamma_\mu N)\,,
\end{align} 
where $Q_i$ denotes a quark doublet of flavour $i$.
In the case of leptons, we will once again treat the three flavour copies of $\mathcal{O}_{\rm LN}$ as operators with independent coefficients.
In the case of quarks, we assume that the coefficients are generation independent.

Finally, it is also possible to construct an independent operator with a scalar structure by employing the totally antisymmetric tensor, $\epsilon$:
\begin{equation}
        \mathcal{O}_{\rm LNL\ell}^{\alpha\beta} = \frac{C_{\rm LNL\ell}^{\alpha\beta}}{\Lambda^2} (\overline{L_\alpha} N)\epsilon (\overline{L_\alpha} \ell_\beta)\,,
        \label{eq:op_4lep}
\end{equation}
The fact that none of these operators mediates CC-like processes makes them difficult to probe. For instance, most bounds on HNL mixing cannot be reinterpreted for this purpose, as they rely on the production of HNLs in the decays of charged mesons. Even the constraints from DELPHI, based on HNL production in $Z$ decays, cannot be used. 

Nevertheless, $\mathcal{O}_{\rm uu}$, ${O}_{\rm dd}$ and $\mathcal{O}_{\rm QN}$ contribute to invisible decay channels for neutral mesons, with the $\pi^0$ and the $\Upsilon$(1S) being the most relevant. 
These operators can also induce quark scatterings in which the only visible signal (aside from the HNLs) is a single jet, produced mainly by a gluon emitted by any of the quarks. 
Such monojet searches have been performed in the LHC. Ref.~\cite{Alcaide:2019pnf} recasted the experimental limits into bounds on the corresponding Wilson coefficients. 
As assumed in Ref.~\cite{Alcaide:2019pnf}, the HNL mass can be neglected in this process for the range of masses we explore, corresponding to a horizontal line in our plots.

We find no observables able to constrain the operators involving two muons or two taus. In the case of two electrons, LEP monophoton searches and supernova cooling arguments apply analogously to previous sections, as electrons are present in the possible production of HNLs. 
As discussed before, a rescaling is required to account for the different cross section each operator would lead to (for details, see App.~\ref{app:supernova}). 
The operators $\mathcal{O}_{\rm ee}$ and $\mathcal{O}_{\rm LN}$ have different chiral structures; however, the bounds on their corresponding Wilson coefficients turn out to be the same. 
The same observables also constrain $C_{\rm LNL\ell}^{\rm \alpha \beta}$, provided $\alpha=\beta=e$. 
If $\alpha\neq\beta$, $\ell_\alpha\to\ell_\beta \nu N$ decays are mediated by this operator. 
We perform a $\chi^2$ fit of the corresponding rates to the measured $\mu$ and $\tau$ leptonic decays, obtaining bounds for $C_{\rm LNL\ell}^{\rm e\mu}$, $C_{\rm LNL\ell}^{\rm e\tau}$ and $C_{\rm LNL\ell}^{\rm \mu\tau}$.

\begin{figure}[t!]
\includegraphics[width=\columnwidth]{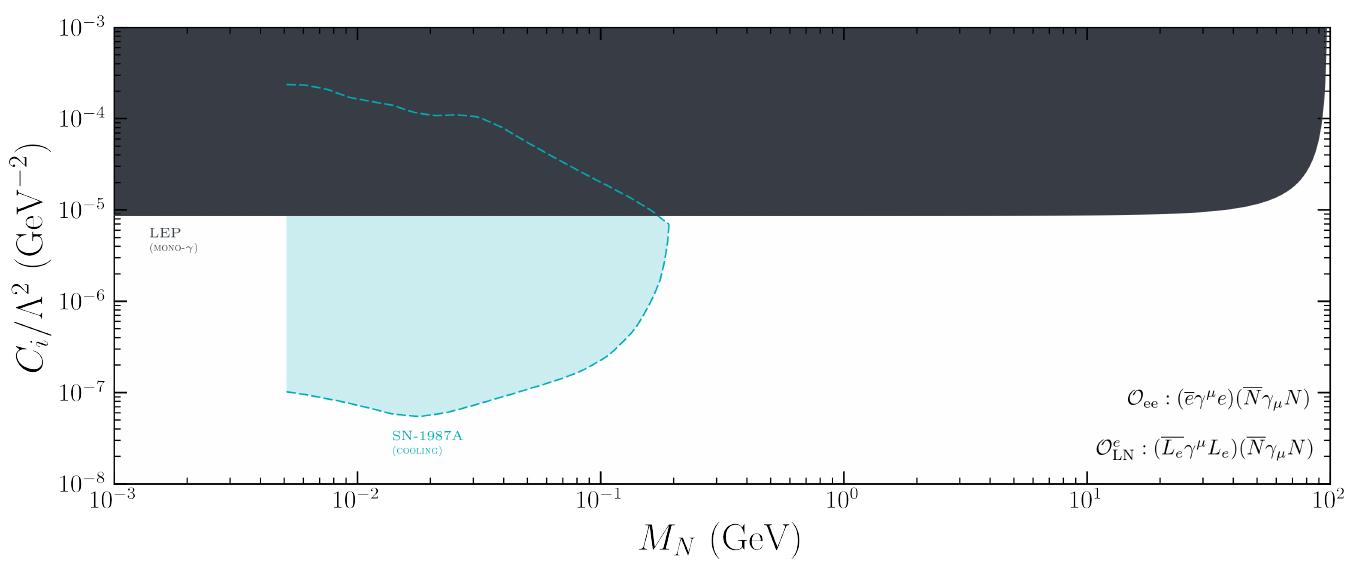}
\includegraphics[width=\columnwidth]{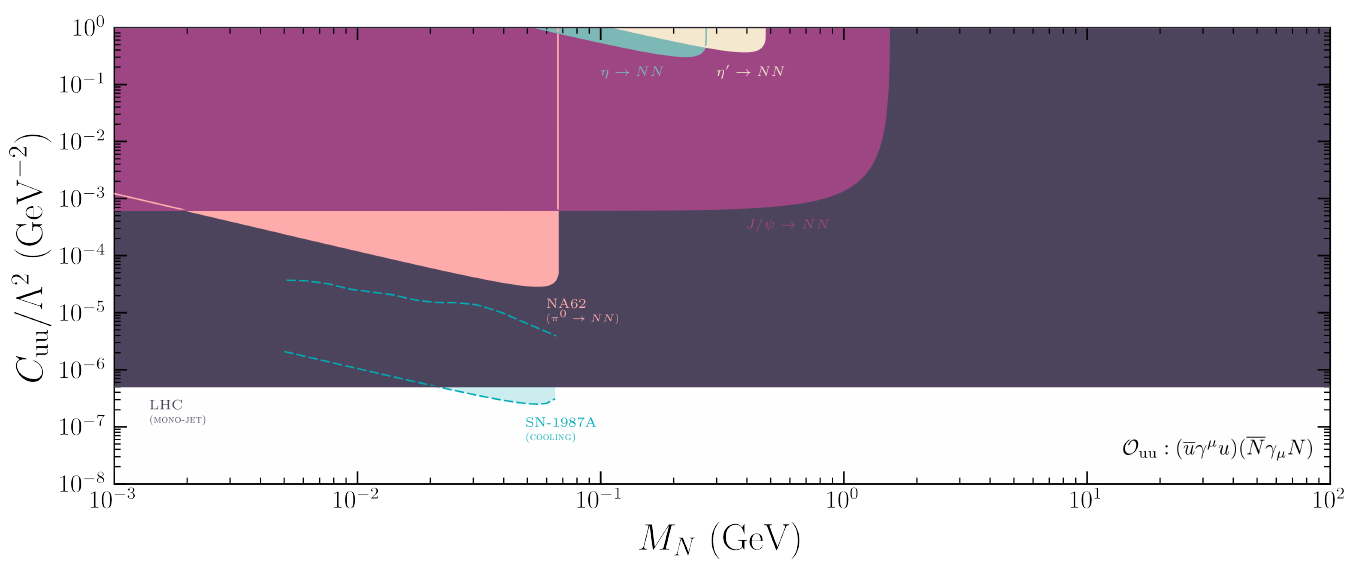}
\includegraphics[width=\columnwidth]{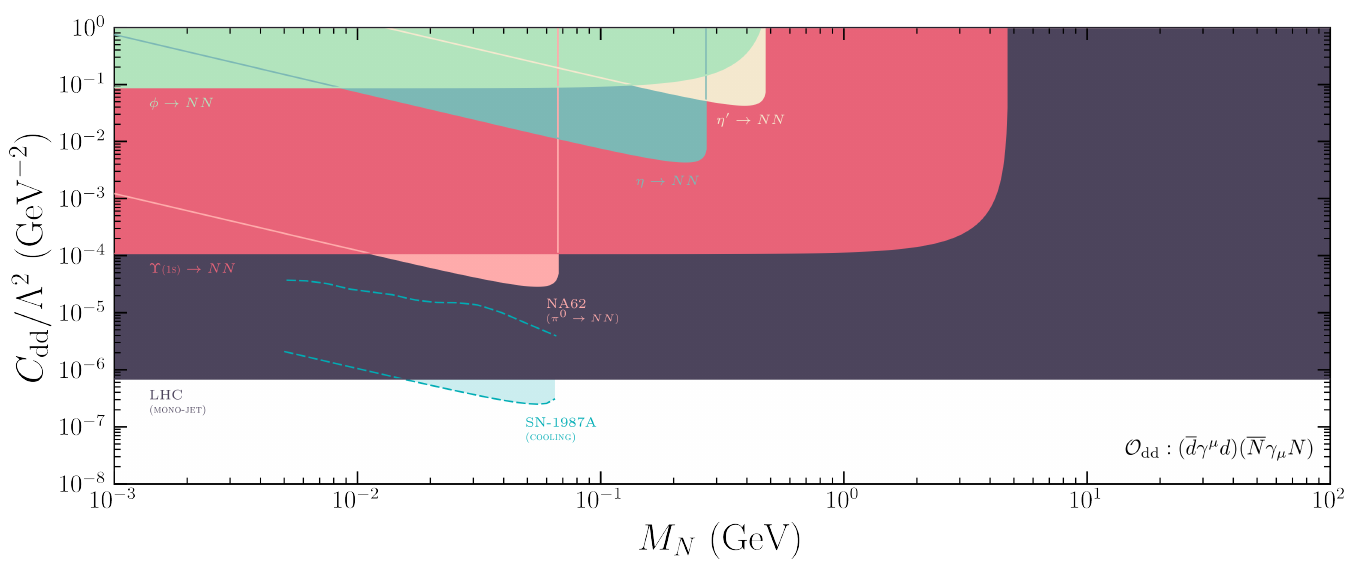}
\caption{90\% CL constraints on some of the Wilson coefficients of the four-fermion NC operators $\mathcal{O}_{\rm ff}$ and $\mathcal{O}_{\rm LN}$ (Eqs.~\ref{eq:Off} and~\ref{eq:OLL} respectively), as a function of the heavy neutrino mass. The bounds in the top panel apply equally to both $O_{\rm ee}$ and $O_{\rm LN}^e$.}
\label{fig:4ferm_neut_ff}
\end{figure}

\begin{figure}[t!]
\includegraphics[width=\columnwidth]{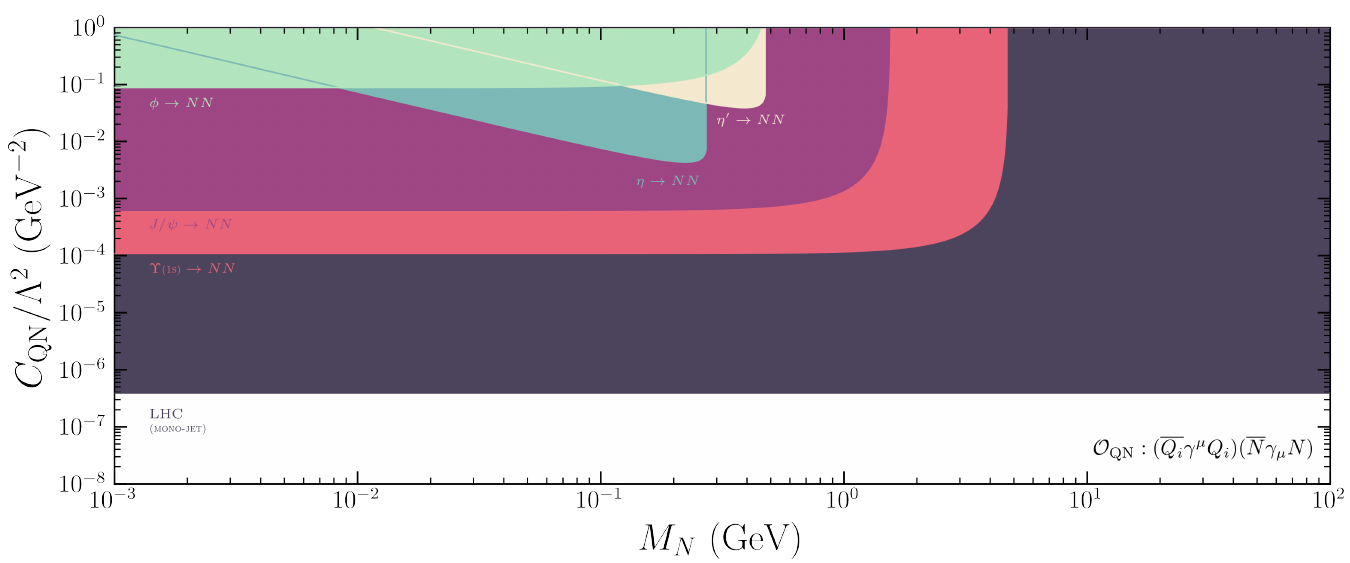}
\caption{90\% CL constraints on the Wilson coefficient of the four-fermion NC operator $\mathcal{O}_{\rm QN}$ (Eq.~\ref{eq:OQN}), as a function of the heavy neutrino mass.}
\label{fig:4ferm_neut_left}
\end{figure}

\begin{figure}[t!]
\includegraphics[width=\columnwidth]{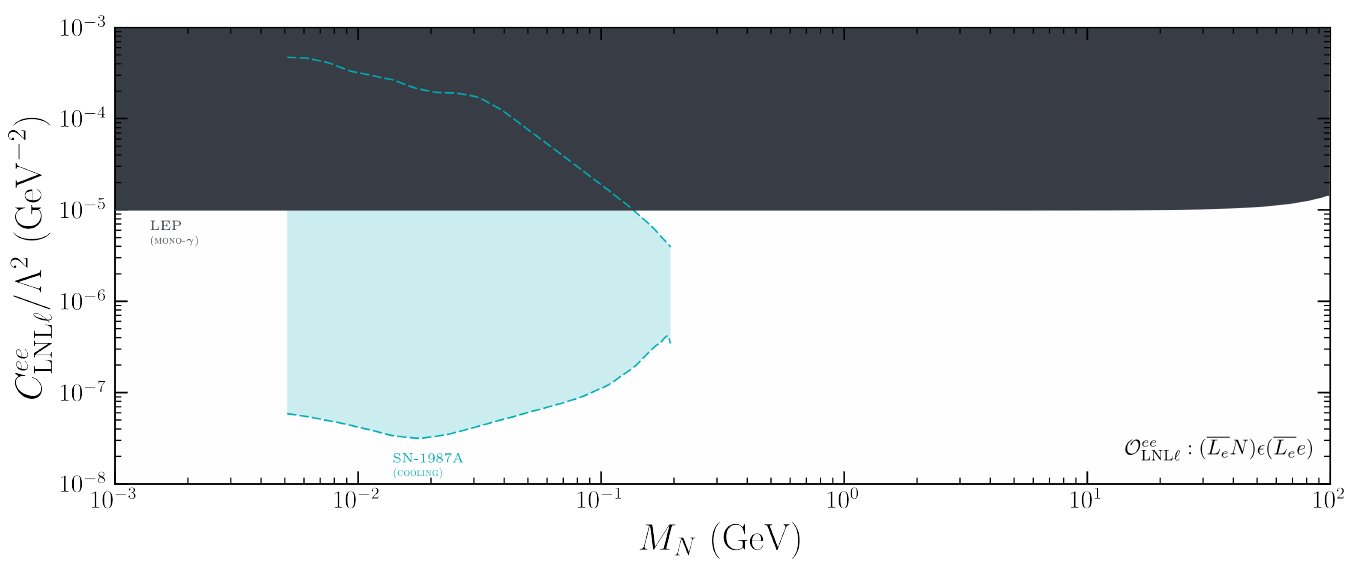}
\includegraphics[width=\columnwidth]{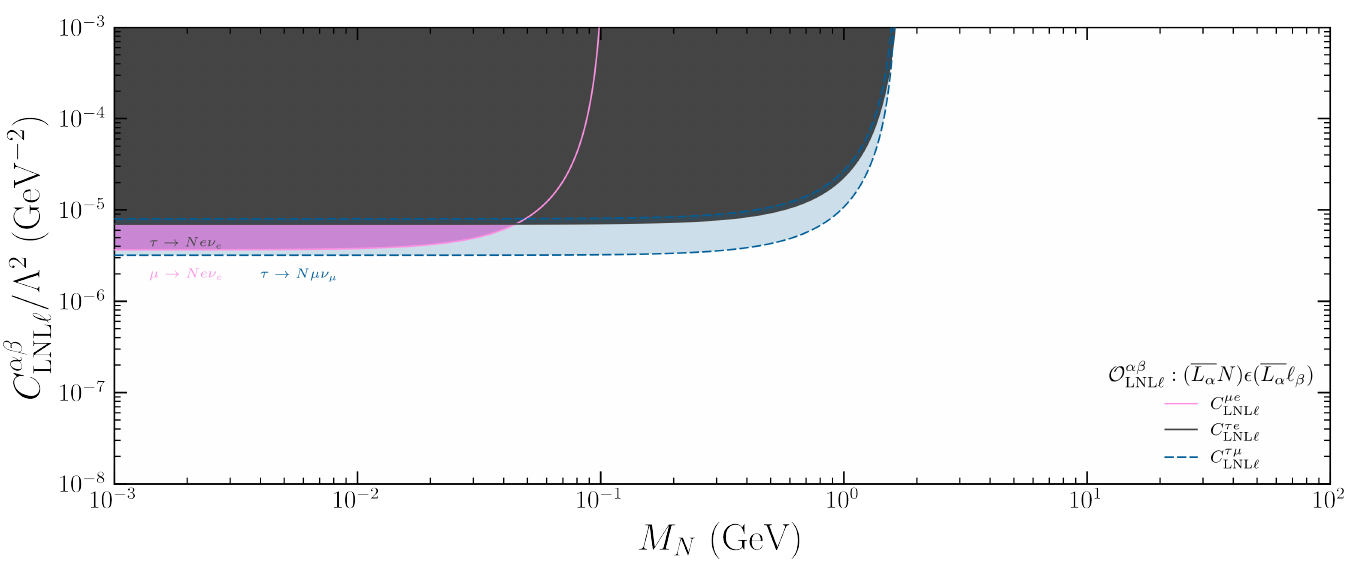}
\caption{90\% CL constraints on some of the Wilson coefficients of the four-fermion NC operator $\mathcal{O}_{\rm LNL\ell}$ (Eq.~\ref{eq:op_4lep}), as a function of the heavy neutrino mass.}
\label{fig:4ferm_neut_scalar}
\end{figure}

Figs.~\ref{fig:4ferm_neut_ff},~\ref{fig:4ferm_neut_left} and~\ref{fig:4ferm_neut_scalar} show the bounds on the mentioned Wilson coefficients. The main limits arise from invisible meson decay and monojet searches in the case of quarks, monophoton searches and supernova cooling for electrons and muon and tau decays for the other lepton flavours.

\subsection{Charged currents}

Three types of four-fermion operators appear at dimension 6 mediating CC-like interactions. The first one involves a right-handed charged lepton, a right-handed down-type quark of flavour $i$, and a right-handed up-type quark of flavour $j$:
\begin{equation}
    \mathcal{O}_{\rm duN\ell}^{\alpha} = \frac{C_{\rm duN\ell}^{\alpha}}{\Lambda^2}\sum_{i,j}\mathcal{Z}_{ij}^{\rm duN\ell}(\overline{d_i} \gamma^\mu u_j) (\overline{N} \gamma_\mu \ell_\alpha)\,.
        \label{eq:CCFFvectorRR}
\end{equation}
In the flavour-blind case, $\mathcal{Z}_{ij}^{\rm duN\ell}=1$, while, under the flavour alignment hypothesis, an insertion of the corresponding CKM matrix element $V_{ij}$ would be needed:
\begin{equation}
    \mathcal{Z}_{ij}^{\rm duN\ell}=V_{ij}^*\,.
\end{equation}
It is also possible to write a scalar coupling of the form 
\begin{equation}\label{eq:CCFFscalarLR_1}
\,.
\end{equation}
As before, under the flavour alignment hypothesis, we assume this operator is flavour universal and diagonal in the flavour basis, and, upon rotating to the mass basis, the CKM mixing matrix will control the degree of flavour violation.
\begin{equation}
\mathcal{Z}^{\rm LNQd}_{ij}=V_{ij}\,,
\end{equation}
while in the flavour-blind scenario $\mathcal{Z}^{\rm LNQd}_{ij}=1$. Note that exchanging the down-type quark and the HNL fields leads to a different operator,
\begin{equation}
        \mathcal{O}_{\rm LdQN}^\alpha = \frac{C_{\rm LdQN}^\alpha}{\Lambda^2}\sum_{ij}\mathcal{Z}^{\rm LdQN}_{ij} (\overline{L_\alpha} d_i)\epsilon (\overline{Q_j} N)\,,
\end{equation}
which shares the flavour coefficients of the previous operator. 
However, through a Fierz transformation, it is possible to rewrite this operator as 
\begin{equation}
\label{eq:opldqn}
       \mathcal{O}_{\rm LdQN}^\alpha = \frac{C_{\rm LdQN}^\alpha}{\Lambda^2}\sum_{ij}\mathcal{Z}^{\rm LdQN}_{ij}\left[ \frac{1}{2}(\overline{Q_j} d_i)\epsilon (\overline{L_\alpha} N) + \frac{1}{8} (\overline{Q_j}\sigma_{\mu\nu} d_i)\epsilon (\overline{L_\alpha}\sigma^{\mu\nu} N)\right]\,.
\end{equation}
Finally, the last independent operator takes the form 
\begin{equation}\label{eq:CCFFscalarLR_2}
    \mathcal{O}_{\rm QuNL}^\alpha = \frac{C_{\rm QuNL}^\alpha}{\Lambda^2}\sum_{ij}\mathcal{Z}^{\rm QuNL}_{ij}(\overline{Q_i} u_j)(\overline{N} L_\alpha)\,,
\end{equation}
with 
\begin{equation}
\mathcal{Z}^{\rm QuNL}_{ij}=V_{ij}^*\,
\end{equation}
in the case of flavour alignment, and $\mathcal{Z}^{\rm QuNL}_{ij}=1$ in the flavour-blind scenario.

The set of operators in Eqs.~\ref{eq:CCFFvectorRR},~\ref{eq:CCFFscalarLR_1} and~\ref{eq:CCFFscalarLR_2} mediate CCs involving quarks and HNLs, producing interactions between charged mesons and HNLs. 
These processes are the main source of bounds on the standard HNL mixing, so the corresponding limits can be translated into constraints on the Wilson coefficients. 
The interactions mediated by the effective operators exhibit different Lorentz structures than the SM, so they yield HNL production and decay rates different from those mediated by standard mixing. 

Thus, as for the previous analyses, a rescaling of the existing bounds is necessary, to account both for the absence of NCs and for the different production and decay rates of the HNL. This procedure was recently advocated and applied to this type of operator in Ref.~\cite{Beltran:2023nli} for some example observables.

\begin{figure}[t!]
\includegraphics[width=\columnwidth]{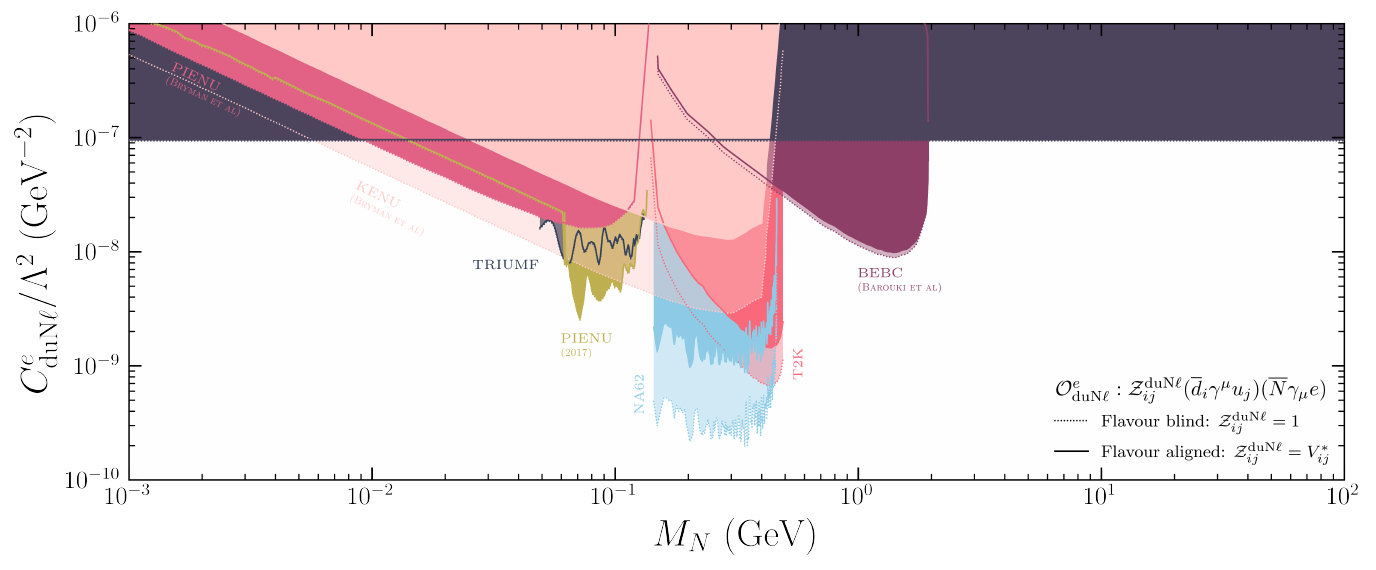}
\includegraphics[width=\columnwidth]{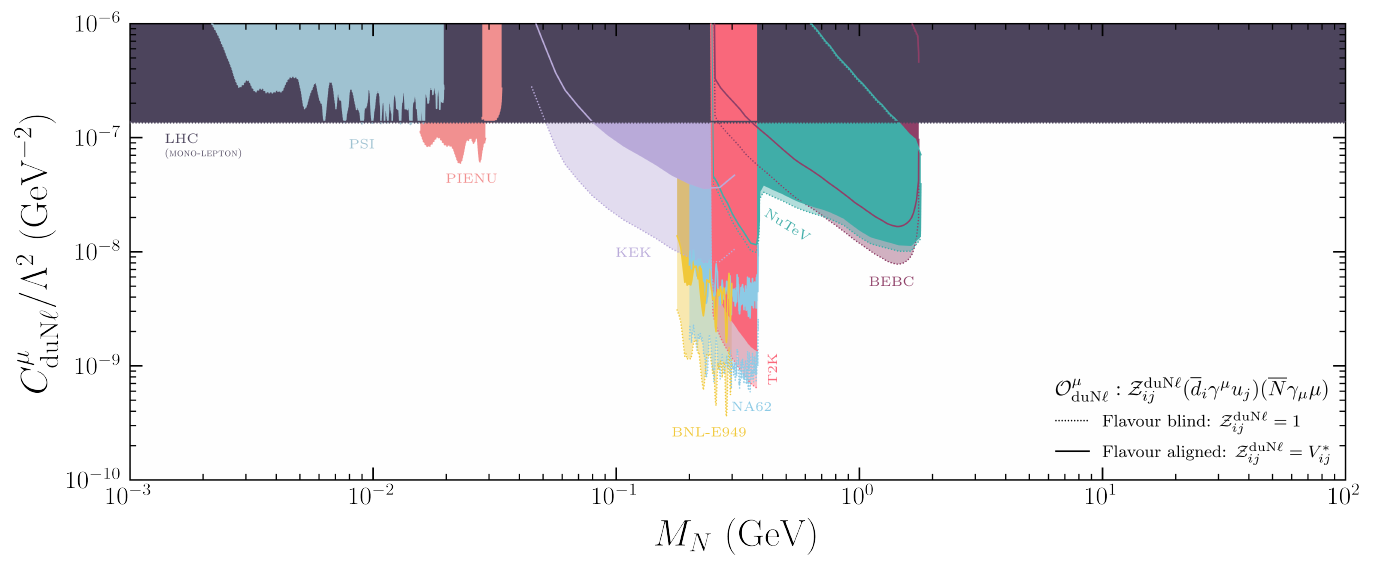}
\includegraphics[width=\columnwidth]{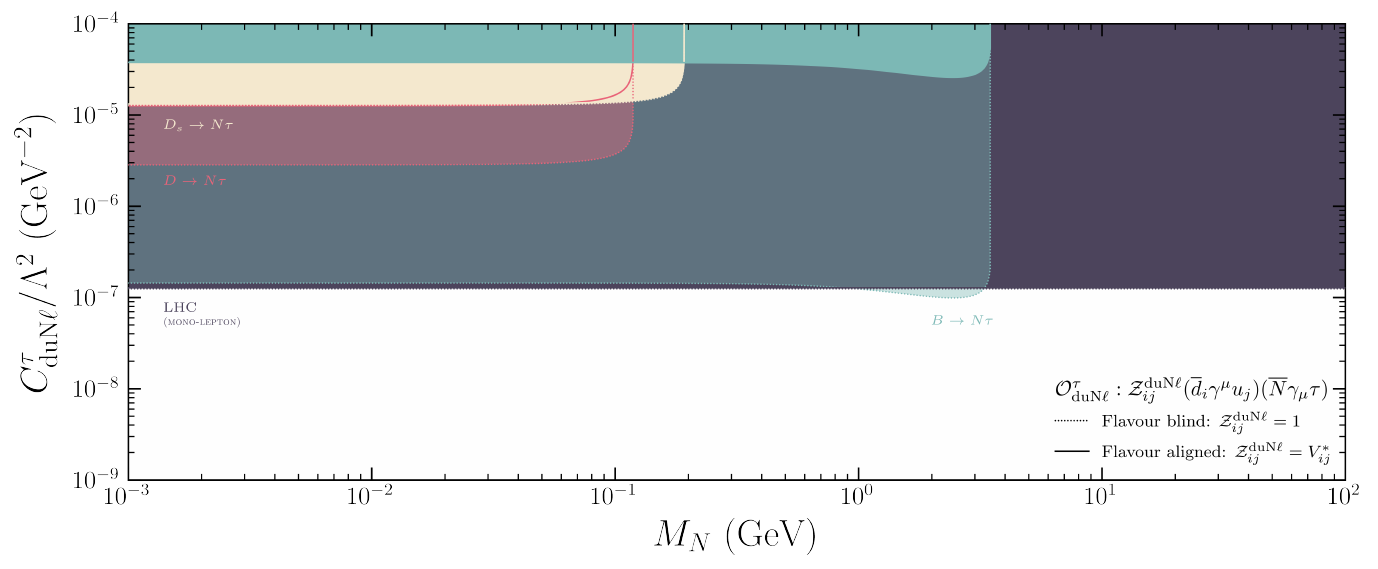}
\caption{90\% CL constraints on the Wilson coefficients of the four-fermion CC operator $\mathcal{O}_{\rm duN\ell}^\alpha$ (Eq.~\ref{eq:CCFFvectorRR}), as a function of the heavy neutrino mass. Dotted light (solid dark) regions represent the flavour blind (aligned) scenario. See text for details.}
\label{fig:4ferm_dune}
\end{figure}

\begin{figure}[t!]
\includegraphics[width=\columnwidth]{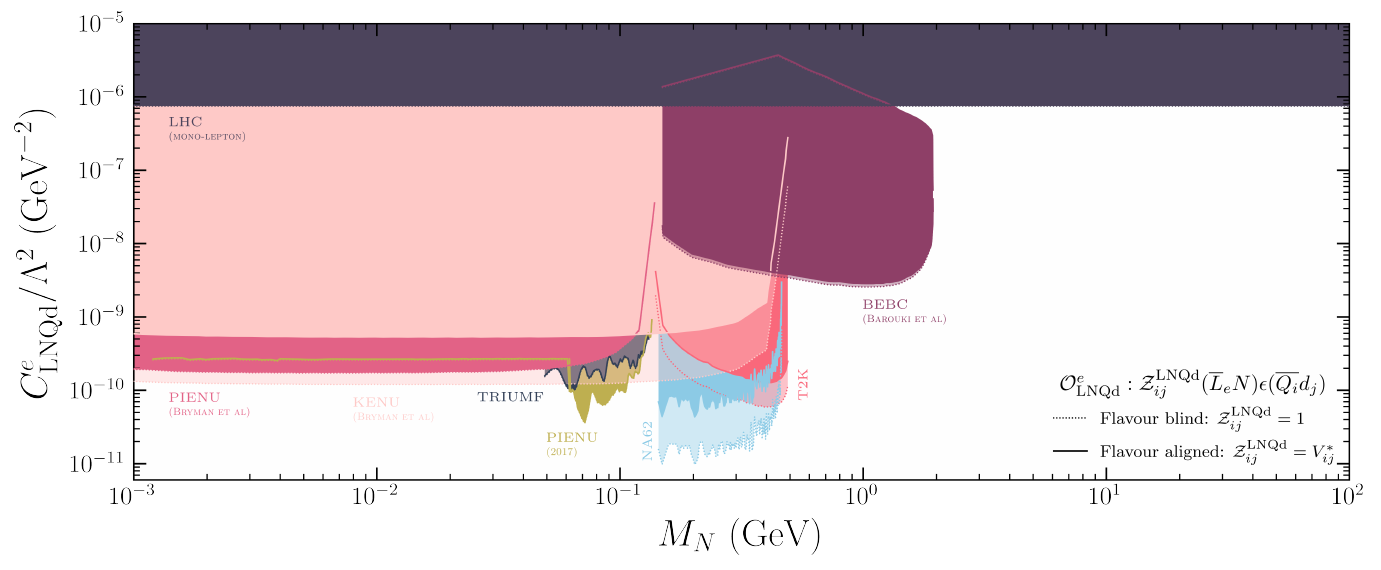}
\includegraphics[width=\columnwidth]{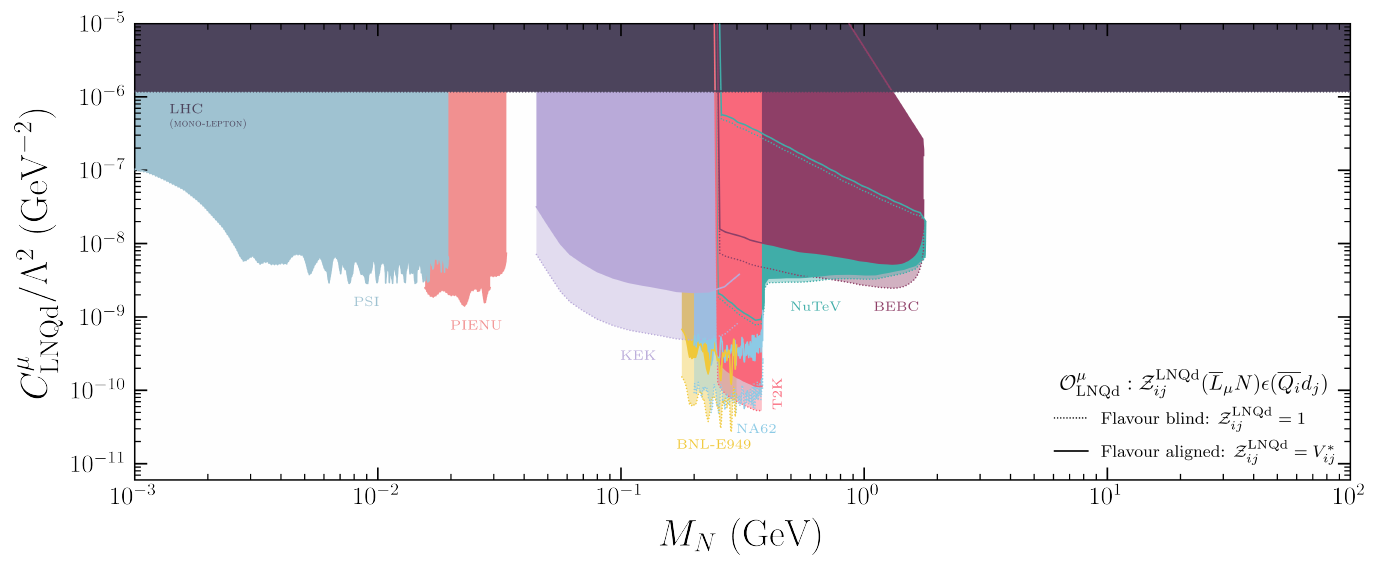}
\includegraphics[width=\columnwidth]{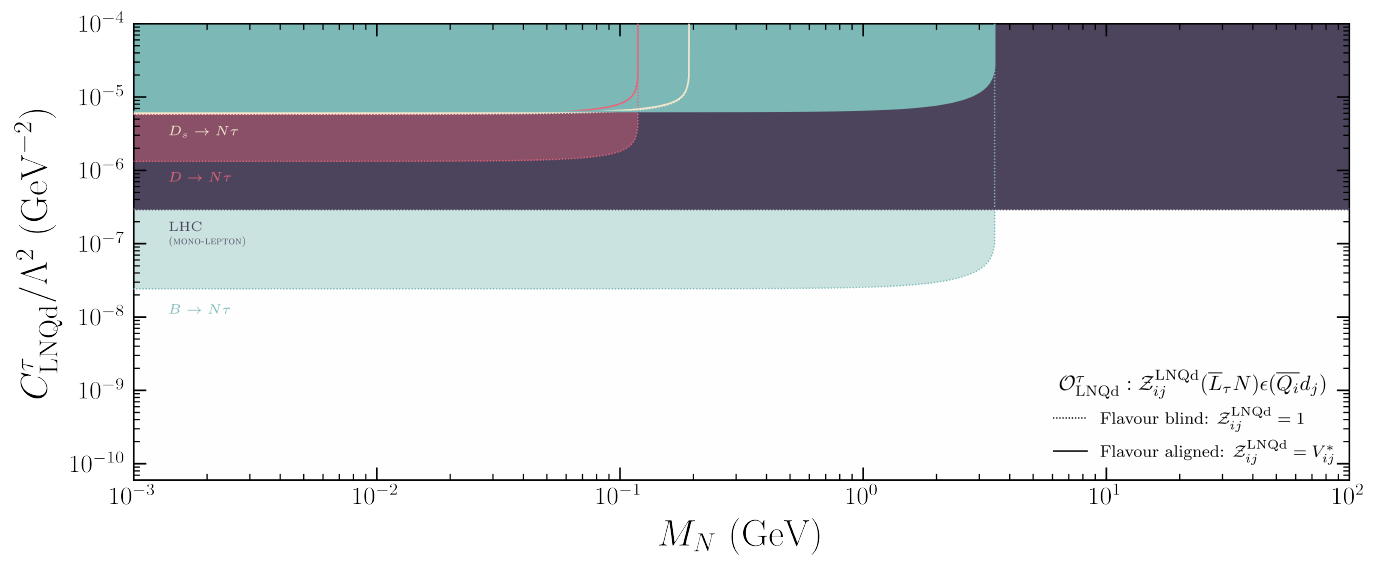}
\caption{90\% CL constraints on the Wilson coefficients of the four-fermion CC operator $\mathcal{O}_{\rm LNQd}$ (Eq.~\ref{eq:CCFFscalarLR_1}), as a function of the heavy neutrino mass. We display in separate panels the bounds relevant for each lepton flavour. Dotted light (solid dark) regions represent the flavour blind (aligned) scenario. See text for details.}
\label{fig:4ferm_lnqd}
\end{figure}
All these operators could also induce monolepton processes in colliders, in which the final state consists of a single observed lepton and missing energy (carried by the HNL). Searches for these signatures have been performed at the LHC and were recasted to constrain the Wilson coefficients of these operators in Ref.~\cite{Alcaide:2019pnf}. We directly employ their bounds, which are independent on the mass of the HNL (as the typical energies at the LHC are much larger than the masses we consider).

\begin{figure}[t!]
\includegraphics[width=\columnwidth]{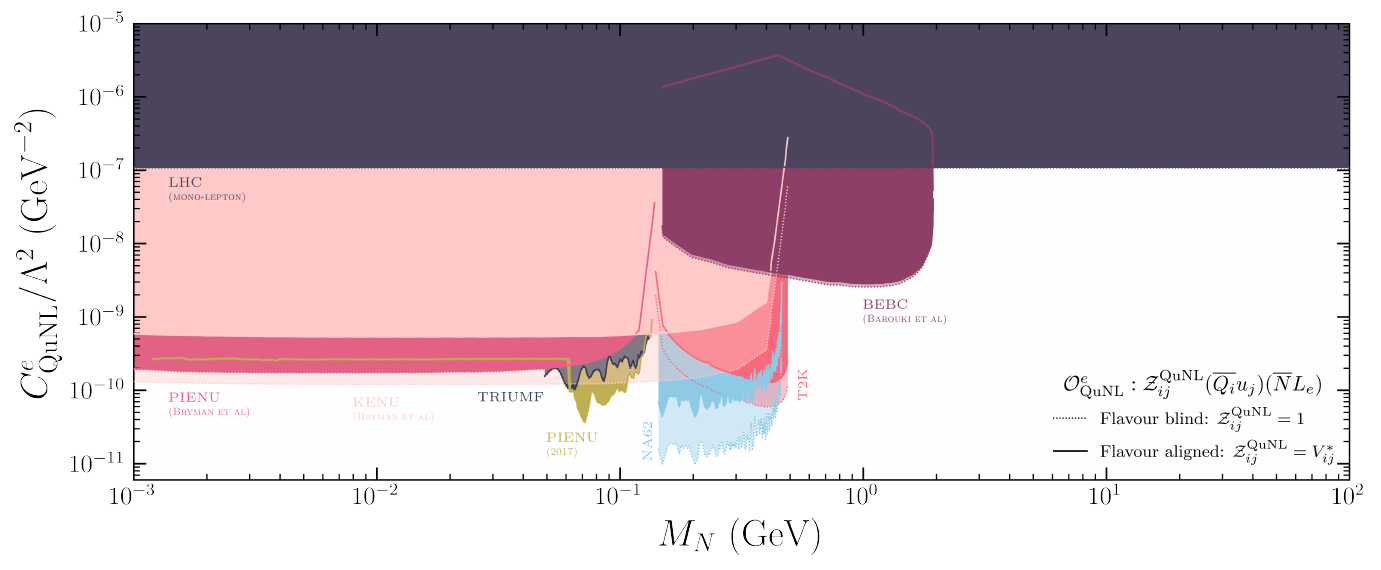}
\includegraphics[width=\columnwidth]{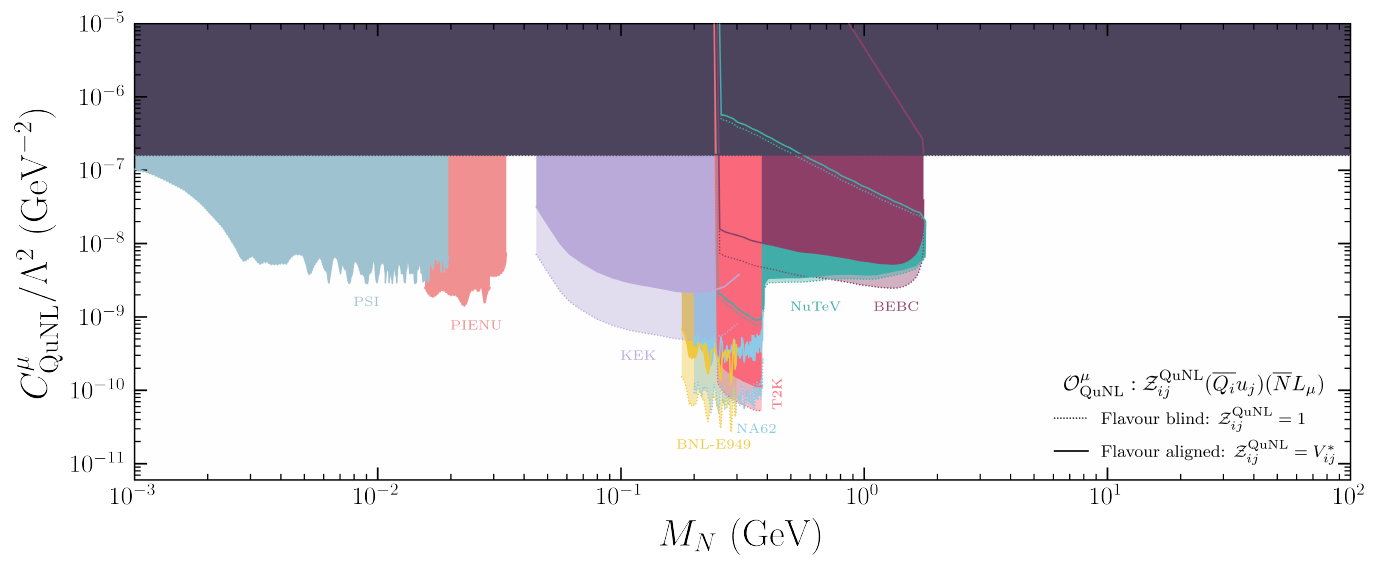}
\includegraphics[width=\columnwidth]{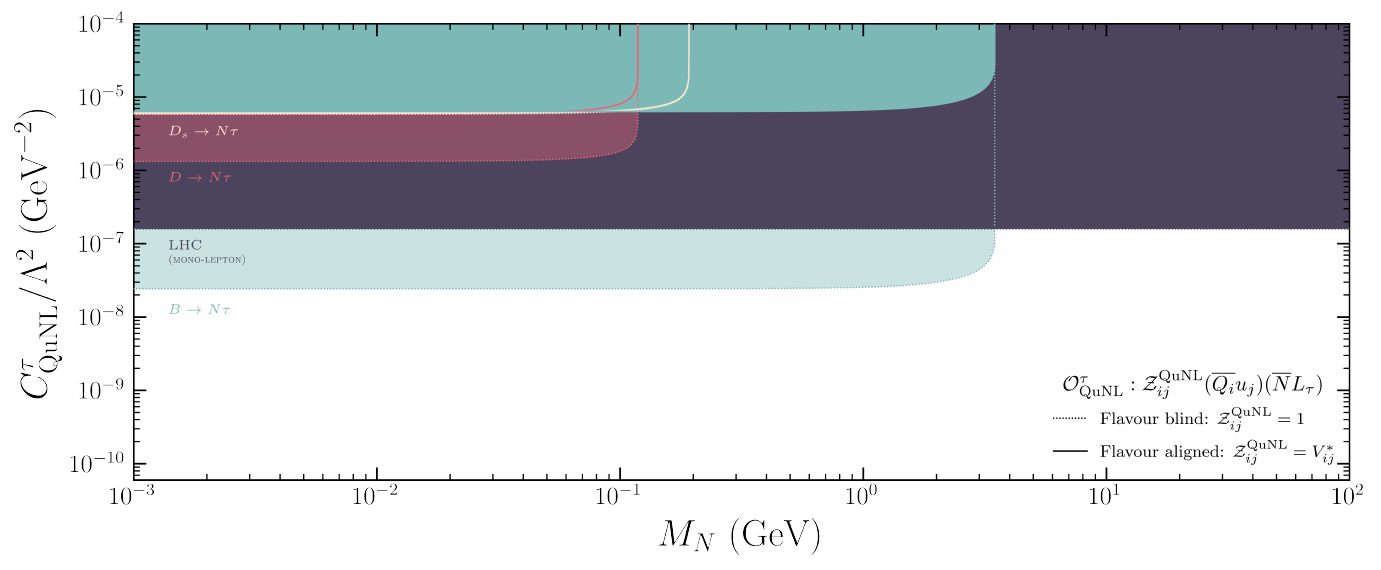}
\caption{90\% CL constraints on the Wilson coefficient of the four-fermion operators $\mathcal{O}_{\rm QuNL}$ (Eq.~\ref{eq:opldqn}), as a function of the heavy neutrino mass. We display in separate panels the bounds relevant for each lepton flavour. Dotted light (solid dark) regions represent the flavour blind(aligned) scenario. See text for details.}
\label{fig:4ferm_qunl}
\end{figure}

We will not discuss the phenomenology associated with $\mathcal{O}_{\rm QuNL}$, as it is identical to that of $\mathcal{O}_{\rm LNQd}$. This can be inferred from Eq.~\ref{eq:opldqn}. 
The term with a tensor structure does not contribute to interactions with mesons. 
On the other hand, the term with a scalar Lorentz structure is identical to $\mathcal{O}_{\rm LNQd}$ (modulo a factor of 2), thus yielding no different phenomenology. The bounds on $C_{\rm LdQN}$ can be directly read from those on $\mathcal{O}_{\rm LNQd}$ by rescaling them by a factor of two. 

The operators $\mathcal{O}_{\rm LNQd}$, $\mathcal{O}_{\rm LdQN}$ and $\mathcal{O}_{\rm QuNL}$ also mediate NC-like processes involving an HNL, a light neutrino, and a $q\bar{q}$ pair. 
This would induce invisible decays of neutral mesons. Imposing the experimental limits on those processes could also constrain the corresponding Wilson coefficients. However, these bounds are always considerably looser than those arising from charged currents. As the limits would apply to the same Wilson coefficients, we will focus on those derived from CC-like processes.

Figs.~\ref{fig:4ferm_dune},~\ref{fig:4ferm_lnqd} and~\ref{fig:4ferm_qunl} contain the bounds on the Wilson coefficients of these operators. 
Each of these figures displays two different regions: in the ones defined by the dotted lines with lighter colors, flavour blindness is assumed ($\mathcal{Z}_{ij}= 1$), whereas in the regions delimited by the solid lines with darker colors we adopt the flavour alignment prescription.

\section{Summary}
\label{sec:conclus}
Effective field theories are a very useful tool in order to describe the effects of a possibly unknown high-energy theory on low-energy processes. They usually rely on integrating out heavy particles, generating an infinite tower of non-renormalizable operators that describe the dynamics of the light degrees of freedom. 

One of the most successful bottom-up approaches is the SMEFT, built out of the SM particle content and respecting its fundamental symmetries. The Wilson coefficients of its operators contain information of possible BSM effects, being eventually matched to the parameters of a full theory.

As argued previously, there are good arguments that motivate the existence of heavy neutrinos at the electroweak scale or below. If this is the case, these particles should be included as light degrees of freedom, along with the rest of SM particles, yielding the $\nu$SMEFT. We have reviewed the dimension-6 operators that include at least one HNL, summarizing the main low-energy processes they can mediate, and employing current experimental data to constrain the corresponding Wilson coefficients. We have assumed a simplified scenario, in which only a single effective operator is active at a time. This approach generically offers conservative bounds, serving as a basis for a more ambitious analysis in the future.

We find that most limits on heavy neutrino mixing can be reinterpreted as bounds on the effective operators, especially those describing CC-like interactions. Operators that include two HNLs, that in general mediate NC-like processes, are harder to constrain; monophoton searches, decays of the $Z$ boson and neutral mesons, and supernovae cooling bounds offer the best options. All the limits have been derived as a function of the heavy neutrino mass, and can be found in \gitlink.

\setupfakechapters
\chapter*{Conclusions}
\addcontentsline{toc}{part}{Conclusions}
\fancyhead[RO]{\scshape \color{lightgray}Conclusions}
\fancyhead[LE]{\scshape \color{lightgray}Conclusions}
\lettrine[depth=1, loversize=0.55,lraise=-0.4]
    {T}{}he Standard Model of particle physics has proven to be one of the greatest scientific achievements ever, describing very successfully the fundamental components of our Universe and the interactions between them. Once some free parameters are fixed by experimental inputs, the SM can offer predictions for virtually any process in particle physics. An immense number of experimental tests have shown that practically all these predictions are correct, up to unprecedented accuracies. 

Despite its tremendous success, the SM cannot be the ultimate theory of fundamental particles and interactions, as it is poses several quandaries. Some of them, such as the hierarchy problem, the flavour puzzle or the strong CP problem, do not indicate an explicit failure of the theory, but rather some unnatural or fine-tuned features. Other issues stem from empirical observations that the SM cannot explain, such as the presence of dark matter in the Universe. 

In this thesis we have tried to tackle the problem of neutrino masses, that constitutes one of the main contradictions between the predictions of the SM and experimental measurements. These particles have been observed to change flavour in an oscillatory way when propagating; such a phenomenon can only occur if neutrinos are massive particles, in contrast to what the SM predicts. It is clear that this theory needs to be extended in order to include some mechanism that explains neutrino masses. Besides, further experimental efforts are also needed to ascertain some properties of neutrinos, that remain unknown due to their non-vanishing mass. Chief among them are their Dirac or Majorana nature, their possible role in breaking CP symmetry or their absolute mass scale.

We have focused on the role of heavy neutrinos in the answer to these mysteries. These hypothetical new particles are present in most of the neutrino mass models that have appeared through the years, exhibiting different characteristics. For instance, the type I seesaw, arguably the simplest proposal, includes heavy neutrinos, that do not partake in the gauge interactions of the SM. In this framework, the mass of light neutrinos is inversely proportional to that of the new states, suggesting the latter to be very heavy. This philosophy has been the starting point for many, more sophisticated models, which often predict new fundamental forces or enlarged spectrums. 

The interest on heavy neutrinos relies not only on generating neutrino masses, but also on their potential to provide solutions to other open problems of the SM. For instance, if they exhibit some particular characteristics, heavy neutrinos are suitable DM candidates, which the SM lacks. Also, if they are Majorana fermions, neutrinos can possibly explain the baryon asymmetry of the Universe via leptogenesis. 

From the experimental point of view, heavy neutrinos are very challenging to probe. As their mass is not determined from first principles, many different strategies must be employed in order to cover a wide range of energies, from the eV scale, tested in oscillations, to the TeV, only reachable at colliders. It is also not clear how these new particles couple to the SM; at least, heavy neutrinos will inherit the interactions of light neutrinos via mixing, offering a way to produce them or observe their decay products. The lack of any signal up to date shows that, if they exist, these particles must interact very feebly with the SM, or be too heavy to be produced. This translates into very strong bounds on the corresponding couplings, usually parametrized as mixing elements.

Here we have concentrated on heavy neutrinos in the MeV-GeV range, which are interesting both from the theoretical and experimental sides. New neutrinos below the electroweak scale are predicted by many models, usually based on an approximate lepton number symmetry that protects light neutrino masses, explaining their smallness without the need for extremely heavy new states. Experimentally, MeV-GeV masses are perfectly accessible at current facilities; in fact, the strongest limits on heavy neutrino mixing are found in this range. 

In this thesis we have explored models that, via the introduction of heavy neutrinos, try to solve several problems in particle physics. First, we have studied a setup in which lepton number is spontaneously broken by a new scalar. This breaking implies a suppression that accounts for small neutrino masses, even with heavy states at the GeV scale. Besides, the corresponding Goldstone boson, the Majoron, couples to light neutrinos, possibly softening the Hubble tension, a longstanding problem in current cosmology. We find that two lepton number assignments are able to provide all these features with no need for fine-tuning, and also satisfying terrestrial and astrophysical bounds. However, one of the two choices is disfavoured by cosmological limits, as it predicts heavy neutrinos that would decay in the early Universe, altering the predictions of BBN. Interestingly, we find that these models could be embedded in a larger framework, based on Minimal Flavour Violation, that attempts to solve the flavour puzzle by predicting the masses of the fermions. It has also been shown how a Peccei-Quinn symmetry can be present in this realization, offering a solution to the strong CP problem via the introduction of an axion. 

These last ideas were further explored in a second model, where Peccei-Quinn and lepton number were identified into a common symmetry group. A single Goldstone boson, the \say{Majoraxion}, is present in this case. We have shown how these new symmetries can dynamically explain an inverse seesaw texture for neutrinos, once two sets of heavy states are introduced. In the usual low-scale seesaw spirit, the observed neutrino masses are reproduced, while keeping the new particles close to the electroweak scale. We adopted an MFV approach, that, on its own, lacks predictive power due to the several spurions present in the neutrino sector. We therefore identified three scenarios in terms of simplified symmetry groups and transformation properties of the heavy neutrinos, which lead to different phenomenologies. In the first case, charged lepton flavour violation is predicted, but the heavy neutrinos can be kept below the TeV thanks to extra freedom in the Yukawas. The second scenario is more restricted, and the new particles must be much heavier in order to satisfy the strong constraints posed by flavour change. Finally, the third case is much simpler, and it does not predict flavour violation. Observables such as the $W$ mass or the effective number of neutrinos constrain the heavy neutrinos to be at the TeV scale.

These model-building attempts are examples of how heavy neutrinos may connect some of the main open questions in particle physics, granting common solutions to neutrino masses and, for instance, to the strong CP problem of the flavour puzzle. Although this requires non-minimal setups, their corresponding phenomenology can be probed by current experiments, mainly due to the relatively light new neutrinos.

Following a different approach, we have turned to explore ways to probe heavy neutrinos in a rather model-independent manner. Each neutrino mass mechanism predicts different interactions for these particles, so it is impossible to perform searches dedicated to test particular models. Experiments instead probe neutrino mixing without assuming any particular setup, leaving that quantity and the heavy neutrino mass as independent, free parameters. Some further simplifications are often performed, usually assuming that the mixing occurs dominantly with one lepton flavour. We have reviewed the main experimental limits that are available up to date in the MeV-GeV range. The strongest bounds are set close to the pion and kaon mass, as the decay of these mesons allows to produce neutrinos copiously. 

It is clear that the interactions between neutrinos and mesons are key in the phenomenology for this range of masses. This topic has been widely studied in the literature, although some discrepancies could be found, especially in the treatment of vector mesons. Here we have derived carefully the effective theory that describes this kind of couplings. In order to do so, we have integrated out the weak gauge bosons and introduced the necessary hadronic matrix elements. Employing this effective theory, we have obtained the branching ratios of a heavy neutrino into all possible channels, as a function of its mass. We have confirmed that decays into mesons are quite sizable for heavy neutrinos at the GeV scale. This low-energy theory was made public in a FeynRules model, that allows for full event generation at the differential level. 

We have made use of this tool to estimate the sensitivity to heavy neutrinos at the DUNE Near Detector. In this environment, the protons that hit the target generate a large number of charged mesons, that may decay into heavy neutrinos. These can travel a certain distance and decay back into SM particles inside the ND, leaving visible signatures. Employing the fluxes of charged mesons, we have computed the number of events of different channels as a function of the heavy neutrino mass and mixing, computing the sensitivity to the latter. Here we have not performed a detailed analysis of the background, so our results are only an estimation. However, we find that the DUNE ND can reach unprecedented sensitivities, especially close to the kaon mass; in fact, mixings as small as those predicted by the type I seesaw could be probed. This is partly thanks to the boost of the heavy neutrinos, that enhances the detector acceptance; such an effect can only be appreciated in a full simulation of the neutrino flux that takes their mass into account.

Finally, we have explored how heavy neutrinos can be implemented in an effective field theory, that allows to describe more exotic interactions and their potential impact on low-energy observables. In particular, we have studied the $\nu$SMEFT, which extends the Standard Model Effective Field Theory by introducing heavy neutrinos, which are gauge singlets of the SM. We have focused on dimension-6 operators, reviewing those that include at least one heavy neutrino and analyzing the observable processes that they could mediate. We have made use of the corresponding experimental information to set bounds on the relevant Wilson coefficients as a function of the heavy neutrino mass. As some simplifications are needed in this approach, we assumed that only one effective operator is active at a time, and that the mixing with the light states is negligible. We find that the limits on heavy neutrino mixing can be recasted for many effective operators, actually constituting some of the most stringent bounds. Other observables, such as monophoton searches, supernovae cooling or $Z$ boson decay can also pose relevant constraints. 

These model-independent studies have made manifest that heavy neutrinos at the MeV-GeV range offer the most promising prospects to be experimentally tested. In contrast to the extremely heavy states predicted by the original seesaws, these lighter neutrinos could be produced in low-energy experiments, leaving clear signals. Current and future facilities may probe extremely small mixings, but can also be sensitive to heavy neutrinos in other observables, that may be described by effective operators. The theoretical motivation for these somewhat light states is quite sound, relying on approximate lepton number symmetries.

In summary, this thesis has aimed at highlighting the role of heavy neutrinos as protagonists in the solutions to open problems of particle physics. From a theoretical perspective, they can be at the center of new models, that offer common explanations to neutrino masses and other issues, such as the flavour puzzle or the strong CP problem. On the experimental side, there are interesting chances to probe heavy neutrinos at current and future facilities, either via mixing with the light states or through more exotic interactions. A positive signal of heavy neutrinos would point towards the mechanism behind neutrino masses, possibly unraveling other mysteries of particle physics.

\chapter*{Conclusiones}
\addcontentsline{toc}{part}{Conclusiones}
\fancyhead[RO]{\scshape \color{lightgray}Conclusiones}
\fancyhead[LE]{\scshape \color{lightgray}Conclusiones}
\lettrine[depth=1, loversize=0.55,lraise=-0.4]
    {E}{}l Modelo Estándar de la física de partículas ha demostrado ser uno de los mayores hitos en la historia de la ciencia, describiendo muy satisfactoriamente los componentes fundamentales de nuestro universo y las interacciones entre ellos. Una vez fijados algunos parámetros libres gracias a información experimental, el SM proporciona predicciones para virtualmente cualquier proceso en física de partículas. Muchísimas pruebas experimentales han mostrado que prácticamente todas estas predicciones son correctas, alcanzando niveles de precisión sin precedentes.

Pese a su tremendo éxito, el SM no puede ser la teoría definitiva de las partículas e interacciones fundamentales, ya que presenta varias incógnitas importantes. Algunas, tales como el problema de la jerarquía, el puzle del sabor o el problema CP fuerte, no apuntan a fallos explícitos de la teoría, sino más bien a características poco naturales o a situaciones de ajuste fino. Otros enigmas se deben a observaciones empíricas que el SM no puede explicar, como la presencia de materia oscura en el universo.

En esta tesis se ha intentado atacar el problema de las masas de los neutrinos, que constituye una de las principales contradicciones entre las predicciones del SM y las medidas experimentales. Se ha observado que estas partículas, al propagarse, cambian de sabor siguiendo un patrón oscilatorio; este fenómeno sólo puede ocurrir si los neutrinos son partículas masivas, al contrario de lo que predice el SM. Es evidente que esta teoría ha de ser ampliada para incluir algún mecanismo que explique la masa de los neutrinos. Además, se necesita más información experimental que aclare algunas propiedades de los neutrinos, que aún se desconocen debido a su masa no nula. Las principales son su naturaleza Dirac o Majorana, su posible papel en la ruptura de la simetría CP o la escala absoluta de su masa.

El papel que juegan los neutrinos pesados en las respuestas a estos misterios ha sido el punto central de esta tesis. Estas hipotéticas nuevas partículas están presentes en la mayoría de modelos de masas de neutrinos que han aparecido a lo largo de los años, presentando características diferentes. Por ejemplo, el \textit{seesaw} tipo I, probablemente la propuesta más sencilla, incluye neutrinos pesados, que no participan de las interacciones \textit{gauge} del SM. En este marco teórico, la masa de los neutrinos ligeros es inversamente proporcional a la de los nuevos estados, sugiriendo que estos últimos son muy pesados. Esta idea ha sido el punto de partida de numerosos nuevos modelos, más sofisticados, que suelen predecir nuevas fuerzas fundamentales o espectros más complejos. 

El interés de los neutrinos pesados no sólo se debe a la generación de masas de neutrinos, sino también a su potencial para proporcionar soluciones a otros problemas abiertos del SM. Por ejemplo, si presentan ciertas características, los neutrinos pesados pueden ser candidatos viables de materia oscura, no presentes en el SM. Además, si son fermiones de Majorana, los neutrinos pueden explicar la asimetría bariónica del universo mediante leptogénesis.

Desde un punto de vista experimental, el estudio de los neutrinos pesados es todo un reto. Como su masa no está determinada por ningún principio fundamental, se necesitan distintas estrategias para cubrir un amplio rango de energías, desde la escala del eV, estudiada en oscilaciones, al TeV, sólo alcanzable en colisionadores. Tampoco está claro cómo estas nuevas partículas se acoplan al SM; al menos, los neutrinos pesados heredarían las interacciones de los ligeros mediante mezcla, ofreciendo una manera de producirlos o de observar los productos de su desintegración. La ausencia de señales hasta la fecha demuestra que, de existir, estas partículas han de interaccionar muy débilmente con el SM, o bien ser demasiado pesadas para ser producidas. Esto se traduce en límites muy fuertes sobre los correspondientes acoplos, habitualmente parametrizados como elementos de mezcla.

Se ha puesto el énfasis en neutrinos pesados en el rango del MeV-GeV, que resultan interesantes tanto desde el aspecto teórico como desde el experimental. Nuevos neutrinos por debajo de la escala electrodébil aparecen en numerosos modelos, normalmente basados en una simetría aproximada de número leptónico que protege las masas de los neutrinos ligeros, explicando su pequeñez sin necesidad de nuevos estados extremadamente pesados. Experimentalmente, masas del orden del MeV-GeV están perfectamente al alcance de la tecnología actual; de hecho, los límites más fuertes sobre la mezcla de neutrinos pesados se obtienen en este rango. 

En esta tesis se han explorado modelos que intentan resolver varios problemas en física de partículas mediante la introducción de neutrinos pesados. Primero, se ha estudiado una situación en la que el número leptónico está espontáneamente roto por un nuevo escalar. Esta ruptura implica una supresión que da cuenta de las pequeñas masas de los neutrinos, incluso con estados pesados a la escala del GeV. Además, el correspondiente bosón de Goldstone, el Majoron, se acopla a los neutrinos ligeros, posiblemente suavizando la tensión de Hubble, un problema persistente en la cosmología actual. Se han encontrado dos asignaciones de número leptónico que presentan todas estas características sin necesidad de ajuste fino, y satisfaciendo los límites terrestres y astrofísicos. Sin embargo, existen restricciones cosmológicas que apuntan en contra de una de estas dos elecciones, ya que predice neutrinos que se desintegrarían en el universo temprano, alterando las predicciones de BBN. Resulta interesante que estos modelos pueden incorporarse a un marco más amplio, basado en Minimal Flavour Violation, que intenta resolver el puzle del sabor, prediciendo las masas de los fermiones. También se ha mostrado cómo una simetría de Peccei-Quinn puede estar presente dentro de este enfoque, ofreciendo una solución al problema CP fuerte a través de la introducción de un axión.

Se ha profundizado en estas últimas ideas en un segundo modelo, donde Peccei-Quinn y número leptónico han sido unificadas en un grupo de simetría común. Un único bosón de Goldstone, el \say{Majoraxion}, está presente en este caso. Se ha mostrado cómo estas nuevas simetrías pueden explicar dinámicamente una textura de \textit{seesaw} inverso para los neutrinos, una vez que se introducen dos grupos de estados pesados. Con la filosofía habitual de los seesaw a baja escala, las medidas de masas de neutrinos ligeros pueden reproducirse manteniendo las nuevas partículas cerca de la escala electrodébil. Se ha mantenido un enfoque de MFV, que, de por sí, carece de poder predictivo debido a los diversos espuriones que aparecen en el sector de los neutrinos. Por tanto, se han identificado tres situaciones en términos de grupos de simetría simplificados y de las propiedades de transformación de los neutrinos pesados, que dan lugar a distintas fenomenologías. En el primer caso, se genera violación de sabor leptónico, pero los neutrinos pesados pueden mantenerse por debajo del TeV debido a una libertad adicional en los Yukawas. El segundo escenario está más restringido, y las nuevas partículas deben ser mucho más pesadas para satisfacer los límites dados por cambio de sabor. Finalmente, el tercer caso es mucho más simple, y no predice violación de sabor. Observables tales como la masa del $W$ o el número efectivo de neutrinos fuerzan a los neutrinos pesados a estar cerca del TeV. 

Los modelos que se han presentado constituyen ejemplos de cómo los neutrinos pesados pueden conectar algunas de las principales incógnitas en física de partículas, proporcionando soluciones comunes a las masas de los neutrinos y, entre otros, al problema CP fuerte o al puzle del sabor. Aunque para ello se requieren escenarios no mínimos, su correspondiente fenomenología está al alcance de los experimentos actuales, principalmente debido a las masas relativamente ligeras de los nuevos neutrinos.

Desde otro punto de vista, se ha explorado cómo testar neutrinos pesados sin necesidad de basarse en ningún modelo concreto. Cada mecanismo de masas de neutrinos predice diferentes interacciones de estas partículas, por lo que resulta imposible llevar a cabo búsquedas dedicadas a teorías específicas. Los experimentos se decantan por estudiar la mezcla de los neutrinos sin asumir ningún marco teórico en especial, manteniendo esta cantidad y la masa de los neutrinos pesados como parámetros libres e independientes. Se suelen realizar incluso más simplificaciones, como asumir que la mezcla se da dominantemente con un único sabor leptónico. Se han revisado los principales límites experimentales disponibles hasta la fecha en el rango del MeV-GeV. Las cotas más fuertes se encuentran cerca de las masas del pión y del kaón, ya que la desintegración de estos mesones permite producir una gran cantidad de neutrinos.

Está claro que las interacciones entre neutrinos y mesones son fundamentales para la fenomenología en este rango de masas. Este tema ha sido estudiado ampliamente en la literatura, aunque pueden encontrarse algunas discrepancias, especialmente en el tratamiento de los mesones vectoriales. Se ha derivado cuidadosamente la teoría efectiva que describe este tipo de interacciones. Para ello, se ha integrado sobre los bosones \textit{gauge} débiles y se han introducido los elementos de matriz hadrónicos necesarios. Empleando esta teoría efectiva, se han obtenido las tasas de desintegración de un neutrino pesado a todos los posibles canales en función de su masa. Se ha confirmado que las desintegraciones a mesones son considerables para neutrinos en la escala del GeV. Esta teoría de bajas energías fue publicada en un modelo FeynRules, que permite la generación de eventos a nivel diferencial.

Se ha utilizado esta herramienta para estimar la sensibilidad a neutrinos pesados en el Detector Cercano de DUNE. En este experimento, los protones que impactan en el objetivo generan un gran número de mesones cargados, que pueden desintegrarse en neutrinos pesados. Estos pueden viajar una cierta distancia y decaer en partículas del SM dentro del ND, dando lugar a señales visibles. Empleando los flujos de mesones cargados, se ha calculado el número de eventos de diferentes canales como función de la masa y mezcla del neutrino pesado, calculando la sensibilidad a este último parámetro. No se ha llevado a cabo un estudio detallado del fondo, de manera que los resultados son meramente una estimación. Sin embargo, se ha comprobado que el DUNE ND puede alcanzar sensibilidades sin precedentes, especialmente cerca de la masa del kaón; de hecho, se podrían testar mezclas tan pequeñas como las predichas por el \textit{seesaw} tipo I. Esto se debe en parte al \textit{boost} de los neutrinos pesados, que aumenta la aceptancia del detector; tal efecto sólo puede apreciarse en una simulación completa de los flujos de neutrinos, que tiene en cuenta su masa.

Finalmente, se ha explorado cómo los neutrinos pesados pueden implementarse en una teoría de campos efectiva, la cual permite describir interacciones más exóticas y su potencial impacto en observables de baja energía. En particular, se ha estudiado el $\nu$SMEFT, que extiende el SMEFT introduciendo neutrinos pesados, que son singletes del SM. Se ha centrado el análisis en operadores de dimensión 6, revisando aquellos que incluyen al menos un neutrino pesado y analizando qué procesos observables podrían generar. Se ha utilizado la correspondiente información experimental para poner cotas a los coeficientes de Wilson relevantes en función de la masa del neutrino pesado. Algunas simplificaciones son necesarias en este enfoque: se ha asumido que sólo un operador efectivo está activo a la vez, y que la mezcla con los estados ligeros es despreciable. Se ha comprobado que los límites en la mezcla de neutrinos pesados pueden reinterpretarse para muchos operadores efectivos, constituyendo algunas de las cotas más restrictivas. Otros observables, como las búsquedas de monofotones, el enfriamiento de supernovas o la desintegración del bosón $Z$, también suponen límites importantes.

Estos análisis, que no dependen de modelos concretos, han puesto de manifiesto que los neutrinos en el rango del MeV-GeV presentan las mejores perspectivas en cuanto a su observación empírica. Al contrario que los estados extremadamente pesados predichos por los \textit{seesaws} originales, estos neutrinos más ligeros pueden producirse en experimentos de baja energía, dejando señales claras. Experimentos presentes y futuros pueden testar mezclas extremadamente pequeñas, además de ser sensibles a neutrinos pesados a través de otros observables, que pueden ser descritos por operadores efectivos. La motivación teórica para estos estados relativamente ligeros es muy robusta y se basa en simetrías aproximadas de número leptónico.

En resumen, esta tesis ha tratado de resaltar el papel de los neutrinos pesados como protagonistas en la solución a diversos problemas abiertos en física de partículas. Desde una perspectiva teórica, pueden constituir la base de nuevos modelos, que ofrecen explicaciones comunes a la masa de los neutrinos y a otras cuestiones, como el puzle del sabor o el problema CP fuerte. En el aspecto experimental, hay interesantes posibilidades de testar neutrinos pesados en el presente y futuro cercano, bien mediante su mezcla o a través de interacciones más exóticas. Una señal de neutrinos pesados apuntaría al mecanismo responsable de generar masas de neutrinos, posiblemente descifrando otros misterios de la física de partículas. 

\PB

\setupappendixparts
\setupappendixchapters

\appendix

\part*{Appendices}
\fancyhead[LE]{\scshape \color{lightgray}Appendices}
\PB
\chapter{Heavy neutrino production and decay}
\fancyhead[RO]{\scshape \color{lightgray}A. Heavy neutrino production and decay} 
In this appendix we summarize some channels of heavy neutrino production and decay that have been employed to obtain the results of this thesis.
\section{Production of heavy neutrinos from meson decays via mixing}
\label{app:hnl_prod_mixing}

Here we provide the expressions for the production of a heavy neutrino via meson decays, which constitute one of the main sources of HNLs in the MeV-GeV range. We have computed them employing the Feynman rules derived in Ch.~\ref{sec:mesons}, assuming that the HNLs only exhibit weaker-than-weak interactions, mediated by mixing.

\subsection{Two-body leptonic decays}

The generic expression for the leptonic decay of a charged pseudoscalar meson $P$ of mass $m_P$ is given by~\cite{Gorbunov:2007ak,Atre:2009rg,Helo:2010cw,Abada:2013aba,Bondarenko:2018ptm}
\begin{align}
\Gamma(P^\pm \rightarrow N \ell^\pm_\alpha)= \displaystyle \dfrac{G_F^2 m_P^3}{8\pi}f_P^2\vert U_{\alpha 4}\vert^2 \vert V_{q q'}\vert^2 \lambda^{1/2}(1, y_N^2, y_\alpha^2) \left(y_N^2+y_\alpha^2-\left(y_N^2-y_\alpha^2 \right)^2\right) \,,
\label{eq:width_P_N_l}
\end{align}
where $G_F$ is Fermi\textquotesingle s constant, $V_{qq^\prime}$ is the relevant CKM element and the values of $f_P$ for the most relevant mesons are given in Tab.~\ref{tab:Fpi}. We have defined $ y_N \equiv M_N/m_P$, $ y_\alpha \equiv m_\alpha/m_P$, and $\lambda(a,b,c) = a^2 + b^2 + c^2 - 2ab -2bc -2ac$ is the Källen function. 

Neutral pseudoscalars can also produce heavy neutrinos, either in pairs or along a light neutrino. However, these channels are completely invisible, playing little role in the phenomenology. 

\subsection{Three-body semileptonic decays}

The decay width for the semileptonic decay of a parent pseudoscalar meson $P$ into a daughter pseudoscalar $D$, a charged lepton and a heavy neutrino is given by~\cite{Gorbunov:2007ak,Helo:2010cw,Abada:2013aba,Bondarenko:2018ptm} 
\begin{align}
\Gamma(P \rightarrow D N \ell^\pm_\alpha)=\displaystyle \dfrac{G_F^2 m_P^5}{64\pi^3} C_D^2 \vert U_{\alpha 4}\vert^2 \vert V_{q q'}\vert^2 \left(I_1^{PD} +I_2^{PD} +I_3^{PD} \right)  \,,
\label{eq:width_P_D_N_l}
\end{align}
where $C_D=1$ unless the daughter meson is a $\pi^0$, where $C_D = \frac{1}{\sqrt{2}}$. The integrals $I_i^{PD}$ are expressed in terms of the form factors $f^{PD}_+(q^2)$ and $f^{PD}_0(q^2)$:  
\begin{align}
I_1^{PD}&= \int_{(y_\alpha+y_N)^2}^{(1-y_D)^2}{\frac{dz}{3z^3}\left| f^{PD}_+\left( z m_P^2 \right)\right|^2 \lambda(1,y^2_D,z)^{3/2}\lambda(z,y^2_4,y_\alpha^2)^{3/2}}  \,, \\
I_2^{PD}&= \int_{(y_\alpha+y_N)^2}^{(1-y_D)^2}{\frac{dz}{2z^3}\left| f^{PD}_+\left( z m_P^2 \right)\right|^2 \lambda(1,y^2_D,z)^{3/2}\lambda(z,y^2_4,y_\alpha^2)^{1/2}g(z)}  \,, \\
I_3^{PD}&= \int_{(y_\alpha+y_N)^2}^{(1-y_D)^2}{\frac{dz}{2z^3}\left| f^{PD}_0\left( z m_P^2 \right)\right|^2 \lambda(1,y^2_D,z)^{1/2}\lambda(z,y^2_4,y_\alpha^2)^{1/2}g(z)\left(1-y_D^2 \right)^2}  \,, 
\label{eq:integrals_P_D_N_l}
\end{align}
where $y_D \equiv m_D/m_P$ and
\begin{equation}
g(z)=z\left(y_N^2+y_\alpha^2 \right)-\left( y_N^2-y_\alpha^2 \right)^2 \, .
\end{equation}
\subsubsection{\fontfamily{put}\selectfont Form factors}
\label{subsec:formfactors}
Several parametrizations for the hadronic form factors are available in the literature, most
of which are given in terms of $f_+$ and $f_0$. The former was defined together with $f_-$ in Eq.~\ref{eq:fplus_fminus}, while the latter can be related to $f_+$ and $f_-$ via
\begin{equation}
f_0(q^2)=f_+(q^2)+\frac{q^2}{m_D^2-m_P^2}f_-(q^2) \, .
\end{equation}
The semileptonic decays we will be mostly interested in are $K\rightarrow \pi n \ell$ and $D\rightarrow K n \ell$ (where $n$ generically denotes a neutrino mass eigenstate), as they may be relevant for HNL production in beam dumps, depending on the neutrino mass\footnote{The FeynRules model that contains the effective theory derived in this thesis also includes semileptonic decays of $B_{(s)}$ mesons, into pions, kaons and $D_{(s)}$. We also employ a pole parametrization for the form factors of these channels, with different coefficients but no conceptual differences.}. For the former we employ a linear parametrization, as in Ref.~\cite{Bijnens:1994me}, according to which 
\begin{equation}
f_{+,0}^{K\pi}(q^2)=f_+^{K\pi}(0)\left[1+\lambda^{K\pi}_{+,0}\frac{q^2}{m_{\pi^+}^2} \right] \, .
\end{equation}
Conversely, in the case of the $D\rightarrow Kn\ell$ decay we make use of a \say{pole} parametrization~\cite{Lubicz:2017syv}:
\begin{align}
f_+^{DK}(q^2) & =  \frac{f_+^{DK}(0)+c_+^{DK}(z-z_0)(1+\frac{z+z_0}{2})}{1-\frac{q^2}{m^2_{D_s^*}}} \, , \label{eq:fpDK} \\
f_0^{DK}(q^2) & =  f_+^{DK}(0)+c_0^{DK}(z-z_0)\left(1+\frac{z+z_0}{2} \right) \, ,
\end{align}
where 
\begin{align}
z & =  \frac{\sqrt{t_+-q^2}-\sqrt{t_+-t_0}}{\sqrt{t_+-q^2}+\sqrt{t_+-t_0}}\,, \\
z_0 & = z(q^2=0)\,,
\end{align}
with
\begin{align}
t_+ & =  (m_D+m_P)^2, \\
t_0 & =  (m_D+m_P)\left(\sqrt{m_D}-\sqrt{m_P} \right) ^2\,.
\end{align}
The values used for the form factor parameters are summarized in Tabs.~\ref{tab:paramtableD} and~\ref{tab:paramtableK}.
\begin{table}[b!]
\begin{center}

\begin{tabular}{ccc} \toprule
    $f_+^{DK}(0)$& $c_+^{DK}$&  $c_0^{DK}$ \\ \midrule
    $0.7647$&$-0.066$&$-2.084$ \\ \bottomrule
\end{tabular}

\end{center}
\caption{Parameters entering our form factor definitions for the semileptonic $D\rightarrow K n \ell$ decays, extracted from Ref.~\cite{Lubicz:2017syv}.
\label{tab:paramtableD} }
\end{table}

\begin{table}[b!]
\begin{center}

\begin{tabular}{cccc} \toprule\vspace{1mm}
    &$f^{\mathrm{PD}}_+(0)$~\cite{FlavourLatticeAveragingGroup:2019iem} &$\lambda^{\mathrm{PD}}_+$~\cite{ParticleDataGroup:2018ovx}&$\lambda^{\mathrm{PD}}_0$~\cite{ParticleDataGroup:2018ovx} \\ \toprule\vspace{1.5mm}
     $K^\pm\pi^0$ &\multirow{2}{*}{$0.9749$}&$0.0297$&$0.0195$ \\
     $K^0\pi^\pm$ &&$0.0282$&$0.0138$  \\\bottomrule
\end{tabular}
\end{center}
\caption{Parameters entering our form factor definitions for the semileptonic $K\rightarrow \pi n \ell$ decays. Note that we make use of different parameters for the decays of charged and neutral kaons, following Ref.~\cite{ParticleDataGroup:2018ovx}. 
\label{tab:paramtableK} }
\end{table}

Note that the momentum dependence of the form factors is not trivial to implement in FeynRules, as this software cannot include energy-dependent parameters. In order to surpass this issue, we chose two implementations. As a simpler option, we included average form factors, computed by fitting the total width of the corresponding channels. These average form factors are obtained as a function of the HNL mass. As a more sophisticated alternative, we provided a Python script that modified the  generated UFO, in order to include the energy dependence in the vertices. This option allows to generate the correct differential decay widths, relevant, for instance, for angular distributions.

\section{Decays of heavy neutrinos into SM particles via mixing}
\label{app:hnl_decay_mixing}

Here we provide expressions for the decay widths of a heavy neutrino into SM particles, assuming that the HNL inherits the interactions of the SM neutrinos, with the corresponding mixing suppression. Once again, we focus in the MeV-GeV regime, in which the main decay channels are those into mesons and leptons. These rates are computed employing the Feynman rules derived in Ch.~\ref{sec:mesons}. Throughout this section, we will neglect the masses of the light neutrinos for simplicity.
\label{sec:decays}

\subsection{Two-body decays}
The generic expression for the heavy neutrino decay width into a neutral pseudoscalar meson $P$ is given by
\begin{equation}
\Gamma(N\rightarrow P \nu)= \displaystyle\sum_{j} \dfrac{G_F^2 M_N^3}{32\pi}f_P^2\vert C_{4j}\vert^2\left(1-x_P^2\right)^2 \,,
\label{eq:width_N_P_nu}
\end{equation}
where we have defined $x_P \equiv m_P/M_N$. Recall that in the case of the $\eta$ and $\eta^\prime$ mesons, effective decay constants are employed, as they are not interaction eigenstates (see Sec.~\ref{sec:neutral-pseudo}). Note that the sum  over $j$ runs over the three light neutrino mass eigenstates, since they cannot be individually identified. However, at leading order in $U_{\alpha 4}$, this is equivalent to a sum running over the three active flavours, since 
\begin{equation}
 \sum_j \vert C_{4j}\vert^2 = \sum_{j,\alpha,\beta} U^*_{\alpha 4} U_{\alpha j} U_{\beta 4} U^*_{\beta j} = \sum_{\alpha,\beta} U^*_{\alpha 4} U_{\beta 4} (\delta_{\alpha \beta} - U_{\alpha 4} U^*_{\beta 4}) \simeq \sum_\alpha \vert U_{\alpha 4}\vert^2 \, .
\end{equation}
On the other hand, the decay width into a charged pseudoscalar meson $P^\pm$ is given by
\begin{equation}
\Gamma(N\rightarrow P^\pm \ell^\mp_\alpha) = 
\dfrac{G_F^2 M_N^3}{16\pi} f_P^2\vert U_{\alpha 4}\vert^2 \vert V_{q q'}\vert^2
\lambda^{1/2}(1, x_P^2, x_\alpha^2) \left[1-x_P^2 - x_\alpha^2 \left(2+ x_P^2 - x_\alpha^2 \right) \right] \,,
\label{eq:width_N_P_l}
\end{equation}
where $x_\alpha  \equiv m_\alpha/M_N$.

In the case of neutral vector mesons, the decay width reads:
\begin{equation}
\Gamma(N\rightarrow V \nu)= \displaystyle\sum_{j} \dfrac{G_F^2 M_N^3}{32\pi m_V^2}f_V^2 g_V^2 \vert C_{4j}\vert^2\left(1+ 2 x_V^2\right)\left(1-x_V^2\right)^2 \,,
\label{eq:width_N_V_nu}
\end{equation}
with $x_V \equiv m_V/M_N$, and where we have again summed over all light neutrinos in the final state. The expressions for $g_V$ in terms of the weak mixing angle are provided in Tab.~\ref{tab:gV}.
\begin{table}[ht!]
\begin{center}

\begin{tabular}{ccc} \toprule\vspace{1mm}
    $N\rightarrow \rho^0 \nu$ & $N\rightarrow \omega \nu$ &$N\rightarrow \phi \nu$ \\ \midrule\vspace{1.5mm}
    $1-2 s_w^2$  & $-\dfrac{2 s_w^2}{3}$ & $-\sqrt{2}\left(\dfrac{1}{2}-\dfrac{2 s_w^2}{3}\right)$  \\ \bottomrule
\end{tabular}
\end{center}
\caption{Couplings $g_V$ entering the heavy neutrino decay widths into neutral vector mesons, see Eq.~\ref{eq:width_N_V_nu}.}
\label{tab:gV}
\end{table}

On the other hand, for the decays into charged vector mesons we get
\begin{align}
\Gamma(N\rightarrow V^\pm \ell^\mp_\alpha)  =&\,  \dfrac{G_F^2 M_N^3}{16\pi m_{V^\pm}^2}f_V^2\vert U_{\alpha 4}\vert^2 \vert V_{q q'}\vert^2  \lambda^{1/2}(1, x_V^2, x_\alpha^2) \times \nonumber 
 \left[\left(1-x_V^2\right)\left(1+2x_V^2\right)\right.
 \\
 &\,\left.+ x_\alpha^2\left( x_V^2 + x_\alpha^2 - 2 \right)\right] 
\,.
\label{eq:width_N_V_l}
\end{align}
As mentioned in Sec.~\ref{sec:vect_meson_decay_const}, these channels are a source of disagreement in the literature. An extra factor 2 with respect to our results is present in Ref.~\cite{Gorbunov:2007ak}, both for neutral and charged vectors. The former case is particulary delicate, as we find different dependences on the weak mixing angle with respect to Refs.~\cite{Bondarenko:2018ptm,Gorbunov:2007ak,Atre:2009rg}. We find some disagreements in the corresponding branching ratios with respect to the results in Ref.~\cite{Ballett:2019bgd}, possibly due to discrepancies in the effective vertices and in the decay constants. 

\subsection{Three-body decays}

Heavy neutrinos may also decay into three body final states, either purely leptonically or semileptonically. The latter include $N \to \pi^+ \pi^0 \ell^- $, $N \to \pi^0 \pi^0 \nu$ and $N \to K^+ \pi^0 \ell^- $. However, their respective contributions are dominated by $N \to \rho^+ \ell^- $, $N \to \rho^0 \nu$ and $N \to K^{*,+}  \ell^- $ respectively, already included in the previous section. This can be seen from the data from tau decays, since the hadronic matrix elements involved in the semileptonic decays would be the same. Indeed, the branching ratio of $\tau^- \to \nu \pi^- \pi^0$ is $25.49\%$, while the contribution which does not correspond to $\tau^- \to \nu \rho^-$ is negligible: $\left(3.0 \pm 3.2 \right)\cdot 10^{-3}$~\cite{ParticleDataGroup:2018ovx}. We will thus review here only the three-body purely leptonic decays $N \to \ell \ell \nu$ and $N \to \nu \nu \nu$, taken from Refs.~\cite{Gorbunov:2007ak,Atre:2009rg,Helo:2010cw,Bondarenko:2018ptm}. 

The invisible decay of the heavy neutrino reads~\cite{Gorbunov:2007ak,Atre:2009rg,Helo:2010cw,Bondarenko:2018ptm}
\begin{equation}
\Gamma(N\rightarrow \nu \nu \nu) = \displaystyle\sum_{j} \vert C_{4j}\vert^2\dfrac{G_F^2 M_N^5}{192\pi^3 } \, ,
\label{eq:width_N_nu_nu_nu}
\end{equation}
where we have summed over all possible light neutrinos in the final state.

For the three-body decays involving charged leptons in the final state, we will distinguish between two cases. If the heavy neutrino decays into two leptons of the same flavour $\beta$, there are both $W$- and $Z$-mediated diagrams contributing to the amplitude. The total decay width can be expressed as~\cite{Gorbunov:2007ak,Atre:2009rg,Helo:2010cw,Bondarenko:2018ptm}
\begin{equation}
\Gamma(N\rightarrow \nu \ell^-_\beta \ell^+_\beta) =  
\sum_{\alpha} \vert U_{\alpha 4} \vert^2 \dfrac{G_F^2 M_N^5}{192\pi^3}
\left[ \left( C_1 + 2 s_w^2 \delta_{\alpha \beta} \right) f_1(x_\beta) + \left( C_2 + s_w^2 \delta_{\alpha \beta} \right)  f_2(x_\beta) \right] \,,
\label{eq:width_N_la_la}
\end{equation}
where 
\begin{align}
C_1&= \frac{1}{4} \left( 1- 4 s_w^2 + 8 s_w^4 \right), \\ C_2 &= \frac{1}{2} \left(-s_w^2+2 s_w^4 \right) \, , 
\end{align}
and we have defined the functions
\begin{align}
f_1(x) & =  (1 - 14 x^2 - 2 x^4 - 12 x^6) \sqrt{1 - 4 x^2} +  12 x^4 (x^4 - 1) L (x) \, , \\
f_2(x) & =  4 \left[x^2 (2 + 10 x^2 - 12 x^4) \sqrt{1 - 4 x^2} + 6 x^4 (1 - 2 x^2 + 2 x^4) L(x)\right] \, , 
\end{align}
with
\begin{equation}
L(x) = \ln{\left(\dfrac{1 - 3 x^2 - (1 - x^2) \sqrt{1 - 4 x^2}}{x^2 (1 + \sqrt{1 - 4 x^2})}\right)}\,.
\label{eq:log_fun}
\end{equation}
On the other hand, the decay of the heavy neutrino into two leptons of different flavour is only mediated by the $W$ interaction. In the limit in which one of the charged lepton masses can be neglected, the corresponding decay width simplifies to~\cite{Gorbunov:2007ak, Bondarenko:2018ptm}
\begin{align}
\Gamma(N\rightarrow \nu \ell^-_\alpha \ell^+_\beta) & \simeq & 
\vert U_{\alpha 4}\vert^2 \dfrac{G_F^2 M_N^5}{192\pi^3}
\left(1 - 8 x_M^2 + 8 x_M^6 - x_M^8 - 12 x_M^4 \ln(x_M^2)\right)\, ,
\label{eq:width_N_la_lb}
\end{align}
where $x_M=\mathrm{max}\left\lbrace x_\alpha,x_\beta\right\rbrace $. Note that this expression corresponds to a Dirac neutrino decay; for Majorana neutrinos there would be a second contribution, proportional to $ |U_{\beta 4}|^2$, since there are two diagrams allowed, each of them controlled by a different mixing matrix element.

\subsection{Decays to 4 or more bodies}
\label{subsec:multimesons}

Finally, for HNL masses above 1~GeV, the appropriate description of the hadronic final states transitions from the effective theory described in Ch.~\ref{sec:mesons} to quark production in the final state, with subsequent hadronization, more suitable for perturbative QCD. For reference, the tau lepton, with its $1.78$~GeV mass, is precisely at the transition region. Indeed, it shows a $10.8 \%$ and $25.5 \%$ branching ratio to the $\nu_\tau \pi^-$ and $\nu_\tau \rho^-$ channels respectively. But also a $9.3 \%$ branching ratio to both $\nu_\tau \pi^- 2 \pi^0 $ and to $\nu_\tau 2 \pi^- \pi^+$, and even $4.6 \%$ and $1.0 \%$ branching ratios to $\nu_\tau 2 \pi^- \pi^+ \pi^0$ and $\nu_\tau \pi^- 3\pi^0$ respectively~\cite{ParticleDataGroup:2018ovx}. These last decay modes, with three or more mesons in the final state, are more suitably described from the underlying quark interactions, with a subsequent correction to account for the hadronization process:
\begin{equation}
1 + \Delta_\mathrm{QCD} \equiv \frac{\Gamma \left( \tau \to \nu_\tau + \mathrm{hadrons} \right)}{\Gamma_{tree} \left( \tau \to \nu_\tau + u + \bar{d} \right)+\Gamma_{tree} \left( \tau \to \nu_\tau + u + \bar{s} \right)}\,,
\end{equation}
with~\cite{Gorishnii:1990vf}
\begin{equation}
\Delta_\mathrm{QCD} = \frac{\alpha_s}{\pi}+5.2\frac{\alpha_s^2}{\pi^2}+26.4\frac{\alpha_s^3}{\pi^3}.
\label{eq:QCDcorrection}
\end{equation}
We adopt the same approach as Ref.~\cite{Bondarenko:2018ptm} (see also Refs.~\cite{SHiP:2018xqw,Bondarenko:2019yob}) and use Eq.~\ref{eq:QCDcorrection} to account for the hadronization of the HNL decays $N \to \ell_\alpha u \bar{d}$ and $N \to \ell_\alpha u \bar{s}$ for HNL masses above 1~GeV. We also apply the same correction to the NC decays $N \to \nu q \bar{q}$, with $q=u,d,s$. However, we add a phase space suppression factor $\sqrt{1- 4 m_K^2/M_N^2}$ for the $N\to \nu s \bar{s}$ channel. Otherwise, this channel would be overestimated for $M_N =1$~GeV, where the phase space prevents two kaons in the final state. For the running of $\alpha_s$ we follow the dedicated review in Ref.~\cite{ParticleDataGroup:2018ovx}. The difference between these fully inclusive hadronic final states and the HNL decays to specific mesons discussed above will provide an estimate of the HNL decays to 3 or more mesons. We have tested this procedure for the tau decays and reached good agreement with its tabulated branching ratios.

\section{Heavy neutrino production and decay in \boldmath{$\nu$}SMEFT}
\label{app:EFT_widths}

In general, the effective operators considered in Ch.~\ref{sec:eft} lead to heavy neutrino production and decay with rates different from those given by usual mixing, described above. These new computations are needed in order to derive the limits on the corresponding Wilson coefficients, and are summarized in the following.

\subsection{Higgs-dressed dimension-5 operator} This operator, given in Eq.~\ref{eq:dim5op_Higgs}, yields an invisible decay of the Higgs into two HNLs, given by the rate
\begin{equation}
    \Gamma_{\rm Higgs}^{d=5}=\frac{\left| C_{\rm Higgs}^{d=5}\right| ^2}{\Lambda^2}\frac{v^2 M_h}{8\pi}(1-2y_N^2)\sqrt{1-4y_N^2}\,,
\end{equation}
where $y_N\equiv M_N/M_h$.

\subsection{Higgs-dressed mixing}
This operator (Eq.~\ref{eq:higgsdressed}) opens an invisible decay channel for the Higgs boson into an HNL and a light neutrino, with a rate given by
\begin{equation}
    \Gamma_{\rm LNH}(H\to \nu N)=\frac{\left| C_{\rm LNH}\right|^2}{\Lambda^4}\frac{9v^4M_h}{64\pi}(1-y_N^2)^2\,,
\end{equation}
where we have neglected the light neutrino mass. Note that no flavour index has been specified, as this decay affects all three flavour copies of the operator. The squared Wilson coefficient in the equation above stands for the sum of all three squared coefficients: $|C_{\rm LNH}|^2=\sum_\alpha|C_{\rm LNH}^\alpha|^2$.

\subsection{Neutral bosonic current} A vertex between a $Z$ boson and two HNLs is introduced (Eq.~\ref{eq:OHN}), thus mediating invisible decays for the $Z$ and for neutral mesons, both pseudoscalar and vector. The corresponding rates are
\begin{align}
 \Gamma_{\rm HN}(Z\to NN) &= \frac{\left| C_{\rm HN}\right|^2}{\Lambda^4}\frac{G_FM_Z^3v^4}{12\sqrt{2}\pi}\left(1-4y_N^2\right)^{3/2}\,,
    \\
    \Gamma_{\rm HN}(P\to NN) &= \frac{\left| C_{\rm HN}\right|^2}{\Lambda^4}\frac{G_F^2M_N^2m_{P}v^4f_{P}^2}{4\pi}\sqrt{1-4y_N^2}\,,
    \\
    \Gamma_{\rm HN}(V\to NN) &= \frac{\left| C_{\rm HN}\right|^2}{\Lambda^4}\frac{G_F^2m_{V}v^4f_{V}^2g_V^2}{24\pi}\sqrt{1-4y_N^2}\,,
\end{align} 
where $y_N$ is the ratio of the masses of the HNL and its parent particle. See Tab.~\ref{tab:Fpi} for the values of the most relevant decay constants\footnote{Recall that effective decay constants are employed for the $\eta$ and $\eta^\prime$ mesons, as they are not interaction eigenstates (see Sec.~\ref{sec:neutral-pseudo}).} and Tab.~\ref{tab:gV} for the $g_V$ couplings. For the heavier mesons we employ $f_{J/\psi}=1.8$ GeV$^2$~\cite{Hwang:1997ie} and $f_{\Upsilon}=9.2$ GeV$^2$~\cite{Merlo:2019anv}. Note that, for the $\Upsilon$(1S), $g_\Upsilon=\sqrt{2}\left(\frac{1}{2}-\frac{2}{3}s_w^2\right)$, as its quark content is analogous to that of the $\phi$.

\subsection{Charged bosonic current}  This operator, given in Eq.~\ref{eq:OHNe}, mediates the most relevant processes for HNL production, mainly charged meson decays. 
Furthermore, there is a direct relation between its Wilson coefficient and the effective mixing they induce, given by Eq.~\ref{eq:effectiveCCmixing}. Thus, the usual expressions for most decay rates involving HNLs, shown above in Apps.~\ref{app:hnl_prod_mixing} and~\ref{app:hnl_decay_mixing}, apply for this effective operator.

However, the absence of NCs affects some HNL decay channels, which, in the case of standard mixing, involve both NC- and CC-like diagrams. This is the case for the $N\to\nu\ell\ell$ processes. The corresponding rate mediated by this operator is analogous to that of the decay $N\to\nu qq^\prime$ in the case of standard mixing~\cite{Bondarenko:2018ptm}, and reads
\begin{equation}
    \Gamma_{\rm HN\ell}(N\to\nu\ell_\alpha^\pm\ell_\alpha^\mp)=\frac{\left| C_{\rm HN\ell}^\alpha\right|^2}{\Lambda^4}\frac{G_F^2v^4M_N^5}{384\pi^3}I(x_\alpha)\,,
\end{equation}
where
\vspace{-4.5mm}
\begin{equation}
    I(x_\alpha)\equiv 12\int_{x_\alpha^2}^{(1-x_\alpha)^2}\frac{ds}{s}(s-x_\alpha^2)(1+x_\alpha^2-s)\sqrt{\lambda(s,0,x_\alpha^2)\lambda\left(1,s,x_\alpha^2\right)}\,.
\end{equation}

\subsection{Neutral four-fermion interactions}
The operators $\mathcal{O}_{\rm uu}$, $\mathcal{O}_{\rm dd}$ (Eq.~\ref{eq:Off}) and $\mathcal{O}_{\rm QN}$ (Eq.~\ref{eq:OQN}) mediate invisible decays of neutral mesons into two HNLs. Although the chiralities of the quark fields are different, the expressions for the widths will be the same, since the vector contribution vanishes. They are given by
\begin{align}
    \Gamma_{\rm NC4F}(P\to NN) &= \frac{\vert C\vert^2}{\Lambda^4}\frac{\xi_P^2M_N^2m_P}{4\pi}\sqrt{1-4y_N^2}\,,
    \label{eq:NC4F_pseudo_width}\\
    \Gamma_{\rm NC4F}(V\to NN) &= \frac{\vert C\vert^2}{\Lambda^4}\frac{\xi_V^2m_V}{24\pi}\sqrt{1-4y_N^2}\,.\label{eq:NC4F_vec_width}
\end{align} 
The couplings $\xi_P$ and $\xi_V$ account for the overlap of the quark content of the involved meson and the quark structure of the operator under consideration. They are summarized in Tab.~\ref{tab:neut_mesons} for the most relevant mesons.

\begin{table}[b!]
\centering

\begin{tabular}{cccc} \toprule\vspace{1mm}
    & $\mathcal{O}_{\rm uu}$& $\mathcal{O}_{\rm dd}$& $\mathcal{O}_{\rm QN}$ \\ \toprule\vspace{1.5mm}
    $\pi^0$&$\frac{f_\pi}{2\sqrt{2}}$&$\frac{-f_\pi}{2\sqrt{2}}$&0 \\\vspace{1.5mm}
    $\eta$&$\frac{1}{2\sqrt{6}}\left(c_8f_8-\sqrt{2}s_0f_0\right)$&$\frac{-1}{2\sqrt{6}}\left(c_8f_8+2\sqrt{2}s_0f_0\right)$&$\frac{-\sqrt{3}}{2}s_0f_0$  \\\vspace{1.5mm}
    $\eta^\prime$&$\frac{1}{2\sqrt{6}}\left(s_8f_8+\sqrt{2}c_0f_0\right)$&$\frac{1}{2\sqrt{6}}\left(-s_8f_8+2\sqrt{2}c_0f_0\right)$&$\frac{\sqrt{3}}{2}c_0f_0$ \\\vspace{1.5mm}
    $\Phi$&0&$\frac{f_\Phi}{2}$&$\frac{f_\Phi}{2}$\\ \vspace{1.5mm}
    $J/\psi$&$\frac{f_{J/\psi}}{2}$&0&$\frac{f_{J/\psi}}{2}$ \\\vspace{1.5mm}
    $\Upsilon$(1S)&0&$\frac{f_\Upsilon}{2}$&$\frac{f_\Upsilon}{2}$\\ \bottomrule
\end{tabular}

\caption{Couplings $\xi_P$ and $\xi_V$ controlling the interactions between neutral mesons and HNLs mediated by NC-like four-fermion operators (Eqs.~\ref{eq:NC4F_pseudo_width} and~\ref{eq:NC4F_vec_width}). $s_{0(8)}$ and $c_{0(8)}$ stand for the sine and cosine of the mixing angles $\theta_{0(8)}$.}
\label{tab:neut_mesons}
\end{table}

A four-lepton interaction is induced by the operator 
$\mathcal{O}_{\rm LNL\ell}^{\alpha\beta}$ (Eq.~\ref{eq:op_4lep}) inducing muon and tau decay into lighter leptons. The corresponding width is given by
\begin{equation}
    \Gamma_{\rm LNL\ell}(\ell_\alpha\to\ell_\beta\nu N)=\frac{\vert C_{\rm LNL\ell}^{\alpha\beta}\vert^2}{\Lambda^4}\frac{1}{256\pi^3m_\alpha^3}\int_{(m_\beta + M_N)^2}^{m_\alpha^2}dm_{12}^2\int_{m_{\rm 23,min}^2}^{m_{\rm 23,max}^2}dm_{23}^2\,\left|\mathcal{M}\right|^2\,,
\end{equation}
where
\begin{equation}
    \left|\mathcal{M}\right|^2=\frac{1}{2}\left(m_\alpha^2+m_\beta^2-m_{12}^2-m_{23}^2\right)\left(m_{12}^2+m_{23}^2-M_N^2\right)\,,
\end{equation}
and 
\begin{align}
    m_{\rm 23,max(min)}^2 &= \frac{1}{2m_{12}^2}\left[m_\alpha^2\left(m_\beta^2+m_{12}^2-M_N^2\right)+m_{12}^2\left(m_\beta^2+M_N^2-m_{12}^2\right)\right.
    \nonumber
    \\
    &\left.\pm\left(m_\alpha^2-m_{12}^2\right)\sqrt{m_{12}^4-2m_{12}^2\left(m_\beta^2+M_N^2\right)+\left(m_\beta^2-M_N^2\right)^2}\right]\,.
\end{align}

\subsection{Charged four-fermion interactions}
The operators $\mathcal{O}_{\rm duN\ell}$, $\mathcal{O}_{\rm LNQd}$ and $\mathcal{O}_{\rm QuNL}$ (Eqs.~\ref{eq:CCFFvectorRR},~\ref{eq:CCFFscalarLR_1} and~\ref{eq:CCFFscalarLR_2} respectively) induce charged meson decays into HNLs and vice versa (we only display here the pseudoscalar case, as it is the main HNL production source). 
The operator $\mathcal{O}_{\rm duN\ell}$  provides the same rates as in the standard mixing case (see Apps.~\ref{app:hnl_prod_mixing} and~\ref{app:hnl_decay_mixing}) but controlled by the \say{effective mixing}
\begin{equation}
    U_{\rm duN\ell} =\frac{1}{4G_FV_{ij}}\mathcal{Z}^{\rm duN\ell}_{ij}\frac{C_{\rm duN\ell}}{\Lambda^2}\,,
\end{equation}
where $i,j$ are the flavours of the quarks composing the meson, and $V_{ij}$ is the corresponding CKM element. 

The operators $\mathcal{O}_{\rm LNQd}$ and $\mathcal{O}_{\rm QuNL}$ exhibit the same decay widths for mesons into HNLs and vice versa:
\begin{align}
    \Gamma (N\to P^\pm\ell_\alpha^\mp) &= \frac{\left| C^\alpha\mathcal{Z}_{ij}\right|^2}{\Lambda^4}\frac{f_P^2m_P^4}{256M_N(m_i+m_j)^2}\left(M_N^2+m_\alpha^2-m_P^2\right)\sqrt{\lambda\left(1,x_\alpha^2,x_P^2\right)}\,,
    \\
     \Gamma (P^\pm\to N\ell_\alpha^\pm) &= \frac{\left| C^\alpha\mathcal{Z}_{ij}\right|^2}{\Lambda^4}\frac{f_P^2m_P^3}{128(m_i+m_j)^2}\left(m_P^2-m_\alpha^2-M_N^2\right)\sqrt{\lambda\left(1,y_\alpha^2,y_N^2\right)}\,,
\end{align}
where $m_i$ and $m_j$ are the masses of the quarks that compose the meson under consideration.

\chapter{Rescaling procedures for $\nu$SMEFT}
\label{app:recasting}
\fancyhead[RO]{\scshape \color{lightgray}B. Rescaling procedures for $\nu$SMEFT}
\section{Meson decays and decay-in-flight searches}
\label{app:rescaling}

Some of the effective operators considered in Ch.~\ref{sec:eft} induce interactions akin to those generated by mixing between heavy and active neutrinos. 
While the standard mixing allows all the processes that could be possible in the SM, by simply replacing a light neutrino with a heavy one, the new operators may only generate a subset of those. 
We discuss how to reinterpret constraints on the mixing as constraints on the Wilson coefficients of the dimension-6 operators. A similar rescaling procedure was also advocated in Ref.~\cite{Beltran:2023nli}.

\subsection{Charged bosonic current}
The operator in Eq.~\ref{eq:OHNe} induces an effective CC interaction.
In analogy to the standard mixing, in Eq.~\ref{eq:effectiveCCmixing} we defined the variable $U^{\rm CC}_{\alpha 4} \equiv \frac{C_{\rm HN\ell}^\alpha v^2}{\sqrt{2}\Lambda^2}$,
which can be related to the standard mixing, $U_{\alpha 4}$, in an experiment-dependent way.

The bounds set by peak search experiments apply equally to $U_{\alpha 4}$ and $U^{\rm CC}_{\alpha 4}$, as the HNLs are always produced via CCs, and their decay is not observed. 
In this case, the rescaling is trivial.
Constraints obtained from HNL production in $Z$ boson decay, such as the ones by DELPHI, would not apply. 

For decay-in-flight searches, HNL production is identical to the standard mixing case, as it takes place primarily through charged meson decays.\footnote{In principle, the leptonic $\ell_\alpha \to \ell_\beta \nu_\beta N$ decays also produce HNLs via $|U_{\alpha 4}|^2$. 
However, these channels are subdominant to the meson decays, $M\to \ell_\alpha N$, except in the mass region $m_{M} - m_\alpha < M_N < m_\alpha - m_{\beta}$.
Nevertheless, as we assume single-flavour dominance, the HNL CC decays are unobservable in this case, since the HNL cannot decay into its parent lepton, $N \, \slashed{\to} \, \ell_\alpha \pi$.}
The decays, however, are modified, and we take into account that NC channels are no longer available.
By equating the number of events, the limits on the effective mixing can be obtained by
\begin{equation}\label{eq:rescaling_beamdump}
     |U^{\rm CC}_{\alpha 4}|^4 = |U_{\alpha 4}|^4 \frac{ \sum_X\hat{\Gamma}^{\rm mixing}(N\to X)}{\sum_X \hat{\Gamma}^{\rm CC}(N\to X)}\,,
\end{equation}
where $\Gamma^{\rm mixing}(N \to X)$ is the decay width of the HNL into the signal final state $X$ (e.g., $X = \nu e^+e^-, e^+ \pi^-, \dots$) induced by the standard mixing scenario, and $\Gamma^{\rm CC}(N \to X)$ is the one induced by the charged bosonic current.
The hat indicates that the decay width is deprived of its $|U_{\alpha 4}|^2$ mixing factor or EFT coefficient $C^2/\Lambda^4$, such that $\hat{\Gamma}^{\rm mixing} = \Gamma^{\rm mixing}/|U_{\alpha 4}|^2$ and $\hat{\Gamma}^{\rm CC} = \Gamma^{\rm CC}/|U_{\alpha 4}^{\rm CC}|^2$. An analogous rescaling applies to collider constraints from ATLAS and CMS, where the HNL is produced in $W^\pm$ decays, and the decay channels vary depending on whether NC was considered in the original search. 
In both cases, the Wilson coefficient will then be obtained by applying consecutively this ratio and the rescaling in Eq.~\ref{eq:effectiveCCmixing}.

Note that in the case of $\tau$ flavour dominance, decay-in-flight searches are insensitive to this operator, as the only possible decays of the HNL involve a tau lepton in the final state, which is either kinematically forbidden or not directly searched for.

\subsection{Charged current four-fermion interactions}
In a similar fashion to the charged bosonic current, the operators $\mathcal{O}_{\rm duN\ell}$, $\mathcal{O}_{\rm LNQd}$ and $\mathcal{O}_{\rm QuNL}$ (Eqs.~\ref{eq:CCFFvectorRR},~\ref{eq:CCFFscalarLR_1} and~\ref{eq:CCFFscalarLR_2} respectively) induce a subset of the interactions contained in the standard mixing case. 
As these operators involve quarks, the only processes they can mediate are mesons decay into neutrinos and vice versa. 
In that case, searches at CHARM, DELPHI, ATLAS, and CMS do not apply, as they tag HNL decays to fully leptonic final states.
Furthermore, the rates of HNL production and decay can be modified due to the different Lorentz structures of the operators. 

To illustrate the rescaling procedure, let us take the operator in Eq.~\ref{eq:CCFFscalarLR_1}, namely $\mathcal{O}_{\rm LNQd}=\frac{C_{\rm LNQd}}{\Lambda^2}\mathcal{Z}^{\rm LNQd}_{ij}(\overline{L}N)\epsilon(\overline{Q_i}d_j)$.
Here, $\mathcal{Z}_{ij}^{\rm LNQd}$ stands for the flavour coefficient. 
If flavour alignment is assumed, $\mathcal{Z}_{ij}^{\rm LNQd}=V_{ij}$.
As usual, the necessary rescaling is achieved by equating the hypothetical number of events between the standard mixing case and the $\nu$SMEFT operator case. 
In the case of a peak search in meson decays, only the production decay rates are relevant. If the decay is, say, a pion into an electron and a neutrino, we can obtain a bound on the Wilson coefficient by means of
\begin{equation}
    \frac{|C_{\rm LNQd}^e|^2}{\Lambda^4}=\frac{|U_{e 4}|^2}{\left|\mathcal{Z}_{du}^{\rm LNQd}\right|^2}\frac{\hat{\Gamma}^{\rm mixing}(\pi\to eN)}{\hat{\Gamma}^{\rm LNQd}(\pi\to e N)}\,.
\end{equation}
This relation can be easily generalized for other meson decays and operators.

In the case of decay-in-flight searches, the decay rates of the HNL need to be taken into account. 
For instance, let us consider the same operator as above, $\mathcal{O}_{\rm LNQd}$, in an electron flavour dominance scenario.
In a search for both $N\to \pi \ell$ and $N\to \ell \ell\nu$ decays in flight from HNLs produced by kaon decays at the target, only the former channel is sensitive to $C_{\rm LNQd}^e$.
The limit on the Wilson coefficient can then be extracted as
\begin{equation}
    \frac{|C_{\rm LNQd}^e|^4}{\Lambda^8}=\frac{|U_{e 4}|^4}{\left|\mathcal{Z}^e_{su}\mathcal{Z}^e_{du}\right|^2}
    \frac{\hat{\Gamma}^{\rm mixing }(K\to e N)}{\hat{\Gamma}^{\rm LNQd}(K\to eN)}
    \frac{\hat{\Gamma}^{\rm mixing}(N\to\pi e)+\hat{\Gamma}^{\rm mixing}(N\to \ell\ell\nu)}{\hat{\Gamma}^{\rm LNQd}(N\to\pi e)}\,.
\end{equation}
The generalization to other lepton flavours and operators is straightforward and takes into account the different mixing assumptions and decay widths.

\section{Supernova cooling}
\label{app:supernova}

Supernova cooling arguments provide upper and lower bounds in the coupling of light particles. 
If the new states couple too feebly to the SM, their production is not enough to cool the supernova efficiently. 
On the other hand, if their interactions with the SM are too strong, they would be trapped inside the supernova, keeping the energy inside the system. 
In the intermediate regime between these two cases, the new particles can escape from the supernova, cooling it efficiently. 

Ref.~\cite{DeRocco:2019jti} constrains the parameter space of a dark matter particle with a four-fermion vector coupling to the SM electromagnetic current,
\begin{equation}\label{eq:FFvectorDM}
    \mathcal{O}_\mathrm{vec}=\frac{C_\mathrm{vec}}{\Lambda^2}\overline{\chi}\gamma^\mu\chi J^\mathrm{em}_\mu\,.
\end{equation}
As HNLs play an identical role to DM fermions in the cooling of the supernova, their bounds can be readily translated as bounds on our operators, upon suitable rescaling,
\begin{equation}
    \frac{|C_i|^2}{\Lambda^4} = \frac{|C_j|^2}{\Lambda^4}\frac{\sigma_j}{\sigma_i}\,,
\end{equation}
where $C_i$ and $\sigma_i$ are, respectively, the Wilson coefficient of a particular operator $\mathcal{O}_i$ and the cross section of a process mediated by this operator. The rescaling factor is then obtained for $\mathcal{O}_i \in \{\mathcal{O}_{\rm HN},\mathcal{O}_{\rm HN\ell}, \mathcal{O}_{\rm ff},\mathcal{O}_{\rm LN},\mathcal{O}_{\rm LNL\ell}\}$ (see Sec.~\ref{sec:4ferm_neut_curr}), and $\mathcal{O}_j=\mathcal{O}_{\rm vec}$. 

To compute the lower bounds, we employ the cross section for the scattering of the new fermion off the SM particles present in the supernova, whereas to compute the upper bound we employ the cross section for the production of the fermion. The particular processes will differ depending on the considered operator. 

HNLs are mainly produced in $e^+e^-$ annihilations. Thus, $\mathcal{O}_{\rm QN}$ cannot mediate this channel, and the operators involving charged leptons can only do so for electron flavour dominance (their muon and tau flavour copies cannot be constrained by supernova cooling arguments). The upper bound on the Wilson coefficient of the corresponding operator is obtained by simply rescaling the HNL production cross section:
\begin{equation}
    \frac{|C_{i}|^2}{\Lambda^4}=\frac{|C_{\rm vec}|^2}{\Lambda^4} \frac{\overline{\sigma}_{\rm vec}(e^+e^-\to NN)}{\overline{\sigma}_{i}(e^+e^-\to NN)}\,.
\end{equation}
Note that an alternative efficient channel for HNL production is $\gamma\gamma\to\pi^0\to NN$. In fact, supernova cooling arguments provide a strong bound on the invisible $\pi^0$ decay. Imposing this limit yields upper limits for the Wilson coefficients of $\mathcal{O}_{\rm uu}$, $\mathcal{O}_{\rm dd}$ and $\mathcal{O}_{\rm HN}$. 
In the latter case, the mentioned constraints will compete with those given by $e^+e^-$ annihilation, whereas they are the only source of upper bounds for $\mathcal{O}_{\rm uu}$ and $\mathcal{O}_{\rm dd}$. 
Note that the quark content of the operator $\mathcal{O}_{\rm QN}$ is orthogonal to that of the $\pi^0$, so this operator is unable to mediate its decay into HNLs, and thus cannot be constrained by supernova cooling arguments.

The scattering processes that may affect the HNL will depend on the effective operator under consideration. The vectorial operator considered in Ref.~\cite{DeRocco:2019jti} mediates scatterings off electrons and protons; $\mathcal{O}_{\rm HN\ell}$, $\mathcal{O}_{\rm ee}$, $\mathcal{O}_{\rm LN}^{\rm e}$ and $\mathcal{O}_{\rm LNL\ell}^{\rm e}$ only induce $Ne\to Ne$ processes, while $\mathcal{O}_{\rm HN}$ yields HNL scatterings off electrons, protons and neutrons. The upper limits on the corresponding Wilson coefficient can be found through
\begin{equation}
    \frac{|C_{i}|^2}{\Lambda^4}=\frac{|C_\mathrm{vec}|^2}{\Lambda^4} \frac{\overline{\sigma}_\mathrm{vec}(e^-N\to e^-N)+\overline{\sigma}_{\rm vec}(p N \to p N)}{\sum_X\overline{\sigma}_{i}(NX\to NX)}\,,
\end{equation}
where the sum in $X$ comprises all the possible HNL scatterings mediated by the particular operator $\mathcal{O}_{i}$.

Of course, all these cross sections are energy dependent. The temperatures inside a supernova, and thus the energies at which these processes take place, vary considerably as a function of the distance with respect to the centre of the supernova. The abundances of electrons, positrons, protons and neutrons also exhibit radial dependences, which determine how likely their interactions are. In order to capture these effects, we compute averaged cross sections (denoted by $\overline{\sigma}$ in the equations above), integrated over the temperature profile of the supernova and convoluted with the abundance of the involved SM particles:
\begin{equation}
\overline{\sigma}=\frac{\int_0^{100 \text{ km}}n(r)T(r)\sigma(T(r))\dd r }{\int_0^{100\text{ km}} T(r)\dd r}\,,   
\end{equation}
where $T(r)$ is the temperature and $n(r)$ is the number density of a given particle at a distance $r$ from the centre of the supernova. $\sigma$ is the energy-dependent cross section of each process (mediated by the operator under study), in which each particle is assigned the energy corresponding to the temperature. We extract the supernova temperature and abundance profiles from Ref.~\cite{DeRocco:2019jti}. 

\section{Monophoton searches}
\label{app:monophoton}

At LEP, searches for $e^+e^-$ collisions with a single photon recoiling against invisible particles were used to derive limits on dark matter particles and HNLs. 
Ref.~\cite{Fox:2011fx} places limits on DM production through effective four-fermion operators using these searches. From a collider perspective, HNLs and DM behave identically, as missing energy, so this information also allows to constrain some of our effective operators, which are able to mediate $e^+e^-\to \gamma NN$ processes. Such is the case of the operators $\mathcal{O}_{\rm ee}$, $\mathcal{O}_{\rm LN}^{\rm e}$, $\mathcal{O}_{\rm HN}$, $\mathcal{O}_{\rm HN\ell}^{\rm e}$ and $\mathcal{O}_{\rm LNL\ell}^{\rm ee}$ (see Sec.~\ref{sec:4ferm_neut_curr}). 

To recast their limits on our Wilson coefficients, we account for the different Lorentz structures of the operators. In particular, one of the operators considered in Ref.~\cite{Fox:2011fx} is a vectorial four-fermion coupling, similar to the one in Eq.~\ref{eq:FFvectorDM}. The bounds on our operators are then related to those on the vectorial one by
\begin{equation}
    \frac{|C_{i}|^2}{\Lambda^4}=\frac{|C_\mathrm{vec}|^2}{\Lambda^4}\frac{\sigma_\mathrm{vec}}{\sigma_{i}}\,,
\end{equation}
where $i$ runs through the operators mentioned above. To compute this ratio, we calculate the $e^+e^- \to \gamma N N$ cross section, as described below.

The $2 \to 3$ phase space is parametrized by the energy of one of the HNLs, $E_N$, the energy of the photon, $E_\gamma$, the relative azimuthal angle between them, $\phi_{N\gamma}$, and the polar angle of the photon, $\theta_\gamma$. 
The total cross section is given by
\begin{equation}
    \sigma=\frac{1}{256\pi^4E_\mathrm{CM}^2}\int_0^{2\pi}d\phi_{\gamma N}\int_{-1}^1 d\cos{\theta_\gamma}\int_0^{E_\gamma^\mathrm{max}}dE_\gamma\int_{E_N^\mathrm{min}}^{E_N^\mathrm{max}}dE_N\vert\mathcal{M_\mathrm}\vert^2\,,
\end{equation}
where $\vert\mathcal{M}\vert^2$ is the corresponding matrix element and $E_\mathrm{CM} = 200$~GeV is the center-of-mass energy~\cite{Fox:2011fx}. 
The amplitude is determined by each particular operator and can be expressed in terms of the products of the four-momenta of the particles involved.
It depends on the relative polar angle between the photon and the neutrino, $\theta_{\gamma N}$, and the polar angle of the neutrino, $\theta_N$. 
The former is fixed once the energies of both particles are determined,  
\begin{equation}
    \cos{\theta_{\gamma N}}=\frac{E_\mathrm{CM}^2-2E_\mathrm{CM}(E_N+E_\gamma)+2E_NE_\gamma}{2E_\gamma\sqrt{E_N^2-M_N^2}}\,,
\end{equation}
while the latter is geometrically determined in terms of the angles in the CM,
\begin{equation}
\cos{\theta_N}=\cos{\theta_\gamma}\cos{\theta_{\gamma N}}-\sin{\theta_\gamma}\sin{\theta_{\gamma N}}\cos{\phi_{\gamma N}}\,.
\end{equation}
The energy of the photon can vary from $0$ to $E_\gamma^\mathrm{max} = (E_\mathrm{CM}^2-4M_N^2)/2 E_\mathrm{CM}$, where the latter is obtained when the HNLs travel in the same direction, opposite to that of the photon.
Finally, the maximum (minimum) energy the HNL can carry depends on the energy of the photon and is obtained by solving $\cos{\theta_{\gamma N}}= -1$ $(\cos{\theta_{\gamma N}}=1)$, which gives
\begin{equation}
    E_N^\mathrm{max(min)}=\frac{1}{2}\left(E_\mathrm{CM}-E_\gamma\pm \frac{E_\gamma \sqrt{E_\mathrm{CM}^2-4M_N^2-2E_\gamma E_\mathrm{CM}}}{\sqrt{E_\mathrm{CM}^2-2E_\gamma E_\mathrm{CM}}}\right)\,.
\end{equation}
We integrate over the physical kinematical range considering the cuts employed in the analysis by LEP, requiring the photon energy to be at least a $6\%$ of that of the electron and its polar angle to be between $45^\circ$ and $135^\circ$.

Note that the case of the operator $\mathcal{O}_{\rm LNL\ell}^{\rm ee}$ is slightly different, as it gives rise to a final state with a light neutrino and an HNL, instead of two HNLs. However, the recasting procedure is very similar, with some differences in the kinematics described above. The expressions can be easily generalized by assuming that one HNL is massless.

\setupbibliochapters
\PB

\fancyhead[LE]{}
\fancyhead[RO]{}
\bibliographystyle{JHEP}


\providecommand{\href}[2]{#2}\begingroup\raggedright\endgroup
\PB
\incmultigraph{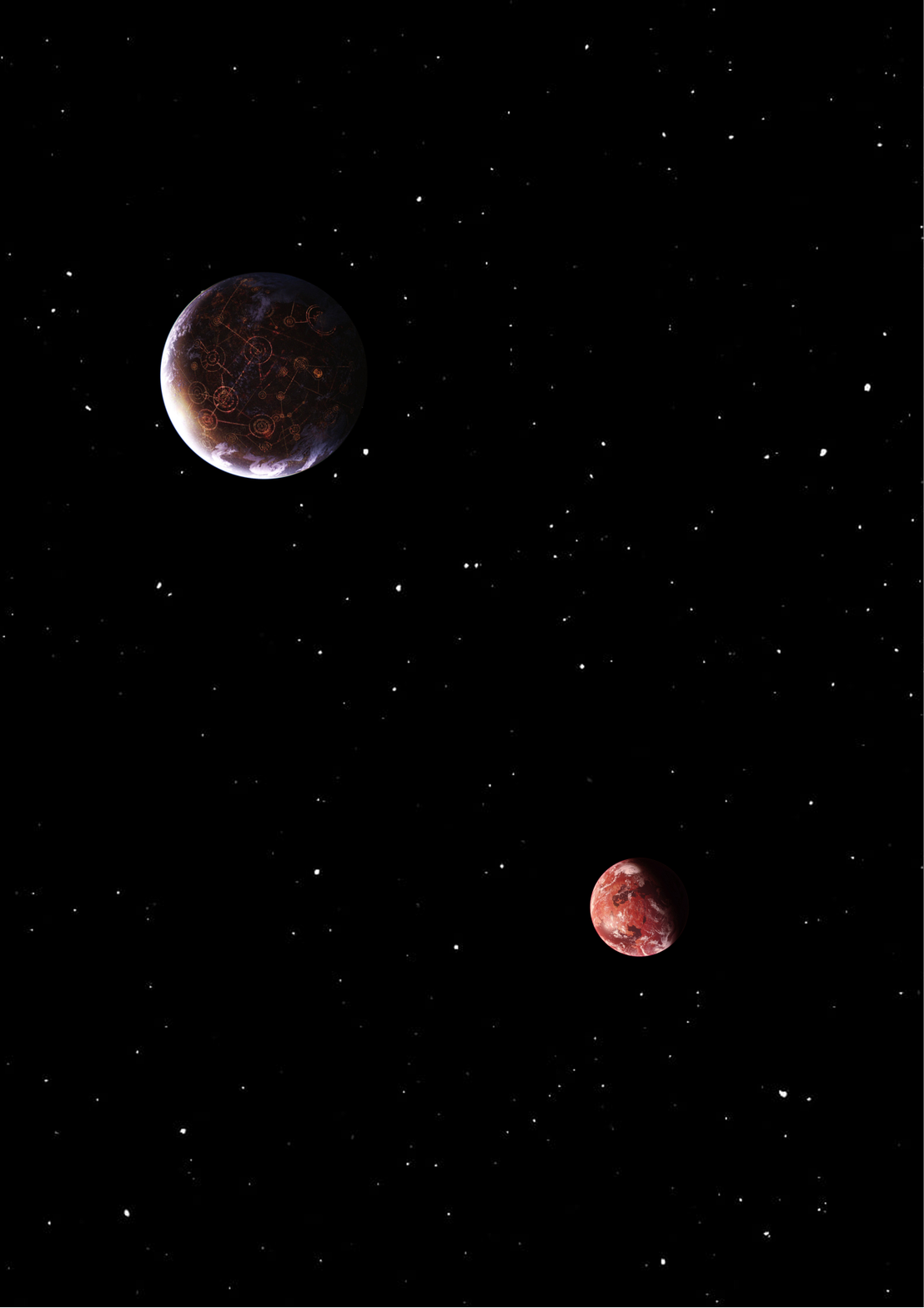}{1}
\end{document}